\def\VersionLong{}
\ifdefined\VersionLong%
	\newcommand{\LongVersion}[1]{#1}
	\newcommand{\ShortVersion}[1]{}
\else
	\newcommand{\LongVersion}[1]{}
	\newcommand{\ShortVersion}[1]{#1}
\fi

\documentclass[acmsmall,screen,nonacm]{acmart}
\setcopyright{cc}\setcctype[4.0]{by}
\acmDOI{}
\acmISBN{}
\acmVolume{}
\acmNumber{}
\acmArticle{}

\usepackage{listings}
\usepackage{amsmath}
\usepackage{bcpproof}
\usepackage{bcprules}\typicallabel{T-Hoge}
\usepackage{multicol}
\usepackage{makecell}
\usepackage{stmaryrd}
\usepackage{thmtools}
\usepackage{soul}

\declaretheorem[name=Remark,style=acmdefinition,numbered=no]{remark}
\newcommand{\proofitem}[1]{\noindentparagraph*{#1}}
\usepackage{comment}

\usepackage{comment}
\ifdefined\VersionLong%
        \includecomment{LongVersionBlock}
        \excludecomment{ShortVersionBlock}
\else
        \excludecomment{LongVersionBlock}
        \includecomment{ShortVersionBlock}
\fi

\ifdefined\VersionWithComments%
	\usepackage{draftwatermark}
	\SetWatermarkText{draft}
	\SetWatermarkScale{3}
	\SetWatermarkColor[gray]{0.9}
\fi

\usepackage[svgnames,table]{xcolor}
\usepackage{mdframed}
\usepackage{colortbl}
\definecolor{darkblue}{rgb}{0.0,0.0,0.6}
\definecolor{darkgreen}{rgb}{0, 0.5, 0}
\definecolor{darkpurple}{rgb}{0.7, 0, 0.7}
\definecolor{darkblue}{rgb}{0, 0, 0.7}

\renewcommand\UrlFont{\color{blue}\rmfamily}
\hypersetup{draft=false}
\usepackage{xurl}

\newcommand{\rocqdecl}[2]{%
  {\urlstyle{tt}\nolinkurl{#2} (\nolinkurl{#1})}%
}
\newcommand{\rocqnote}[1]{%
  \par\nobreak\smallskip
  {\normalfont\small
    \renewcommand{\UrlFont}{\ttfamily}%
    \noindent #1\par}%
}
\newcommand{\rocqref}[3][Rocq]{%
  \rocqnote{#1: \rocqdecl{#2}{#3}.}%
}

\usepackage[capitalise,english,nameinlink]{cleveref} %
\crefname{figure}{Figure}{Figures}
\Crefname{figure}{Figure}{Figures}
\crefalias{AlgoLine}{line}
\crefname{line}{\text{line}}{\text{lines}} %
\crefname{item}{\text{item}}{\text{items}} %
\crefname{thm}{Theorem}{Theorems}
\crefname{example}{Example}{Examples}
\crefname{assumption}{Assumption}{Assumptions}
\crefname{algorithm}{Algorithm}{Algorithms}
\crefname{appendix}{Appendix}{Appendices}

\ifdefined\VersionWithComments%
	\usepackage[colorinlistoftodos,textsize=footnotesize]{todonotes}
	\setuptodonotes{inline}
\newcommand{\wantcite}[1][]{\todo{cite\IfBlankTF{#1}{}{: #1}}}
\newcommand{\TODO}[1]{\todo{#1}}
\newcommand{\gennote}[3]{\todo[linecolor=#2,backgroundcolor=#2!25,bordercolor=#2]{#3: #1}}
\newcommand{\ym}[1]{{\gennote{#1}{green}{YM}}}
\newcommand{\yf}[1]{{\gennote{#1}{blue}{YF}}}
\newcommand{\ks}[1]{{\gennote{#1}{purple}{KS}}}
\newcommand{\ai}[1]{{\gennote{#1}{red}{AI}}}
\newcommand{\note}[1]{{\gennote{#1}{cyan}{Note}}}
\else
\usepackage[disable]{todonotes}
\newcommand{\wantcite}[1][]{}
\newcommand{\TODO}[1]{}
\newcommand{\gennote}[3]{}
\newcommand{\ym}[1]{}
\newcommand{\yf}[1]{}
\newcommand{\ks}[1]{}
\newcommand{\ai}[1]{}
\newcommand{\note}[1]{}
\fi

\ifdefined\VersionFinal

\else
  \newenvironment{revision}{}{}
\fi

\newcommand{\ottnt}[1]{\mathit{#1}}
\newcommand{\ottmv}[1]{\mathit{#1}}
\newcommand{\ottkw}[1]{\mathbf{#1}}
\newcommand{\ottsym}[1]{#1}

\DeclareMathOperator{\SEQ}{;}

\DeclareMathOperator{\LET}{\mathbf{let}}
\DeclareMathOperator{\IN}{\mathbf{in}}
\DeclareMathOperator{\WRITE}{:=}

\newcommand{\NULL}{\mathbf{null}}
\DeclareMathOperator\IFNP{\mathbf{ifnp}}

\DeclareMathOperator\IF{\mathbf{if}}
\DeclareMathOperator\ALLOC{\mathbf{alloc}}

\DeclareMathOperator\THEN{\mathbf{then}}
\DeclareMathOperator\ELSE{\mathbf{else}}
\DeclareMathOperator{\COL}{:}
\newcommand\ASSERT{\mathbf{assert}}

\newcommand{\consort}{\textsc{ConSORT}\xspace}

\newcommand{\DOM}{\mathit{dom}}

\newcommand\tuple[1]{\left\langle{#1}\right\rangle}

\newcommand\TINT{\mathbf{int}}
\DeclareMathOperator\TREF{\mathbf{ref}}

\newcommand\set[1]{\left\{{#1}\right\}}

\DeclareMathOperator{\produces}{\Rightarrow}

\DeclareMathOperator{\ra}{\rightarrow}

\DeclareMathOperator{\subt}{\le}

\newcommand{\fv}{\mathbf{FV}}

\newcommand\ALIAS{\mathbf{alias}}

\newcommand\sem[1]{\left\llbracket{#1}\right\rrbracket}

\newcommand\fml[1]{\textit{fml}({#1})}
\newcommand\fmlT[2]{\textit{fml}_{#2}({#1})}
\newcommand\Empty[1]{\textit{Empty}({#1})}

\newif\ifdraftComments
\draftCommentsfalse
\def\mkDraftFn#1#2{
  \expandafter\def\csname #1\endcsname##1{\ifdraftComments\textcolor{#2}{[#1: ##1]}\marginpar[$\longrightarrow$]{$\longleftarrow$}\fi}
}
\mkDraftFn{JT}{blue}
\mkDraftFn{AI}{red}
\mkDraftFn{KS}{purple}
\mkDraftFn{NK}{cyan}
\mkDraftFn{RS}{teal}
\usepackage{xspace}
\makeatletter
\def\needcite@with[#1]{\ifdraftComments\textcolor{blue}{[citation needed #1]}\else\empty\fi\xspace}
\def\needcite@bare{\ifdraftComments\textcolor{blue}{[citation needed]}\else\empty\fi\xspace}
\def\needcite{\@ifnextchar[{\needcite@with}{\needcite@bare}}
\makeatother

\lstdefinelanguage{Imp}{
  keywords=[0]{ifnp,then,else,alias,assert,mkref,let,in,null,return,if},
  morecomment=[l]{//},
  morecomment=[s]{/*}{*/}
}
\definecolor{comment-green}{rgb}{0,0.6,0}
\newcommand{\LLBRAKET}{\llbracket}
\newcommand{\RRBRAKET}{\rrbracket}

\newcommand\VARS{\mathbf{Var}}

\newcommand\CONOWN{\mathbf{ConOwn}}
\newcommand\OWNALL{\mathbf{Own}}
\newcommand\OWN{\mathbf{own}}
\newcommand\SAT{\mathbf{SAT}}
\newcommand\SATV{\mathbf{SATv}}
\newcommand\defeq{\overset{\text{\tiny def}}{=}}
\usepackage{xcolor}              %
\usepackage{listings}
\lstdefinelanguage{Rocq}{
  sensitive=true,
  morekeywords={
    Definition,Fixpoint,CoFixpoint,
    Inductive,CoInductive,Record,Variant,
    Theorem,Lemma,Corollary,Proposition,Fact,Remark,
    Proof,Qed,Defined,Admitted,Abort,
    Require,Import,Export,From,
    Module,Module Type,End,
    Section,Context,Variable,Variables,
    Hypothesis,Hypotheses,
    Arguments,Notation,Infix,
    Instance,Class,
    Program,Equations,
    forall,fun,match,with,end,let,in,
    if,then,else,return,as,
    Type,Prop,Set
  },
  morecomment=[s]{(*}{*)},
  morestring=[b]"
}

\lstdefinestyle{rocq}{
  language=Rocq,
  basicstyle=\small\ttfamily,
  keywordstyle=\bfseries,
  commentstyle=\itshape,
  columns=fullflexible,
  keepspaces=true,
  showstringspaces=false,
  breaklines=true,
  mathescape=true,
  xleftmargin=1em,
  aboveskip=0.7\baselineskip,
  belowskip=0.7\baselineskip
}
\usepackage{booktabs}
\ifdefined\VersionFinal{}
\else
\fi
\title{Ownership Refinement Types for Pointer Arithmetic and Nested Arrays}

\author{Yusuke Fujiwara}
\orcid{0009-0004-8423-0782}
\affiliation{ \institution{Kyoto University} \city{Kyoto} \country{Japan}}
\email{yfujiwara@fos.kuis.kyoto-u.ac.jp}

\author{Yusuke Matsushita}
\orcid{0000-0002-5208-3106}
\affiliation{ \institution{Kyoto University} \city{Kyoto} \country{Japan}}
\email{ymat@fos.kuis.kyoto-u.ac.jp}

\author{Kohei Suenaga}
\affiliation{ \institution{Kyoto University} \city{Kyoto} \country{Japan}}
\email{kohei.suenaga@acm.org}
\orcid{0000-0002-7466-8789}

\author{Atsushi Igarashi}
\affiliation{ \institution{Kyoto University} \city{Kyoto} \country{Japan}}
\orcid{0000-0002-5143-9764}
\email{igarashi@kuis.kyoto-u.ac.jp}



\begin{document}

\begin{abstract}
Tanaka et al.\ proposed a type system for verifying functional correctness
 properties of programs that use arrays and pointer arithmetic.
Their system extends \consort---a type system combining fractional
ownership and refinement types for imperative program
verification---with support for pointer arithmetic.
Their idea was to extend fractional ownership so that it can depend on an array index.
Their formulation, however, does not handle nested arrays, which are essential for
representing practical data structures such as matrices.
We extend Tanaka et al.'s type system to support nested arrays by allowing
ownership to depend on the indices of enclosing arrays. We formalize the
extended calculus in Rocq using a mixed embedding and provide a machine-checked
proof of type soundness.
We have implemented a verifier based on the proposed type system and
demonstrated that it can verify the correctness of programs that manipulate nested arrays, 
which were beyond the reach of Tanaka et al.

\end{abstract}

\maketitle

\section{Introduction}
\label{sec:intro}
Type-based automated program verification has been a popular methodology for
making software reliable.
 Among various type systems proposed so far, recent years have seen significant
 success of \emph{refinement type
 systems}~\cite{HybridTypeChecking,LiquidTypeTutorial}---type systems that can constrain values using predicates called \emph{refinement predicates}---to guarantee properties that cannot be expressed with base types alone.
For example, a refinement type system can express the type $\{\nu \COL \TINT \mid \nu > 0\}$ of positive integers using the refinement predicate $\nu > 0$, which is more expressive than its simple type $\TINT$.
Refinement type systems have been applied to various types of programs including functional programs~\cite{LiquidTypeTutorial,10.1145/3158100},
object-oriented programs~\cite{sun_et_al:LIPIcs.ECOOP.2024.39}, and
smart contracts~\cite{10.1007/978-3-030-72013-1_14,icon}.

However, applying refinement types to imperative programs with pointers and aliasing has been less popular than for other kinds of programs.
A naive way of introducing refinement types for an imperative language would be to incorporate reference types.
For example, such a type system would have type $\{\nu \COL \TINT \mid \nu > 0\} \TREF$ for pointers to positive integers.
The type system would be designed to be flow-sensitive: typing rules would be designed as if the type of each variable were updated by each statement.
For example, if $x$ has type $\{\nu \COL \TINT \mid \nu > 0\} \TREF$ just before a statement $x := -1$ that updates the memory cell pointed to by $x$ to $-1$, then the type of $x$ just after this statement would be $\{\nu \COL \TINT \mid \nu = -1\} \TREF$.

One major challenge in this approach is the \emph{strong update} problem, explained below.
Suppose pointers $x$ and $y$ both have type $\{\nu \COL \TINT \mid \nu > 0\} \TREF$ just before a statement $x := -1$.
Then, the type of $x$ after this statement is updated to a type like $\{\nu \COL \TINT \mid \nu = -1\} \TREF$.
If $x$ and $y$ point to the same memory cell, retaining the original type of
$y$ would incorrectly assert that the cell still contains a positive integer.
Updating only the type of $x$ is therefore unsound.

\ks{}
To address this issue, Toman et al.~\cite{DBLP:conf/esop/TomanSSI020} proposed a type system \textsc{ConSORT}.
Their type system, instead of conducting static must-alias analysis, constrains aliases based on types.
Concretely, their reference type is augmented with information called \emph{(fractional) ownership}, with which their type system permits the following mutually exclusive cases for each memory cell:
\begin{itemize}
  \item A single pointer can read and update the cell and may carry
    non-trivial refinement information about its contents; all other aliases
    carry only trivial refinement information about that cell.
  \item Multiple pointers may read the cell and carry non-trivial refinement
    information about its contents, but none can update it.
\end{itemize}
With these constraints, the type system does not need to handle aliasing in a strong update, since it is guaranteed that there are no aliases reading non-trivial information from the updated memory cell.

Later, Tanaka et al.~\cite{DBLP:conf/pepm/TanakaSK24} extended \textsc{ConSORT}
 to support pointer arithmetic, enabling the verification of programs that manipulate integer arrays and perform pointer arithmetic.
 Their key idea is to extend the notion of ownership to \emph{(fractional)
 ownership functions} from array indices to fractional ownerships.
With this extension, their type system can reason about the properties of array elements in an index-sensitive way.
Pointer arithmetic is reflected as index-shifting operations over the
ownership function at the type system's level.

\begin{figure}[tb]
  \centering
  \begin{lstlisting}[backgroundcolor=,escapechar=\!]
initArray (l, n, p){
  if l <= 0 then { 0 }
  else {
    p := n; // Updating the memory cell pointed to by p to n 
    let q = p + 1 in 
    let x = initArray(l-1, n, q) in 0 
  }
}

initMatrix(n, pp)
{ 
  if n <= 0 then { 0 } else {
    let s = alloc n : int ref in !\label{motiv:line:alloc}!
    let d = initArray(n, n, s) in !\label{motiv:line:callInitArray}!
    pp := s; !\label{motiv:line:updatepp}!
    let d2 = initMatrix(n-1, pp+1) in 0 !\label{motiv:line:fun_ty_goal}!
  }
}
  \end{lstlisting}
  \caption{A motivating example.}
  \Description{Two recursive functions initialize a jagged matrix.
  initArray writes the value n into l consecutive integer cells, advancing
  its pointer after each write.  For positive n, initMatrix allocates a row
  of length n, fills it with n, stores its pointer in the current outer
  cell, and recurses on the next outer cell with n decreased by one.
  Successive rows therefore have decreasing lengths and element values.}
  \label{fig:motivatingExample}
\end{figure}

In this paper, we extend Tanaka et al.'s type system to handle another kind of programs that they do not support: programs manipulating \emph{nested arrays}, which frequently arise in programs that manipulate matrices and tensors.
\Cref{fig:motivatingExample} is the motivating example that we use to illustrate our contribution.
The function \texttt{initMatrix} takes an integer $n$ and a pointer $\mathit{pp}$ to an array of length $n$, and allocates an integer array to each element of the array pointed to by $pp$ so that $pp$ points to a matrix represented by a two-dimensional array; concretely, for each $0 \le i < n$, the cell at $\mathit{pp} + i$ stores a pointer to an array of length $n-i$, all of whose elements are set to $n-i$.
To this end, if $n > 0$, \texttt{initMatrix} allocates an array of length $n$, binds $s$ to a pointer to it (\cref{motiv:line:alloc}), calls $\texttt{initArray}(n,n,s)$ that initializes all the elements of the array pointed to by $s$ of length $n$ to $n$ (\cref{motiv:line:callInitArray}), writes $s$ to the memory cell pointed to by $\mathit{pp}$ (\cref{motiv:line:updatepp}), and recursively calls $\texttt{initMatrix}(n-1, \mathit{pp} + 1)$ (\cref{motiv:line:fun_ty_goal}).

Verifying the above specification of \texttt{initMatrix} using the idea of ownership functions by Tanaka et al.\ requires the ownership of an inner array to be dependent not only on its index but also on the index of the outer array.
However, Tanaka et al.'s type system does not support such dependency; hence, their type system cannot handle the program in \cref{fig:motivatingExample}.

Our extension, based on Tanaka et al.’s type system, introduces \emph{(fractional) ownership terms} to describe a function from array indices to ownerships.
Intuitively, the ownership of the inner array of the matrix pointed to by $\mathit{pp}$ after the execution of \texttt{initMatrix} must satisfy the following: Via pointer $\mathit{pp}$, the memory
 cell $(*(\mathit{pp} + x_2)) + x_1$ can be updated if $0 \le x_2 \le n - 1 \land 0 \le x_1 \le n - x_2 - 1$, where
$x_1$ represents the index of an inner array and
$x_2$ represents the index of an outer array.
Using an ownership term, this constraint is described as
$(0 \le x_2 \le n - 1 \land 0 \le x_1 \le n - x_2 - 1)  \produces  1$; here, $ \produces  1$ expresses the ownership for reading and updating the memory cell.
Note that this ownership term for an inner array refers to the index $x_2$ of the outer array.

\subsection*{Contributions}
We formally define an imperative language and its type system equipped with ownership terms.
We also prove the soundness of the type system: a well-typed program does not cause assertion failures, out-of-bounds array accesses, or null-pointer dereferences. 
We also design a type inference procedure and implement a prototype verifier based on the procedure.
We apply our verifier to several programs that manipulate nested arrays, and we
demonstrate that it successfully and efficiently verifies the functional
correctness of these programs, which was not possible with Tanaka et al.'s type system.
\begin{revision}
Although our motivating examples primarily use two-dimensional arrays for clarity of presentation, our type system and implementation support nested arrays of arbitrary depth. In particular, we demonstrate this generality by verifying programs that manipulate three- and four-dimensional arrays (\cref{sec:experiments}).
\end{revision}

\subsection*{Relationship to the Conference Version}
This article extends our ECOOP~2026 paper~\cite{fujiwara_et_al:LIPIcs.ECOOP.2026.6}.
An earlier full version~\cite{conferenceFullVersion} included detailed
proofs.  The principal addition here is a complete Rocq formalization of
the calculus and its type soundness proof.  \cref{sec:rocq} explains the mixed embedding
of programs, types, and logical predicates, and the treatment of
well-formedness and variable binding.
The formalization also led to corrections to the definition of type
emptiness, ownership accounting, and rule premises, together with revisions
to the prose proofs.  We present the soundness argument in
\cref{sec:soundness} and identify corresponding Rocq results in the detailed
proofs in \cref{sec:proofs}.
We also expand the account of the implementation and the benchmark programs
in \cref{sec:implementation,sec:experiments}.

\subsection*{Organization}
The rest of this paper is organized as follows.
\Cref{sec:targetLanguage} defines the target imperative language and its operational semantics. 
\Cref{sec:typing} presents the proposed type system and states its soundness. 
\Cref{sec:typeInference} describes a template-based type inference procedure.
\Cref{sec:implementation} explains the implementation's restrictions
and syntax extensions.
\Cref{sec:experiments} reports on the experimental results, comparing the performance of our implemented verifier and evaluating the verification capabilities for programs involving nested arrays. 
\Cref{sec:rocq} describes the mechanization of the calculus and its
type soundness proof in Rocq.
\Cref{sec:discussion} discusses the applicability of our approach to mainstream programming languages and possible extensions to more general heap structures.
\Cref{sec:relatedWork} discusses related work.
\Cref{sec:conclusion} concludes the paper.
\ifdefined\VersionLong
    We omit some definitions and proofs; they can be found in Appendix.
  \else
    We omit some definitions and proofs, which will be found in a full version of the paper~\cite{extendedVersion}.
  \fi

\section{Target Language}
\label{sec:targetLanguage}
\subsection{Syntax}
\label{sec:syntax}

\Cref{fig:SyntaxOfSourceLanguage} shows the syntax of the target
language.  
The set of variables $\VARS$ is ranged over by $w$, $x$, $y$, and $z$.  The
metavariable $n$ represents integer constants; $f$ represents function
names; $\varphi$ represents first-order logic formulae over integers.
The metavariable $e$ represents expressions; $\ottnt{fd}$ represents
function definitions; $D$ represents a finite set of function
definitions; $\ottnt{P}$ represents programs; and $\tau^{-}$ represents
simple types, consisting of integer and reference types.  An expression of
the form $ \LET  x  =   \ldots   \IN  e_{{\mathrm{1}}} $ binds $x$ in $e_{{\mathrm{1}}}$.  We write
$ [  y_{{\mathrm{1}}}  /  x_{{\mathrm{1}}}  , \ldots,  y_{\ottmv{n}}  /  x_{\ottmv{n}}  ] $ for simultaneous capture-avoiding
substitution of variable $y_i$ for variable $x_i$ (for
$1 \leq i \leq n$).  A substitution is also denoted by $\theta$.

\begin{figure}
  \[
    \begin{array}{llll}
      \tau^{-} & ::= &  \TINT \mid  \tau^{-}  \TREF  \\[.3em]
      e & ::= & x \mid  \LET  x  =  n  \IN  e  \mid  \LET  x  =  \ottkw{null}  \IN  e 
               \mid  \LET  x  =   y  \mathop{  -  }  z   \IN  e  \\
        & \hspace{.8em} \mid &  \IFNP  x  \THEN  e_{{\mathrm{0}}}  \ELSE  e_{{\mathrm{1}}}  \mid  \LET  x  =   f (  y_{{\mathrm{1}}} ,\ldots, y_{\ottmv{n}}  )   \IN  e   \mid  \LET  x  =  e_{{\mathrm{0}}}  \IN  e_{{\mathrm{1}}}  \\
        & \hspace{.8em} \mid &  \LET  x  =   \ALLOC  y   \ottsym{:}    \tau^{-}  \TREF    \IN  e  \mid  \LET  x  =   \ast  y   \IN  e  
          \mid     x  \WRITE  y   \SEQ  e  \mid  \LET  x  =   y   \boxplus   z   \IN  e  \\
        & \hspace{.8em} \mid &   \ALIAS(  x  =  y   \boxplus   z  )   \SEQ  e \mid   \ALIAS(  x  = \ast  y  )   \SEQ  e  \mid   \ASSERT( \varphi )   \SEQ  e   \\[.3em]
      \ottnt{fd} & ::= & f  \mapsto   (  x_{{\mathrm{1}}}  , \ldots ,  x_{\ottmv{n}}  )  \, e \\[.3em]
      D & ::= & \ottsym{\{}  \ottnt{fd_{{\mathrm{1}}}}  \ottsym{,} \, ... \, \ottsym{,}  \ottnt{fd_{\ottmv{n}}}  \ottsym{\}} \\[.3em]
      \ottnt{P} & ::= &  \tuple{ D ,  e } 
    \end{array}
  \]
  \caption{Syntax of the target language.}
  \Description{A grammar defines integer and reference simple types,
  expressions, function definitions, and programs.  Expressions include
  variables, let bindings, integer subtraction, conditionals, function
  calls, array allocation, dereference, assignment, pointer arithmetic,
  two forms of alias checks, and assertions.  A program consists of a
  finite collection of function definitions and a main expression.}
  \label{fig:SyntaxOfSourceLanguage}
\end{figure}

In this language, constants and the value of every intermediate computation are named with $\LET$.
We informally explain the meaning of expression forms.
\begin{itemize}
\item
  $ \IFNP  x  \THEN  e_{{\mathrm{0}}}  \ELSE  e_{{\mathrm{1}}} $ evaluates
  $e_{{\mathrm{0}}}$ if $x$ is not a positive integer; it evaluates $e_{{\mathrm{1}}}$
  otherwise.
  \item
  $ \LET  x  =   f (  y_{{\mathrm{1}}} ,\ldots, y_{\ottmv{n}}  )   \IN  e $ calls function
  $f$ with actual arguments $y_1, \dots, y_n$, binds $x$ to the value
  returned from $f$, and evaluates $e$.  Without loss of
  generality, we assume $y_i \ne y_j$ if $i \ne j$.
  For readability, examples omit some integer-copy bindings before calls.
  When necessary, we first copy an integer argument into a fresh variable
  and pass the copy instead.  Such copies make repeated arguments distinct
  and allow facts about the original integer variables to be retained
  after the call.  The latter use can require a copy even when an integer
  argument occurs only once.
  \item
  $ \LET  x  =   \ALLOC  y   \ottsym{:}    \tau^{-}  \TREF    \IN  e $ allocates an array of
  size $y$, binds $x$ to the pointer to the zeroth memory cell of the
  array, and evaluates $e$.
  We assume that the simple type of $x$ (i.e., $ \tau^{-}  \TREF )$ is
  annotated.
  As shown below, the type annotation determines the initial values stored in the allocated array.
  \begin{revision}
    If a nested $\mathbf{ref}$ type is specified as $\tau^{-}$, then $x$ is bound to a pointer to an array of length $y$ whose elements
    are initialized to $ \NULL $.
    Each cell is expected to be initialized with a pointer to
    another (nested) array before it is accessed.
  \end{revision}
  If $y$ is not positive, an empty array will be allocated and a
  dangling pointer will be returned.  However, the type system
  prevents such a pointer from being accessed.
  \item
  $ \LET  x  =   \ast  y   \IN  e $ reads from the memory cell pointed to by the pointer
  $y$, binds $x$ to the read value, and evaluates
  $e$.
  \item
   $  x  \WRITE  y   \SEQ  e $ updates the memory cell pointed to by
  $x$ with the value $y$ and evaluates $e$.
\item  $ \LET  x  =   y   \boxplus   z   \IN  e $ is for pointer
  arithmetic; it binds $x$ to the pointer obtained by shifting $y$ by
  offset $z$ and evaluates $e$.
\item 
\begin{revision}
$  \ALIAS(  x  =  y   \boxplus   z  )   \SEQ  e $ (resp., $  \ALIAS(  x  = \ast  y  )   \SEQ  e $) evaluates $e$ if $x =  y   \boxplus   z $
  (resp., $x =  \ast  y $) holds; otherwise, it raises an exception
  indicating that the alias relationship does not hold.  
  Operationally, this expression works as a runtime check for the alias relationship; if the alias relationship does not hold, execution is safely terminated by raising an exception.
  Statically, this expression serves as a hint to the type checker: it declares the programmer's intended aliasing relationship between pointers; the type checker is allowed to redistribute the ownership between the pointers $x$ and $ y   \boxplus   z $ (resp., $ \ast  y $); see \rn{T-AliasAddPtr} and \rn{T-AliasDeref} in \cref{sec:typesystem} for details.
As in previous work on fractional ownership~\cite{DBLP:conf/pepm/TanakaSK24,DBLP:conf/esop/TomanSSI020,DBLP:conf/aplas/SuenagaK09,DBLP:conf/oopsla/SuenagaFI12}, 
our type system guarantees the absence of errors due to runtime assertion failures, out-of-bounds accesses, and null-pointer dereferences; an incorrect alias expression leads to the explicit error state $\mathbf{AliasFail}$ rather than to undefined behavior
(see \cref{sec:opsem} and \cref{thm:soundness}).
\end{revision}
\item $  \ASSERT( \varphi )   \SEQ  e $ evaluates $e$ if the formula $\varphi$
  holds; otherwise, the evaluation gets stuck due to an assertion failure, which means that
  the program goes into an unsafe state.  A well-typed program in our
  type system is guaranteed not to cause this error.
\end{itemize}
A function definition
$\ottnt{fd}$ is of the form $f \mapsto (x_1,\dots,x_n)e$ where $f$ is
  the name of the defined function, $x_1,\dots,x_n$ are the names of
  the parameters, and $e$ is the body of the function.  We assume that
  $x_1,\dots,x_n$ are pairwise distinct.
  Finally, a program $\ottnt{P}$ is a pair of a set of function definitions
$D$ and a main expression $e$.  

\subsection{Operational Semantics}
\label{sec:opsem}

We define the operational semantics of the language.

First, we give the syntax of \emph{pointer values} and \emph{values}, ranged over by $pv$ and $\ottnt{v}$, respectively.
\[
  a \in \mathcal{A} \qquad pv ::= \ottsym{(}  a  \ottsym{,}  i  \ottsym{)} \mid  \NULL  \qquad \ottnt{v} ::= n \mid pv 
\]
An \emph{address} is represented by a pair $\ottsym{(}  a  \ottsym{,}  i  \ottsym{)}$ of a base address
$a$, which is an element of a fixed denumerable set $\mathcal{A}$, and an
offset $i$, which is an integer.  Two addresses $\ottsym{(}  a  \ottsym{,}  i  \ottsym{)}$ and $\ottsym{(}  a'  \ottsym{,}   i '   \ottsym{)}$
are equal if and only if $a = a'$ and $i =  i ' $.
A \emph{pointer value} $pv$ is either an address or the null pointer $ \NULL $.
Unlike other type systems using fractional ownership~\cite{DBLP:conf/aplas/SuenagaK09},
in which $ \NULL $ is used to represent the end of a linked list,
our type system gives $ \NULL $ zero ownership, preventing
access to the null pointer statically.  A \emph{value} $\ottnt{v}$ is either
a pointer value or an integer.

The small-step operational semantics of our language is given as a
rewriting relation between \emph{configurations}, ranged over by $C$, of the form
$ \tuple{ \ottnt{H} ,  \ottnt{R} ,  e } $ or an error state $ \mathbf{AliasFail} $.
Here, $R$ is a map from a finite subset of $\VARS$ to the set of values,
modeling a register file,
and $H$ is a finite partial map from the set of addresses to the set of values,

We write $ \DOM( \ottnt{R} ) $ and $ \DOM( \ottnt{H} ) $ for the domain of $R$ and $H$, respectively.
We consider only heaps satisfying the following heap-domain invariant:
\[
  (a,i)\in\DOM(H)
  \quad\Longrightarrow\quad
  0\leq i \mathrel{\land} (a,0)\in\DOM(H).
\]
This restriction concerns locations allocated in the heap; pointer values may
have negative offsets.  The empty heap satisfies the invariant, and every
reduction rule below preserves it: assignment does not change the heap domain,
while allocation either adds no locations or adds a block beginning at offset
zero.  Consequently, $(a,0)\notin\DOM(H)$ implies that no address with base
$a$ belongs to $\DOM(H)$.
If $\set{x_{{\mathrm{1}}}, \ldots, x_{\ottmv{n}}} \cap  \DOM( \ottnt{R} )  = \emptyset$, we write $R\set{x_{{\mathrm{1}}} \mapsto \ottnt{v_{{\mathrm{1}}}}, \ldots, x_{\ottmv{n}} \mapsto \ottnt{v_{\ottmv{n}}}}$ for the map obtained by
extending $R$ with new bindings that map $x_{\ottmv{i}}$ to $\ottnt{v_{\ottmv{i}}}$ (for $1 \leq i \leq n$).
Similarly, if $\set{(a,0), \ldots, (a,n)} \cap  \DOM( \ottnt{H} )  = \emptyset$, we write $ \ottnt{H}  \{ (  a  ,   0   ) \mapsto   \ottnt{v_{{\mathrm{0}}}}  , \ldots, (  a  ,   n   ) \mapsto   \ottnt{v_{\ottmv{n}}}   \} $ for an extension of $H$ with
$(a,i) \mapsto v_i$ (for $0 \leq i \leq n$) and, if $ \ottsym{(}  a  \ottsym{,}  n  \ottsym{)}  \in   \DOM( \ottnt{H} )  $, we write $\ottnt{H}  \ottsym{\{}  \ottsym{(}  a  \ottsym{,}   n   \ottsym{)}  \hookleftarrow   \ottnt{v}   \ottsym{\}}$ for the heap that is
identical to $H$ except that the address $\ottsym{(}  a  \ottsym{,}  n  \ottsym{)}$ is now mapped to the
value $v$.

\Cref{fig:opsem} gives the excerpt of the rules to define
the reduction step $ \longrightarrow _{D}$, parameterized by a set of
function definitions $D$.
\ifdefined\VersionLong
The full set of rules is given in \cref{sec:opsemFull}.
\else
The full set of rules is given in the extended version.
\fi
We write $\mathbb{Z}$ for the set of integers.

\begin{figure*}
\footnotesize
{\raggedright\noindent\fbox{$C  \longrightarrow _{  D  }  C'$}\par}
\begin{multicols}{2}
  \infrule[Rs-LetNull]{
     x'  \not\in   \DOM( \ottnt{R} )  
  }{
     \tuple{ \ottnt{H} ,  \ottnt{R} ,   \LET  x  =  \ottkw{null}  \IN  e  }   \longrightarrow _D\\ \tuple{H, R\{ x'  \mapsto  \NULL \},   [  x'  /  x  ]    e }
  }
  \infrule[Rs-AddPtr]{
    \ottnt{R}  \ottsym{(}  y  \ottsym{)} \,  =  \, pv
    \andalso  x'  \not\in   \DOM( \ottnt{R} )  
    \andalso R(z) \in  \mathbb{Z} 
  }{
     \begin{array}{r}  \tuple{ \ottnt{H} ,  \ottnt{R} ,   \LET  x  =   y   \boxplus   z   \IN  e  }     \longrightarrow _{  D  }   \\   \tuple{ \ottnt{H} ,  \ottnt{R}  \ottsym{\{}  x'  \mapsto  pv  \boxplus  \ottnt{R}  \ottsym{(}  z  \ottsym{)}  \ottsym{\}} ,    [  x'  /  x  ]    e  }  \end{array} 
  }
  \end{multicols}
  \infrule[Rs-MkArrayIntref]{
     \ottsym{(}  a  \ottsym{,}  0  \ottsym{)}  \not\in   \DOM( \ottnt{H} )  
    \andalso  x'  \not\in   \DOM( \ottnt{R} )  
    \andalso H' =  \ottnt{H}  \{ (  a  ,   0   ) , \ldots, (  a  ,  \ottnt{R}  \ottsym{(}  y  \ottsym{)} \,  -  \,  1   ) \mapsto   0   \} 
    \andalso R(y) \in  \mathbb{Z} 
  }{
      \tuple{ \ottnt{H} ,  \ottnt{R} ,   \LET  x  =   \ALLOC  y   \ottsym{:}     \TINT   \TREF    \IN  e  }     \longrightarrow _{  D  }     \tuple{ \ottnt{H'} ,  \ottnt{R}  \ottsym{\{}  x'  \mapsto  \ottsym{(}  a  \ottsym{,}   0   \ottsym{)}  \ottsym{\}} ,    [  x'  /  x  ]    e  }  
  }
  \infrule[Rs-MkArrayNestedref]{
     \ottsym{(}  a  \ottsym{,}  0  \ottsym{)}  \not\in   \DOM( \ottnt{H} )  
    \andalso  x'  \not\in   \DOM( \ottnt{R} )  
    \andalso \ottnt{H'}  =   \ottnt{H}  \{ (  a  ,   0   ) , \ldots, (  a  ,  \ottnt{R}  \ottsym{(}  y  \ottsym{)} \,  -  \,  1   ) \mapsto  \ottkw{null}  \} 
    \andalso R(y) \in  \mathbb{Z} 
  }{
      \tuple{ \ottnt{H} ,  \ottnt{R} ,   \LET  x  =   \ALLOC  y   \ottsym{:}    \ottsym{(}   \tau^{-}  \TREF   \ottsym{)}  \TREF    \IN  e  }     \longrightarrow _{  D  }     \tuple{ \ottnt{H'} ,  \ottnt{R}  \ottsym{\{}  x'  \mapsto  \ottsym{(}  a  \ottsym{,}   0   \ottsym{)}  \ottsym{\}} ,    [  x'  /  x  ]    e  }  
  }
  \begin{multicols}{2}
  \infrule[Rs-AliasAddPtr]{
    \ottnt{R}  \ottsym{(}  y  \ottsym{)} \,  =  \, pv \andalso
    R(z) \in  \mathbb{Z}  \andalso
    \ottnt{R}  \ottsym{(}  x  \ottsym{)} = pv  \boxplus  \ottnt{R}  \ottsym{(}  z  \ottsym{)}
  }{
      \tuple{ \ottnt{H} ,  \ottnt{R} ,    \ALIAS(  x  =  y   \boxplus   z  )   \SEQ  e  }     \longrightarrow _{  D  }     \tuple{ \ottnt{H} ,  \ottnt{R} ,  e }  
  }
  \infrule[Rs-AliasAddPtrFail]{
    \ottnt{R}  \ottsym{(}  y  \ottsym{)} \,  =  \, pv \andalso
    R(z) \in  \mathbb{Z}  \andalso
    \ottnt{R}  \ottsym{(}  x  \ottsym{)} \neq pv  \boxplus  \ottnt{R}  \ottsym{(}  z  \ottsym{)}
  }{
      \tuple{ \ottnt{H} ,  \ottnt{R} ,    \ALIAS(  x  =  y   \boxplus   z  )   \SEQ  e  }     \longrightarrow _{  D  }     \mathbf{AliasFail}  
  }
  \pagebreak
  \infrule[Rs-AliasDeref]{
    \ottnt{H}  \ottsym{(}  \ottnt{R}  \ottsym{(}  y  \ottsym{)}  \ottsym{)} \,  =  \, \ottnt{R}  \ottsym{(}  x  \ottsym{)}
  }{
      \tuple{ \ottnt{H} ,  \ottnt{R} ,    \ALIAS(  x  = \ast  y  )   \SEQ  e  }     \longrightarrow _{  D  }     \tuple{ \ottnt{H} ,  \ottnt{R} ,  e }  
  }
  \infrule[Rs-AliasDerefFail]{
    R(y) =  \NULL  \mbox{ or }
    \ottnt{H}  \ottsym{(}  \ottnt{R}  \ottsym{(}  y  \ottsym{)}  \ottsym{)} \, \neq \, \ottnt{R}  \ottsym{(}  x  \ottsym{)}
  }{
      \tuple{ \ottnt{H} ,  \ottnt{R} ,    \ALIAS(  x  = \ast  y  )   \SEQ  e  }     \longrightarrow _{  D  }     \mathbf{AliasFail}  
  }
  \infrule[Rs-Assert]{
    \models  \ottsym{[}  \ottnt{R}  \ottsym{]} \, \varphi
  }{
      \tuple{ \ottnt{H} ,  \ottnt{R} ,    \ASSERT( \varphi )   \SEQ  e  }     \longrightarrow _{  D  }     \tuple{ \ottnt{H} ,  \ottnt{R} ,  e }  
  }
  \end{multicols}
  \caption{Operational semantics (excerpt).}
  \Description{Reduction rules operate on a heap, register file, and
  expression.  Null binding and pointer arithmetic add fresh register
  bindings.  Allocation creates a fresh array initialized with zeros for
  integer elements or null pointers for reference elements.  Alias checks
  either continue unchanged or produce AliasFail, according to whether the
  claimed pointer equality holds.  An assertion continues when its formula
  holds under the register file.}
\label{fig:opsem}
\end{figure*}

The rules formalize the informal meaning described in
\cref{sec:syntax} in a straightforward manner.
\begin{itemize}
\item The rules for evaluating an expression $e$ that binds a variable
  $x$ 
  take a fresh
  variable $x'$, extend the register file $R$ with the value of the right-hand side,
  and rename the occurrences of $x$ in the body of $e$ to $x'$ to
  avoid the collision of bound variable names.
\item In the rule \rn{Rs-AddPtr} for pointer arithmetic expressions $ \LET  x  =   y   \boxplus   z   \IN  e $,
  we use the meta-level operator $pv  \boxplus   j $, defined by:
  $\ottsym{(}  a  \ottsym{,}  i  \ottsym{)}  \boxplus  j = \ottsym{(}  a  \ottsym{,}  i  \ottsym{+}  j  \ottsym{)}$ and $ \NULL   \boxplus  j =  \NULL $.
  Adding an offset to $ \NULL $ results in $ \NULL $.
  
\item The allocation rules \rn{Rs-MkArrayIntref} and \rn{Rs-MkArrayNestedref} extend
  the heap as follows.
  In both rules $a$ is a fresh base address and the heap $H$ is extended to $H'$ so that
  $(a,0), \dots, (a, R(y) - 1)$ are all mapped to initial values that depend on $\tau^{-}$.
  If $\tau^{-}$ is $ \TINT $, the initial values are $0$ (\rn{Rs-MkArrayIntref});
  otherwise the initial values are $ \NULL $ (\rn{Rs-MkArrayNestedref}).
  \begin{revision}
  In the latter case, each cell is intended to be initialized later with
  a pointer to another (nested) array before it is accessed.
  \end{revision}
  If $\ottnt{R}  \ottsym{(}  y  \ottsym{)} \leq 0$, we regard
  $\set{(a,0) \mapsto 0, \dots, (a, R(y) - 1) \mapsto 0}$ as the empty
  set, hence $\ottnt{H'} = \ottnt{H}$: Although a fresh address is returned, no memory cell will be
  allocated.
  The operational semantics in Tanaka et
  al.~\cite{DBLP:conf/pepm/TanakaSK24} does not include a rule for nested array allocation.

\item The two rules \rn{Rs-AliasAddPtr} and \rn{Rs-AliasDeref} deal with \textbf{alias} expressions.
  If the two pointers compared are equal, the execution proceeds to
  $e$ without changing $\ottnt{R}$ or $\ottnt{H}$.  Otherwise, the configuration goes to
  $ \mathbf{AliasFail} $, which is a special erroneous configuration.
  
\item The rule \rn{Rs-Assert} requires $\ottsym{[}  \ottnt{R}  \ottsym{]} \, \varphi$ to be valid, where
  \(
    \ottsym{[}  \ottnt{R}  \ottsym{]} \, \varphi =
    [R(x_1)/x_1, \ldots, R(x_n)/x_n]\,\varphi.
  \)
  Here, $x_1,\ldots,x_n$ enumerate the variables in $ \DOM( \ottnt{R} ) $
  whose values in $\ottnt{R}$ are integers.
  In a well-typed program, all free variables in $\varphi$
  have integer type.
  If $\ottsym{[}  \ottnt{R}  \ottsym{]} \, \varphi$ is not valid, the program gets stuck.

\end{itemize}

A program $ \tuple{ D ,  e } $ starts its execution from the configuration
$ \tuple{  \emptyset  ,   \emptyset  ,  e } $ and reduces by using $ \longrightarrow _{D}$.  The
execution of an (untyped) program (1) terminates at a configuration of
the form $ \tuple{ \ottnt{H} ,  \ottnt{R} ,  x } $, representing successful termination, (2)
diverges, (3) abnormally terminates at $ \mathbf{AliasFail} $, or (4) gets
stuck.  There are several reasons for an execution to get stuck.
\begin{enumerate}
\item Unbound variable errors, which occur when the referenced variable is not in $ \DOM( \ottnt{R} ) $.
\item Simple type errors, such as applying subtraction to a pointer, conditional branching on a pointer,
  dereferencing an integer, and so on.
\item Dereferencing the null pointer.
  It occurs if $\ottnt{R}  \ottsym{(}  x  \ottsym{)}$ is $ \NULL $ for a given
  $ \tuple{ \ottnt{H} ,  \ottnt{R} ,   \LET  y  =   \ast  x   \IN  e  } $ or $ \tuple{ \ottnt{H} ,  \ottnt{R} ,    x  \WRITE  y   \SEQ  e  } $.
\item An out-of-bounds array access.  It occurs when $\ottnt{R}  \ottsym{(}  x  \ottsym{)} =  \ottsym{(}  a  \ottsym{,}  i  \ottsym{)}  \not\in   \DOM( \ottnt{H} )  $, for a given
  $ \tuple{ \ottnt{H} ,  \ottnt{R} ,   \LET  y  =   \ast  x   \IN  e  } $ or $ \tuple{ \ottnt{H} ,  \ottnt{R} ,    x  \WRITE  y   \SEQ  e  } $.
\item An assertion failure.  This happens when $ \tuple{ \ottnt{H} ,  \ottnt{R} ,    \ASSERT( \varphi )   \SEQ  e  } $ does not satisfy $\models \ottsym{[}  \ottnt{R}  \ottsym{]} \, \varphi$.
\end{enumerate}
Our type system guarantees that well-typed programs do not get stuck.

\section{Type System}
\label{sec:typing}
In this section, we give our type system to prevent assertion failures
and null/dangling pointer access and state a type soundness theorem.
The type system is based on Tanaka et al.~\cite{DBLP:conf/pepm/TanakaSK24}; we give a more
precise formalization of ownership functions, in particular, how
they depend on the variables in context.

\subsection{Types}
\label{sec:syntax of types}

\begin{figure}
  \[
    \begin{array}{lllll}
      \tau & \text{(types)} & ::= &  \{  \nu  :   \TINT    \mid   \varphi  \}  \mid  \Pi x .( \tau  \TREF^{\hspace{0.5pt} r })  \\
      r & \text{(ownership terms)} & ::= &  \mathbf{0}  \mid   \varphi   \produces    q    ,  r  \\
      \Gamma & \text{(type environments)} & ::= &  x_{{\mathrm{1}}} \COL \tau_{{\mathrm{1}}} ,\ldots, x_{\ottmv{n}} \COL \tau_{\ottmv{n}}  \\
      \sigma & \text{(function types)} & ::= &   \tuple{ \Gamma }\ra\tuple{ \Gamma'  \mid  \tau }  \quad (\text{where } \DOM( \Gamma )  =  \DOM( \Gamma' ) )
    \end{array}
  \]
  \caption{Syntax of types.}
  \Description{Types are refined integers or array references with an
  element type and an ownership term.  Each reference binds an array index
  in its element type and ownership term.  Ownership terms are conditional
  rational permissions with a zero default.  Type environments map
  variables to types; function types specify input and output environments
  with the same domain, together with a return type.}
  \label{fig:SyntaxOfTypes}
\end{figure}

The syntax of \emph{types}, ranged over by $\tau$, \emph{ownership
  terms}, ranged over by $r$, and \emph{function types},
ranged over by $\sigma$ is shown in \cref{fig:SyntaxOfTypes}.
An integer type $ \{  \nu  :   \TINT    \mid   \varphi  \} $ represents integers satisfying a
refinement predicate $\varphi$.
Precisely, $ \{  \nu  :   \TINT    \mid   \varphi  \} $ stands for a subset of integers
$i$ that satisfy $\ottsym{[}   i   \ottsym{/}  \nu  \ottsym{]}  \varphi$.
We often write $ \TINT $ for $ \{  \nu  :   \TINT    \mid   \top  \} $.

A type of the form $ \Pi x .( \tau  \TREF^{\hspace{0.5pt} r }) $ represents pointers to arrays
of type $\tau$ with an \emph{ownership term} $r$, explained below.
The variable $x$, which stands for an array index, is bound in $\tau$ and $r$.
Thus, the element type and ownership term can depend on the array index.
Moreover, if a reference type is nested, the inner type can depend on outer array indices.
We write $|\tau|$ for the simple type obtained from $\tau$ in an obvious manner.
\begin{revision}
For example, let $\tau_{\mathit{ex}} :=  \Pi x_{{\mathrm{1}}} .(  \Pi x_{{\mathrm{2}}} .(  \Pi x_{{\mathrm{3}}} .(  \{  \nu  :   \TINT    \mid   \varphi  \}   \TREF^{\hspace{0.5pt} r_{{\mathrm{3}}} })   \TREF^{\hspace{0.5pt} r_{{\mathrm{2}}} })   \TREF^{\hspace{0.5pt} r_{{\mathrm{1}}} }) $
be a type for three-dimensional arrays of integers with ownership terms $r_{{\mathrm{1}}}$, $r_{{\mathrm{2}}}$, and $r_{{\mathrm{3}}}$,
where $x_1$ is the index of the outermost array, $x_2$ that of the middle array,
and $x_3$ that of the innermost array.
The index $x_1$ may appear in $r_{{\mathrm{1}}}$, $r_{{\mathrm{2}}}$, $r_{{\mathrm{3}}}$, and $\varphi$;
$x_2$ may appear in $r_{{\mathrm{2}}}$, $r_{{\mathrm{3}}}$, and $\varphi$; and
$x_3$ may appear in $r_{{\mathrm{3}}}$ and $\varphi$.
\end{revision}

As in previous work~\cite{DBLP:conf/pepm/TanakaSK24, DBLP:conf/esop/TomanSSI020}, the type system keeps
track of pointer \emph{ownership}, which is represented by a rational
number $q$ between 0 and 1 (that is, $(0 \le q \le 1)$).  The rational number
denotes the permissions for dereferencing (reading) and writing to the
element.
The ownership expresses permissions granted to pointers as follows:
\begin{itemize}
    \item $q = 1$: allows both dereferencing and writing to the element (read-write permission).
    \item $0 < q < 1$: allows dereferencing but not writing (read-only permission).
    \item $q = 0$: allows neither reading nor writing (no
      access).\footnote{%
        Toman et al.~\cite{DBLP:conf/esop/TomanSSI020} \emph{do} allow reading even if
        $q=0$.  Their type system does not support pointer arithmetic
        or prevent null pointer access.  Thus, every pointer access is
        safe (except for the null pointer).  The $0$ ownership is
        different from positive ownership in that the refinement predicate of the element pointed to by a pointer with ownership $0$ is regarded as $\top$, because the ownership being 0 means that there may be a
        writable alias with ownership 1.}
\end{itemize}

An ownership term, which takes the form
$  \varphi_{{\mathrm{1}}}   \produces    q_{{\mathrm{1}}}  , \ldots,  \varphi_{\ottmv{n}}   \produces    q_{\ottmv{n}}    ,   \mathbf{0}  $ where each $q_{\ottmv{i}}$ is a
rational number such that $0 \leq q_{\ottmv{i}} \leq 1 $, represents the
ownership of the pointer to each element of an array.
Intuitively, it means ``if $\varphi_{{\mathrm{1}}}$ is true, then the ownership is $q_{{\mathrm{1}}}$;
or else if $\varphi_{{\mathrm{2}}}$ is true, then the ownership is $q_{{\mathrm{2}}}$, \ldots, and if none of them is true, then the ownership is $0$''.
We often omit $ \mathbf{0} $ at the end.
For example, the type %
$ \Pi x .(  \{  \nu  :   \TINT    \mid   \nu \,  >  \, x  \}   \TREF^{\hspace{0.5pt} \ottsym{(}   \varphi   \produces    1    \ottsym{)} }) $ where $\varphi = 0 \leq x \leq 2$
means that the pointer has full ownership of the first three elements
but no ownership of the other elements, and that
the $i$-th element is an integer greater than $i$.
\begin{revision}
For example, in the above type $\tau_{\mathit{ex}}$ for three-dimensional arrays,
if $r_{{\mathrm{1}}} = (x_1 \in [0,9] \Rightarrow 1)$,
$r_{{\mathrm{2}}} = (x_2 \in [0, x_1] \Rightarrow 1)$, and
$r_{{\mathrm{3}}} = (x_3 \in [0, x_2] \Rightarrow 1)$,
then the pointer has full ownership at indices $i_1, i_2, i_3$,
where $i_1, i_2, i_3$ range over the outermost, middle, and innermost indices,
and satisfy $0 \leq i_3 \leq i_2 \leq i_1 \leq 9$.
\end{revision}

\subsubsection*{Type Environments}
A \emph{type environment}, denoted
by $\Gamma$, is a sequence of type declarations $ x_{\ottmv{i}} \COL \tau_{\ottmv{i}} $ of pairwise
distinct variables.  Given a type environment $x_{{\mathrm{1}}}  \ottsym{:}  \tau_{{\mathrm{1}}}  \ottsym{,} \, .. \, \ottsym{,}  x_{\ottmv{n}}  \ottsym{:}  \tau_{\ottmv{n}}$,
all the variables are bound in $\tau_{{\mathrm{1}}}, \ldots, \tau_{\ottmv{n}}$.  Thus, even
mutual dependency between variables is allowed.
\ifdefined\VersionLong{}%
As defined in \cref{sec:type-well-formedness}, a well-formed type environment allows only integer variables to appear in the refinement predicates and ownership terms of other types.
\fi
We regard a type environment $x_{{\mathrm{1}}}  \ottsym{:}  \tau_{{\mathrm{1}}}  \ottsym{,} \, .. \, \ottsym{,}  x_{\ottmv{n}}  \ottsym{:}  \tau_{\ottmv{n}}$ as a function that
maps each $x_{\ottmv{i}}$ to $\tau_{\ottmv{i}}$.  We write $ \DOM( \Gamma ) $ for
$\set{x_{{\mathrm{1}}}, \ldots, x_{\ottmv{n}}}$. If $ x  \not\in   \DOM( \Gamma )  $, we write
$\Gamma  \ottsym{,}   x \COL \tau $ to add a new declaration $ x \COL \tau $ to $\Gamma$.  We write
$\Gamma  \ottsym{[}  x  \ottsym{:}  \tau  \ottsym{]}$ to signify $ \Gamma  (  x  )   =  \tau$ and, if $ x  \in   \DOM( \Gamma )  $,
write $ \Gamma  \left[  x \hookleftarrow \tau  \right] $ for a type environment $\Gamma'$ such that
$ \DOM( \Gamma' )   =   \DOM( \Gamma ) $ and $ \Gamma'  (  x  )   =  \tau$ and $ \Gamma'  (  y  )   =   \Gamma  (  y  ) $ for
$y \neq x$.

\subsubsection*{Function Types and Function Type Environments}

A \emph{function type} $\sigma$ is of the form
\(
   \tuple{ \Gamma }\ra\tuple{ \Gamma'  \mid  \tau } 
  \)
(where $ \DOM( \Gamma )   =   \DOM( \Gamma' ) $).
If
$\Gamma = x_{{\mathrm{1}}}  \ottsym{:}  \tau_{{\mathrm{1}}}  \ottsym{,} \, .. \, \ottsym{,}  x_{\ottmv{n}}  \ottsym{:}  \tau_{\ottmv{n}}$ and
$\Gamma' = x_{{\mathrm{1}}}  \ottsym{:}  \tau'_{{\mathrm{1}}}  \ottsym{,} \, .. \, \ottsym{,}  x_{\ottmv{n}}  \ottsym{:}  \tau'_{\ottmv{n}}$,
it means that the
function takes $x_{{\mathrm{1}}}, \ldots, x_{\ottmv{n}}$ of types
$\tau_{{\mathrm{1}}}, \ldots, \tau_{\ottmv{n}}$, respectively, as arguments, returns $\tau$,
and changes the types of the arguments to $\tau'_{{\mathrm{1}}}, \ldots, \tau'_{\ottmv{n}}$
as a side effect.  Formal arguments are named and can appear in
$\tau_{\ottmv{i}}, \tau'_{\ottmv{i}}$, and $\tau$.
For example, the function type
\[
   \tuple{ x_{{\mathrm{1}}}  \ottsym{:}   \TINT   \ottsym{,}  x_{{\mathrm{2}}}  \ottsym{:}   \Pi z .(  \TINT   \TREF^{\hspace{0.5pt} r })  }\ra\tuple{ x_{{\mathrm{1}}}  \ottsym{:}   \TINT   \ottsym{,}  x_{{\mathrm{2}}}  \ottsym{:}   \Pi z .(  \{  \nu  :   \TINT    \mid   \nu \,  =  \, x_{{\mathrm{1}}}  \}   \TREF^{\hspace{0.5pt} r })  \mid  \TINT  } 
\]
where $r = (0 \leq z < x_{{\mathrm{1}}} \Rightarrow 1)$ means that the function takes
an integer $x_{{\mathrm{1}}}$ and an integer array whose length is $x_{{\mathrm{1}}}$, returns an integer,
and updates the elements of the array with $x_{{\mathrm{1}}}$.

A \emph{function type environment}, denoted by $\Theta$,
is a finite mapping from function names to function types.

\subsubsection*{Type Well-Formedness}

We enforce well-formedness conditions on types.  Roughly speaking, a
type $\tau$ is well-formed under type environment $\Gamma$,
written $\Gamma  \vdash   \tau  \mbox{ ok} $, if every
predicate $\varphi$ in $\tau$ refers only to integer variables in
context and if the ownership of an array element is zero, its element
type is \emph{empty}.  An empty type has zero ownership at every reference
level and integer refinements that hold for every integer.
The zero-ownership/emptiness implication must be valid for all valuations,
without assuming the refinements in $\Gamma$; the environment determines
only which integer variables may occur in the type.
For example, the type
$\tau =  \Pi x .(  \Pi y .(  \TINT   \TREF^{\hspace{0.5pt} r_{{\mathrm{0}}} })   \TREF^{\hspace{0.5pt} r }) $ where $r = (0 \leq x \leq 1  \produces  1)$
and $r_{{\mathrm{0}}} = (0 \leq x \leq 1 \land 0 \leq y \leq 1  \produces  1)$
is well-formed but
$\tau' =  \Pi x .(  \Pi y .(  \TINT   \TREF^{\hspace{0.5pt} r'_{{\mathrm{0}}} })   \TREF^{\hspace{0.5pt} r }) $
with $r'_{{\mathrm{0}}} = (0 \leq x \leq 2 \land 0 \leq y \leq 1  \produces  1)$
is not.
In $\tau$, the ownership of the element of an inner array is 1 only if
the outer array element is accessible (i.e., $x$ is either 0 or 1), whereas,
in $\tau'$, the ownership of an element of the third inner array is 1 even if
the outer array is not accessible, violating the emptiness condition.
\ifdefined\VersionLong{}%
Formally, type well-formedness is defined by the rules in
\cref{sec:type-well-formedness}.%
\fi\footnote{\interfootnotelinepenalty=10000
Throughout our illustrative examples, including tables and program annotations,
conditions enforcing emptiness are implicit: inner ownership is zero whenever
an enclosing ownership is zero, and integer refinements impose no restriction
at zero-ownership indices.
For example,
$ \Pi x .(  \{  \nu  :   \TINT    \mid   \nu \,  >  \, x  \}   \TREF^{\hspace{0.5pt} \ottsym{(}   \varphi   \produces    1    \ottsym{)} }) $
abbreviates
$ \Pi x .(  \{  \nu  :   \TINT    \mid   \varphi  \implies  \nu \,  >  \, x  \}   \TREF^{\hspace{0.5pt} \ottsym{(}   \varphi   \produces    1    \ottsym{)} }) $
for $\varphi = 0 \leq x \leq 2$.}

A type environment $\Gamma$ is well-formed if each type declared in
$\Gamma$ is well-formed under $\Gamma$ itself.
A function type $ \tuple{ \Gamma }\ra\tuple{ \Gamma'  \mid  \tau } $ is well-formed if
both $\Gamma$ and $\Gamma'$ are well-formed,
$| \Gamma  (  x  ) | = | \Gamma'  (  x  ) |$ for every $ x  \in   \DOM( \Gamma )  $, and
$\tau$ is well-formed under $\Gamma'$.
Thus, a function may change the ownership and refinements of its parameters,
but not their simple types.
A function type environment is well-formed if each function type in it is
well-formed.

We will assume that types, type environments, function types,
function type environments, and assertions are well-formed in judgments
introduced later and omit well-formedness conditions from derivation rules.
For an assertion $  \ASSERT( \varphi )   \SEQ  e $, every free variable of $\varphi$
must be bound to an integer type in the input type environment of its
typing judgment.  The type environment under which a given type is
well-formed is obvious in most cases---we will note when it is not clear.

\begin{remark}[Comparison with Tanaka et al.]
In Tanaka et al.~\cite{DBLP:conf/pepm/TanakaSK24}, an array type is decorated with an
\emph{ownership function} $r$, which is a (seemingly set-theoretic) \emph{function} from integers to
rational numbers (in the interval $[0, 1]$).
The way ownership information is represented there, however, obscures
how reference types may depend on array indices and other variables,
especially when they are nested.
For example, the type of \lstinline{pp} on \cref{motiv:line:fun_ty_goal} in \cref{fig:motivatingExample} is
$%
   \Pi x_{{\mathrm{1}}} .(  \Pi x_{{\mathrm{2}}} .(  \{  \nu  :   \TINT    \mid   \varphi  \}   \TREF^{\hspace{0.5pt} r_{{\mathrm{2}}} })   \TREF^{\hspace{0.5pt} r_{{\mathrm{1}}} }) %
$
where the ownership term $r_{{\mathrm{2}}}$ of the inner array type is $0 \leq x_{{\mathrm{2}}} < \texttt{n} - x_{{\mathrm{1}}}  \produces  1$
since the length of the array pointed to by $\mathtt{pp} \boxplus i$ is $\mathtt{n}-i$.
Because $r_{{\mathrm{2}}}$ depends on $x_{{\mathrm{1}}}$, it would need to be a
\emph{family} of functions, rather than a function.
\end{remark}

\subsection{Auxiliary Judgments and Encoding into Verification Logic}

We often encode auxiliary judgments into formulae within the logic
used for verification. For example, we write $\Gamma  \models  \varphi$, which means
that formula $\varphi$ is valid under $\Gamma$. This relation is
formally defined as follows: $\Gamma  \models  \varphi \iff \models   \fml{ \Gamma }   \implies  \varphi$,
where the function $ \fml{ \Gamma } $ is defined by:
\begin{gather*}
   \fml{  x_{{\mathrm{1}}} \COL \tau_{{\mathrm{1}}} ,\ldots, x_{\ottmv{n}} \COL \tau_{\ottmv{n}}  }  =  \fmlT{ \tau_{{\mathrm{1}}} }{ x_{{\mathrm{1}}} }  \land \cdots \land  \fmlT{ \tau_{\ottmv{n}} }{ x_{\ottmv{n}} }  \\
   \fmlT{  \{  \nu  :   \TINT    \mid   \varphi  \}  }{ y }  = \ottsym{[}  y  \ottsym{/}  \nu  \ottsym{]}  \varphi \qquad
   \fmlT{  \Pi x .( \tau'  \TREF^{\hspace{0.5pt} r })  }{ y }  =  \top .
\end{gather*}
The function $ \fml{ \Gamma } $ translates refinements on integer variables into propositions, ignoring
variables of reference types.

We also encode auxiliary judgments for pointwise comparison and addition of ownership functions,
such as $\Gamma  \models  r_{{\mathrm{1}}} \, \le \, r_{{\mathrm{2}}}$ and $\Gamma  \models  r_{{\mathrm{1}}}  \ottsym{+}  r_{{\mathrm{2}}} \,  =  \, r_{{\mathrm{3}}}$ into the verification logic.
Intuitively, the comparisons $r_{{\mathrm{1}}} \, \le \, r_{{\mathrm{2}}}$ and $r_{{\mathrm{1}}} \,  =  \, r_{{\mathrm{2}}}$ for ownership terms are defined as
a pointwise extension of comparison over rational numbers.  We denote
the first-order formula expressing $r_{{\mathrm{1}}} \, \le \, r_{{\mathrm{2}}}$ by
$ \fml{ r_{{\mathrm{1}}}   \le   r_{{\mathrm{2}}} } $.  We omit the full definition for brevity.  For
example, if $r_{{\mathrm{1}}} = (  \varphi_{{\mathrm{1}}}   \produces    q_{{\mathrm{11}}}    ,    true    \produces    q_{{\mathrm{12}}}   )$ and
$r_{{\mathrm{2}}} = (  \varphi_{{\mathrm{2}}}   \produces    q_{{\mathrm{21}}}    ,    true    \produces    q_{{\mathrm{22}}}   )$, then $ \fml{ r_{{\mathrm{1}}}   \le   r_{{\mathrm{2}}} } $ is
\begin{gather*}
\ottsym{(}  \varphi_{{\mathrm{1}}}  \wedge  \varphi_{{\mathrm{2}}}  \implies  q_{{\mathrm{11}}} \, \le \, q_{{\mathrm{21}}}  \ottsym{)}  \wedge  \ottsym{(}   \neg  \varphi_{{\mathrm{1}}}   \wedge  \varphi_{{\mathrm{2}}}  \implies  q_{{\mathrm{12}}} \, \le \, q_{{\mathrm{21}}}  \ottsym{)}  \\
\land \ottsym{(}  \varphi_{{\mathrm{1}}}  \wedge   \neg  \varphi_{{\mathrm{2}}}   \implies  q_{{\mathrm{11}}} \, \le \, q_{{\mathrm{22}}}  \ottsym{)}  \wedge  \ottsym{(}   \neg  \varphi_{{\mathrm{1}}}   \wedge   \neg  \varphi_{{\mathrm{2}}}   \implies  q_{{\mathrm{12}}} \, \le \, q_{{\mathrm{22}}}  \ottsym{)}
\end{gather*}
where $q_{\ottmv{i}\,\ottmv{j}} \, \le \, q_{\ottmv{k}\,\ottmv{l}}$ stands for either $\top$ or $\bot$, depending
on the comparison of the two rational numbers.  (Note that no rational
number will appear in the resulting formula because the comparison is
performed during the encoding.)
Then, $\Gamma  \models  r_{{\mathrm{1}}} \, \le \, r_{{\mathrm{2}}}$ stands for $\models   \fml{ \Gamma }   \implies   \fml{ r_{{\mathrm{1}}}   \le   r_{{\mathrm{2}}} } $.
$ \fml{ r_{{\mathrm{1}}}    =    r_{{\mathrm{2}}} } $ and the encoding of pointwise addition
$ \fml{ r_{{\mathrm{1}}}   \ottsym{+}   r_{{\mathrm{2}}}   =   r_{{\mathrm{3}}} } $ can be defined similarly.

\subsection{Type System}
\label{sec:typesystem}

Our type system consists of several judgment forms.  The main one is
for expression typing of the form: $ \Theta   \mid   \Gamma   \vdash   e  :  \tau   \produces   \Gamma' $.
Intuitively, it means ``under the environments $\Theta$ and $\Gamma$,
the type of expression $e$ is $\tau$, and the initial
type environment $\Gamma$ is updated to $\Gamma'$.''
As we already noted, we assume the type environments and the type are well-formed,
namely $ \vdash   \Gamma  \mbox{ ok} $ and $ \vdash   \Gamma'  \mbox{ ok} $ and $\Gamma'  \vdash   \tau  \mbox{ ok} $.
We often call $\Gamma$ a pre-environment and $\Gamma'$ a post-environment.

\Cref{fig:typing rules 1,fig:typing rules 2} show the
typing rules for expressions.  The rules in \cref{fig:typing rules 1}
are for expressions not related to heap manipulation and
are very similar to those in previous work~\cite{DBLP:conf/pepm/TanakaSK24,DBLP:conf/esop/TomanSSI020}, except
for a few minor notational changes.
Standard typing rules (i.e., \rn{T-Int}, \rn{T-Minus}, and \rn{T-Let}) are omitted
from \cref{fig:typing rules 1} for brevity; see
\ifdefined\VersionLong%
  \cref{sec:typingOmitted}.
\else
the full version.
\fi

\begin{figure*}[t]
  \small
{\raggedright\noindent\fbox{$ \Theta   \mid   \Gamma   \vdash   e  :  \tau   \produces   \Gamma' $}\par}
\typicallabel{T-Null}
\infrule[T-Var]{
  \Gamma  \vdash    \tau_{{\mathrm{1}}}  +  \tau_{{\mathrm{2}}}    \approx   \tau_{{\mathrm{3}}} 
}{
   \Theta   \mid   \Gamma  \ottsym{[}  x  \ottsym{:}  \tau_{{\mathrm{3}}}  \ottsym{]}   \vdash   x  :   \ottsym{(}  \tau_{{\mathrm{1}}}  \ottsym{)}  ^ {= x }    \produces    \Gamma  \left[  x \hookleftarrow \tau_{{\mathrm{2}}}  \right]  
}
\infrule[T-Null]{
 \models   \Empty{  \Pi z .( \tau_{{\mathrm{1}}}  \TREF^{\hspace{0.5pt} r })  }  \andalso
  \Theta   \mid   \Gamma  \ottsym{,}   x \COL  \Pi z .( \tau_{{\mathrm{1}}}  \TREF^{\hspace{0.5pt} r })     \vdash   e  :  \tau   \produces   \ottsym{(}  \Gamma'  \ottsym{,}   x \COL \tau'   \ottsym{)} 
}{
  \Theta   \mid   \Gamma   \vdash    \LET  x  =  \ottkw{null}  \IN  e   :  \tau   \produces   \Gamma' 
}
\infrule[T-If]{
   \Theta   \mid    \Gamma  \left[  x \hookleftarrow  \{  \nu  :   \TINT    \mid    \varphi  \wedge  \nu \, \le \,  0    \}   \right]    \vdash   e_{{\mathrm{0}}}  :  \tau   \produces   \Gamma'  \\
   \Theta   \mid    \Gamma  \left[  x \hookleftarrow  \{  \nu  :   \TINT    \mid    \varphi  \wedge  \nu \,  >  \,  0    \}   \right]    \vdash   e_{{\mathrm{1}}}  :  \tau   \produces   \Gamma' 
}{
   \Theta   \mid   \Gamma  \ottsym{[}  x  \ottsym{:}   \{  \nu  :   \TINT    \mid   \varphi  \}   \ottsym{]}   \vdash    \IFNP  x  \THEN  e_{{\mathrm{0}}}  \ELSE  e_{{\mathrm{1}}}   :  \tau   \produces   \Gamma' 
}
\infrule[T-Call]{
  \Theta  \ottsym{(}  f  \ottsym{)} =  \tuple{  x_{{\mathrm{1}}} \COL \tau_{{\mathrm{1}}} ,\ldots, x_{\ottmv{n}} \COL \tau_{\ottmv{n}}  }\ra\tuple{  x_{{\mathrm{1}}} \COL \tau'_{{\mathrm{1}}} ,\ldots, x_{\ottmv{n}} \COL \tau'_{\ottmv{n}}   \mid  \tau }  \andalso
  \theta  =   [  y_{{\mathrm{1}}}  /  x_{{\mathrm{1}}}  , \ldots,  y_{\ottmv{n}}  /  x_{\ottmv{n}}  ]  \\
   \Theta   \mid    \Gamma  \left[  y_{\ottmv{i}} \hookleftarrow \theta \, \tau'_{\ottmv{i}}  \right]   \ottsym{,}   x \COL \theta \, \tau    \vdash   e  :  \tau'   \produces   \ottsym{(}  \Gamma'  \ottsym{,}   x \COL \tau''   \ottsym{)}  %
}{
   \Theta   \mid   \Gamma  \ottsym{[}  y_{\ottmv{i}}  \ottsym{:}  \theta \, \tau_{\ottmv{i}}  \ottsym{]}   \vdash    \LET  x  =   f (  y_{{\mathrm{1}}} ,\ldots, y_{\ottmv{n}}  )   \IN  e   :  \tau'   \produces   \Gamma' 
}
\infrule[T-Sub]{
  \Gamma  \leq  \Gamma'
  \andalso  \Theta   \mid   \Gamma'   \vdash   e  :  \tau   \produces   \Gamma'' 
  \andalso \Gamma''  \ottsym{,}  \tau  \leq  \Gamma'''  \ottsym{,}  \tau'
}{
   \Theta   \mid   \Gamma   \vdash   e  :  \tau'   \produces   \Gamma''' 
}
\caption{Expression typing rules (1).}
\Description{Five rules specify expression types and changes to the type
environment.  Variable use splits its type between the returned value and
the remaining binding.  Null binding requires an empty reference type.
Conditionals strengthen refinements by their branch conditions.  Calls
instantiate function types with actual arguments.  Subsumption combines
subtyping of the input environment, output environment, and result type.}
\label{fig:typing rules 1}
\end{figure*}

The rule \rn{T-Var} splits a variable's type between its use and its
remaining binding.  To understand ownership splitting, consider
$ \LET  y  =  x  \IN  e_{{\mathrm{0}}} $, which creates an alias
of $x$.  A part of the ownership of $x$ has to be given to
$y$ when typing $e_{{\mathrm{0}}}$.  Such ``split'' of ownership is
expressed as the auxiliary judgment $\Gamma  \vdash    \tau_{{\mathrm{1}}}  +  \tau_{{\mathrm{2}}}    \approx   \tau_{{\mathrm{3}}} $---which
reads ``$\tau_{{\mathrm{3}}}$ is split into $\tau_{{\mathrm{1}}}$ and $\tau_{{\mathrm{2}}}$ under $\Gamma$''
or ``$\tau_{{\mathrm{1}}}$ and $\tau_{{\mathrm{2}}}$ merge into $\tau_{{\mathrm{3}}}$ under $\Gamma$''---for
type addition.  The latter reading will be useful when we discuss the
typing rules for $\ALIAS$ expressions.

We show a type derivation for $ \LET  y  =  x  \IN  e_{{\mathrm{0}}} $ below.
Here, we write $\tau(r) =  \Pi z .(  \TINT   \TREF^{\hspace{0.5pt} r }) $ and omit $\Theta$.
\[
  \frac{
    \displaystyle{\frac{
        \vdots
      }{
      \Gamma, x:\tau(r_{{\mathrm{1}}})  \vdash  x : \tau(r_{{\mathrm{2}}})  \produces  \Gamma, x:\tau(r_{{\mathrm{2}}})
    }}
    \andalso
    \displaystyle{\frac{\vdots}{
        \Gamma, x:\tau(r_{{\mathrm{2}}}), y:\tau(r_{{\mathrm{2}}})  \vdash  e_{{\mathrm{0}}} : \tau(r_{{\mathrm{2}}})  \produces  (\Gamma', y:\tau(r_{{\mathrm{2}}}))
        }}
  }{
    \Gamma, x:\tau(r_{{\mathrm{1}}})  \vdash   \LET  y  =  x  \IN  e_{{\mathrm{0}}}  : \tau(r_{{\mathrm{2}}})  \produces  \Gamma'
  }
\]
where $r_{{\mathrm{1}}} = 0 \leq z \leq 9  \produces  1$ and
$r_{{\mathrm{2}}} = 0 \leq z \leq 9  \produces  0.5$.
Notice that $\Gamma  \models  r_{{\mathrm{2}}}  \ottsym{+}  r_{{\mathrm{2}}} \,  =  \, r_{{\mathrm{1}}}$ holds.

The rule also uses \emph{selfification}, introduced for refinement-type
checking by Ou et al.~\cite{OuTMW04} and subsequently used in Liquid
Types~\cite{DBLP:conf/pldi/RondonKJ08}, to record equality with the source
variable when copying an integer.
The equality-strengthening operation is defined by
\[
\begin{array}{c}
    \{  \nu  :   \TINT    \mid   \varphi  \}   ^ {= x }  =  \{  \nu  :   \TINT    \mid    \varphi  \wedge  \ottsym{(}  \nu \,  =  \, x  \ottsym{)}   \} 
  \hspace{2.5cm}
    \Pi z .( \tau  \TREF^{\hspace{0.5pt} r })   ^ {= x }  =  \Pi z .( \tau  \TREF^{\hspace{0.5pt} r }) .
\end{array}
\]
It leaves reference types unchanged, as in the derivation above.

In \rn{T-Null}, the type of
$x$ is given an \emph{empty} reference type because $ \NULL $ is
inaccessible.  The premise $\models   \Empty{ \tau } $ requires emptiness for every
valuation, independently of the refinements in $\Gamma$.  We require this
independence so that the chosen type remains provably empty when the
surrounding environment loses refinement information.  For example, a
function signature may specify $ y \COL  \{  \nu  :   \TINT    \mid   \nu \,  =  \,  0   \}  $ on input but only
$ y \COL  \TINT  $ on output.  After the call, the typing environment no longer
records that $y$ is zero, even though its value is unchanged.
Thus, establishing emptiness should not rely on facts specific to the point
where the null value is introduced.
As with ownership comparisons,
emptiness is encoded into a logical formula.  The function
$ \Empty{ \tau } $ is defined by induction on $\tau$ as follows:
\begin{align*}
   \Empty{  \{  \nu  :   \TINT    \mid   \varphi  \}  }  & =  \varphi & %
   \Empty{  \Pi z .( \tau  \TREF^{\hspace{0.5pt} r })  }  & =   \fml{ r    =     \mathbf{0}  }  \land  \Empty{ \tau } .
\end{align*}
\begin{revision}
$ \Empty{ \tau } $ gathers conditions recursively if $\tau$ is a nested array type.
\end{revision}

The rule \rn{T-If} is standard in a refinement type system.  Since the
$\THEN$ branch (the $\ELSE$ branch, resp.) is taken only when $x$ is
non-positive (positive, resp.), the predicate $\varphi$ for $x$ is
strengthened by $\nu \, \le \,  0 $ ($\nu \,  >  \,  0 $, resp.).

In rule \rn{T-Call}, substitution $\theta$ renames formal parameter
names in the function type $\Theta  \ottsym{(}  f  \ottsym{)}$ by actual argument names.  This
rule means that the types of arguments $y_{\ottmv{i}}$ change from
$\theta \, \tau_{\ottmv{i}}$ to $\theta \, \tau'_{\ottmv{i}}$ by the function call, $x$ is given
type $\theta \, \tau$, where $\tau$ is the return type.

The rule \rn{T-Sub} is for subsumption, which strengthens the
pre-environment ($\Gamma  \leq  \Gamma'$) and weakens the expression type and
the post-environment.
\begin{revision}
The full rules for subtyping are similar to those of previous work~\cite{DBLP:conf/pepm/TanakaSK24, DBLP:conf/esop/TomanSSI020} and are given in
\ifdefined\VersionLong{}%
  \cref{sec:subtyping}.
\else%
  the extended version.
\fi
We explain two rules here.
The rule \rn{S-Int} for integer types:
\infrule[S-Int]{
   \Gamma  ,  \nu  \colon \TINT   \models  \varphi_{{\mathrm{1}}}  \implies  \varphi_{{\mathrm{2}}}
}{
  \Gamma  \vdash    \{  \nu  :   \TINT    \mid   \varphi_{{\mathrm{1}}}  \}    \leq    \{  \nu  :   \TINT    \mid   \varphi_{{\mathrm{2}}}  \}  
}
requires that the predicate of a subtype has to be stronger than that
of a supertype.
The rule \rn{S-Ref} for reference types:
\infrule[S-Ref]{
  \Gamma  \ottsym{,}   x \COL  \TINT    \models   r _{ 1 }  \, \ge \,  r _{ 2 } 
  \andalso \Gamma  \ottsym{,}   x \COL  \TINT    \vdash   \tau_{{\mathrm{1}}}   \leq   \tau_{{\mathrm{2}}} 
}{
  \Gamma  \vdash    \Pi x .( \tau_{{\mathrm{1}}}  \TREF^{\hspace{0.5pt}  r _{ 1 }  })    \leq    \Pi x .( \tau_{{\mathrm{2}}}  \TREF^{\hspace{0.5pt}  r _{ 2 }  })  
}
is covariant and additionally requires that the ownership term in a subtype has to
be pointwise greater than that in a supertype.  Note that covariance
is safe because if an array is writable through one alias, the other
aliases will be assigned no ownership and cannot observe the update~\cite[Section 6.3]{10.1145/1411204.1411235}.
\end{revision}

\begin{figure*}
  \small
{\raggedright\noindent\fbox{$ \Theta   \mid   \Gamma   \vdash   e  :  \tau   \produces   \Gamma' $}\par}
\typicallabel{}
\infrule[T-MkIntArray]{
  \Gamma  \ottsym{,}   z \COL  \TINT    \models  r \,  =  \, \ottsym{(}   \ottsym{(}    0  \, \le \, z  \wedge  z \, \le \, y  \ottsym{-}   1    \ottsym{)}   \produces    1    \ottsym{)} \andalso
  \\
     \Theta   \mid   \Gamma  \ottsym{[}  y  \ottsym{:}   \{  \nu  :   \TINT    \mid   \varphi  \}   \ottsym{]}  \ottsym{,}   x \COL  \Pi z .(  \{  \nu  :   \TINT    \mid     0  \, \le \, z  \wedge  z \, \le \, y  \ottsym{-}   1    \implies  \nu \,  =  \,  0   \}   \TREF^{\hspace{0.5pt} r })     \vdash   e_{{\mathrm{0}}}  :  \tau   \produces   \ottsym{(}  \Gamma'  \ottsym{,}   x \COL \tau'   \ottsym{)} 
  }{
     \Theta   \mid   \Gamma  \ottsym{[}  y  \ottsym{:}   \{  \nu  :   \TINT    \mid   \varphi  \}   \ottsym{]}   \vdash    \LET  x  =   \ALLOC  y   \ottsym{:}     \TINT   \TREF    \IN  e_{{\mathrm{0}}}   :  \tau   \produces   \Gamma' 
  }
\infrule[T-MkNestedArray]{
    \Gamma  \ottsym{,}   z \COL  \TINT    \models  r \,  =  \, \ottsym{(}   \ottsym{(}    0  \, \le \, z  \wedge  z \, \le \, y  \ottsym{-}   1    \ottsym{)}   \produces    1    \ottsym{)} \andalso
    \models   \Empty{ \tau_{{\mathrm{0}}} }  \\
     \Theta   \mid   \Gamma  \ottsym{[}  y  \ottsym{:}   \{  \nu  :   \TINT    \mid   \varphi  \}   \ottsym{]}  \ottsym{,}   x \COL  \Pi z .( \tau_{{\mathrm{0}}}  \TREF^{\hspace{0.5pt} r })     \vdash   e_{{\mathrm{0}}}  :  \tau   \produces   \ottsym{(}  \Gamma'  \ottsym{,}   x \COL \tau'   \ottsym{)}  \andalso
    |\tau_{{\mathrm{0}}}| =  \tau^{-}  \TREF   %
  }{
     \Theta   \mid   \Gamma  \ottsym{[}  y  \ottsym{:}   \{  \nu  :   \TINT    \mid   \varphi  \}   \ottsym{]}   \vdash    \LET  x  =   \ALLOC  y   \ottsym{:}    \ottsym{(}   \tau^{-}  \TREF   \ottsym{)}  \TREF    \IN  e_{{\mathrm{0}}}   :  \tau   \produces   \Gamma' 
  }
\infrule[T-Deref]{
  \Gamma  \ottsym{,}   z \COL  \{  \nu  :   \TINT    \mid   \nu \,  =  \,  0   \}    \models  r \,  >  \,  \mathbf{0}  \andalso
  \Gamma  \ottsym{,}   z \COL  \{  \nu  :   \TINT    \mid   \nu \,  =  \,  0   \}    \vdash     \tau'  +  \tau  _{ x }    \approx    \tau _{ y }   \\
  \Gamma  \ottsym{,}   x \COL  \tau _{ x }    \ottsym{,}   z \COL  \{  \nu  :   \TINT    \mid   \nu \,  =  \,  0   \}    \vdash    \tau' _{ y }    \approx    \ottsym{(}  \tau'  \ottsym{)}  ^ {= x }   \andalso
  \Gamma  \ottsym{,}   x \COL  \tau _{ x }    \ottsym{,}   z \COL  \{  \nu  :   \TINT    \mid   \nu \, \neq \,  0   \}    \vdash    \tau' _{ y }    \approx    \tau _{ y }   \\
   \Theta   \mid    \Gamma  \left[  y \hookleftarrow  \Pi z .(  \tau' _{ y }   \TREF^{\hspace{0.5pt} r })   \right]   \ottsym{,}   x \COL  \tau _{ x }     \vdash   e_{{\mathrm{0}}}  :  \tau   \produces   \ottsym{(}  \Gamma'  \ottsym{,}   x \COL \tau''   \ottsym{)}  %
  }{
     \Theta   \mid   \Gamma  \ottsym{[}  y  \ottsym{:}   \Pi z .(  \tau _{ y }   \TREF^{\hspace{0.5pt} r })   \ottsym{]}   \vdash    \LET  x  =   \ast  y   \IN  e_{{\mathrm{0}}}   :  \tau   \produces   \Gamma' 
  }

\infrule[T-Assign]{
  \Gamma  \ottsym{,}   z \COL  \{  \nu  :   \TINT    \mid   \nu \,  =  \,  0   \}    \models  r \,  =  \,  \mathbf{1}  \andalso
  \Gamma  \vdash     \tau'  +  \tau'  _{ y }    \approx    \tau _{ y }   \\
  \Gamma  \ottsym{,}   z \COL  \{  \nu  :   \TINT    \mid   \nu \,  =  \,  0   \}    \vdash    \tau' _{\ast  x }    \approx    \ottsym{(}  \tau'  \ottsym{)}  ^ {= y }   \andalso
  \Gamma  \ottsym{,}   z \COL  \{  \nu  :   \TINT    \mid   \nu \, \neq \,  0   \}    \vdash    \tau' _{\ast  x }    \approx    \tau _{\ast  x }   \\
   \Theta   \mid     \Gamma  \left[  x \hookleftarrow  \Pi z .(  \tau' _{\ast  x }   \TREF^{\hspace{0.5pt} r })   \right]   \left[  y \hookleftarrow  \tau' _{ y }   \right]    \vdash   e_{{\mathrm{0}}}  :  \tau   \produces   \Gamma' 
  }{
     \Theta   \mid   \Gamma  \ottsym{[}  x  \ottsym{:}   \Pi z .(  \tau _{\ast  x }   \TREF^{\hspace{0.5pt} r })   \ottsym{]}  \ottsym{[}  y  \ottsym{:}   \tau _{ y }   \ottsym{]}   \vdash     x  \WRITE  y   \SEQ  e_{{\mathrm{0}}}   :  \tau   \produces   \Gamma' 
  }

\infrule[T-AddPtr]{
    \Gamma  \vdash     \Pi w .( \tau_{{\mathrm{1}}}  \TREF^{\hspace{0.5pt}  { r }_{ y_{{\mathrm{1}}} }  })   +   \Pi w .   [ (  w  -  z  ) /  w  ]   ( \tau_{{\mathrm{2}}}  \TREF^{\hspace{0.5pt}  { r }_{ x }  })     \approx    \Pi w .( \tau_{{\mathrm{3}}}  \TREF^{\hspace{0.5pt}  { r }_{ y }  })   \\
     \Theta   \mid    \Gamma  \ottsym{[}  z  \ottsym{:}   \{  \nu  :   \TINT    \mid   \varphi  \}   \ottsym{]}  \left[  y \hookleftarrow  \Pi w .( \tau_{{\mathrm{1}}}  \TREF^{\hspace{0.5pt}  { r }_{ y_{{\mathrm{1}}} }  })   \right]   \ottsym{,}   x \COL  \Pi w .( \tau_{{\mathrm{2}}}  \TREF^{\hspace{0.5pt}  { r }_{ x }  })     \vdash   e_{{\mathrm{0}}}  :  \tau   \produces   \ottsym{(}  \Gamma'  \ottsym{,}   x \COL \tau'   \ottsym{)}  %
  }{
     \Theta   \mid   \Gamma  \ottsym{[}  y  \ottsym{:}   \Pi w .( \tau_{{\mathrm{3}}}  \TREF^{\hspace{0.5pt}  { r }_{ y }  })   \ottsym{]}  \ottsym{[}  z  \ottsym{:}   \{  \nu  :   \TINT    \mid   \varphi  \}   \ottsym{]}   \vdash    \LET  x  =   y   \boxplus   z   \IN  e_{{\mathrm{0}}}   :  \tau   \produces   \Gamma' 
  }

\infrule[T-AliasAddPtr]{
    x \, \neq \, y\\
  \begin{array}{r}
    \Gamma  \vdash  \ottsym{(}    \Pi w' .   [ (  w'  -  z  ) /  w'  ]   (  \tau _{\ast  x }   \TREF^{\hspace{0.5pt}  { r }_{ x }  })   +   \Pi w .(  \tau _{\ast  y }   \TREF^{\hspace{0.5pt}  { r }_{ y }  })    \ottsym{)}  \approx  \qquad \qquad \\   \ottsym{(}   \Pi w' .   [ (  w'  -  z  ) /  w'  ]   (  \tau' _{\ast  x }   \TREF^{\hspace{0.5pt}  { r' }_{ x }  })   \ottsym{)}  +   \Pi w .(  \tau' _{\ast  y }   \TREF^{\hspace{0.5pt}  { r' }_{ y }  })  
    \end{array}\\
     \Theta   \mid     \Gamma  \ottsym{[}  z  \ottsym{:}   \{  \nu  :   \TINT    \mid   \varphi  \}   \ottsym{]}  \left[  x \hookleftarrow  \Pi w' .(  \tau' _{\ast  x }   \TREF^{\hspace{0.5pt}  { r' }_{ x }  })   \right]   \left[  y \hookleftarrow  \Pi w .(  \tau' _{\ast  y }   \TREF^{\hspace{0.5pt}  { r' }_{ y }  })   \right]    \vdash   e_{{\mathrm{0}}}  :  \tau   \produces   \Gamma' \\
  }{
     \Theta   \mid   \Gamma  \ottsym{[}  x  \ottsym{:}   \Pi w' .(  \tau _{\ast  x }   \TREF^{\hspace{0.5pt}  { r }_{ x }  })   \ottsym{]}  \ottsym{[}  y  \ottsym{:}   \Pi w .(  \tau _{\ast  y }   \TREF^{\hspace{0.5pt}  { r }_{ y }  })   \ottsym{]}  \ottsym{[}  z  \ottsym{:}   \{  \nu  :   \TINT    \mid   \varphi  \}   \ottsym{]}   \vdash     \ALIAS(  x  =  y   \boxplus   z  )   \SEQ  e_{{\mathrm{0}}}   :  \tau   \produces   \Gamma' 
  }

\infrule[T-AliasDeref]{
  \Gamma  \ottsym{,}   w \COL  \{  \nu  :   \TINT    \mid   \nu \,  =  \,  0   \}    \models  r \,  >  \,  \mathbf{0}  \\
  \Gamma  \ottsym{,}   w \COL  \{  \nu  :   \TINT    \mid   \nu \,  =  \,  0   \}    \vdash   \ottsym{(}    \Pi z' .(  \tau _{\ast  x }   \TREF^{\hspace{0.5pt}  { r }_{ x }  })   +   \Pi z .(  \tau _{\ast \ast  y }   \TREF^{\hspace{0.5pt}  { r }_{ \ast  y }  })    \ottsym{)}   \approx   \ottsym{(}    \Pi z' .(  \tau' _{\ast  x }   \TREF^{\hspace{0.5pt}  { r' }_{ x }  })   +   \Pi z .(  \tau' _{\ast \ast  y }   \TREF^{\hspace{0.5pt}  { r' }_{ \ast  y }  })    \ottsym{)}  \\
  \Gamma  \ottsym{,}   w \COL  \{  \nu  :   \TINT    \mid   \nu \, \neq \,  0   \}    \vdash    \Pi z .(  \tau _{\ast \ast  y }   \TREF^{\hspace{0.5pt}  { r }_{ \ast  y }  })    \approx    \Pi z .(  \tau' _{\ast \ast  y }   \TREF^{\hspace{0.5pt}  { r' }_{ \ast  y }  })  \\
   \Theta   \mid     \Gamma  \left[  x \hookleftarrow  \Pi z' .(  \tau' _{\ast  x }   \TREF^{\hspace{0.5pt}  { r' }_{ x }  })   \right]   \left[  y \hookleftarrow  \Pi w .(  \Pi z .(  \tau' _{\ast \ast  y }   \TREF^{\hspace{0.5pt}  { r' }_{ \ast  y }  })   \TREF^{\hspace{0.5pt} r })   \right]    \vdash   e_{{\mathrm{0}}}  :  \tau   \produces   \Gamma' \\
}{
   \Theta   \mid   \Gamma  \ottsym{[}  x  \ottsym{:}   \Pi z' .(  \tau _{\ast  x }   \TREF^{\hspace{0.5pt}  { r }_{ x }  })   \ottsym{]}  \ottsym{[}  y  \ottsym{:}   \Pi w .(  \Pi z .(  \tau _{\ast \ast  y }   \TREF^{\hspace{0.5pt}  { r }_{ \ast  y }  })   \TREF^{\hspace{0.5pt} r })   \ottsym{]}   \vdash     \ALIAS(  x  = \ast  y  )   \SEQ  e_{{\mathrm{0}}}   :  \tau   \produces   \Gamma' 
}
\infrule[T-Assert]{
  \Gamma  \models  \varphi \andalso
   \Theta   \mid   \Gamma   \vdash   e_{{\mathrm{0}}}  :  \tau   \produces   \Gamma' 
}{
   \Theta   \mid   \Gamma   \vdash     \ASSERT( \varphi )   \SEQ  e_{{\mathrm{0}}}   :  \tau   \produces   \Gamma' 
}
\caption{Expression typing rules (2).}
\Description{Rules for allocation, dereference, assignment, pointer
arithmetic, alias checks, and assertions.  Allocation gives full ownership
within the new array's bounds.  Dereference requires positive ownership
and splits the element type; assignment requires full ownership and
updates the element type.  Pointer arithmetic shifts array indices and
splits ownership.  Alias checks redistribute ownership and refinement
information between known aliases.  Assertions require their formula to
follow from the current type environment.}
\label{fig:typing rules 2}
\end{figure*}

Now, we explain the typing rules related to heap-manipulating constructs.

The rules \rn{T-MkIntArray} and \rn{T-MkNestedArray} are for the
creation of new arrays of integers and (nested) arrays, respectively.
In both rules, the first premise about $r$ (in the type of $x$)
means that, $x$ has full ownership $\Gamma  \models  r \,  =  \,  \mathbf{1} $, i.e., write permission,
for elements at indices $0$ through $y-1$, but no ownership $\Gamma  \models  r \,  =  \,  \mathbf{0} $ for all the other elements.
For integer arrays, the element type asserts that all (accessible) elements are initialized to $0$.
Since nested arrays are initialized with $ \NULL $, their element type
$\tau_{{\mathrm{0}}}$ must satisfy the same unconditional emptiness requirement
$\models   \Empty{ \tau_{{\mathrm{0}}} } $ as in \rn{T-Null}.

The rule \rn{T-Deref} is for dereference $ \LET  x  =   \ast  y   \IN  e $.
The first premise requires the ownership at $z = 0$ is greater
than 0, meaning that the address denoted by $y$ (with offset 0)
can be read.
Since $x$ becomes an alias of $ \ast  y $, the type of the zeroth element
must be split---analogously to variable reference---and distributed between $x$
and $y$ after the dereference.
However, since the types of all elements have to be represented by a
single type expression $ \tau _{ y } $, which depends on the array index
$z$, type splitting is more complicated.
The second premise means that the type $ \tau _{ y } $ (under $z = 0$)
for the zeroth element is split into $\tau'$ and $ \tau _{ x } $, the latter of
which is given to $x$.
The third and fourth premises concern the element types $ \tau' _{ y } $ after
dereferencing.
They roughly mean that the zeroth element type ($ \tau' _{ y } $ with
$z = 0$) is equivalent to $\tau'$ and the other element types
($ \tau' _{ y } $ with $z \neq 0$) remain the same as $ \tau _{ y } $.
We use another judgment $\Gamma  \vdash   \tau_{{\mathrm{1}}}   \approx   \tau_{{\mathrm{2}}} $ to state the two types
are \emph{equivalent}, defined by:
$
  \Gamma  \vdash   \tau_{{\mathrm{1}}}   \approx   \tau_{{\mathrm{2}}}  \iff \Gamma  \vdash   \tau_{{\mathrm{1}}}   \leq   \tau_{{\mathrm{2}}}  \text{ and } \Gamma  \vdash   \tau_{{\mathrm{2}}}   \leq   \tau_{{\mathrm{1}}} .
$
The type $ \ottsym{(}  \tau'  \ottsym{)}  ^ {= x } $ in the third premise uses the same
equality-strengthening operation as \rn{T-Var}: it adds the fact that
$ \ast  y $ is equal to $x$ when the element type is integer.

We show a type derivation using \rn{T-Deref}.
Suppose $y$ is given type
\(
   \Pi z .(  \Pi w .(  \TINT   \TREF^{\hspace{0.5pt} r_{{\mathrm{0}}} })   \TREF^{\hspace{0.5pt} r }) ,
\)
where
\[
  r = \ottsym{(}   \ottsym{(}    0  \, \le \, z  \wedge  z \, \le \,  1    \ottsym{)}   \produces    1    \ottsym{)},
  \qquad
  r_{{\mathrm{0}}} = \ottsym{(}   \ottsym{(}     0  \, \le \, z  \wedge  z \, \le \,  1    \wedge  w \,  =  \,  0    \ottsym{)}   \produces    1    \ottsym{)}.
\]
Thus, $y$ has full ownership of two pointer cells, each of which
provides full ownership of one integer cell.
Consider $ \LET  x  =   \ast  y   \IN  e $.
We can give $x$ half the ownership of the integer cell reached
through the zeroth element, retaining the other half through $y$.
The body $e$ is then checked with
\[
\begin{aligned}
  x & :  \Pi w .(  \TINT   \TREF^{\hspace{0.5pt} r_{{\mathrm{1}}} }) , &
  y & :  \Pi z .(  \Pi w .(  \TINT   \TREF^{\hspace{0.5pt} r'_{{\mathrm{0}}} })   \TREF^{\hspace{0.5pt} r }) ,
\end{aligned}
\]
where
\[
\begin{aligned}
  r_{{\mathrm{1}}} & = (w \,  =  \,  0   \produces  \tfrac{1}{2}), &
  r'_{{\mathrm{0}}} & = ( z \,  =  \,  0   \wedge  w \,  =  \,  0    \produces  \tfrac{1}{2},\quad
                 z \,  =  \,  1   \wedge  w \,  =  \,  0    \produces  1).
\end{aligned}
\]
For the derivation below, write
\[
  \tau_{{\mathrm{0}}} =  \Pi w .(  \TINT   \TREF^{\hspace{0.5pt} r_{{\mathrm{0}}} }) ,\qquad
  \tau_{{\mathrm{1}}} =  \Pi w .(  \TINT   \TREF^{\hspace{0.5pt} r_{{\mathrm{1}}} }) ,\qquad
  \tau'_{{\mathrm{0}}} =  \Pi w .(  \TINT   \TREF^{\hspace{0.5pt} r'_{{\mathrm{0}}} }) .
\]
Omitting $\Theta$, we obtain the following instance of \rn{T-Deref}:
\[
\frac{
  \begin{gathered}
    \Gamma  \ottsym{,}   z \COL  \{  \nu  :   \TINT    \mid   \nu \,  =  \,  0   \}    \models  r \,  >  \,  \mathbf{0} 
    \qquad
    \Gamma  \ottsym{,}   z \COL  \{  \nu  :   \TINT    \mid   \nu \,  =  \,  0   \}    \vdash    \tau_{{\mathrm{1}}}  +  \tau_{{\mathrm{1}}}    \approx   \tau_{{\mathrm{0}}} 
    \\[0.4ex]
    \Gamma  \ottsym{,}   x \COL \tau_{{\mathrm{1}}}   \ottsym{,}   z \COL  \{  \nu  :   \TINT    \mid   \nu \,  =  \,  0   \}    \vdash   \tau'_{{\mathrm{0}}}   \approx   \tau_{{\mathrm{1}}} 
    \qquad
    \Gamma  \ottsym{,}   x \COL \tau_{{\mathrm{1}}}   \ottsym{,}   z \COL  \{  \nu  :   \TINT    \mid   \nu \, \neq \,  0   \}    \vdash   \tau'_{{\mathrm{0}}}   \approx   \tau_{{\mathrm{0}}} 
    \\[0.8ex]
     \Gamma  \left[  y \hookleftarrow  \Pi z .( \tau'_{{\mathrm{0}}}  \TREF^{\hspace{0.5pt} r })   \right]   \ottsym{,}   x \COL \tau_{{\mathrm{1}}} 
     \vdash  e : \tau  \produces  (\Gamma',  x \COL \tau'' )
  \end{gathered}
}{
  \Gamma  \ottsym{[}  y  \ottsym{:}   \Pi z .( \tau_{{\mathrm{0}}}  \TREF^{\hspace{0.5pt} r })   \ottsym{]}  \vdash   \LET  x  =   \ast  y   \IN  e  : \tau  \produces  \Gamma'
}
\]
At $z = 0$, the original inner ownership $1$ is split into
$\tfrac{1}{2} + \tfrac{1}{2}$; at every other index, the element type
is unchanged.
The outer ownership $r$ also remains unchanged: reading the pointer
cell splits ownership of its contents, not ownership of the cell itself.
Since the contents are a pointer, the equality-strengthening operation
leaves the retained type unchanged.

The rule \rn{T-Assign} handles updating an array element. The first
premise requires the ownership at $z = 0$ is 1, meaning that the
address denoted by $x$ (with offset 0) is writable. The second,
third, and fourth premises are similar to those in \rn{T-Deref}. Here,
the type of $y$ is split into $\tau'$ and $ \tau' _{ y } $. The former
is used as the zeroth element type and the latter the type of $y$
after the update.

The rule \rn{T-AddPtr} for pointer arithmetic can be considered as a
generalization of typing for $ \LET  x  =  y  \IN  e_{{\mathrm{0}}} $ discussed above
(except that the type of $y$ is a reference type), in the sense
that the type of $y$ before $\LET$ is split into the type for
$x$ and $y$ after $\LET$. The capture-avoiding substitution
$[w-z/w]$ (for logical terms for variables in formulae) means that the
$n$-th element of $y$ corresponds to the $(n-z)$-th element of
$x$.

The rules \rn{T-AliasAddPtr} and \rn{T-AliasDeref} handle $\ALIAS$
expressions. From the viewpoint of ownership, an $\ALIAS$ expression
merges and re-splits the types for the aliases into two new types by using type
addition.   The judgment $\Gamma  \vdash    \tau_{{\mathrm{1}}}  +  \tau_{{\mathrm{2}}}    \approx    \tau_{{\mathrm{3}}}  +  \tau_{{\mathrm{4}}}  $ stands for
$\Gamma  \vdash    \tau_{{\mathrm{1}}}  +  \tau_{{\mathrm{2}}}    \approx   \tau_{{\mathrm{5}}} $ and $\Gamma  \vdash    \tau_{{\mathrm{3}}}  +  \tau_{{\mathrm{4}}}    \approx   \tau_{{\mathrm{5}}} $ for some $\tau_{{\mathrm{5}}}$.
A typical use is to transfer the predicate of one alias to another: for example, we can
derive
\[
 \Theta   \mid   \Gamma_{{\mathrm{1}}}   \vdash     \ALIAS(  x  =  y   \boxplus   z  )   \SEQ  z   :   \TINT    \produces   \Gamma_{{\mathrm{2}}} 
\]
where
\begin{gather*}
  \Gamma_{{\mathrm{1}}} =  x \COL  \Pi w .(  \{  \nu  :   \TINT    \mid   \nu \,  =  \, w  \}   \TREF^{\hspace{0.5pt} r_{{\mathrm{1}}} })    \ottsym{,}   y \COL  \Pi w .(  \TINT   \TREF^{\hspace{0.5pt}  \mathbf{0}  })    \ottsym{,}   z \COL  \{  \nu  :   \TINT    \mid   z \,  =  \,  1   \}  , \\
  \Gamma_{{\mathrm{2}}} =  x \COL  \Pi w .(  \{  \nu  :   \TINT    \mid   \nu \,  =  \, w  \}   \TREF^{\hspace{0.5pt} r_{{\mathrm{2}}} })    \ottsym{,}   y \COL  \Pi w .(  \{  \nu  :   \TINT    \mid   \nu \,  =  \, w  \ottsym{+}   1   \}   \TREF^{\hspace{0.5pt} r_{{\mathrm{3}}} })    \ottsym{,}   z \COL  \TINT  , \\
  r_{{\mathrm{1}}} = 0 \leq w \leq 9  \produces  1, r_{{\mathrm{2}}} = 0 \leq w \leq 9  \produces  0.5\, \text{ and } r_{{\mathrm{3}}} = -1 \leq w \leq 8  \produces  0.5\ .
\end{gather*}
Here, the refinement for $y$ in the post-environment $\Gamma_{{\mathrm{2}}}$ is
\emph{strengthened} by telling the type system that $y  \boxplus  1$ is an alias of $x$.

Finally, the rule \rn{T-Assert} requires the predicate $\varphi$ to be valid under $\Gamma$.

Function and program typing use three judgments.
The judgment $ \Theta   \vdash   f  \mapsto   (  x_{{\mathrm{1}}}  , \ldots ,  x_{\ottmv{n}}  )  \, e $ means that
the function $f$ is well typed under $\Theta  \ottsym{(}  f  \ottsym{)}$.
The judgment $ \Theta   \vdash   D $ is its pointwise extension.
Finally, $ \vdash    \tuple{ D ,  e }  $ states that the program is well typed:
for some well-formed $\Theta$, $ \Theta   \vdash   D $ holds and the main
expression $e$ is typable under $\Theta$ from the empty type
environment $ \bullet $.
The corresponding rules are the same as those in previous
work~\cite{DBLP:conf/pepm/TanakaSK24, DBLP:conf/esop/TomanSSI020} and are given in
\ifdefined\VersionLong{}%
  \cref{sec:functionTyping}.
\else%
  the extended version.
\fi

\begin{figure*}
\begin{lstlisting}[backgroundcolor=,label = code:typeexam,escapechar=\!]
!\color{comment-green}// pp:$ \Pi x_2 (. \Pi x _{ 1 }  . (\{\nu \COL \TINT \mid  \top  \} \TREF ^ { \,  \mathbf{0}  } )   ) \TREF^{ \, 0 \le x_2 \le \mathtt{n}-1  \produces  1 }$ ! !\label{typingExample:line1}!
let s = alloc n : int ref in!\label{typingExample:line2}!
!\color{comment-green}// s:$ (\Pi x_{1}. \{\nu \COL \TINT \mid 0 \le x_{1} \le  \mathtt{n} - 1 \implies \nu = 0 \} ) \TREF^{ \, 0 \le x_1 \le \mathtt{n}-1  \produces  1 } $!!\label{typingExample:line3}!
let d = initArray(n, n, s) in!\label{typingExample:line4}!
!\color{comment-green}// s:$ (\Pi x_{1}. \{\nu \COL \TINT \mid 0 \le x_{1} \le  \mathtt{n} - 1 \implies \nu = \mathtt{n} \} ) \TREF^{ \, 0 \le x_1 \le \mathtt{n}-1  \produces  1 } $!!\label{typingExample:line5}!
pp := s;!\label{typingExample:line6}!
!\color{comment-green}// pp:$ (\Pi  x_2. (\Pi x_{1}. \{\nu \COL \TINT \mid x_2 = 0  \wedge  0 \le x_{1} \le  \mathtt{n} - 1 \implies \nu = \mathtt{n} \} ) \TREF^{ \, r_{{\mathrm{1}}} }   ) \TREF^{ \, r_{{\mathrm{2}}} } $!!\label{typingExample:line7}!
!\color{comment-green}// $\quad\text{where}\quad r_{{\mathrm{1}}}  = ( x_2 = 0  \wedge  0 \le x_{ 1 } \le \mathtt{n} - 1  \produces  1 ), r_{{\mathrm{2}}}  = ( 0 \le x_2 \le \mathtt{n} - 1  \produces  1 )$!!\label{typingExample:line8}!
!\color{comment-green}// s:$ (\Pi x_{1}. \{\nu \COL \TINT \mid  \top  \} ) \TREF^{ \,  \mathbf{0}  } $!!\label{typingExample:line9}!
let t = pp $ \boxplus $ 1 in!\label{typingExample:line10}!
!\color{comment-green}// pp:$ (\Pi  x_2. (\Pi x_1. \{ \nu \COL \TINT \mid x_2 = 0  \wedge  0 \le x_{1} \le  \mathtt{n} - 1  \implies  \nu  = \mathtt{n} \}   ) \TREF^{ \, r_{{\mathrm{1}}} }   ) \TREF^{ \, r_{{\mathrm{2}}} } $!!\label{typingExample:line11}!
!\color{comment-green}// $\quad\text{where}\quad r_{{\mathrm{1}}}  = ( x_2 = 0  \wedge  0 \le x_1 \le \mathtt{n} - 1  \produces  1 ), r_{{\mathrm{2}}}  = ( x_2 = 0  \produces  1 )$!!\label{typingExample:line12}!
!\color{comment-green}// t:$ (\Pi  x_2. (\Pi x_1. \{ \nu \COL \TINT \mid  \top  \}   ) \TREF^{ \,  \mathbf{0}  }   ) \TREF^{ \, 0 \le x_2 \le \mathtt{n}-2  \produces  1} $!!\label{typingExample:line13}!
let d2 = initMatrix(n-1, t) in !\label{typingExample:line14}!
!\color{comment-green}// t:$ (\Pi  x_2. (\Pi x_1. \{ \nu \COL \TINT \mid 0 \le x_2 \le \mathtt{n} - 2  \wedge  0 \le x_{ 1 } \le \mathtt{n} - x_2 - 2 \implies \nu = \mathtt{n} - 1 - x_2 \}   ) \TREF^{ \, r_{{\mathrm{1}}} }   ) \TREF^{ \, r_{{\mathrm{2}}} } $!!\label{typingExample:line15}!
!\color{comment-green}// $\quad\text{where}\quad r_{{\mathrm{1}}}  = ( 0 \le x_2 \le \mathtt{n} - 2  \wedge  0 \le x_{ 1 } \le \mathtt{n} - x_2 - 2  \produces  1 ), r_{{\mathrm{2}}}  = ( 0 \le x_2 \le \mathtt{n} - 2  \produces  1 )$!!\label{typingExample:line16}!
alias(t = pp + 1); 0!\label{typingExample:line17}!
!\color{comment-green}// pp:$ (\Pi x_2 . ( \Pi x_1 . \{ \nu \COL \TINT \mid 0 \le x_2 \le \mathtt{n} - 1  \wedge  0 \le x_{ 1 } \le \mathtt{n} - x_2 - 1 \implies \nu = \mathtt{n} - x_2 \} ) \TREF^{ \, r_{{\mathrm{1}}} } ) \TREF^{ \, r_{{\mathrm{2}}} } $!!\label{typingExample:line18}!
!\color{comment-green}// $\quad\text{where}\quad r_{{\mathrm{1}}}  = ( 0 \le x_2 \le \mathtt{n} - 1  \wedge  0 \le x_1 \le \mathtt{n} - x_2 - 1  \produces  1 ), r_{{\mathrm{2}}}  = ( 0 \le x_2 \le \mathtt{n} - 1  \produces  1 )$!!\label{typingExample:line19}!
!\color{comment-green}// t:$ (\Pi  x_2. (\Pi x_1. \{ \nu \COL \TINT \mid  \top  \}   ) \TREF^{ \,  \mathbf{0}  }   ) \TREF^{ \,  \mathbf{0} } $!!\label{typingExample:line20}!
\end{lstlisting}
\caption{Typing example.}
\Description{The recursive branch of initMatrix is annotated with pointer
types after each operation.  A newly allocated integer array is initialized
and stored in the first outer cell.  Pointer arithmetic separates this cell
from the remaining outer array, which is initialized by a recursive call.
An alias check then recombines ownership through the original matrix
pointer.  The final type records decreasing row lengths and element values
as functions of the outer index.}
\label{fig:typingexam}
\end{figure*}

\Cref{fig:typingexam} describes how the $\ELSE$ branch of \texttt{initMatrix} in \cref{fig:motivatingExample} is typed to express the intuition explained in \cref{sec:intro}.
\begin{itemize}
\item The pre-environment of the entire expression expresses that
the pointer \texttt{pp} has full ownership to access the length-$n$ outer array,
whereas it has no ownership for the inner array (\cref{typingExample:line1}).
\item The type environment at \cref{typingExample:line3} expresses that, following \rn{T-MkIntArray},
the pointer \texttt{s} has full ownership
to access all the memory cells of the newly created length-$n$ array; the refinement predicate of the type of \texttt{s}
expresses that each memory cell is initialized by $0$.
\item Line \ref{typingExample:line4} initializes \texttt{s} using the function
\texttt{initArray};
\begin{revision}
we assume that \texttt{initArray} is typed
$
\langle \mathtt{l} \COL \TINT , \mathtt{n} \COL \TINT , \mathtt{p} \COL \Pi x _{ 1 }  . (\TINT \TREF ^ { \, r_1 } )\rangle   \rightarrow 
\langle \mathtt{l} \COL \TINT , \mathtt{n} \COL \TINT , \mathtt{p} \COL \Pi x_1 . (\{ \nu \COL \TINT \mid \varphi \}  \TREF^{\, r_2})  \mid  \TINT \rangle
$
elsewhere, where
$r_1 = r_2 =  0 \le x_1 \le \mathtt{l}-1  \produces  1$
and
$\varphi = 0 \le x_1 \le \mathtt{l}-1 \implies \nu = \mathtt{n}$,
reflecting the intuition explained in \cref{sec:intro}.
\end{revision}
The call uses fresh integer copies $y$ and $z$ of $\mathtt n$,
following the convention in \cref{sec:syntax}.
Before the call, \rn{T-Sub} can retain the original binding with refinement
$ \{  \nu  :   \TINT    \mid     \nu \,  >  \,  0   \wedge  y \,  =  \,  n    \wedge  z \,  =  \,  n    \} $, while weakening the actual
copies to $ \TINT $ as required by the contract.
The original binding is untouched by \rn{T-Call}, so its positivity and
the copy equalities remain available to express the returned array type
using $\mathtt n$ and justify the following assignment.
The argument $\mathtt n-1$ at \cref{typingExample:line14} is treated
similarly: after naming it, its equality with $\mathtt n-1$ is retained
in an untouched integer binding before the call.
The type environment at \cref{typingExample:line5} expresses that the ownership term in the type of \texttt{s}
is unchanged by this function call and the refinement predicate expresses that each memory cell
is initialized to $n$ as required.
\item The type environment at
Lines~\ref{typingExample:line7}--\ref{typingExample:line9} shows how
 \rn{T-Assign} can be used to distribute the ownership retained by \texttt{s} between \texttt{pp} and \texttt{s}.
In this example, all the ownership held by \texttt{s} is transferred to \texttt{pp},
resulting in the ownership term $r_1$ at \cref{typingExample:line8}
expressing that \texttt{pp} has the full ownership to access the inner array
pointed to by the zeroth element of the outer array.
\item At \cref{typingExample:line10} where pointer arithmetic is conducted and
\rn{T-AddPtr} is applied, the ownership $0 \le x_2 \le \mathtt{n}-1 \produces 1$
held by \texttt{pp} for the outer array is split to
$x_2 = 0 \produces 1$ and $1 \le x_2 \le \texttt{n}-1 \produces 1$.
The former is kept by \texttt{pp}.
The latter is transferred to \texttt{t} bound to
$\mathtt{pp} \boxplus 1$ after it is shifted by the offset $1$ of this pointer arithmetic,
resulting in $0 \le x_2 \le \mathtt{n}-2 \produces 1$, and hence the type of \texttt{t} at \cref{typingExample:line13}.
\item Then, \texttt{initMatrix} is called recursively with the pointer to a matrix \texttt{t} at \cref{typingExample:line14},
changing the type of \texttt{t} so that it expresses that
\texttt{t} retains full ownership for the outer array of length $\mathtt{n}-1$ and
each inner array at index $x_2$ of length $\mathtt{n} - 1 - x_2$.
Furthermore, the refinement predicate of the type of $t$ at
\cref{typingExample:line15} expresses that the $i$-th inner integer array
is initialized to the value $\mathtt{n}-i-1$ as required.
\item Finally, the $\ALIAS$ expression at \cref{typingExample:line17}
transfers back the ownership held by \texttt{t} to \texttt{pp},
resulting in the type of \texttt{pp} at \cref{typingExample:line18}.
\end{itemize}
\begin{revision}
Combined with the types of the other part of \texttt{initMatrix}, this function is typed as
\begin{gather*}
\langle \mathtt{n} \COL \TINT , \mathtt{pp} \COL \Pi x_2 . (\Pi x _{ 1 }  . (\TINT \TREF ^ { \,  \mathbf{0}  } )    \TREF^{ \, r_1 })  \rangle   \rightarrow 
\langle \mathtt{n} \COL \TINT , \mathtt{pp} \COL \Pi x_2 . ( \Pi x_1 . (\{ \nu \COL \TINT \mid \varphi \}  \TREF^{\, r_2} ) \TREF^{ \, r_3 })  \mid  \TINT \rangle
\end{gather*}
where
$r_1 = r_3 = (0 \le x_2 \le \mathtt{n}-1  \produces  1, \mathbf{0})$
and
$r_2 = (0 \le x_2 \le \mathtt{n} - 1 \land 0 \le x_1 \le  \mathtt{n} - x_2   - 1  \produces  1, \mathbf{0})$
and
$\varphi = (0 \le x_2 \le \mathtt{n} - 1 \land 0 \le x_{ 1 } \le \mathtt{n} - x_2 - 1 \implies \nu = \mathtt{n} - x_2)$.
\end{revision}

\subsection{Soundness}
\label{sec:soundness}

We state a type soundness theorem below.
We write $C \not  \longrightarrow _{D}$ if no $C'$ satisfies
$C  \longrightarrow _{D} C'$.
\begin{restatable}[Soundness]{thm}{soundness}
  \label{thm:soundness}
  If
  $ \vdash    \tuple{ D ,  e }  $ and $ \tuple{  \emptyset  ,   \emptyset  ,  e }   \longrightarrow^* _{D} C \not  \longrightarrow _{D}$, then %
  either (1) $C =  \tuple{ \ottnt{H} ,  \ottnt{R} ,  x } $ for some $R$, $H$, and $x$, or (2) $C =  \mathbf{AliasFail} $.
\rocqref{Metatheory/Soundness.v}{soundness}
\end{restatable}
The theorem excludes assertion failures, null-pointer dereferences, and
out-of-bounds accesses.  Incorrect $\ALIAS$ expressions are detected as
$ \mathbf{AliasFail} $ rather than invalidating these guarantees.

From a practical standpoint, however, $\ALIAS$ expressions do affect
the \emph{verifiability} of a program: an incorrect or missing $\ALIAS$ expression
may prevent the type checker from redistributing ownership appropriately,
causing type inference to fail even for an otherwise correct program.
As reported in \cref{sec:experiments} (\cref{table:experiments}),
the majority of $\ALIAS$ expressions required by our benchmarks are
inserted automatically by our implementation; manual expressions are needed
only in a small number of cases involving non-trivial pointer aliasing patterns.
Notice that this design uses $\ALIAS$ construct as an interface between the type system and
any external alias analysis: \Cref{thm:soundness} does not depend on the particular
alias analysis used to insert or discharge $\ALIAS$ expressions.
Our current implementation uses a simple syntactic automated $\ALIAS$ insertion, but it can be replaced by
more sophisticated alternatives without affecting the metatheory.

The proof connects expression typing to runtime states through satisfaction
of refinements and consistency of ownership.  We follow the argument in the
full version of~\citet[Appendix~A]{DBLP:conf/esop/TomanSSI020}, adapting its runtime
invariants to indexed ownership and null pointers.
Write $\sem{r}_R$ for the value of the first branch of $r$ whose
condition holds under $R$, or zero if no condition holds.

\begin{definition}[Satisfaction and ownership]
Value satisfaction $\SATV$ and environment satisfaction $\SAT$ are defined by
\label{def:satisfaction-ownership}
\begin{align*}
  \mathbf{SATv} ( \ottnt{H} ,  \ottnt{R} ,   n  ,   \{  \nu  :   \TINT    \mid   \varphi  \}  )  &\iff \models  \ottsym{[}  \ottnt{R}  \ottsym{]} \, \ottsym{[}   n   \ottsym{/}  \nu  \ottsym{]}  \varphi,\\
  \mathbf{SATv} ( \ottnt{H} ,  \ottnt{R} ,  \ottkw{null} ,   \Pi z .( \tau  \TREF^{\hspace{0.5pt} r })  )  &\iff
   \forall j.\ \sem{  [  j  /  z  ]    r }_R=0,\\
  \mathbf{SATv} ( \ottnt{H} ,  \ottnt{R} ,  \ottsym{(}  a  \ottsym{,}   k   \ottsym{)} ,   \Pi z .( \tau  \TREF^{\hspace{0.5pt} r })  )  &\iff
   \forall j.\ \sem{  [  j  /  z  ]    r }_R>0 \implies\\
 &\qquad\qquad (a,k+j)\in\DOM(H)\ \land\
    \mathbf{SATv} ( \ottnt{H} ,  \ottnt{R} ,  \ottnt{H}  \ottsym{(}  \ottsym{(}  a  \ottsym{,}   k   \ottsym{+}   j   \ottsym{)}  \ottsym{)} ,   [  j  /  z  ]  \, \tau ) ,\\
  \mathbf{SAT}( \ottnt{H} ,  \ottnt{R} ,  \Gamma )  &\iff \forall x\in\DOM(\Gamma).\ x\in\DOM(R)\ \land\
   \SATV(H,R,R(x),\Gamma(x)).
\end{align*}
Other value/type combinations do not satisfy $\SATV$.
Ownership maps are total functions from addresses to rationals; $\emptyset$
is the zero map, and map operations are pointwise.  Define
\begin{align*}
  \mathbf{own} ( \ottnt{H} ,  \ottnt{R} ,  \ottsym{(}  a  \ottsym{,}   k   \ottsym{)} ,   \Pi z .( \tau  \TREF^{\hspace{0.5pt} r })  ) (a',k')
 &\defeq f(a',k')+
 \begin{cases}
   \sem{  [   k '   \ottsym{-}  k  /  z  ]    r }_R & a=a',\\
   0 & \text{otherwise},
 \end{cases}\\
   & \qquad \text{where} \\
   & \qquad f \defeq
 \sum_{\substack{j\in\mathbb Z \\ (a,k+j)\in\DOM(H)\\
                 \sem{  [  j  /  z  ]    r }_R>0}}
  \mathbf{own} ( \ottnt{H} ,  \ottnt{R} ,  \ottnt{H}  \ottsym{(}  \ottsym{(}  a  \ottsym{,}   k   \ottsym{+}   j   \ottsym{)}  \ottsym{)} ,   [  j  /  z  ]  \, \tau ) ,\\
  \mathbf{own} ( \ottnt{H} ,  \ottnt{R} ,   \ottnt{v}  ,  \tau )  &\defeq \emptyset \quad\text{otherwise},\\
  \mathbf{Own}( \ottnt{H} ,  \ottnt{R} ,  \Gamma )  &\defeq
   \sum_{x\in\DOM(\Gamma)}\OWN(H,R,R(x),\Gamma(x)).
\end{align*}
\end{definition}

$\SATV$ checks integer refinements and follows array contents only at
positively owned offsets.  The index $j$ is relative to $(a,k)$, and the
allocation premise makes $H(a,k+j)$ defined.  Null requires zero outer
ownership.  The map $\OWN$ adds direct array ownership to ownership through
nested contents; its recursive sum is finite because it ranges over
allocated cells.  $\OWNALL$ sums these contributions over all typed variables.

\begin{definition}[Configuration typing]
\label{def:configuration-typing}
Write $ \mathbf{ConOwn}( \ottnt{H} ,  \ottnt{R} ,  \Gamma ) $ if
$\OWNALL(H,R,\Gamma)(a,k)\le1$ for every $(a,k)\in\DOM(H)$.
We write $\vdash_D  \tuple{ \ottnt{H} ,  \ottnt{R} ,  e } \COL\tau\Rightarrow\Gamma'$ if there exist
$\Theta$ and $\Gamma$ such that
\[
  \Theta   \vdash   D ,\qquad  \Theta   \mid   \Gamma   \vdash   e  :  \tau   \produces   \Gamma' ,\qquad
  \mathbf{SAT}( \ottnt{H} ,  \ottnt{R} ,  \Gamma ) ,\qquad  \mathbf{ConOwn}( \ottnt{H} ,  \ottnt{R} ,  \Gamma ) .
\]
\end{definition}

Satisfaction ensures that the current values match their types.  The
ownership bound ensures that full ownership of a cell excludes positive
ownership through any other alias, which justifies strong updates.
The following lemmas establish and maintain configuration typing.

\begin{restatable}[Initialization]{lemma}{initializationlemma}
  \label{lem:init}
  If $ \vdash    \tuple{ D ,  e }  $, then there exist $\tau$ and $\Gamma'$ such that
  $\vdash_D  \tuple{  \emptyset  ,   \emptyset  ,  e }  \COL \tau \Rightarrow \Gamma'$.
\rocqref{Metatheory/Soundness.v}{initial_configuration_typed}
\end{restatable}

\begin{restatable}[Preservation]{lemma}{preservationlemma}
  \label{lem:preservation}
  If $\vdash_D  \tuple{ \ottnt{H} ,  \ottnt{R} ,  e }  \COL \tau \Rightarrow \Gamma'$ and
  $  \tuple{ \ottnt{H} ,  \ottnt{R} ,  e }     \longrightarrow _{  D  }     \tuple{ \ottnt{H'} ,  \ottnt{R'} ,  e' }  $, then
  $\vdash_D  \tuple{ \ottnt{H'} ,  \ottnt{R'} ,  e' }  \COL \tau \Rightarrow \Gamma'$.
\rocqref{Metatheory/Preservation.v}{configuration_preservation}
\end{restatable}

\begin{restatable}[Progress]{lemma}{progresslemma}
  \label{lem:progress}
  If $\vdash_D  \tuple{ \ottnt{H} ,  \ottnt{R} ,  e }  \COL \tau \Rightarrow \Gamma'$, then either
  (1) there exists a configuration $C'$ such that
  $ \tuple{ \ottnt{H} ,  \ottnt{R} ,  e }   \longrightarrow _{D} C'$, or
  (2) $e = x$ for some variable $x$.
\rocqref{Metatheory/Progress.v}{typed_configuration_progress}
\end{restatable}

Initialization and repeated Preservation type every reachable running
configuration; Progress then excludes any irreducible outcome other than a
returned variable or $ \mathbf{AliasFail} $.
\ifdefined\VersionLong
The proofs are given in \cref{sec:proofs}.
\else
The proofs are given in the extended version.
\fi

\section{Type Inference}
\label{sec:typeInference}
We describe the type inference procedure used in the experiments reported in \cref{sec:experiments}.
The type inference procedure consists of the following three steps:
\begin{itemize}
\item Step 1: Simple type inference inferring the simple type of each variable.
\item Step 2: Ownership inference inferring ownership expressions for each reference type.
\item Step 3: Refinement inference inferring refinement predicates for each refinement type.
\end{itemize}
Step 1 is the standard unification-based type inference.

\begin{revision}
The main strategy of Steps 2 and 3 is to generate constraints based on
the syntax-directed typing rules derived from the rules introduced in
\cref{sec:typesystem}. To make type inference tractable,
we use a template-based approach for ownership inference (Step 2).
Using the obtained ownership expressions, we then generate CHC constraints
for refinement inference (Step 3).
\end{revision}
We explain Steps 2 and 3 in detail in the following subsections.

\subsection{Step 2: Ownership Inference}
\label{sec:OwnershipInference}

\subsubsection*{Step 2.1: Generating Ownership Constraints}

Step 2 of our type inference procedure is based on a template-based approach: We
prepare a template for each ownership expression associated with a reference
type, generate a set of constraints based on the typing rules, and solve the
constraints to obtain a concrete ownership expression as a solution.
Concretely, we prepare a template type environment at each location in a
simply-typed program, in which each reference type is accompanied by a template
ownership expression.

In the present implementation, a simple type $\tau^- \TREF$ is
promoted to $\Pi x. (\tau \TREF^{r})$, where $\tau$ is obtained by
promoting $\tau^-$ recursively and choosing the template for $r$ from one of
(1) $x \in [l, h]  \produces  o$, 
(2) $x = 0  \produces  o_1, x \in [1,h]  \produces  o_2$ and
(3) $y=0 \land x \in [l_1, h_1]  \produces  o_1, y \in [1, h_2] \land x \in [l_3, h_3]  \produces  o_2$.
Here, $o$, $o_1$, and $o_2$ are variables representing rational
numbers; $l$ and $h$ are linear combinations of the form
$c_0 + c_1 x_1 + \cdots + c_{k-1} x_{k-1} + c_{k} x$, where $c_i$ are
variables representing integers; $x,x_1,\dots,x_{k-1}$ are
variables of integer type available at this program point.
We explain the variable $y$ in Template (3) later.

Template (2) expresses that a pointer to an array has ownership $o_1$ for the
head element, ownership $o_2$ for the other elements in an interval of indices,
and $0$ for all the other elements.
Template~(2) is useful for expressing an access pattern in which
the head element is treated differently from the other elements.
This pattern frequently occurs in programs that iterate over
an array via a pointer.
We found that using both Templates~(1) and~(2) improves
the performance of the type inference procedure.

For nested reference types (e.g.,
$\Pi x_2. (\Pi x_1. \TINT \TREF^{r_1}) \TREF^{r_2}$), where the outermost
 ownership expression (e.g., $r_2$) is represented using Template (2), the type inference procedure uses Template (3) for the inner ownership expression (e.g., $r_1$), with $y$ being the index variable of the outermost array type (e.g., $x_2$).
 This template enables another access pattern of a program that iterates over a
 nested array where its zeroth row is processed differently from the other rows.

 Then, the type inference procedure generates constraints based on the typing
 rules in a forward-reasoning style: Given an expression $e$ and a
 pre-environment $\Gamma$, the procedure generates a set of constraints $C$ and a
 post-environment $\Gamma'$, potentially involving recursive calls to the procedure if $e$ is a compound expression.
At the beginning of a program or function and for a recursive call, the type inference procedure prepares a pre-environment; the templates for the ownership terms in this pre-environment are chosen from Templates (1--3) based on the following heuristics, which are designed so that Templates (2) and (3) are used only where the zeroth element of an array needs to be treated specially.
\begin{itemize}
  \item Template (1) is used at the beginning of a program or function or on
        array creation, where we do not need special treatment of the head
        element of an array template.
        For a read $ \LET  y  =   \ast  x   \IN  e $ or a write
        $  x  \WRITE  y   \SEQ  e $ involving a reference-valued $y$,
        the type of $x$ in the pre-environment of $e$
        uses Template~(2) for its outermost ownership expression
        and Template~(3) for its inner ownership expressions.
        This distinguishes ownership of the reference stored in the
        head element from ownership of references stored in other elements.
        The type of $y$ in this environment uses Template~(1)
        at every reference level.
        Reads and writes of integer values do not introduce this
        head/tail distinction.

  \item For a pointer-arithmetic expression ($ \LET  x  =   y   \boxplus   z   \IN  e $),
        the type of $x$ in the pre-environment of $e$
        uses Template~(1).
        The type of $y$ retains its template form, except in the
        special case of partitioning by one described in
        \cref{sec:implementation}, where ownership expressions
        using Templates~(2) and~(3) are converted to Template~(1).

    \item For $  \ALIAS(  x  = \ast  y  )   \SEQ  e $, the type of $y$ in the
        pre-environment of the alias expression is assumed to use
        Template~(2) for its outermost ownership expression and
        Template~(3) for its inner ownership expressions, if any.
    \item For $  \ALIAS(  x  =  y   \boxplus   z  )   \SEQ  e $, the ownership expressions
        in the type of $x$ in the pre-environment of the alias expression
        are assumed to use Template~(1) at every reference level.

  \item At the end of a function body and at the end of each branch of an $\IF$
        expression, ownership expressions in reference types are restricted
        to Template~(1) at every reference level.
        The current implementation assumes that these types already have
        this form.

\end{itemize}

Although the above heuristics are not complete in that there exist well-typed
programs that cannot be typed under these restrictions, we will demonstrate in
\cref{sec:experiments} that these heuristics successfully handle the benchmarks
 used in our experiments.

 We remark that the implementation by Tanaka et
 al.~\cite{DBLP:conf/pepm/TanakaSK24} uses only Template (1).
As demonstrated in experimental results in \cref{sec:experiments}, we need to use Templates (2) and (3) to handle frequent access patterns in matrix-manipulating programs.

The generated constraints are solved using an SMT solver to determine
the unknown coefficients and ownership values in the templates.
These values are substituted into the templates to obtain concrete ownership terms
associated with the types at each location.

\newcommand\PP{\mathit{pp}}

\begin{figure*}
\begin{lstlisting}[backgroundcolor=,escapechar=\!]
!\color{comment-green}// pp:$\Pi x_2 .(\Pi x_1. (\TINT \TREF ^{x_1 \in [l^{\PP}_{0,0}, h^{\PP}_{0,0}]  \produces  o^{\PP}_{0,0}}) \TREF ^{x_2 \in [l^{\PP}_{0,1}, h^{\PP}_{0,1}]  \produces  o^{\PP}_{0,1} })$!
!\color{comment-green}// s:$\Pi x_1 . (\TINT \TREF ^{x_1 \in [l^{s}_{0,0}, h^{s}_{0,0}]  \produces  o^{s}_{0,0}})$!  
pp := s; 
!\color{comment-green}// pp:$\Pi x_2. (\Pi x_1. (\TINT \TREF^{r_{1,1}}) \TREF^{r_{1,2}}) $!  
!\color{comment-green}// $\quad\text{where}\quad r_{1,1} = {(x_2 = 0)  \wedge  x_1 \in [l^{\PP}_{1,0}, h^{\PP}_{1,0}]  \produces  o^{\PP}_{1,0}}, (1 \le x_2 \le h^{\PP}_{1,3} )  \wedge  x_1 \in [l^{\PP}_{1,1}, h^{\PP}_{1,1}]  \produces  o^{\PP}_{1,1}$!
!\color{comment-green}// $\quad\phantom{\text{where}\quad}r_{1,2} = {(x_2 = 0)  \produces  o^{\PP}_{1,2} , x_2 \in [1,h^{\PP}_{1,3}]  \produces  o^{\PP}_{1,3} } $!
!\color{comment-green}// s:$ \Pi x_1 . (\TINT \TREF ^{x_1 \in [l^{s}_{1,0}, h^{s}_{1,0}]  \produces  o^{s}_{1,0}}) $!  
let t = pp $ \boxplus $ 1 in 
!\color{comment-green}// pp:$ \Pi x_{{\mathrm{2}}} .(  \Pi x_{{\mathrm{1}}} .(  \TINT   \TREF^{\hspace{0.5pt}  r _{ 2 , 1 }  })   \TREF^{\hspace{0.5pt}  r _{ 2 , 2 }  })  $!  
!\color{comment-green}// $\quad\text{where}\quad r_{2,1} = {(x_2 = 0)  \wedge  x_1 \in [l^{\PP}_{2,0},h^{\PP}_{2,0}]  \produces  o^{\PP}_{2,0}} ,(1 \le x_2 \le h^{\PP}_{2,3} )  \wedge  x_1 \in [l^{\PP}_{2,1},h^{\PP}_{2,1}]  \produces  o^{\PP}_{2,1}$!
!\color{comment-green}// $\quad\phantom{\text{where}\quad}r_{2,2} = {(x_2 = 0)  \produces  o^{\PP}_{2,2} , x_2 \in [1,h^{\PP}_{2,3}]  \produces  o^{\PP}_{2,3} } $!
!\color{comment-green}// t:$ \Pi x_{{\mathrm{2}}} .(  \Pi x_{{\mathrm{1}}} .(  \TINT   \TREF^{\hspace{0.5pt}  r' _{ 2 , 1 }  })   \TREF^{\hspace{0.5pt}  r' _{ 2 , 2 }  })  $!  
!\color{comment-green}// $\quad\text{where}\quad r'_{2,1} = {x_1 \in [l^{t}_{2,0},h^{t}_{2,0}]  \produces  o^{t}_{2,0}}, r'_{2,2} = {x_2 \in [l^{t}_{2,1},h^{t}_{2,1}]  \produces  o^{t}_{2,1} } $!
\end{lstlisting}
\caption{Code snippet with ownership expression templates.}
\Description{The assignment of an array pointer s through pp, followed by
pointer arithmetic producing t, is annotated with ownership templates.
Unknown lower and upper bounds and fractional permissions are introduced
for each program point and reference level.  After assignment, the
templates distinguish outer index zero from the remaining indices, so
that ownership of the stored array can differ from that of other rows.}
\label{fig:templateTypeEnv}
\end{figure*}

For example, consider the following code snippet, where the simple
types for $\PP$ and $s$ are
$(\TINT \TREF) \TREF$ and $\TINT \TREF$, respectively.
\[
\begin{array}{c}
  \PP := s; \LET t = \PP  \boxplus  1 \IN \dots
\end{array}
\]
\Cref{fig:templateTypeEnv} illustrates template construction
using the general head/tail representation, with templates
shown in green.
For $\LET t = \PP  \boxplus  1 \IN \dots$, this presentation
does not show the implementation's special-case conversion
to Template~(1), described in \cref{sec:implementation}.
Under the implementation's partitioning policy, only the
head element remains owned through $\PP$.

Because an assignment to the array occurs, the template of $pp$ after assignment treats the head element separately.
Then, the following constraints are generated from the first statement of the
snippet and \rn{T-Assign}:
\begin{align*}
  &o^{pp}_{0,1} = 1  \wedge  \forall n>0.  l^{pp}_{0,1} \le 0 \le h^{pp}_{0,1} \\
  &\forall n>0. 
   (l^{pp}_{0,1} \le 0  \wedge  \text{max} \{  h^{pp}_{1,3}, 0 \}  \le h^{pp}_{0,1}  
    \wedge  o^{pp}_{1,2} \ge o^{pp}_{0,1}  \wedge  o^{pp}_{1,3} \ge o^{pp}_{0,1} )\\
  &\forall n>0. 
   (\forall x_2. x_2 = 0  \produces  (l^{s}_{0,0} \le l^{pp}_{1,0}   \wedge  h^{pp}_{1,0} \le h^{s}_{0,0} 
    \wedge  o^{pp}_{1,0} \le o^{s}_{0,0}  \wedge  o^{s}_{1,0} = 0) )\\
  &\forall n>0. (\forall x_2. ((1 \le x_2 \le h^{pp}_{1,3} )  \produces  (l^{pp}_{0,0} \le l^{pp}_{1,1} 
   \wedge  h^{pp}_{1,1} \le h^{pp}_{0,0}  \wedge  o^{pp}_{1,1} \le o^{pp}_{0,0}))).
\end{align*}
We remark that, for each nested reference type, we generate \textit{Empty}
 constraints that we mentioned in \cref{sec:syntax of types}.
These constraints must hold for all values of program variables and array
indices, without assuming the surrounding environment's refinements.
For example, for each occurrence of $\Pi x_2. (\Pi x_1. \TINT \TREF^{x_1 \in [l_1, h_1]  \produces  o_1}) \\ \TREF^{x_2 \in [l_2, h_2]  \produces  o_2}$, the following constraints are generated: $(o_2 = 0  \produces  o_1 = 0) \land (x_2 < l_2  \produces  o_1 = 0 \lor h_1 < l_1) \land (x_2 > h_2  \produces  o_1 = 0 \lor h_1 < l_1)$.

\subsubsection*{Step 2.2: Solving Ownership Constraints}

By default, the ownership inference procedure solves the
constraints collected from the entire program to determine
the unknown coefficients and ownership values.
When ownership annotations fully specify function interfaces,
ownership constraints can instead be solved separately for
each function, as described in \cref{sec:implementation}.
The generated constraints are in the form of
$\exists \vec{c}, \vec{d}, \vec{o} . \forall \vec{x} . \varphi $, where
 $\vec{c}, \vec{d}, \vec{o}$ are unknowns representing the coefficients and ownership values in the templates and $\vec{x}$ are program variables and array indices.
The same form of constraints is also generated by the type-inference procedure by Tanaka et al. 
To solve these $\exists\forall$ constraints using an SMT solver, Tanaka et al.\ use the following ``guess-and-check'' method.
Specifically, given a constraint of the form $\exists \vec{c}. \forall \vec{x}. \psi(\vec{c},\vec{x})$,
they first generate a set of random integers $\{\vec{n}_1,\dots,\vec{n}_k\}$ and 
weaken the constraint to $\psi(\vec{c},\vec{n}_1) \land \dots \land \psi(\vec{c},\vec{n}_k)$.
The satisfiability of this weakened constraint is then checked using an SMT solver to obtain a candidate solution $\vec{m}$ for $\vec{c}$; if it is unsatisfiable, then the original constraint is unsatisfiable.
Then, again using an SMT solver, the validity of $\forall \vec{x}. \psi(\vec{m},\vec{x})$ is checked; if it is valid, then $\vec{m}$ is accepted as the solution, and the ownership inference terminates. 
Otherwise, the number of random samples $k$ is increased, and the process is repeated until a correct $\vec{m}$ is found or the problem is found to be unsatisfiable.

Although the overall structure of our procedure is similar to that of Tanaka et al., we improve their procedure as follows.
If a solution $\vec{m}$ for $\vec{c}$ in
 $\psi(\vec{c},\vec{n}_1) \land \dots \land \psi(\vec{c},\vec{n}_k)$ is obtained but $\forall \vec{x}. \psi(\vec{m},\vec{x})$ is not valid, then instead of choosing the next value $\vec{n}_{k+1}$ randomly as they do, we use the obtained counterexample for $\forall \vec{x}. \psi(\vec{m},\vec{x})$ as $\vec{n}_{k+1}$; so $\vec{n}_{k+1}$ satisfies $\neg \psi(\vec{m},\vec{n}_{k+1})$.
By this improvement, we can guarantee progress of the procedure since the new search space $\{ \vec{m} \mid \psi(\vec{m}, \tilde{n}_{1}) \land \ldots \land \psi (\vec{m}, \tilde{n}_{k}) \land \psi (\vec{m}, \tilde{n}_{k+1})\}$ is strictly smaller than the previous search space $\{ \vec{m} \mid \psi(\vec{m}, \tilde{n}_{1}) \land \ldots \land \psi (\vec{m}, \tilde{n}_{k}) \}$.

\subsection{Step 3: Refinement Inference}

Once ownership expressions for each reference type are determined,  
the type inference procedure conducts refinement inference in a manner similar to that of Tanaka et al.
Our procedure extends each simple integer type $\TINT$ with a predicate variable
 $P$ that represents a predicate over integers to create an integer type template $\set{\nu \COL \TINT \mid P}$, generates constraints over the predicate variables as constrained Horn clauses (CHC), and solves the constraints using a CHC solver.
For each predicate variable $P$ in an integer type
$\set{\nu \COL \TINT \mid P}$, $P$ may refer to the variables
$\vec{x}$ consisting of all the integer variables in the type
environment at the program point and $\nu$.

\begin{figure*}
\begin{lstlisting}[backgroundcolor=,escapechar=\!]
!\color{comment-green}// pp:$\Pi x_2 . (\Pi x_1. (\{ \nu  : \TINT \mid P_{0}(n, x_1, x_2,  \nu  ) \} \TREF ^{ \mathbf{0} } ) \TREF ^{x_2 \in [0, n-1]  \produces  1 }) $! 
!\color{comment-green}// s:$\Pi x_1 . (\{ \nu  : \TINT \mid P_{1}(n, x_1,  \nu  ) \}  \TREF ^{x_1 \in [0, n-1]  \produces  1}) $! 
pp := s;
!\color{comment-green}// pp:$\Pi x_2 .(\Pi x_1 . (\{ \nu  : \TINT \mid P_{2}(n, x_1, x_2,  \nu )  \}  \TREF ^{r_{1,1}} ) \TREF ^{r_{1,2}}) $! 
!\color{comment-green}// $\quad\text{where}\quad r_{1,1} = {(x_2 = 0)  \wedge  x_1 \in [0, n-1]  \produces  1},(1 \le x_2 \le n - 1)  \wedge  x_1 \in [0,0]  \produces  0 $!
!\color{comment-green}// $\quad\phantom{\text{where}\quad}r_{1,2} = {(x_2 = 0)  \produces  1 , x_2 \in [1,n-1]  \produces  1 } $!
!\color{comment-green}// s:$\Pi x_1 .(\{\nu : \TINT \mid P_{3}(n, x_1,  \nu  )\}  \TREF ^{x_1 \in [0,0]  \produces  0} )$! 
let t = pp $ \boxplus $ 1 in 
!\color{comment-green}// pp:$\Pi x_2 .(\Pi x_1 .(\{ \nu  : \TINT \mid P_{4}(n, x_1, x_2,  \nu  ) \} \TREF ^{r_{2,1}} ) \TREF ^ {r_{2, 2}}) $! 
!\color{comment-green}// $\quad\text{where}\quad r_{2,1} = (x_2 = 0)  \wedge  x_1 \in [0,n-1]  \produces  1,\quad r_{2,2} = x_2 \in [0,0]  \produces  1$!
!\color{comment-green}// t:$\Pi x_2 .(\Pi x_1 .(\{ \nu  : \TINT \mid P_{5}(n, x_1, x_2,  \nu  ) \}  \TREF ^{ \mathbf{0} } ) \TREF ^{x_2 \in [0,n-2]  \produces  1} ) $!
\end{lstlisting}
\caption{Code snippet with refinement type templates.}
\Description{The assignment through pp and subsequent pointer arithmetic
are annotated with refinement templates after ownership inference.
Predicate variables P0 through P5 describe integer elements at successive
program points.  Their arguments include the array size, relevant array
indices, and the integer value being refined.  Concrete ownership terms
record the transfer from s to the first row and the separation of the
remaining outer array through t.}
\label{fig:refinementTemplate}
\end{figure*}
For an example of refinement type inference, we use the same code
snippet as in the ownership inference section.
\Cref{fig:refinementTemplate} shows the type environments with
refinement type templates associated at each program location in the
code snippet.
For example, from $pp := s$ and \rn{T-Assign}, the following CHC is
generated.  
\begin{align*} 
  &\forall n, x_1, x_2,  \nu . n>0  \wedge  x_2 = 0  \wedge  P_{1}(n, x_1,  \nu  )  \produces  P_{2}(n, x_1, x_2,  \nu  )  \\ 
  &\forall n, x_1, x_2,  \nu .  n>0  \wedge  0 < x_2 \le n - 1  \wedge  P_{0}(n, x_1, x_2,  \nu  )  \produces  P_{2}(n, x_1, x_2,  \nu  ) 
\end{align*}
This constraint is solved using a CHC solver to obtain solutions for $P_1$ and
$P_2$; these are substituted back into the original type templates to obtain a
type derivation of the given program.

\begin{revision}
\subsection{Properties of the Type-Inference Procedure}

This section discusses the properties of our type-inference procedure.
Since the procedure itself is similar to the one by Tanaka et al.,
we discuss the properties in prose rather than mathematical formalisms.

\paragraph{Soundness}
The ownership-inference step generates constraints enforcing the ownership
requirements of the typing rules. After solving these constraints, the
refinement-inference step generates refinement constraints using the inferred
ownership terms. If both stages produce valid solutions, these solutions
instantiate the type templates to yield a typing derivation of the input program.

The refinement inference step follows the standard approach of reducing refinement constraints to constrained Horn clauses (CHCs) and solving them with a CHC solver, as adopted in previous refinement-type-inference procedures~\cite{MukaiKS22,BjornerGMR15,HashimotoUnno15}.
This step is sound provided that the employed CHC solver is sound, meaning that
every model it returns is a correct solution to the CHCs.
This style of argument---reducing soundness of refinement-type inference
to the soundness of the underlying constraint solver---is standard in
refinement-type systems; see, e.g., Rondon~\cite[Appendix A]{Rondon2012Liquid}.

\paragraph{Completeness}
In contrast, neither Step 2 nor Step 3 of the type-inference procedure is complete.
The ownership inference step is based on a fixed finite set of templates
for ownership terms. As is standard in template-based program verification, 
this immediately implies incompleteness: the algorithm may fail to find
ownership terms that lie outside the chosen template language, 
even when the program is well-typed. 
Similarly, the refinement inference step may fail to find a solution since
CHC solving is not complete in general~\cite{DeAngelis2022Analysis}.

\paragraph{Complexity}
It is difficult to characterize the time complexity of the entire procedure
since our type-inference procedure invokes Z3 to solve ownership constraints
and HoIce to solve CHC constraints.
Let $m$ be the size of the input program and $n$ be the maximum nesting depth
of the $\mathbf{ref}$ type constructor appearing in the input program.
Each application of a syntax-directed typing rule generates ownership constraints and CHC constraints of size $O(n)$.
Since the total number of syntax-directed rule applications is 
$O(m)$, the overall number of generated constraints is $O(nm)$.

\end{revision}

\section{Implementation}
\label{sec:implementation}
We implemented the type inference procedure described in
\cref{sec:typeInference}.
The source code is available at
\url{https://github.com/SoftwareFoundationGroupAtKyotoU/nested_array_ConSORT}.
The implementation uses the ownership templates introduced in
\cref{sec:OwnershipInference}, with additional restrictions on
ownership redistribution and template conversion intended to
simplify and accelerate inference.
The implementation also provides additional syntax for annotations, allocation,
and assertions.
We first describe the restrictions on ownership redistribution,
then explain template usage and conversion, and finally present
the syntax extensions and annotation support.

\subsection{Restrictions on Ownership Redistribution}

The implementation restricts ownership redistribution in pointer
arithmetic by explicitly distinguishing array partitioning from
array sharing.
Both forms support constant and variable offsets.
The expression $ \LET  x  =   y   \boxplus   z   \IN  e $ partitions the region
owned through $y$ at offset $z$, measured relative to the location
pointed to by $y$.
All ownership of the suffix beginning at this offset is transferred
to $x$, while ownership of the preceding portion remains with $y$.
The expression $\LET \mathbf{immut} x = y \boxplus z \IN e$
instead shares ownership of the same heap region between $x$ and $y$.

The implementation also fixes ownership redistribution for
array reads and writes.
For a read $ \LET  x  =   \ast  y   \IN  e $, all ownership carried by the
value stored in the head element of $y$ is transferred to $x$.
For a write $  x  \WRITE  y   \SEQ  e $, all ownership held by $y$ is
transferred to the head element of $x$.
These fixed transfers avoid having to infer how ownership is
split between the source and destination.

At alias expressions, ownership is returned in full.
The expression $  \ALIAS(  x  =  y   \boxplus   z  )   \SEQ  e $ returns all ownership
held by $x$ to $y$.
The expression $  \ALIAS(  x  = \ast  y  )   \SEQ  e $ returns all ownership
held by $x$ to the head element of $y$.

\subsection{Ownership Templates}

The implementation also restricts the use and conversion of
ownership templates during constraint generation.
For the partitioning expression $\LET x = y \boxplus 1 \IN e$,
if the outermost ownership expression of $y$ uses Template~(2),
the implementation replaces it with Template~(1) with the interval
$[0,0]$ in the pre-environment of $e$, consistent with the
partitioning policy.
For nested arrays, the inner ownership expressions represented
using Template~(3) are also replaced with expressions using
Template~(1), retaining the case corresponding to outer index $0$.
This removes the need to maintain separate cases for the first
element and the remaining elements in subsequent constraint
generation for $y$.

At alias expressions, the implementation generates constraints
under the following assumptions about ownership templates.
For $  \ALIAS(  x  = \ast  y  )   \SEQ  e $, the type of $y$ is assumed to use
Template~(2) for its outermost ownership expression and
Template~(3) for its inner ownership expressions, if any.
For $  \ALIAS(  x  =  y   \boxplus   z  )   \SEQ  e $, the ownership expressions
in the type of $x$ are assumed to use Template~(1).
The current implementation does not introduce additional template
environments and subtyping constraints to convert other
representations into these forms; supporting such conversions
is left for future work.

For each function definition, the implementation uses Template~(1)
at every reference level in both the input and output types of
reference-valued parameters.
Constraint generation for the function body assumes this
representation for these parameters at function entry and exit.
At function calls, the actual reference arguments are likewise
assumed to already use Template~(1) at every reference level,
and their types after the call are determined by the callee's
output types.

For $\IF$ expressions, reference types at the ends of both the
$\THEN$ and $\ELSE$ branches are assumed to use Template~(1)
at every reference level.
This common template form simplifies the combination of the
type environments obtained from the two branches.

\subsection{Syntax Extensions and Annotation}

The implementation allows users to provide ownership annotations
for input and output types of reference-valued parameters 
and refinement annotations for integer types.
For an element type $\tau$,
\texttt{ref(l, h, o)} denotes
$\Pi x. (\tau \TREF^{x \in [l, h]  \produces  o})$.
Integer parameters that may occur in ownership terms must be
explicitly marked with \texttt{\#}.
When input and output ownership annotations fully specify
function interfaces, the implementation can solve ownership
constraints separately for each function.
Refinement constraints are still collected across the program
and passed together to the CHC solver.

The implementation provides the indeterminate integer literal
\texttt{\_}, with an optional refinement annotation.
The supported forms are $ \LET  x  =   \_   \IN  e $ and
$ \LET  x  =   \_ \colon ( \varphi )   \IN  e $.
The unannotated form binds $x$ to an arbitrary integer with type
$ \{  \nu  :   \TINT    \mid   \top  \} $, while the annotated form restricts the value
to satisfy $\varphi$.
For example,
$ \LET  x  =  0  \IN   \LET  y  =   \_ \colon ( y \,  >  \, x )   \IN  e  $
binds $y$ to an arbitrary positive integer.
This construct also allows array sizes to be left unspecified:
$ \LET  y  =   \_ \colon ( y \,  >  \,  0  )   \IN   \LET  x  =   \ALLOC  y   \ottsym{:}     \TINT   \TREF    \IN  e  $
creates an array $x$ of arbitrary positive length.

Refinement annotations on allocation expressions constrain the
initial values of array elements.
For example,
$ \LET  y  =  10  \IN   \LET  x  =   \ALLOC  y   \ottsym{:}     \{  \nu  :   \TINT    \mid   \ottsym{(}  \nu \,  =  \,  1   \ottsym{)}  \}   \TREF    \IN  e  $
creates an array of length $10$ whose elements are all initialized
to $1$.
An element refinement such as $\nu > 1$ allows each element
to be initialized to any integer greater than $1$.

Allocation refinements may also depend on array indices.
The implementation uses the fixed names \texttt{i1}, \texttt{i2},
and so on for indices at successive array layers, numbered from
the innermost layer outward.
Thus, for a two-dimensional array, \texttt{i1} denotes the column
index and \texttt{i2} denotes the row index.
For example,
\[
\LET y = 10 \IN
\LET x = \ALLOC y \ottsym{:}
  \{ \nu : \TINT \mid
     (\texttt{i1} = 0) \implies (\nu = 1) \} \TREF
\IN e
\]
creates an integer array of length $10$ whose first element is
initialized to $1$, while the initial values of the remaining
elements are unconstrained.

The implementation also supports direct array indexing in
$\ASSERT$ expressions.
For example, $\ASSERT(x[0][y] = z); e$ asserts that the element
in row $0$ and column $y$ of $x$ equals $z$.
Combined with indeterminate integers and refinement annotations,
this syntax can express properties over a range of indices.
For example,
\[
\LET y = \_ \colon
  (y \ge 0 \mathbin{\&\&} y \le 9)
\IN \ASSERT(x[0][y] = z); e
\]
checks that every element in columns $0$ through $9$ of row $0$
equals $z$, since $y$ may be any integer in that range.

To reduce the annotation burden on programmers, the implementation
automatically inserts $\ALIAS$ expressions.
For a binding $ \LET  x  =   \ast  y   \IN  e $, if its body contains a
subsequent binding through $y$, of the form
$ \LET  z  =   \ast  y   \IN  e_{{\mathrm{1}}} $ or $\LET z = y \boxplus n \IN e_1$,
the implementation inserts $\ALIAS(x = *y)$ immediately before
that binding.
This returns ownership from $x$ to the head element of $y$,
making it available for redistribution to $z$ together with
ownership already held through $y$.

If the body contains no such binding, the corresponding $\ALIAS$
expression is inserted at the end of the body, where the scope
of $x$ ends.
The same placement is used when $x$ is introduced by pointer
arithmetic, as in $\LET x = y \boxplus n \IN e$.
These rules cover common patterns in which ownership redistribution
is needed and the inserted $\ALIAS$ expression is valid.
Tanaka et al.~\cite{DBLP:conf/pepm/TanakaSK24} implement a similar
mechanism for automatically inserting $\ALIAS$ expressions for
non-nested arrays.

\section{Experiments}
\label{sec:experiments}

Using the implementation described in \cref{sec:implementation}, we conducted experiments to evaluate the effectiveness of our verifier.
We aim to address the following research questions through our experiments:
\begin{description}
\item[RQ1] Can our verifier effectively analyze programs containing
  nested arrays, a feature not supported by Tanaka et al.'s verifier,
  and produce useful verification results?
\item[RQ2] Can our verifier handle the same set of programs verifiable by Tanaka et al.'s implementation without significant overhead?
\end{description}

All the experiments were conducted on a machine with an Apple M2 CPU
and 24 GB of RAM.  Our implementation is written in
OCaml 4.14.1; we used Z3 4.14.1~\cite{z3} as an SMT solver and HoIce
1.10.0~\cite{hoice} as a CHC solver.
\todo{}
\todo{}

\ks{}

\subsection{RQ1: Verification of Programs with Nested Arrays}

\subsubsection{Benchmarks}

To answer RQ1, we wrote programs that manipulate matrices---one of the main use cases of nested arrays in practice---in our target
language and verified them using our verifier.

We used the following benchmark programs in the experiments to answer RQ1.
\begin{description}
\item[Init-Matrix] is a variant of our motivating example in \cref{fig:motivatingExample}, which initializes a rectangular matrix with ones. It verifies that every element in the initialized matrix is one.
\item[Indexed-Matrix] initializes a matrix with ones, with row lengths decreasing by one from one row to the next. 
It checks that selected elements of the initialized matrix are one.
\item[Indexed-Value] constructs a $10 \times 10$ matrix 
such that the zeroth row is filled with $10$, the 1st row with $9$, and so on, with the value decreasing as the row index increases.
  It verifies that for any arbitrary position $(i, j)$ in the generated matrix, the value is always equal to $10 - i$.
\item[Sum-Matrix] adds all the elements in a matrix of non-negative integers.
  It verifies that the resulting sum is non-negative.
\item[Copy-Matrix] copies the elements of one matrix to another.
  It verifies that the copied matrix is identical to the original matrix.
\item[Add-Matrix] takes two integer matrices of the same shape and computes their element-wise sum into a third matrix.
  It verifies that each element in the resulting matrix is the sum of the corresponding elements from the two input matrices.
\item[Trace-Matrix] calculates the trace of a matrix of non-negative integers and verifies that the result is non-negative.
\item[Trans-Matrix] computes the transpose of a matrix and verifies that each element is correctly positioned.
\item[Swap] swaps two rows within a matrix and verifies that the values are correctly updated.
\item[Eta-Equ-Sum] computes the sum of all elements in a matrix using two different methods---one that sums elements of each row from left to right before aggregating the row sums, and another that sums elements of each row from right to left before aggregating the row sums---and verifies that both methods yield the same result.
\item[Eta-Equ-Trace] computes the trace of a matrix using two different methods---one that sums elements from the top-left to the bottom-right and another from the bottom-right to the top-left---and verifies that both methods yield the same result.
\item[Lower-Triangle] generates a lower triangular matrix and verifies that all elements above the main diagonal are zero.
\item[Compare-Element] performs an element-wise comparison of two different arrays.
\item[Row-Add] creates an array containing the sum of each row from a matrix.
\item[Boomerang] reads a value from a matrix and immediately writes it back to the same position.
\item[Share-Add-Matrix] generates two pointers to the same matrix and computes their element-wise sum into a third matrix. 
It verifies that each element in the resulting matrix is twice the value in the original matrix.
\end{description}

\begin{revision}
This benchmark suite was designed to systematically cover typical programming patterns that arise when manipulating matrix-like data structures with pointer arithmetic and aliasing. The benchmarks vary along three main dimensions: (i) array shape (rectangular vs.\ non-rectangular), (ii) traversal order (row-major, column-major, and diagonal-like traversals), and (iii) the number of matrices and aliasing/ownership patterns involved (single-matrix updates vs.\ coordinated updates of multiple matrices with shared read access). 

Concretely, several benchmarks operate on \emph{rectangular} matrices, where every row has the same length; typical examples include programs that initialize or update all elements of a dense matrix (e.g., \textsc{Init-Matrix}, \textsc{Sum-Matrix}, \textsc{Add-Matrix}, \textsc{Trans-Matrix}), whose ownership terms do not need to distinguish rows by length. In contrast, other benchmarks use \emph{non-rectangular} shapes such as lower-triangular matrices (e.g., \textsc{Indexed-Matrix}); in these programs, the length of each inner array depends on the outer index, so the ownership terms for inner arrays must depend on outer indices, directly stressing the main generalization of our system over Tanaka et al.'s work.

The suite also covers a range of access and aliasing patterns. Row-major traversal appears in initialization and accumulation programs (such as \textsc{Init-Matrix}, \textsc{Indexed-Value}, \textsc{Sum-Matrix}, \textsc{Copy-Matrix}, \textsc{Row-Add}, \textsc{Share-Add-Matrix}), which exercise ownership templates tailored to pointer iteration where the head element is treated differently from the remaining elements. Column-major traversal is represented by transpose-like programs (e.g., \textsc{Trans-Matrix}), where the access pattern repeatedly visits the same column across different rows, while diagonal-style traversals (e.g., \textsc{Trace-Matrix} and variants) visit entries whose indices satisfy constraints such as $i = j$ or $i \le j$. 

The benchmarks also differ in the number of matrices manipulated simultaneously: some programs manipulate a single matrix in place (\textsc{Init-Matrix}, \textsc{Indexed-Matrix}, \textsc{Lower-Triangle}, \textsc{Boomerang}), whereas others operate on two or more matrices (\textsc{Copy-Matrix}, \textsc{Add-Matrix}, \textsc{Compare-Element}), requiring the system to track ownership of multiple nested arrays with correlated shapes. 

Finally, examples such as \textsc{Share-Add-Matrix} use explicit aliasing of pointers to express shared read access to source matrices while accumulating results into a separate target matrix. This benchmark stresses the interaction between $\ALIAS$ expressions, ownership splitting/merging, and nested-array shapes. 
Furthermore, the suite is intended to cover a representative range of nested-array usage patterns in low-level matrix code, and none of these nested-array benchmarks can be verified by the previous verifier of Tanaka et al.
\end{revision}

\begin{table}[t]
  \caption{The results of the experiments for RQ1. Times are presented in seconds.
    The rightmost three columns show the number of $\ALIAS$ expressions in each benchmark program,
    broken down into those inserted manually by the programmer
    and those inserted automatically by the verifier.}
  \label{table:experiments}
  \centering
  \footnotesize
  \begin{tabular}{l|c|r|r|r}\hline
                                & \text{Total time (ownership/refinement)} & \multicolumn{3}{c}{\textbf{\# of $\ALIAS$}} \\
    \textbf{Name}               & {\textbf{Ours}}                          & \textbf{Manual} & \textbf{Auto} & \textbf{Total} \\
    \hline\hline
    \textbf{Init-Matrix(2D)}        & 5.093 (2.412+2.680)                      & 0  & 2  & 2  \\
    \textbf{Init-Matrix(3D)}        & 205.544 (16.437+189.106)                 & 0  & 3  & 3  \\
    \textbf{Indexed-Matrix(2D)}     & 36.895 (0.433+36.462)                    & 0  & 2  & 2  \\
    \textbf{Indexed-Matrix(3D)}     & 17.232 (13.311+3.920)                    & 0  & 3  & 3  \\
    \textbf{Indexed-Matrix(4D)}     & 350.784 (114.746+236.037)                & 0  & 4  & 4  \\
    \textbf{Indexed-Value}      & 3.336 (2.943+0.393)                          & 0  & 2  & 2  \\
    \textbf{Sum-Matrix}         & 48.244 (31.694+16.549)                       & 0  & 5  & 5  \\
    \textbf{Copy-Matrix}        & 131.380 (80.184+51.195)                      & 0  & 7  & 7  \\
    \textbf{Add-Matrix}         & 208.473 (191.750+16.722)                     & 0  & 10 & 10 \\
    \textbf{Trace-Matrix}       & 54.533 (10.845+43.688)                       & 1  & 4  & 5  \\
    \textbf{Trans-Matrix}       & 163.251 (79.337+83.913)                  & 1  & 4  & 5  \\
    \textbf{Eta-Equ-Sum}        & 19.318 (17.646+1.671)                    & 0  & 6  & 6  \\
    \textbf{Eta-Equ-Trace}      & 18.331 (16.563+1.767)                    & 3  & 3  & 6  \\
    \textbf{Lower-Triangle}     & 26.132 (3.053+23.079)                    & 0  & 2  & 2  \\
    \textbf{Swap}               & 75.411 (73.636+1.774)                    & 4  & 4  & 8  \\
    \textbf{Compare-Element}    & 16.550 (2.110+14.439)                    & 0  & 2  & 2  \\
    \textbf{Row-Add}            & 28.302 (4.563+23.739)                    & 0  & 4  & 4  \\
    \textbf{Boomerang}          & 17.674 (14.742+2.931)                    & 0  & 4  & 4  \\
    \textbf{Share-Add-Matrix}   & 376.867 (29.328+347.538)                 & 3  & 5  & 8  \\
    \hline
  \end{tabular}
\end{table}

\subsubsection{Results}

\Cref{table:experiments} presents the time spent on verifying each program with our implementation.
In addition to the total verification time, we also provide a breakdown of time spent on ownership inference and refinement type verification.
Depending on the benchmark, the total verification time varies from around 3 seconds to about 6 minutes; none exceeded the time or memory limits, despite the modest hardware used.

\begin{revision}
\newcommand{\br}[1]{#1}

To illustrate the variety of ownership terms used by the benchmarks, we present the ownership terms inferred for the post-type of the main matrix pointer in representative programs.  \Cref{tab:ownership-terms} summarizes the results; we write $r_{\mathit{outer}}$ for the ownership term associated with the outermost $\mathbf{ref}$ type and $r_{\mathit{inner}}$ for the one associated with the next inner $\mathbf{ref}$ type (and $r_{\mathit{mid}}$ for the middle level in 3D examples).

\begin{table}[t]
\caption{Inferred ownership terms for the matrix pointer after
initialization / main computation.
$x_1$ and $x_2$ denote the inner and outer array indices, respectively;
$x_3$ denotes the outermost index in 3D examples;
$n$, $n_1$, $n_2$, and $m$ are function parameters in scope.
T1/T2/T3 indicate which type of template introduced in \cref{sec:OwnershipInference} was used.}
\label{tab:ownership-terms}
\centering
\footnotesize
\begin{tabular}{@{}l l l@{}}
\toprule
Program & $r_{\mathit{outer}}$ & $r_{\mathit{inner}}$ \\
\midrule
\textsc{Indexed-Matrix(2D)}
  & $\br{0 \le x_2 \le n{-}1} \mapsto 1$ \;(T1)
  & $\br{0 \le x_2 \le n{-}1 \land 0 \le x_1 \le n{-}x_2{-}1} \mapsto 1$ \;(T3) \\
  \multicolumn{3}{@{}l}{\quad\itshape
    Inner ownership depends on the outer index $x_2$.} \\
\addlinespace
\textsc{Sum-Matrix}
  & $\br{x_2 = 0} \mapsto 1,$ 
                               & $\br{0 \le x_1 \le n_1{-}1} \mapsto 1$ \;(T1) \\
  & $\br{1 \le x_2 \le n_2{-}1} \mapsto 1$ \;(T2) \\
  \multicolumn{3}{@{}l}{\quad\itshape
    Outer ownership uses T2: head row is dereferenced before advancing the pointer.} \\
\addlinespace
\textsc{Trans-Matrix}
  & $\br{0 \le x_2 \le n_2{-}1} \mapsto 1$ \;(T1)
  & $\br{0 \le x_1 \le n_1{-}1} \mapsto 1$ \;(T1) \\
  \multicolumn{3}{@{}l}{\quad\itshape
    Uniform rectangular matrix; column-major traversal.} \\
\addlinespace
\textsc{Share-Add-Matrix}
  & $\br{0 \le x_2 \le n_2{-}1} \mapsto 0.1$ \;(T1)
  & $\br{0 \le x_1 \le n_1{-}1} \mapsto 0.1$ \;(T1) \\
  \multicolumn{3}{@{}l}{\quad\itshape
    Fractional (read-only) sharing.} \\
\addlinespace
  & $r_{\mathit{outer}}$ & $r_{\mathit{mid}}$ / $r_{\mathit{inner}}$ \\
\cmidrule(l){2-3}
\textsc{Indexed-Matrix}
  & $\br{0 \le x_3 \le m{-}1} \mapsto 1$ \;(T1)
  & $\br{0 \le x_2 \le m{-}x_3{-}1} \mapsto 1$ \;(T3) \\
\quad(3D)
  & & $\br{0 \le x_1 \le m{-}x_3{-}x_2{-}1} \mapsto 1$ \;(T3) \\
  \multicolumn{3}{@{}l}{\quad\itshape
    Inner ownership depends on $x_3$; innermost depends on both
    $x_3$ and $x_2$.} \\
\bottomrule
\end{tabular}
\end{table}

Several observations are worth noting.
First, all three templates are exercised across the benchmark suite.
Template~1 suffices when rows have uniform length and uniform
access permissions, as in \textsc{Trans-Matrix}.
Template~2 is required whenever pointer iteration treats the head
element differently from the tail, which arises in every program that
traverses a matrix via recursive pointer arithmetic (e.g.,
\textsc{Init-Matrix}, \textsc{Sum-Matrix}, \textsc{Copy-Matrix}).
Template~3, which introduces a dependency on an outer index variable,
is required for non-rectangular matrices and for the inner ownership
terms in all programs where the length of an inner array varies
with the outer index (e.g., \textsc{Indexed-Matrix}).

Second, several benchmarks require ownership terms that depend on variables other than array indices. For example, \cref{fig:typingexam3d} shows how the 3D version of \textsc{Indexed-Matrix} is typed. In the final type of \texttt{ppp} after alias recombination, the ownership term for the innermost array is inferred as $\br{0 \le x_1 \le m - x_3 - x_2 - 1} \mapsto 1$, where $m$ is the function parameter. This demonstrates that the generality of our ownership terms, allowing dependency on outer indices \emph{and} on variables in the type environment, is essential for verifying programs with higher-dimensional nested arrays.

Third, the \textsc{Share-Add-Matrix} benchmark demonstrates a qualitatively different use of ownership terms: the inferred ownership is $0.1$ (read-only) rather than $1$ (read-write), reflecting that two aliases to the same matrix share read access while a third matrix receives full write access.  This pattern exercises the fractional ownership aspect of our system jointly with nested-array reasoning.
\end{revision}

\begin{revision}
  We also evaluated the effectiveness of our automated $\ALIAS$ insertion mechanism.
The rightmost three columns of \Cref{table:experiments} show the number of $\ALIAS$ expressions
required in each benchmark program from \cref{table:experiments},
distinguishing between those that were manually written by the programmer
and those that were automatically inserted by the verifier
using the mechanism described above.
In 14 out of 19 benchmarks, no manual insertion of $\ALIAS$ expressions was needed at all;
even in the remaining benchmarks, the number of manually inserted $\ALIAS$ expressions
is at most 4.
These results demonstrate that the automated insertion mechanism
covers the majority of $\ALIAS$ expressions required in practice,
substantially reducing the annotation burden on the programmer.
\end{revision}

Based on these results, our conclusion to RQ1 is affirmative:
Our verifier effectively analyzes programs containing nested
  arrays, a feature not supported by Tanaka et al.'s verifier, and
  produces useful verification results.

\begin{revision}
\begin{figure*}[t]
\begin{lstlisting}[backgroundcolor=,escapechar=\!]
!\color{comment-green}// iM:$\langle \mathtt{m}:  \TINT ,\; \mathtt{r}:  \Pi x_{{\mathrm{2}}} .(  \Pi x_{{\mathrm{1}}} .(  \TINT   \TREF^{\hspace{0.5pt}  \mathbf{0}  })   \TREF^{\hspace{0.5pt} r_{{\mathrm{2}}} }) \rangle  \rightarrow \langle \mathtt{m}:  \TINT ,\; \mathtt{r}:  \Pi x_{{\mathrm{2}}} .(  \Pi x_{{\mathrm{1}}} .(  \{  \nu  :   \TINT    \mid   \varphi  \}   \TREF^{\hspace{0.5pt} r_{{\mathrm{1}}} })   \TREF^{\hspace{0.5pt} r_{{\mathrm{2}}} })  \mid  \TINT \rangle$!
!\color{comment-green}// $\quad\text{where}\quad r_1 = (0 \le x_2 \le \mathtt{m}-1  \wedge  0 \le x_1 \le \mathtt{m}-x_2-1  \produces  1),\quad r_2 = (0 \le x_2 \le \mathtt{m}-1  \produces  1)$!
!\color{comment-green}// $\quad\phantom{\text{where}\quad}\varphi = (0 \le x_2 \le \mathtt{m}-1  \wedge  0 \le x_1 \le \mathtt{m}-x_2-1 \implies \nu=1)$!

iTm(m, ppp)
!\color{comment-green}// ppp:$ \Pi x_3 . ( \Pi x _{ 2 }  . (  \Pi x _{ 1 }  . (\{\nu \COL \TINT \mid  \top  \} \TREF ^ { \,  \mathbf{0}  } ) \TREF ^ { \,  \mathbf{0}  } )    \TREF^{ \, x_3 \in [ 0,  \mathtt{m}-1 ]  \produces  1 })$ ! 
{ 
  if m > 0 then { 
    let r = alloc m : int ref ref in
!\color{comment-green}// r:$ (\Pi x_{2}. ( \Pi x _{ 1 }  . (\{\nu \COL \TINT \mid  \top  \} \TREF ^ { \,  \mathbf{0}  } )    \TREF^{ \,x_2 \in [ 0,  \mathtt{m}-1 ]  \produces  1 })) $!
    let d = iM(m, r) in
!\color{comment-green}// r:$ (\Pi x_2 . ( \Pi x_1 . (\{ \nu \COL \TINT \mid  0 \le x_2 \le \mathtt{m} - 1  \wedge  0 \le x_{ 1 } \le \mathtt{m} - x_2 - 1 \implies \nu = 1 \}  \TREF^{ \, r_{{\mathrm{1}}} } ) \TREF^{ \, r_{{\mathrm{2}}} } ))$!
!\color{comment-green}// $\quad\text{where}\quad r_{{\mathrm{1}}}  = x_1 \in [0, \mathtt{m} - x_2 - 1 ] \produces  1 , r_{{\mathrm{2}}}  = ( x_2 \in [ 0, \mathtt{m}-1 ]  \produces  1 )$!
    ppp := r;
!\color{comment-green}// ppp:$ ( \Pi x_3. (\Pi  x_2. (\Pi x_{1}. (\{\nu \COL \TINT \mid  x_3 = 0  \wedge   0 \le x_2 \le \mathtt{m} - 1  \wedge  0 \le x_{ 1 } \le \mathtt{m} - x_2 - 1 \implies \nu = 1 \}  \TREF^{ \, r_{{\mathrm{1}}} }   ) \TREF^{ \, r_{{\mathrm{2}}} }) \TREF^{ \, r_{{\mathrm{3}}} }))$!
!\color{comment-green}// $\quad\text{where}\quad r_{{\mathrm{1}}}  = ( x_3 = 0  \wedge  x_1 \in [0, \mathtt{m} - x_2 - 1 ]  \produces  1 ), r_{{\mathrm{2}}}  = ( x_3 = 0  \wedge  x_2 \in [ 0, \mathtt{m}-1 ]  \produces  1 ), r_{{\mathrm{3}}} = ( x_3 \in [ 0, \mathtt{m}-1 ]  \produces  1)$!
!\color{comment-green}// r:$ (\Pi x_{2}. ( \Pi x _{ 1 }  . (\{\nu \COL \TINT \mid  \top  \} \TREF ^ { \,  \mathbf{0}  } )    \TREF^{ \,  \mathbf{0}  } ))$!
    let q = ppp $ \boxplus $ 1 in
!\color{comment-green}// ppp:$ ( \Pi x_3. (\Pi  x_2. (\Pi x_{1}. \{\nu \COL \TINT \mid  x_3 = 0  \wedge   0 \le x_2 \le \mathtt{m} - 1  \wedge  0 \le x_{ 1 } \le \mathtt{m} - x_2 - 1 \implies \nu = 1 \} ) \TREF^{ \, r_{{\mathrm{1}}} }   ) \TREF^{ \, r_{{\mathrm{2}}} }) \TREF^{ \, r_{{\mathrm{3}}} }$!
!\color{comment-green}// $\quad\text{where}\quad r_{{\mathrm{1}}}  = ( x_1 \in [0, \mathtt{m} - x_2 - 1 ]  \produces  1 ), r_{{\mathrm{2}}}  = ( x_2 \in [ 0, \mathtt{m}-1 ]  \produces  1 ), r_{{\mathrm{3}}}  = (x_3 = 0  \produces  1 )$!
!\color{comment-green}// q:$ (\Pi  x_3. (\Pi x_2. (\Pi x_1. (\{ \nu \COL \TINT \mid  \top  \}    \TREF^{ \,  \mathbf{0}  }   ) \TREF^{ \,  \mathbf{0}  }   ) \TREF^{ \, x_3 \in [0 , \mathtt{m}-2]  \produces  1})) $!
    let d2 = iTm(m-1, q) in
!\color{comment-green}// q:$\begin{aligned}[t]& (\Pi  x_3. (\Pi x_2. (\Pi x_1. (\{ \nu \COL \TINT \mid\\&\qquad 0 \le x_3 \le \mathtt{m} - 2  \wedge  0 \le x_{ 2 } \le \mathtt{m} - x_3 - 2  \wedge  0 \le x_{ 1 } \le \mathtt{m} - x_3 - x_2 - 2 \implies \nu = \mathtt{1} \}    \TREF^{ \, r_{{\mathrm{1}}} }   ) \TREF^{ \, r_{{\mathrm{2}}} }   ) \TREF^{ \, r_{{\mathrm{3}}} })) \end{aligned}$!
!\color{comment-green}// $\quad\text{where}\quad r_{{\mathrm{1}}}  = (x_1 \in [0 , \mathtt{m} - x_3 - x_2 - 2]  \produces  1 ), r_{{\mathrm{2}}}  = (x_2 \in [0 , \mathtt{m} - x_3 - 2]  \produces  1 ), r_{{\mathrm{3}}}  = ( x_3 \in [0 , \mathtt{m} - 2]  \produces  1 )$!
    alias(q=ppp+1);0
!\color{comment-green}// ppp:$\begin{aligned}[t]& (\Pi  x_3. (\Pi x_2. (\Pi x_1. (\{ \nu \COL \TINT \mid\\&\qquad 0 \le x_3 \le \mathtt{m} - 1  \wedge  0 \le x_{ 2 } \le \mathtt{m} - x_3 - 1  \wedge  0 \le x_{ 1 } \le \mathtt{m} - x_3 - x_2 - 1 \implies \nu = \mathtt{1} \}    \TREF^{ \, r_{{\mathrm{1}}} }   ) \TREF^{ \, r_{{\mathrm{2}}} }   ) \TREF^{ \, r_{{\mathrm{3}}} } ))\end{aligned}$!
!\color{comment-green}// $\quad\text{where}\quad r_{{\mathrm{1}}}  = (x_1 \in [0 , \mathtt{m} - x_3 - x_2 - 1]  \produces  1 ), r_{{\mathrm{2}}}  = (x_2 \in [0 , \mathtt{m} - x_3 - 1]  \produces  1 ), r_{{\mathrm{3}}}  = ( x_3 \in [0 , \mathtt{m} - 1]  \produces  1 )$!
!\color{comment-green}// q:$ (\Pi  x_3. (\Pi x_2. (\Pi x_1. (\{ \nu \COL \TINT \mid  \top  \}    \TREF^{ \,  \mathbf{0}  }   ) \TREF^{ \,  \mathbf{0}  }   ) \TREF^{ \,  \mathbf{0} })) $!
  } else { 0 }
}
\end{lstlisting}
\caption{Typing example of IndexedMatrix (3D).}
\Description{A recursive function initializes a three-dimensional jagged
array with ones.  Annotated types track allocation of a two-dimensional
slice, its initialization, its assignment through the outer pointer, and
a recursive call on the remaining slices. The helper's function type
specifies initialization of a two-dimensional jagged array with ones.
An alias check recombines
ownership.  The final type makes the middle dimension's bound depend on
the outer index and the innermost dimension's bound depend on both
enclosing indices.}
\label{fig:typingexam3d}
\end{figure*}

\end{revision}

\subsection{RQ2: Comparison with Prior Work}

To address RQ2, we compared our verifier with the implementation by Tanaka et al.~\cite{DBLP:conf/pepm/TanakaSK24} using the same benchmark suite they employed.
Concretely, we used the following benchmark programs in our experiments to answer RQ2:
\begin{description}
    \item[Init-$n$] initializes an array of length $n$ with $0$ and verifies that the array is successfully initialized.  %
    \item[Sum] computes the sum of an array.  It verifies that the sum of a constant array is computed correctly.
    \item[Sum-Back] computes the sum of an array starting from the end. It verifies that the sum of a constant array is computed correctly.
    \item[Sum-Both] computes the sum of an array by alternately summing elements from the beginning and the end. It verifies that the sum of a constant array is computed correctly.
    \item[Sum-Div] computes the sum of an array by splitting the array into two regions,
    recursively computing the sum of each region, and adding the obtained results.
    It verifies that the sum is computed correctly.
    \item[Copy-Array] copies the elements of one array to another. It verifies that the copied array is identical to the original array.
    \item[Add-Array] computes the element-wise sum of two arrays of the same size and verifies that the resulting array contains the correct sums.
\end{description}
\newcommand{\splitnum}[2]{\makecell{#1\\#2}}
\begin{table}[t]
  \caption{The results of the experiments for RQ2. Times are presented in seconds. The fastest time for each benchmark is shown in bold.}
  \label{table:experimentsArray}
  \centering
  \footnotesize
  \begin{tabular}{lc|c|c|c}\hline
                        & \multicolumn{3}{c}\textit{Total time(ownership/refinement)} \\
                        & \multicolumn{2}{c}{\textbf{Tanaka+}}                                      & \multicolumn{2}{c}{\textbf{Ours}}                                   \\         
    z3 version          & 4.11.2                              & 4.14.1                              & 4.11.2                            & 4.14.1                          \\ 
    \hline\hline
    \textbf{Init-10}    & \splitnum{\textbf{0.255}}{\textbf{(0.052+0.202)}}     & \splitnum{0.262}{(0.041+0.220)}     & \splitnum{0.590}{(0.193+0.397)}   & \splitnum{0.493}{(0.102+0.390)} \\
    \textbf{Init-1000}  & \splitnum{3.689}{(3.466+0.223)}     & \splitnum{157.486}{(0.299+157.186)} & \splitnum{\textbf{1.102}}{\textbf{(0.684+0.417)}}   & \splitnum{2.274}{(0.332+1.942)} \\
    \textbf{Sum}        & \splitnum{1.962}{(1.797+0.165)}     & \splitnum{\textbf{0.266}}{\textbf{(0.138+0.128)}}     & \splitnum{1.056}{(0.202+0.854)}   & \splitnum{0.337}{(0.110+0.227)} \\
    \textbf{Sum-Back}   & \splitnum{0.471}{(0.134+0.337)}     & \splitnum{0.836}{(0.509+0.326)}     & \splitnum{0.335}{(0.224+0.111)}   & \splitnum{\textbf{0.224}}{\textbf{(0.119+0.105)}} \\
    \textbf{Sum-Both}   & \splitnum{1.078}{(0.575+0.503)}     & \splitnum{\textbf{0.739}}{\textbf{(0.310+0.428)}}     & \splitnum{0.916}{(0.702+0.214)}   & \splitnum{0.919}{(0.182+0.736)} \\
    \textbf{Sum-Div}    & \splitnum{0.789}{(0.361+0.427)}     & \splitnum{0.780}{(0.341+0.439)}     & \splitnum{0.777}{(0.630+0.146)}   & \splitnum{\textbf{0.647}}{\textbf{(0.492+0.155)}} \\
    \textbf{Copy-Array} & \splitnum{30.585}{(28.648+1.937)}   & \splitnum{61.470}{(61.203+0.266)}   & \splitnum{1.180}{(0.622+0.558)}   & \splitnum{\textbf{1.036}}{\textbf{(0.318+0.717)}} \\
    \textbf{Add-Array}  & \splitnum{328.066}{(327.652+0.413)} & timeout($> 10^3$)                  & \splitnum{\textbf{38.205}}{\textbf{(2.848+35.357)}} & \splitnum{338.428}{(1.315+337.112)} \\
    \hline
  \end{tabular}
\end{table}

\cref{table:experimentsArray} shows the results of the experiments comparing our verifier with that of Tanaka et al.
Since the implementation by Tanaka et al.\ was tested with an older version of Z3 (4.11.2), we conducted our experiments using two different Z3 versions.
The verifier by Tanaka et al.\ fails to verify \textbf{Add-Array} due to a timeout when using Z3 4.14.1, so we used an older version (4.11.2) for that benchmark.
\todo{}
We can observe that our verifier successfully verifies all the benchmarks that Tanaka et al.'s verifier can handle.
Regarding the verification time for each benchmark, we observe that our verifier incurs considerably less overhead overall than Tanaka et al.'s verifier.
For \textbf{Copy-Array} and \textbf{Add-Array}, ownership inference by our verifier is significantly faster than by Tanaka et al.'s verifier.
\ks{}
As is evident from the experimental results, the time required for ownership inference is highly dependent on the version of Z3. 
Furthermore, although we did not change the HoIce version in this experiment, 
the refinement type inference for Init-1000 from Tanaka et al.\ is also significantly affected by the version of Z3. 
We conjecture that differences in the Z3 version affect ownership-inference results, 
and it is likely that the resulting refinement constraints were incompatible with the CHC solver (HoIce), causing the significant time increase.
\todo{%
}

Based on these results, our conclusion to RQ2 is affirmative: Our
  verifier can handle the same set of programs that are verifiable by
  the implementation from the prior work by Tanaka et al.\ without much
  overhead.

\section{Mechanization in Rocq}
\label{sec:rocq}
We have mechanized the calculus and proved \cref{thm:soundness} in
Rocq.  The proof sources and build instructions are available on
Zenodo~\cite{rocqArtifact}.\footnote{Source paths in the code annotations
are relative to \texttt{rocq/theories/}.}
This section explains how programs, types, and logical predicates
are represented, with particular attention to well-formedness, type
relations, and variable binding.

\subsection{Mixed Embedding}
\label{sec:rocq-mixed}

We use a \emph{mixed embedding}~\cite{LiWeirich22}: the structure of
programs and types is deeply embedded, while the predicates in integer
types, assertions, and ownership terms are shallowly embedded as Rocq
predicates.
This lets us reuse Rocq's logical connectives and proofs without defining a
formula syntax and its interpretation.  The cost is that variable
dependencies and substitution in the shallow components must be expressed
semantically, as discussed in \cref{sec:rocq-semantics,sec:rocq-binding}.
The combination of deeply embedded programs and proof systems with
shallowly embedded assertions is often found in program
verification~\cite{WangCaoTao24}.

The expression language is an inductive datatype.  Expression
variables use a \emph{locally nameless
  representation}~\cite{Chargueraud12}: bound occurrences are
represented by de Bruijn indices, while free occurrences retain names.
Free variable identifiers have type \texttt{var}, represented by
natural numbers.  The type \texttt{term\_var} distinguishes bound
occurrences (\texttt{TermBound index}) from free occurrences
(\texttt{TermFree name}, with \texttt{name : var}).  Binding
constructors therefore do not carry source-level binder names.  The
following excerpt shows variable occurrences, integer bindings
$ \LET  x  =  n  \IN  e $, general expression bindings
$ \LET  x  =  e_{{\mathrm{1}}}  \IN  e_{{\mathrm{2}}} $, and assertions $  \ASSERT( \varphi )   \SEQ  e $; the
remaining constructors are omitted.  In the code, \texttt{Z} denotes
the type of integers, and \texttt{Prop} is Rocq's sort of
propositions; the type \texttt{term\_assertion}, which represents shallowly
embedded predicates, will be explained shortly.
\begin{lstlisting}[style=rocq]
(* Language/Syntax.v:257-271 (excerpt) *)
Inductive expr : Type :=
| EVar (x : term_var)
| ELetInt (n : Z) (body : expr)
(* ... *)
| ELetExpr (bound body : expr)
(* ... *)
| EAssert (test : term_assertion) (body : expr).
\end{lstlisting}
As a running example, consider $ \LET  x  =  0  \IN    \ASSERT( x \,  =  \,  0  )   \SEQ  x  $.
Both uses of $x$ in the body refer to the nearest lexical binder,
identified by \texttt{TermBound 0}.  It is expressed
by
\lstinline[style=rocq]|ELetInt 0
where the ellipsis will be revealed shortly.

The \texttt{term\_assertion} carried by \texttt{EAssert} illustrates the
shallow part of the expression encoding.  Instead of a formula datatype with
a separate interpretation function, an assertion carries its semantic meaning
directly: a predicate returning a proposition in Rocq's \texttt{Prop}.
Its interface is defined as follows.
\begin{lstlisting}[style=rocq]
(* Language/Syntax.v:16-19 *)
Inductive value : Type :=
| VInt (n : Z)
| VPtr (b : base) (offset : Z)
| VNull.

(* Base/Names.v:72 *)
Definition term_environment (A : Type) : Type := term_var -> option A.

(* Language/Syntax.v:94-95 *)
Definition term_assertion : Type := term_environment value -> Prop.
\end{lstlisting}
An assertion describes a property of the variable values supplied by its
environment.
The environment maps each term-variable occurrence to an optional
runtime value: an integer (\texttt{VInt}), an address (\texttt{VPtr}), or
null (\texttt{VNull}).  The type \texttt{base} represents base addresses,
encoded as natural numbers.
The example expression above is represented by
\begin{lstlisting}[style=rocq]
ELetInt 0%Z
  (EAssert (fun env => env (TermBound 0%nat) = Some (VInt 0%Z))
           (EVar (TermBound 0%nat)))
\end{lstlisting}
A \texttt{term\_assertion} receives no heap, so it cannot inspect heap
cells, whereas---the way it is defined---it can depend on pointer
variables as well as integer variables.  The typing rule, however,
requires the predicate to depend only on integer-typed variables in
the current type environment, enforced by well-formedness.
We will discuss well-formedness in \cref{sec:rocq-semantics}.

The Rocq code below shows one use of \texttt{term\_assertion}.
\begin{lstlisting}[style=rocq]
(* Language/Syntax.v:112-114 *)
Definition term_assertion_holds_runtime
    (test : term_assertion) (R : register) : Prop :=
  test (runtime_term_environment R).

(* Language/Semantics.v:56-58 *)
Inductive configuration : Type :=
| Running (H : heap) (R : register) (e : expr)
| AliasFail.

(* Language/Semantics.v:65-66, 144-146 (excerpt) *)
Inductive step (D : function_defs) :
    configuration -> configuration -> Prop :=
(* ... *)
| RS_Assert : forall H R test body,
    term_assertion_holds_runtime test R ->
    (Running H R (EAssert test body)) ---> D (Running H R body).
\end{lstlisting}
The inductive type \texttt{step D C1 C2} (also denoted by \lstinline[style=rocq]|C1 ---> D C2|) corresponds to
the paper's reduction judgment $  \tuple{ \ottnt{H} ,  \ottnt{R} ,  e }     \longrightarrow _{  D  }     \tuple{ \ottnt{H'} ,  \ottnt{R'} ,  e' }  $.
The type of the constructor \texttt{RS\_Assert} represents that
$  \ASSERT( \varphi )   \SEQ  e $ proceeds to the body $e$ when $\models  \ottsym{[}  \ottnt{R}  \ottsym{]} \, \varphi$ holds.
Thus, the premise applies the assertion predicate to
\texttt{runtime\_term\_environment R}, which interprets named free
occurrences by register lookup.  The rule is relational: it requires evidence
that the predicate holds, not a computation of its truth value.

The following definitions show how these predicates are combined with the
deep structure of ownership terms and types.

\noindent\begin{minipage}{\linewidth}
\begin{lstlisting}[style=rocq]
(* Logic/Refinement.v:10, 12, 17-18 *)
Definition valuation : Type := var -> option Z.
Definition index_stack : Type := list Z.
Definition refinement : Type := valuation -> index_stack -> Z -> Prop.

(* Logic/Refinement.v:158-162 *)
Record guard : Type := Guard {
  guard_holds : valuation -> index_stack -> Prop;
  guard_decides : forall rho indices,
    {guard_holds rho indices} + {~ guard_holds rho indices}
}.

(* Logic/Ownership.v:11-13 *)
Inductive ownership : Type :=
| OwnZero
| OwnWhen (test : guard) (q : Q) (otherwise : ownership).

(* Typing/Types.v:8-10 *)
Inductive ty : Type :=
| TyInt (predicate : refinement)
| TyRef (element : ty) (permission : ownership).
\end{lstlisting}
\end{minipage}

Refinements in types are also shallow predicates, but use a different
interface.  Unlike an expression assertion, an integer refinement constrains
a distinguished integer value (denoted by $\nu$).
Its arguments are a \texttt{valuation}, an \texttt{index\_stack}, and the integer being refined.  The
valuation maps named program variables (\texttt{var}) to optional integers, \texttt{var -> option Z}; the index
stack records array-index values (represented by variables bound by $\Pi$) in innermost-first order.
When interpreting a reference's ownership term and element type at index
$i$, we prepend $i$ to the stack; predicates access an enclosing array-index
variable's value by its position in this list.
For example, at an inner array index of $5$ within an outer array index of
$2$, the stack is \texttt{[5; 2]}.
The refinement saying that an integer equals a fixed integer, say
\texttt{42}, is represented by
\lstinline[style=rocq]|fun _ _ nu => nu = 42

The type \texttt{ownership} of ownership terms is fairly
straightforward, except for predicates $\varphi$ represented by a Rocq
record of type \texttt{guard}.  \texttt{OwnWhen test q otherwise}
denotes \texttt{q} when the predicate represented by \texttt{test}
holds and otherwise uses the remaining term; \texttt{OwnZero}
represents $ \mathbf{0} $.  The type \texttt{guard} pairs a predicate
\texttt{guard\_holds} with a decision function
\texttt{guard\_decides}.  The type \lstinline[style=rocq]|{P} + {~P}|
is an executable two-way choice carrying evidence of \texttt{P} in one
branch and of its negation in the other.  For a given valuation and
index stack, this function supplies the choice needed by ownership
interpretation.  This decision function is used to define the
interpretation of \texttt{ownership} as a (total) function returning a
rational number (Rocq type \texttt{Q}).

The different types for predicates reflect where they occur.
Assertions are part of expression syntax and must represent dependencies on
both free and bound variables.
Program-variable occurrences in types, however, are always named free
variables.  Array-index variables remain bound by $\Pi$ and are
represented separately through \texttt{index\_stack}.
Thus, \texttt{valuation} used to define
\texttt{refinement} and \texttt{ownership} uses \texttt{var}, rather
than \texttt{term\_var}.  Assertions and type refinements need no
decision function as in \texttt{guard\_decides}, because their
satisfaction is used as a proposition in the soundness proof.

By using this machinery, the inductive definition for types is
straightforward: \texttt{TyInt} carries a shallow predicate, whereas
\texttt{TyRef} retains the inductive structure needed to reason about
nested references.

\subsection{Well-Formedness and Type Relations}
\label{sec:rocq-semantics}

The judgments on types combine structural checks with semantic conditions
on refinements and ownership terms.  We first formalize which variables a
shallow predicate may depend on, then use this notion of scoping to define
type and environment well-formedness.  Type relations compare refinements
and ownership denotations under the current environment.

\paragraph{Semantic scoping.}
For a syntactic formula, well-formedness can check that its free variables
belong to a permitted set.  For a shallow predicate, we instead require that
its truth be unchanged whenever its environments agree on the permitted
variables and indices.  We call this condition \emph{semantic scoping}.
For instance, the assertion in our running example must be unaffected by
changes to bindings other than the one for its \texttt{let}-bound variable.
This semantic account of dependency is related to \emph{unrestriction} in
Isabelle/UTP~\cite[Section~5.4]{FosterZeydaWoodcock15}, which expresses that
changing a variable's value does not affect a predicate.  Here, the dependency
bound consists of permitted named variables and a prefix of the array-index
stack.  For integer refinements, this condition is defined as follows.

\noindent\begin{minipage}{\linewidth}
\begin{lstlisting}[style=rocq]
(* Logic/Refinement.v:14-15, 27-32 (reformatted) *)
Definition valuation_agree
    (xs : list var) (rho1 rho2 : valuation) : Prop :=
  forall x, In x xs -> rho1 x = rho2 x.

Definition refinement_scoped
    (xs : list var) (depth : nat) (p : refinement) : Prop :=
  forall rho1 rho2 indices1 indices2 nu,
    valuation_agree xs rho1 rho2 ->
    firstn depth indices1 = firstn depth indices2 ->
    p rho1 indices1 nu <-> p rho2 indices2 nu.
\end{lstlisting}
\end{minipage}
Here, \texttt{xs} lists the named variables on which \texttt{p} may depend,
and \texttt{depth} represents the number of nested $\Pi$s surrounding \texttt{p}.%
\footnote{\texttt{firstn depth} selects the prefix of the index stack containing
the permitted array-index values.}
Agreement on these inputs must preserve
truth for each fixed value of the distinguished integer \texttt{nu}.

Similarly, we define
\texttt{term\_assertion\_scoped} for assertions
and \texttt{ownership\_scoped} for ownership terms.

\paragraph{Type and environment well-formedness.}
Type well-formedness is defined using the following auxiliary predicate.

\begin{lstlisting}[style=rocq]
(* Typing/Types.v:190-202 *)
Fixpoint ty_wf_names
    (integer_names : list var) (depth : nat) (t : ty) : Prop :=
  match t with
  | TyInt p => refinement_scoped integer_names depth p
  | TyRef element r =>
      ownership_wf r /\
      ownership_scoped integer_names (S depth) r /\
      ty_wf_names integer_names (S depth) element /\
      (forall rho indices i,
        length indices = depth ->
        ownership_denote r rho (i :: indices) == 0 ->
        ty_empty_at element rho (i :: indices))
  end.
\end{lstlisting}
The parameter \texttt{integer\_names} lists the names of
integer-typed variables, and \texttt{depth} counts the enclosing array-index
binders.  An integer type \texttt{TyInt p} must satisfy
\texttt{refinement\_scoped integer\_names depth p}: its predicate may depend
only on those names and array indices.
For a reference type, it checks four obligations: the
rational numbers in its ownership term lie in $[0,1]$; the predicates in that
term have the permitted scope;
its element type is well-formed with one additional array index; and its
element type is empty wherever ownership is zero, for every valuation
\texttt{rho}.\footnote{\texttt{==} denotes rational equality.}
The \texttt{TyRef} constructor binds one additional array index, so both
its ownership term and element type
are scoped at \texttt{S depth}.  The final conjunct uses
\texttt{ty\_empty\_at} to express emptiness: an integer refinement must hold for
every integer, and a reference must have zero ownership and recursively
empty elements at every index.
To check type well-formedness, \texttt{ty\_wf Gamma t}, which represents $\Gamma  \vdash   \tau  \mbox{ ok} $, calls
\texttt{ty\_wf\_names} with the names of the integer-typed bindings in
\texttt{Gamma} and depth zero.  Thus, as in \rn{WF-Ref}, the environment
supplies variable scope but no assumptions for the emptiness implication.
The type may still depend on these integer variables; its
zero-ownership/emptiness implication must hold for all their values.
Here, a type environment \texttt{Gamma} has Rocq type \texttt{tyenv}, represented
as a list of variable--type pairs.
The predicate \texttt{env\_wf Gamma} requires each variable name to occur at
most once in \texttt{Gamma} and checks each binding's type under the whole
environment.

\paragraph{Type relations.}

Subtyping, type equivalence, and type addition are defined by
recursion on the structure of types, using semantic comparisons of
integer refinements and ownership terms.  For example, \texttt{subtype
  Gamma t1 t2}, which represents $\Gamma  \vdash   \tau_{{\mathrm{1}}}   \leq   \tau_{{\mathrm{2}}} $, is defined as
follows:
\begin{lstlisting}[style=rocq]
(* Typing/Relations.v:16-26 (notation expanded) *)
Fixpoint subtype (Gamma : tyenv) (t1 t2 : ty) : Prop :=
  match t1, t2 with
  | TyInt p1, TyInt p2 =>
      forall rho indices nu, env_formula Gamma rho -> p1 rho indices nu -> p2 rho indices nu
  | TyRef e1 r1, TyRef e2 r2 =>
      ownership_le (env_assumption Gamma) r2 r1 /\ subtype Gamma e1 e2
  | _, _ => False
  end.
\end{lstlisting}
The integer case requires implication under the current environment's
refinements: \texttt{env\_formula Gamma rho} says that \texttt{rho}
satisfies those refinements, and the implication must hold for every
index stack and distinguished integer value.  The reference case uses
\texttt{ownership\_le} to require that the denotation of \texttt{r2}
be no greater than that of \texttt{r1}, pointwise under the same
environment assumption, packaged as \texttt{env\_assumption Gamma}.
The recursive call relates the element types.  Type equivalence and
addition are defined similarly.

\subsection{Binding and Typing}
\label{sec:rocq-binding}

Expression typing combines structural rules over programs with semantic
conditions on their shallow predicates.  Binding constructs must also
account for the choice of a fresh name.  We first explain
opening~\cite[Section~3.1]{Chargueraud12} and substitution, then illustrate
expression typing with the assertion and let rules.  Lexical expression
binders and array-index binders remain distinct: the former use
\texttt{term\_var}, whereas the latter supply integer values through
\texttt{index\_stack}.

\paragraph{Opening and substitution.}
Opening makes the body of a nameless binder available for typing or
evaluation by replacing its bound occurrences with a named free variable.
In our running example, $ \LET  x  =  0  \IN    \ASSERT( x \,  =  \,  0  )   \SEQ  x  $,
\rn{Rs-LetInt} binds zero to a fresh register name \texttt{fresh} and opens
the let body with that name.  The intended effect on the encoded expression
is shown below:

\noindent\begin{minipage}{\linewidth}
\begin{lstlisting}[style=rocq]
(* Illustrative example; opening is defined in
   Language/Syntax.v:377-378. *)
(* Before the binding step *)
ELetInt 0%Z
  (EAssert (fun env => env (TermBound 0%nat) = Some (VInt 0%Z))
           (EVar (TermBound 0%nat)))

(* After opening the body with fresh : var *)
EAssert (fun env => env (TermFree fresh) = Some (VInt 0%Z))
        (EVar (TermFree fresh))
\end{lstlisting}
\end{minipage}
Although \texttt{TermBound 0} is visible in the displayed \texttt{fun},
an assertion is represented as a Gallina function, not as formula syntax.
The opening operation cannot traverse an arbitrary function's body
and replace that occurrence syntactically.

We define the opening function \texttt{open\_expr\_at} so that, when opening
an assertion, it changes the environment supplied to the original predicate.
In the running example, given the new predicate's argument \texttt{env},
we apply the original predicate to an environment that returns
\texttt{env (TermFree fresh)} for \texttt{TermBound 0} and agrees with
\texttt{env} elsewhere:

\noindent\begin{minipage}{\linewidth}
\begin{lstlisting}[style=rocq]
(* Illustrative example; environment reindexing is defined in
   Base/Names.v:74-77. *)
fun env =>
  (fun env' => env' (TermBound 0%nat) = Some (VInt 0%Z))
    (fun u => match u with
              | TermBound 0%nat => env (TermFree fresh)
              | _ => env u
              end)
\end{lstlisting}
\end{minipage}
The application reduces to
\texttt{env (TermFree fresh) = Some (VInt 0\%Z)}, yielding exactly the
opened predicate shown above without inspecting the original function.

Substitution here is only variable-for-variable: it replaces named free
variables with other named free variables, not with values.  It uses the
same environment-reindexing idea as opening, while leaving bound indices
unchanged.  Substitution in type refinements and predicates in ownership
terms likewise changes the environment supplied to the predicate.

Pointer arithmetic changes the array-index values supplied to predicates.
For example, if $x = y  \boxplus  2$, a cell at offset $5$ from $y$
is at offset $3$ from $x$.  When relating their types, we therefore
interpret the predicates for $x$ with index $3$ where those for $y$
use index $5$.  In Rocq, this changes the corresponding entry of the
\texttt{index\_stack} supplied to the predicates.  Under a nested reference,
a stack such as \texttt{[j; 5]} becomes \texttt{[j; 3]}, leaving the inner
array index \texttt{j} unchanged.  Like opening, this transformation changes
the argument supplied to a shallow predicate rather than inspecting its body.

\paragraph{Expression typing.}
The encoding of \rn{T-Assert} illustrates how the paper's implicit
well-formedness requirements become explicit premises in Rocq.
The following excerpt of the inductive relation \texttt{has\_type}
shows this rule, with the auxiliary premise
\texttt{typing\_judgment\_wf} expanded into its three components.

\noindent\begin{minipage}{\linewidth}
\begin{lstlisting}[style=rocq]
(* Typing/Rules.v:378-379, 567-574 (excerpt;
   typing_judgment_wf from lines 329-331 unfolded) *)
Inductive has_type (Theta : ftyenv) :
    tyenv -> expr -> ty -> tyenv -> Prop :=
(* ... *)
| T_Assert : forall Gamma Gamma' test body t,
    expr_scoped (dom Gamma) 0 (EAssert test body) ->
    term_assertion_scoped (env_integer_names Gamma) 0 test ->
    assertion_valid Gamma test ->
    Theta ; Gamma |- body : t => Gamma' ->
    (env_wf Gamma /\ env_wf Gamma' /\ ty_wf Gamma' t) ->
    Theta ; Gamma |- (EAssert test body) : t => Gamma'.
\end{lstlisting}
\end{minipage}
The notation \lstinline[style=rocq]/Theta ; Gamma |- e : t => Gamma'/
corresponds to $ \Theta   \mid   \Gamma   \vdash   e  :  \tau   \produces   \Gamma' $.
The first premise requires the expression to be scoped in \texttt{Gamma};
the second restricts the assertion to integer-typed variable names.
The premise \texttt{assertion\_valid Gamma test} requires the assertion to
follow from the integer-environment formula $ \fml{ \Gamma } $, represented by
\texttt{env\_formula}, while the next premise types the body.
The final conjunction requires well-formed input and output environments
and a returned type well-formed under the output environment.
Recording these conditions at every typing node makes the paper's implicit
well-formedness convention available in proofs by induction on typing.

\paragraph{Fresh names and binder typing.}
We use \emph{cofinite quantification}~\cite{Chargueraud12} to type locally
nameless binders.  Typing environments remain named, so a binder rule checks
the body after opening it with a fresh name.  The body-typing premise must
hold for every name outside a finite exclusion set; for example,
\texttt{T\_Let} excludes the names in the intermediate typing environment.
In the proof of Preservation, we can therefore instantiate this premise
with the fresh name selected by the reduction rule.  The following excerpt of \texttt{T\_Let}
makes this quantification explicit:

\noindent\begin{minipage}{\linewidth}
\begin{lstlisting}[style=rocq]
(* Typing/Rules.v:435-445 (excerpt;
   scoping and well-formedness premises omitted) *)
| T_Let : forall Gamma Gamma1 Gamma' bound body bound_ty t (discarded : var -> ty),
    Theta ; Gamma |- bound : bound_ty => Gamma1 ->
    (forall fresh_name,
      ~ In fresh_name (dom Gamma1) ->
      Theta ; (extend fresh_name bound_ty Gamma1) |-
        (open_expr_one fresh_name body) : t =>
        (extend fresh_name (discarded fresh_name) Gamma')) ->
    Theta ; Gamma |- (ELetExpr bound body) : t => Gamma'
\end{lstlisting}
\end{minipage}
Here, \texttt{extend x t Gamma} adds the binding $ x \COL \tau $ to
\texttt{Gamma}, and \texttt{open\_expr\_one x body} opens the body with
\texttt{x}.  For an opening name \texttt{x}, \texttt{discarded x} is the
type remaining for the local variable after the body finishes.  The body's
output environment therefore consists of \texttt{Gamma'} extended with
\texttt{x : discarded x}; the conclusion drops this local binding and
retains \texttt{Gamma'}.

In our mechanization, we choose to represent the remaining local type by a
family \texttt{discarded : var -> ty}, indexed by the opening name.  This
representation choice accommodates name-dependent refinements within the
binding discipline described above.  For example, a remaining integer type
with refinement $\nu \,  =  \, x$ must instead use $\nu \,  =  \, y$ when the same
body is opened with $y$.  The family supplies the appropriate local type
for each fresh name in the cofinite premise.  By contrast, the returned
type \texttt{t} and the outer output environment \texttt{Gamma'} are fixed
across that premise and do not depend on the choice of opening name.

\subsection{Checked Results and Scope}
\label{sec:rocq-results}

The Rocq development verifies \cref{thm:soundness} by formalizing
the initialization, Preservation, and Progress argument outlined in
\cref{sec:soundness}.  Detailed proofs and annotations identifying the
corresponding Rocq definitions and lemmas are given in \cref{sec:proofs}.
The verification concerns type soundness of the calculus: typing premises
involving shallow predicates are proof obligations in Rocq, while type
inference, SMT translation, and solver correctness are outside the verified
scope.

\paragraph{AI-assisted development.}
We used OpenAI Codex interactively to assist with developing and revising
the Rocq formalization, including definitions, lemma statements, and proof
scripts, and to identify and repair discrepancies in the prose proofs.
The development proceeded under author supervision through iterative
discussion, revision, and machine checking.  The resulting proofs were
checked by Rocq, and the main soundness theorem has no additional axiomatic
assumptions.  The authors remain responsible for the formalized statements,
their correspondence with the paper, and all claims in this article.

\begin{revision}
\section{Discussion}
\label{sec:discussion}
\paragraph{Applicability to mainstream languages}
Our target language is a C-like imperative language with explicit
pointer arithmetic and heap allocation.  Applying our type system is
most natural for \emph{type-safe subsets} of C, programs that do not
perform arbitrary pointer casts; many numerical and scientific programs---in which nested arrays are heavily used---already satisfy
these restrictions.
The \textbf{alias} construct can be
seen as a dynamically checked annotation that replaces the need
for a separate must-alias analysis.
Our method can be used to verify functional correctness properties for such programs.

Beyond C, our system has a natural connection to modern typed
languages.  For example, in OCaml, mutable arrays and references
are first-class values (although it does not support pointer arithmetic).
Our method could be used to verify
functional correctness of OCaml programs in which mutable arrays and references
are involved by accommodating ownership.
In this regard, Cameleer~\cite{Pereira2021Cameleer},
a functional-correctness verifier for OCaml, 
already supports mutable references~\cite{Pereira2021Cameleer};
however, their backend Why3 poses a restriction that all the aliases of a reference
and an array must be known statically. Our ownership-based approach could provide
a more flexible alternative to Cameleer's approach.

For Rust, the borrow checker already enforces a strict ownership discipline at the language level, and mature verification tools such as Prusti~\cite{DBLP:journals/pacmpl/Astrauskas0PS19} and Creusot~\cite{DBLP:conf/icfem/DenisJM22} leverage this discipline for functional correctness verification. Our work addresses a different setting, languages without built-in ownership guarantees, where aliasing must be tracked explicitly. Nevertheless, the idea of index-dependent ownership terms could complement Rust verification tools when reasoning about unsafe code blocks that perform raw pointer arithmetic on nested arrays, a direction we leave for future work.

\paragraph{Towards more general heap structures}
Although the present work focuses on nested arrays, our formulation of
ownership terms can be applied to more general data structures.
Ownership terms are, in essence, predicate-guarded maps from indices to fractional
permissions, and this mechanism is not inherently tied to arrays.
For example, pairs and tuples can be handled by extending ownership expressions
to allow mentioning the position of each component.
C-like structures also can be handled by extending ownership expressions
to handle the fields of a structure.
However, further extensions for heap-allocated data structures such as linked lists
and trees would be more challenging, as they require introducing
recursive types and suitably extended ownership terms.

\end{revision}

\section{Related Work}
\label{sec:relatedWork}
\paragraph{\consort and its extensions}

Toman et al.~\cite{DBLP:conf/esop/TomanSSI020} have proposed and implemented \consort, an ownership refinement type system for verifying the functional correctness of imperative programs that involve mutable references and aliasing.
Their type system employs fractional ownership types with refinement, which provide flow-sensitive aliasing information through types and enable strong updates to be performed soundly.
Tanaka et al.~\cite{DBLP:conf/pepm/TanakaSK24} extended \consort to support arrays and pointer arithmetic.

Our type system is a further extension of Tanaka et al.'s to support nested arrays, in which the ownership of a pointer to an inner array can depend on the index of that array in the outer array.
We also implemented a prototype type inference tool that supports nested arrays, whereas the implementation by Tanaka et al.\ only supports one-dimensional arrays.

\paragraph{Refinement types}

Refinement types~\cite{DBLP:conf/pldi/FreemanP91}, including liquid types~\cite{DBLP:conf/pldi/RondonKJ08,Rondon2012Liquid,LiquidTypeTutorial}, are types constrained with extra information for values, typically expressed as logical predicates.
They can be seen as a lightweight form of dependent types~\cite{DBLP:conf/popl/XiP99,DBLP:journals/jfp/Xi07}.
They have been applied to automated verification for various programming paradigms, including higher-order non-deterministic languages~\cite{10.1145/3158100}, class-based functional object-oriented languages~\cite{sun_et_al:LIPIcs.ECOOP.2024.39}, and smart contracts~\cite{10.1007/978-3-030-72013-1_14,DBLP:journals/ngc/NishidaSCKFSI22,icon}.
The studies mentioned above primarily target a functional or smart-contract language, but not an imperative heap-manipulating language like we target.

\paragraph{Ownership refinement types}

Recently, ownership refinement types, i.e., ownership types with refinements, have been studied for automated verification of imperative memory-manipulating programs.
\consort by Toman et al.~\cite{DBLP:conf/esop/TomanSSI020} is one of the earliest systems in this line.

RefinedC~\cite{DBLP:conf/pldi/SammlerLKMD021} is a verification framework for C programs based on ownership refinement types.
It automates foundational proof for C programs in Rocq, where typing and proof rules are mechanized and proved sound with respect to an idealized semantics of C.
RefinedC is based on the Iris framework~\cite{DBLP:conf/popl/JungSSSTBD15,DBLP:journals/jfp/JungKJBBD18} for separation logic and thus highly expressive.
However, it does not provide very powerful automation, as it does not use external backend solvers outside of Rocq.
\consort and its extensions including our work, on the other hand, are designed to be more lightweight and automated.

\paragraph{Rust verification}

There are several recent studies on automated verification of Rust programs using ownership refinement types.
RefinedRust~\cite{DBLP:journals/pacmpl/GaherSJKD24} is a foundational automated verification framework for Rust based on RefinedC~\cite{DBLP:conf/pldi/SammlerLKMD021}.
Like RefinedC, it is highly expressive but does not provide as powerful automation as ours.
Flux~\cite{DBLP:journals/pacmpl/LehmannGVJ23} and Thrust~\cite{DBLP:journals/pacmpl/OgawaSU25} are frameworks for automated verification of Rust programs by annotating Rust's ownership types with refinement.
They provide nice automation through external backend SMT or CHC solvers.
However, ownership is enforced by Rust and cannot depend on dynamic values, unlike our work.

Verus~\cite{DBLP:journals/pacmpl/LattuadaHCBSZHPH23,DBLP:conf/sosp/0001HB0CLSACHHL24} is a well-engineered framework for automated verification of Rust programs, through a wrapper language that extends Rust's ownership types with ghost state and refinement using an SMT solver.
It can be used for verifying even unsafe Rust and has been actively used for various verification projects.
However, Verus combines Rust's borrow checker and manual annotations for ownership management, and thus does not provide strong automation for ownership, especially regarding fractions, quite unlike our work.

There are also Rust verifiers that translate Rust programs into functional programs~\cite{Electrolysis,DBLP:conf/icfem/DenisJM22,DBLP:journals/pacmpl/HoP22} or logical formulas~\cite{DBLP:conf/esop/0002T020,DBLP:journals/toplas/MatsushitaTK21} leveraging Rust's ownership types, providing an interface different from refinement types.
However, they are quite bound to Rust's ownership principles, unlike our work.

\paragraph{Index-dependent types for arrays}

The idea of making element types depend on the index in an array has long been known and used, such as in \textsc{Dsolve} \cite{DBLP:conf/pldi/KawaguchiRJ09} and Lean 3's \texttt{d\_array}.
Still, the type system of Tanaka et al.~\cite{DBLP:conf/pepm/TanakaSK24} and ours are peculiar in that they combine that idea with ownership types and that they support pointer arithmetic on array pointers.

RefinedC~\cite{DBLP:conf/pldi/SammlerLKMD021} also has a library for an array type whose element types can differ depending on the index, but interestingly the array type takes a list of element types instead of a map from indices to element types.
A list of element types is isomorphic to a pair of a length and a map from indices up to the length to element types.
Still, this representation does not support ownership at negative indices, unlike Tanaka et al.'s system and ours.

\paragraph{Fractional ownership}

The idea of using fractional numbers $\{r \mid 0 \le r \le 1\}$ to represent pointer permission was first introduced by Boyland~\cite{DBLP:conf/sas/Boyland03}.
It can be understood as a relaxation of a coarse ownership analysis using integer-valued permissions in $\{0,1\}$ used in \cite{DBLP:conf/pldi/HeineL03}.

The idea of using fractions has then been incorporated into permission accounting in separation logic~\cite{DBLP:conf/popl/BornatCOP05}, with extensions to concurrent settings~\cite{DBLP:conf/cav/BrotherstonCHW20}.
Related type-theoretic treatments of resource usage appear in capability calculi~\cite{10.1145/292540.292564,10.1145/1411204.1411235}, though these use non-fractional capabilities.
The fractional-permission approach has been applied to memory deallocation~\cite{DBLP:conf/aplas/SuenagaK09,DBLP:conf/aplas/SonobeSI14}, race freedom~\cite{DBLP:conf/pldi/Terauchi08}, concurrent resource deallocation~\cite{DBLP:conf/oopsla/SuenagaFI12}, authenticity~\cite{DBLP:conf/aplas/KikuchiK07,DBLP:conf/esop/KikuchiK09,DBLP:conf/atva/DahlKSH11}, and functional-correctness verification~\cite{DBLP:conf/vmcai/NakayamaMSSK24,DBLP:conf/pepm/UenoT0T21}, with a fraction attached to a reference variable, a region, etc.

\consort and its extensions, including our present work, belong to the fractional-permission vein, attaching fractions to refinement-typed pointers.
Our work is quite special in the treatment of fractional ownership for nested arrays.

\paragraph{Separation logic}

Separation logic~\cite{DBLP:conf/popl/IshtiaqO01,DBLP:conf/csl/OHearnRY01,DBLP:conf/lics/Reynolds02} is a program logic that enable modular reasoning about heap states and other resources by using non-persistent assertions that have ownership of resources.
There are various verification tools for imperative languages based on separation logic, Smallfoot~\cite{DBLP:conf/fmco/BerdineCO05}, VeriFast~\cite{DBLP:conf/nfm/JacobsSPVPP11}, Viper~\cite{DBLP:conf/vmcai/0001SS16}, Prusti~\cite{DBLP:journals/pacmpl/Astrauskas0PS19}, Gobra~\cite{DBLP:conf/cav/WolfACOPM21}, Steel~\cite{DBLP:journals/pacmpl/SwamyRFMAM20,DBLP:journals/pacmpl/FromherzRSGMMR21}, and Pulse~\cite{DBLP:journals/pacmpl/EbnerMRDFRS25}.
Separation logic has also served for static analysis and bug finding in Infer~\cite{DBLP:journals/jacm/CalcagnoDOY11,DBLP:journals/cacm/DistefanoFLO19} and CN~\cite{DBLP:journals/pacmpl/PulteMSMSK23,DBLP:journals/pacmpl/AamerBKKEMMKPPS26}.
The Iris framework~\cite{DBLP:conf/popl/JungSSSTBD15,DBLP:journals/jfp/JungKJBBD18} provides a higher-order concurrent separation logic mechanized in Rocq, and it has served as foundations for ownership types~\cite{DBLP:journals/pacmpl/0002JKD18} and verification tools~\cite{DBLP:conf/pldi/MatsushitaDJD22,DBLP:journals/pacmpl/HanceEMD26}.
\consort and its extensions, including ours, are not as expressive and flexible as verification based on separation logic, but provide a lightweight automated approach that naturally extends an existing, standard type system with ownership and refinement.

\section{Conclusion}
\label{sec:conclusion}
In this paper, we proposed an extension to a type system featuring ownership and refinement types 
to support index-sensitive nested arrays, and we formally proved its soundness with mechanization in Rocq. 
We also implemented a verifier based on this type system and confirmed, 
through comparison with Tanaka et al.'s ConSORT implementation, that our verifier's capabilities are comparable. 
Furthermore, we successfully verified non-trivial properties of several programs involving nested arrays, 
a task not achievable with the previous work.

One limitation of our current implementation of the verifier is that it requires type annotations on function arguments beyond simple types, even though ownership/predicate inference is
theoretically possible without annotations.
Improving the type-inference procedure in this regard is left for future work.

\begin{revision}
Another practical concern is the need for $\ALIAS$ expressions.
Our prototype already automates the insertion of $\ALIAS$ expressions
at syntactically determined points, and this mechanism eliminates
the need for manual insertion of $\ALIAS$ expressions in the majority of our benchmarks
(\cref{table:experiments}).
Nevertheless, certain patterns, particularly those involving
non-trivial aliasing across function boundaries or
multiple simultaneous redistributions, still require manual insertion.
Extending the automated insertion to cover a wider range of aliasing patterns,
as well as exploring more flexible ownership splitting strategies
that would reduce the number of $\ALIAS$ expressions needed in the first place,
remain important directions for future work.
We are looking at the direction of applying recent
progress on demand-driven must-alias analysis,
such as Boomerang~\cite{DBLP:conf/ecoop/SpathDAB16}
and its sparsification-based accelerations~\cite{DBLP:conf/icst/KarakayaB23},
to discharge $\ALIAS$ annotations statically.
\end{revision}

\begin{acks}
We appreciate the ECOOP reviewers' constructive comments.
We are also grateful to
Naoki Kobayashi,
Tsubasa Matsumoto,
Ken Sakayori,
and Izumi Tanaka
for valuable comments and discussions on this work.
Takashi Suwa helped us test the artifact.
This work is partially supported by
JSPS KAKENHI Grant Number 20H05703, Japan, %
JST CREST Grant Number JPMJCR2012, Japan, JSPS KAKENHI Grant Number 25H01113, Japan, %
the Hakubi Project at Kyoto University for the second author, and JSPS KAKENHI Grant Number JP24KJ0133, Japan. %
The second author is hosted by MPI-SWS as a JSPS Overseas Research Fellow at the time of submission.
\end{acks}

\bibliography{references}

\ifdefined\VersionLong%
\appendix
\crefalias{section}{appendix}
\crefalias{subsection}{appendix}
\crefalias{subsubsection}{appendix}
\onecolumn
\section{Full Definition}
\label{sec:def}

\subsection{Syntax}

  \[
    \begin{array}{llll}
      \tau^{-} & ::= &  \TINT \mid  \tau^{-}  \TREF  \\
      e & ::= & x \mid  \LET  x  =  n  \IN  e  \mid  \LET  x  =  \ottkw{null}  \IN  e 
               \mid  \LET  x  =   y  \mathop{  -  }  z   \IN  e  \\
        & \mid &  \IFNP  x  \THEN  e_{{\mathrm{0}}}  \ELSE  e_{{\mathrm{1}}}  \mid  \LET  x  =   f (  y_{{\mathrm{1}}} ,\ldots, y_{\ottmv{n}}  )   \IN  e   \mid  \LET  x  =  e_{{\mathrm{0}}}  \IN  e_{{\mathrm{1}}}  \\
        & \mid &  \LET  x  =   \ALLOC  y   \ottsym{:}    \tau^{-}  \TREF    \IN  e  \mid  \LET  x  =   \ast  y   \IN  e 
          \mid     x  \WRITE  y   \SEQ  e  \mid  \LET  x  =   y   \boxplus   z   \IN  e  \\
        & \mid &   \ALIAS(  x  =  y   \boxplus   z  )   \SEQ  e \mid   \ALIAS(  x  = \ast  y  )   \SEQ  e  \mid   \ASSERT( \varphi )   \SEQ  e   \\
      \ottnt{fd} & ::= & f  \mapsto   (  x_{{\mathrm{1}}}  , \ldots ,  x_{\ottmv{n}}  )  \, e \\
      D & ::= & \ottsym{\{}  \ottnt{fd_{{\mathrm{1}}}}  \ottsym{,} \, ... \, \ottsym{,}  \ottnt{fd_{\ottmv{n}}}  \ottsym{\}} \\
      \ottnt{P} & ::= &  \tuple{ D ,  e } 
    \end{array}
  \]

\subsection{Operational Semantics}
\label{sec:opsemFull}

\infrule[Rs-LetInt]{
   x'  \not\in   \DOM( \ottnt{R} )  
}{
    \tuple{ \ottnt{H} ,  \ottnt{R} ,   \LET  x  =  n  \IN  e  }     \longrightarrow _{  D  }     \tuple{ \ottnt{H} ,  \ottnt{R}  \ottsym{\{}  x'  \mapsto   n   \ottsym{\}} ,    [  x'  /  x  ]    e  }  
}
\infrule[Rs-LetNull]{
   x'  \not\in   \DOM( \ottnt{R} )  
}{
   \tuple{ \ottnt{H} ,  \ottnt{R} ,   \LET  x  =  \ottkw{null}  \IN  e  }   \longrightarrow _D \tuple{H, R\{ x'  \mapsto  \NULL \},   [  x'  /  x  ]    e }
}
\infrule[Rs-Minus]{
   x'  \not\in   \DOM( \ottnt{R} )  
  \andalso R(y), R(z) \in  \mathbb{Z} 
}{
   \tuple{ \ottnt{H} ,  \ottnt{R} ,   \LET  x  =   y  \mathop{  -  }  z   \IN  e  }   \longrightarrow _D \tuple{H, R\{ x'  \mapsto  R(y) - R(z) \},   [  x'  /  x  ]    e }
}
\infrule[Rs-IfTrue]{
  \ottnt{R}  \ottsym{(}  x  \ottsym{)} \, \le \,  0 
}{
    \tuple{ \ottnt{H} ,  \ottnt{R} ,   \IFNP  x  \THEN  e_{{\mathrm{0}}}  \ELSE  e_{{\mathrm{1}}}  }     \longrightarrow _{  D  }     \tuple{ \ottnt{H} ,  \ottnt{R} ,  e_{{\mathrm{0}}} }  
}
\infrule[Rs-IfFalse]{
  \ottnt{R}  \ottsym{(}  x  \ottsym{)} \,  >  \,  0 
}{
    \tuple{ \ottnt{H} ,  \ottnt{R} ,   \IFNP  x  \THEN  e_{{\mathrm{0}}}  \ELSE  e_{{\mathrm{1}}}  }     \longrightarrow _{  D  }     \tuple{ \ottnt{H} ,  \ottnt{R} ,  e_{{\mathrm{1}}} }  
}
\infrule[Rs-LetVar]{
   x'  \not\in   \DOM( \ottnt{R} )  
}{
    \tuple{ \ottnt{H} ,  \ottnt{R} ,   \LET  x  =  y  \IN  e  }     \longrightarrow _{  D  }     \tuple{ \ottnt{H} ,  \ottnt{R}  \ottsym{\{}  x'  \mapsto  \ottnt{R}  \ottsym{(}  y  \ottsym{)}  \ottsym{\}} ,    [  x'  /  x  ]    e  }  
}
\infrule[Rs-Deref]{
   x'  \not\in   \DOM( \ottnt{R} )   \andalso
  \ottnt{R}  \ottsym{(}  y  \ottsym{)} =  \ottsym{(}  a  \ottsym{,}  i  \ottsym{)}  \in   \DOM( \ottnt{H} )  
}{
    \tuple{ \ottnt{H} ,  \ottnt{R} ,   \LET  x  =   \ast  y   \IN  e  }     \longrightarrow _{  D  }     \tuple{ \ottnt{H} ,  \ottnt{R}  \ottsym{\{}  x'  \mapsto  \ottnt{H}  \ottsym{(}  \ottnt{R}  \ottsym{(}  y  \ottsym{)}  \ottsym{)}  \ottsym{\}} ,    [  x'  /  x  ]    e  }  
}
\infrule[Rs-Assign]{
  \ottnt{R}  \ottsym{(}  x  \ottsym{)} =  \ottsym{(}  a  \ottsym{,}  i  \ottsym{)}  \in   \DOM( \ottnt{H} )  
}{
    \tuple{ \ottnt{H} ,  \ottnt{R} ,    x  \WRITE  y   \SEQ  e  }     \longrightarrow _{  D  }     \tuple{ \ottnt{H}  \ottsym{\{}  \ottsym{(}  a  \ottsym{,}   i   \ottsym{)}  \hookleftarrow  \ottnt{R}  \ottsym{(}  y  \ottsym{)}  \ottsym{\}} ,  \ottnt{R} ,  e }  
}
\infrule[Rs-AddPtr]{
  \ottnt{R}  \ottsym{(}  y  \ottsym{)} \,  =  \, pv
  \andalso  x'  \not\in   \DOM( \ottnt{R} )  
  \andalso R(z) \in  \mathbb{Z} 
}{
    \tuple{ \ottnt{H} ,  \ottnt{R} ,   \LET  x  =   y   \boxplus   z   \IN  e  }     \longrightarrow _{  D  }     \tuple{ \ottnt{H} ,  \ottnt{R}  \ottsym{\{}  x'  \mapsto  pv  \boxplus  \ottnt{R}  \ottsym{(}  z  \ottsym{)}  \ottsym{\}} ,    [  x'  /  x  ]    e  }  
}
\infrule[Rs-Call]{
   f  \mapsto   (  x_{{\mathrm{1}}}  , \ldots ,  x_{\ottmv{n}}  )  \, e  \in  D 
}{
    \tuple{ \ottnt{H} ,  \ottnt{R} ,   \LET  x  =   f (  y_{{\mathrm{1}}} ,\ldots, y_{\ottmv{n}}  )   \IN  e'  }     \longrightarrow _{  D  }     \tuple{ \ottnt{H} ,  \ottnt{R} ,   \LET  x  =    [  y_{{\mathrm{1}}}  /  x_{{\mathrm{1}}}  , \ldots,  y_{\ottmv{n}}  /  x_{\ottmv{n}}  ]    e   \IN  e'  }  
}
\infrule[Rs-MkArrayIntref]{
   \ottsym{(}  a  \ottsym{,}  0  \ottsym{)}  \not\in   \DOM( \ottnt{H} )  
  \andalso  x'  \not\in   \DOM( \ottnt{R} )  
  \andalso H' =  \ottnt{H}  \{ (  a  ,   0   ) , \ldots, (  a  ,  \ottnt{R}  \ottsym{(}  y  \ottsym{)} \,  -  \,  1   ) \mapsto   0   \} 
  \andalso R(y) \in  \mathbb{Z} 
}{
    \tuple{ \ottnt{H} ,  \ottnt{R} ,   \LET  x  =   \ALLOC  y   \ottsym{:}     \TINT   \TREF    \IN  e  }     \longrightarrow _{  D  }     \tuple{ \ottnt{H'} ,  \ottnt{R}  \ottsym{\{}  x'  \mapsto  \ottsym{(}  a  \ottsym{,}   0   \ottsym{)}  \ottsym{\}} ,    [  x'  /  x  ]    e  }  
}
\infrule[Rs-MkArrayNestedref]{
   \ottsym{(}  a  \ottsym{,}  0  \ottsym{)}  \not\in   \DOM( \ottnt{H} )  
  \andalso  x'  \not\in   \DOM( \ottnt{R} )   \\
  \andalso \ottnt{H'}  =   \ottnt{H}  \{ (  a  ,   0   ) , \ldots, (  a  ,  \ottnt{R}  \ottsym{(}  y  \ottsym{)} \,  -  \,  1   ) \mapsto  \ottkw{null}  \} 
  \andalso R(y) \in  \mathbb{Z} 
}{
    \tuple{ \ottnt{H} ,  \ottnt{R} ,   \LET  x  =   \ALLOC  y   \ottsym{:}    \ottsym{(}   \tau^{-}  \TREF   \ottsym{)}  \TREF    \IN  e  }     \longrightarrow _{  D  }     \tuple{ \ottnt{H'} ,  \ottnt{R}  \ottsym{\{}  x'  \mapsto  \ottsym{(}  a  \ottsym{,}   0   \ottsym{)}  \ottsym{\}} ,    [  x'  /  x  ]    e  }  
}
\infrule[Rs-AliasAddPtr]{
  \ottnt{R}  \ottsym{(}  y  \ottsym{)} \,  =  \, pv \andalso
  R(z) \in  \mathbb{Z}  \andalso
  \ottnt{R}  \ottsym{(}  x  \ottsym{)} = pv  \boxplus  \ottnt{R}  \ottsym{(}  z  \ottsym{)}
}{
    \tuple{ \ottnt{H} ,  \ottnt{R} ,    \ALIAS(  x  =  y   \boxplus   z  )   \SEQ  e  }     \longrightarrow _{  D  }     \tuple{ \ottnt{H} ,  \ottnt{R} ,  e }  
}
\infrule[Rs-AliasAddPtrFail]{
  \ottnt{R}  \ottsym{(}  y  \ottsym{)} \,  =  \, pv \andalso
  R(z) \in  \mathbb{Z}  \andalso
  \ottnt{R}  \ottsym{(}  x  \ottsym{)} \neq pv  \boxplus  \ottnt{R}  \ottsym{(}  z  \ottsym{)}
}{
    \tuple{ \ottnt{H} ,  \ottnt{R} ,    \ALIAS(  x  =  y   \boxplus   z  )   \SEQ  e  }     \longrightarrow _{  D  }     \mathbf{AliasFail}  
}
\infrule[Rs-AliasDeref]{
  \ottnt{H}  \ottsym{(}  \ottnt{R}  \ottsym{(}  y  \ottsym{)}  \ottsym{)} \,  =  \, \ottnt{R}  \ottsym{(}  x  \ottsym{)}
}{
    \tuple{ \ottnt{H} ,  \ottnt{R} ,    \ALIAS(  x  = \ast  y  )   \SEQ  e  }     \longrightarrow _{  D  }     \tuple{ \ottnt{H} ,  \ottnt{R} ,  e }  
}
\infrule[Rs-AliasDerefFail]{
  R(y) =  \NULL  \mbox{ or }
  \ottnt{H}  \ottsym{(}  \ottnt{R}  \ottsym{(}  y  \ottsym{)}  \ottsym{)} \, \neq \, \ottnt{R}  \ottsym{(}  x  \ottsym{)}
}{
    \tuple{ \ottnt{H} ,  \ottnt{R} ,    \ALIAS(  x  = \ast  y  )   \SEQ  e  }     \longrightarrow _{  D  }     \mathbf{AliasFail}  
}
\infrule[Rs-Assert]{
  \models  \ottsym{[}  \ottnt{R}  \ottsym{]} \, \varphi
}{
    \tuple{ \ottnt{H} ,  \ottnt{R} ,    \ASSERT( \varphi )   \SEQ  e  }     \longrightarrow _{  D  }     \tuple{ \ottnt{H} ,  \ottnt{R} ,  e }  
}
\infrule[Rs-Context]{
    \tuple{ \ottnt{H} ,  \ottnt{R} ,  e }     \longrightarrow _{  D  }     \tuple{ \ottnt{H'} ,  \ottnt{R'} ,  e' }  
}{
    \tuple{ \ottnt{H} ,  \ottnt{R} ,   \LET  x  =  e  \IN  e_{{\mathrm{0}}}  }     \longrightarrow _{  D  }     \tuple{ \ottnt{H'} ,  \ottnt{R'} ,   \LET  x  =  e'  \IN  e_{{\mathrm{0}}}  }  
}
\infrule[Rs-ContextFail]{
    \tuple{ \ottnt{H} ,  \ottnt{R} ,  e }     \longrightarrow _{  D  }     \mathbf{AliasFail}  
}{
    \tuple{ \ottnt{H} ,  \ottnt{R} ,   \LET  x  =  e  \IN  e_{{\mathrm{0}}}  }     \longrightarrow _{  D  }     \mathbf{AliasFail}  
}

\subsection{Type Well-Formedness}
\label{sec:type-well-formedness}

\infrule[WF-Phi]{
  \forall x  \in  \ottkw{FV} \, \ottsym{(}  \varphi  \ottsym{)} \backslash \{  \nu  \}.
  | \Gamma  (  x  ) | =  \TINT 
}{
   \Gamma   \vdash   \varphi  \mbox{ ok} 
}
\infrule[WF-Int]{
   \Gamma   \vdash   \varphi  \mbox{ ok} 
}{
  \Gamma  \vdash    \{  \nu  :   \TINT    \mid   \varphi  \}   \mbox{ ok} 
}
\infrule[WF-Ref]{
  \models   \fml{ r    =     \mathbf{0}  }   \implies   \Empty{ \tau }   \andalso
  \Gamma  \ottsym{,}   z \COL  \TINT    \vdash   \tau  \mbox{ ok}  \andalso
   \Gamma  \ottsym{,}   z \COL  \TINT     \vdash   r  \mbox{ ok} 
}{
  \Gamma  \vdash    \Pi z .( \tau  \TREF^{\hspace{0.5pt} r })   \mbox{ ok} 
}
\infrule[WF-Env]{
  \forall x  \in  dom(\Gamma) . \Gamma  \vdash    \Gamma  (  x  )   \mbox{ ok} 
}{
   \vdash   \Gamma  \mbox{ ok} 
}
\infrule[WF-OwnPhi]{
   \Gamma   \vdash   r  \mbox{ ok}  \andalso q  \in  [0,1] \andalso    \Gamma   \vdash   \varphi  \mbox{ ok} 
}{
   \Gamma   \vdash   \ottsym{(}    \varphi   \produces    q    ,  r   \ottsym{)}  \mbox{ ok} 
}
\infrule[WF-OwnZero]{}{
   \Gamma   \vdash    \mathbf{0}   \mbox{ ok} 
}
\infrule[WF-FunType]{
   \DOM( \Gamma )  =  \DOM( \Gamma' )  \andalso
  \forall x  \in   \DOM( \Gamma ) .\ | \Gamma  (  x  ) | = | \Gamma'  (  x  ) | \\
   \vdash   \Gamma  \mbox{ ok}  \andalso
   \vdash   \Gamma'  \mbox{ ok}  \andalso
  \Gamma'  \vdash   \tau  \mbox{ ok} 
}{
   \vdash    \tuple{ \Gamma }\ra\tuple{ \Gamma'  \mid  \tau }   \mbox{ ok} 
}
\infrule[WF-FunEnv]{
  \forall f  \in  dom(\Theta) .  \vdash   \Theta  \ottsym{(}  f  \ottsym{)}  \mbox{ ok} 
}{
   \vdash   \Theta  \mbox{ ok} 
}

\subsection{Rules for Subtyping, Type Equivalence, and Type Addition}
\label{sec:subtyping}

\infrule[S-Int]{
   \Gamma  ,  \nu  \colon \TINT   \models  \varphi_{{\mathrm{1}}}  \implies  \varphi_{{\mathrm{2}}}
}{
  \Gamma  \vdash    \{  \nu  :   \TINT    \mid   \varphi_{{\mathrm{1}}}  \}    \leq    \{  \nu  :   \TINT    \mid   \varphi_{{\mathrm{2}}}  \}  
}
\infrule[S-Ref]{
  \Gamma  \ottsym{,}   x \COL  \TINT    \models   r _{ 1 }  \, \ge \,  r _{ 2 } 
  \andalso \Gamma  \ottsym{,}   x \COL  \TINT    \vdash   \tau_{{\mathrm{1}}}   \leq   \tau_{{\mathrm{2}}} 
}{
  \Gamma  \vdash    \Pi x .( \tau_{{\mathrm{1}}}  \TREF^{\hspace{0.5pt}  r _{ 1 }  })    \leq    \Pi x .( \tau_{{\mathrm{2}}}  \TREF^{\hspace{0.5pt}  r _{ 2 }  })  
}
\infrule[S-TyEnv]{
   \forall    x  \in   \DOM( \Gamma' )   .  \Gamma  \vdash    \Gamma  (  x  )    \leq    \Gamma'  (  x  )   
}{
  \Gamma  \leq  \Gamma'
}
\infrule[S-Res]{
  \Gamma  \leq  \Gamma'
  \andalso \Gamma  \vdash   \tau   \leq   \tau' 
}{
  \Gamma  \ottsym{,}  \tau  \leq  \Gamma'  \ottsym{,}  \tau'
}
\infrule[TEq-Sub]{
  \Gamma  \vdash   \tau_{{\mathrm{1}}}   \leq   \tau_{{\mathrm{2}}} 
  \andalso \Gamma  \vdash   \tau_{{\mathrm{2}}}   \leq   \tau_{{\mathrm{1}}} 
}{
  \Gamma  \vdash   \tau_{{\mathrm{1}}}   \approx   \tau_{{\mathrm{2}}} 
}
\infrule[TA-Int]{
  \Gamma  \models  \varphi_{{\mathrm{1}}}  \wedge  \varphi_{{\mathrm{2}}}  \iff  \varphi_{{\mathrm{3}}}
}{
  \Gamma  \vdash     \{  \nu  :   \TINT    \mid   \varphi_{{\mathrm{1}}}  \}   +   \{  \nu  :   \TINT    \mid   \varphi_{{\mathrm{2}}}  \}     \approx    \{  \nu  :   \TINT    \mid   \varphi_{{\mathrm{3}}}  \}  
}
\infrule[TA-Ref]{
  \Gamma  \ottsym{,}   x \COL  \TINT    \models  r_{{\mathrm{1}}}  \ottsym{+}  r_{{\mathrm{2}}} \,  =  \, r_{{\mathrm{3}}}
  \andalso \Gamma  \ottsym{,}   x \COL  \TINT    \vdash    \tau_{{\mathrm{1}}}  +  \tau_{{\mathrm{2}}}    \approx   \tau_{{\mathrm{3}}} 
}{
  \Gamma  \vdash     \Pi x .( \tau_{{\mathrm{1}}}  \TREF^{\hspace{0.5pt} r_{{\mathrm{1}}} })   +   \Pi x .( \tau_{{\mathrm{2}}}  \TREF^{\hspace{0.5pt} r_{{\mathrm{2}}} })     \approx    \Pi x .( \tau_{{\mathrm{3}}}  \TREF^{\hspace{0.5pt} r_{{\mathrm{3}}} })  
}
\infrule[TA-PlusPlus]{
  \Gamma  \vdash    \tau_{{\mathrm{1}}}  +  \tau_{{\mathrm{2}}}    \approx   \tau_{{\mathrm{5}}} 
  \andalso \Gamma  \vdash    \tau_{{\mathrm{3}}}  +  \tau_{{\mathrm{4}}}    \approx   \tau_{{\mathrm{5}}} 
}{
  \Gamma  \vdash    \tau_{{\mathrm{1}}}  +  \tau_{{\mathrm{2}}}    \approx    \tau_{{\mathrm{3}}}  +  \tau_{{\mathrm{4}}}  
}

\subsection{Typing Rules for Expressions}
\label{sec:typingOmitted}

\leavevmode
\typicallabel{T-Null}
\infrule[T-Var]{
  \Gamma  \vdash    \tau_{{\mathrm{1}}}  +  \tau_{{\mathrm{2}}}    \approx   \tau_{{\mathrm{3}}} 
}{
   \Theta   \mid   \Gamma  \ottsym{[}  x  \ottsym{:}  \tau_{{\mathrm{3}}}  \ottsym{]}   \vdash   x  :   \ottsym{(}  \tau_{{\mathrm{1}}}  \ottsym{)}  ^ {= x }    \produces    \Gamma  \left[  x \hookleftarrow \tau_{{\mathrm{2}}}  \right]  
}
\infrule[T-Int]{
     \Theta   \mid   \Gamma  \ottsym{,}   x \COL  \{  \nu  :   \TINT    \mid   \nu \,  =  \,  n   \}     \vdash   e  :  \tau   \produces   \ottsym{(}  \Gamma'  \ottsym{,}   x \COL \tau'   \ottsym{)} 
\andalso
  }{
     \Theta   \mid   \Gamma   \vdash    \LET  x  =  n  \IN  e   :  \tau   \produces   \Gamma' 
  }
\infrule[T-Null]{
 \models   \Empty{  \Pi z .( \tau_{{\mathrm{1}}}  \TREF^{\hspace{0.5pt} r })  }  \andalso
  \Theta   \mid   \Gamma  \ottsym{,}   x \COL  \Pi z .( \tau_{{\mathrm{1}}}  \TREF^{\hspace{0.5pt} r })     \vdash   e  :  \tau   \produces   \ottsym{(}  \Gamma'  \ottsym{,}   x \COL \tau'   \ottsym{)} 
}{
  \Theta   \mid   \Gamma   \vdash    \LET  x  =  \ottkw{null}  \IN  e   :  \tau   \produces   \Gamma' 
}
\infrule[T-Minus]{
   \Theta   \mid   \Gamma  \ottsym{[}  y  \ottsym{:}   \{  \nu  :   \TINT    \mid   \varphi  \}   \ottsym{]}  \ottsym{[}  z  \ottsym{:}   \{  \nu  :   \TINT    \mid   \varphi'  \}   \ottsym{]}  \ottsym{,}   x \COL  \{  \nu  :   \TINT    \mid   \nu \,  =  \, y  \ottsym{-}  z  \}     \vdash   e  :  \tau   \produces   \ottsym{(}  \Gamma'  \ottsym{,}   x \COL \tau'   \ottsym{)} 
  }{
     \Theta   \mid   \Gamma  \ottsym{[}  y  \ottsym{:}   \{  \nu  :   \TINT    \mid   \varphi  \}   \ottsym{]}  \ottsym{[}  z  \ottsym{:}   \{  \nu  :   \TINT    \mid   \varphi'  \}   \ottsym{]}   \vdash    \LET  x  =   y  \mathop{  -  }  z   \IN  e   :  \tau   \produces   \Gamma' 
  }
\infrule[T-If]{
   \Theta   \mid    \Gamma  \left[  x \hookleftarrow  \{  \nu  :   \TINT    \mid    \varphi  \wedge  \nu \, \le \,  0    \}   \right]    \vdash   e_{{\mathrm{0}}}  :  \tau   \produces   \Gamma'  \\
   \Theta   \mid    \Gamma  \left[  x \hookleftarrow  \{  \nu  :   \TINT    \mid    \varphi  \wedge  \nu \,  >  \,  0    \}   \right]    \vdash   e_{{\mathrm{1}}}  :  \tau   \produces   \Gamma' 
}{
   \Theta   \mid   \Gamma  \ottsym{[}  x  \ottsym{:}   \{  \nu  :   \TINT    \mid   \varphi  \}   \ottsym{]}   \vdash    \IFNP  x  \THEN  e_{{\mathrm{0}}}  \ELSE  e_{{\mathrm{1}}}   :  \tau   \produces   \Gamma' 
}
\infrule[T-Call]{
  \Theta  \ottsym{(}  f  \ottsym{)} =  \tuple{  x_{{\mathrm{1}}} \COL \tau_{{\mathrm{1}}} ,\ldots, x_{\ottmv{n}} \COL \tau_{\ottmv{n}}  }\ra\tuple{  x_{{\mathrm{1}}} \COL \tau'_{{\mathrm{1}}} ,\ldots, x_{\ottmv{n}} \COL \tau'_{\ottmv{n}}   \mid  \tau }  \andalso
  \theta  =   [  y_{{\mathrm{1}}}  /  x_{{\mathrm{1}}}  , \ldots,  y_{\ottmv{n}}  /  x_{\ottmv{n}}  ]  \\
   \Theta   \mid    \Gamma  \left[  y_{\ottmv{i}} \hookleftarrow \theta \, \tau'_{\ottmv{i}}  \right]   \ottsym{,}   x \COL \theta \, \tau    \vdash   e  :  \tau'   \produces   \ottsym{(}  \Gamma'  \ottsym{,}   x \COL \tau''   \ottsym{)}  %
}{
   \Theta   \mid   \Gamma  \ottsym{[}  y_{\ottmv{i}}  \ottsym{:}  \theta \, \tau_{\ottmv{i}}  \ottsym{]}   \vdash    \LET  x  =   f (  y_{{\mathrm{1}}} ,\ldots, y_{\ottmv{n}}  )   \IN  e   :  \tau'   \produces   \Gamma' 
}
\infrule[T-Let]{
     \Theta   \mid   \Gamma   \vdash   e_{{\mathrm{0}}}  :  \tau'   \produces   \Gamma'  \andalso
     \Theta   \mid   \Gamma'  \ottsym{,}   x \COL \tau'    \vdash   e_{{\mathrm{1}}}  :  \tau   \produces   \ottsym{(}  \Gamma''  \ottsym{,}   x \COL \tau''   \ottsym{)}  \andalso  x \not \in (\Gamma', \tau)
  }{
     \Theta   \mid   \Gamma   \vdash    \LET  x  =  e_{{\mathrm{0}}}  \IN  e_{{\mathrm{1}}}   :  \tau   \produces   \Gamma'' 
  }
\infrule[T-MkIntArray]{
  \Gamma  \ottsym{,}   z \COL  \TINT    \models  r \,  =  \, \ottsym{(}   \ottsym{(}    0  \, \le \, z  \wedge  z \, \le \, y  \ottsym{-}   1    \ottsym{)}   \produces    1    \ottsym{)} \andalso
  \\
    {\thickmuskip=2mu
     \Theta   \mid   \Gamma  \ottsym{[}  y  \ottsym{:}   \{  \nu  :   \TINT    \mid   \varphi  \}   \ottsym{]}  \ottsym{,}   x \COL  \Pi z .(  \{  \nu  :   \TINT    \mid     0  \, \le \, z  \wedge  z \, \le \, y  \ottsym{-}   1    \implies  \nu \,  =  \,  0   \}   \TREF^{\hspace{0.5pt} r })     \vdash   e_{{\mathrm{0}}}  :  \tau   \produces   \ottsym{(}  \Gamma'  \ottsym{,}   x \COL \tau'   \ottsym{)} }
  }{
     \Theta   \mid   \Gamma  \ottsym{[}  y  \ottsym{:}   \{  \nu  :   \TINT    \mid   \varphi  \}   \ottsym{]}   \vdash    \LET  x  =   \ALLOC  y   \ottsym{:}     \TINT   \TREF    \IN  e_{{\mathrm{0}}}   :  \tau   \produces   \Gamma' 
  }
\infrule[T-MkNestedArray]{
    \Gamma  \ottsym{,}   z \COL  \TINT    \models  r \,  =  \, \ottsym{(}   \ottsym{(}    0  \, \le \, z  \wedge  z \, \le \, y  \ottsym{-}   1    \ottsym{)}   \produces    1    \ottsym{)} \andalso
    \models   \Empty{ \tau_{{\mathrm{0}}} }  \\
     \Theta   \mid   \Gamma  \ottsym{[}  y  \ottsym{:}   \{  \nu  :   \TINT    \mid   \varphi  \}   \ottsym{]}  \ottsym{,}   x \COL  \Pi z .( \tau_{{\mathrm{0}}}  \TREF^{\hspace{0.5pt} r })     \vdash   e_{{\mathrm{0}}}  :  \tau   \produces   \ottsym{(}  \Gamma'  \ottsym{,}   x \COL \tau'   \ottsym{)}  \andalso |\tau_{{\mathrm{0}}}| =  \tau^{-}  \TREF   %
  }{
     \Theta   \mid   \Gamma  \ottsym{[}  y  \ottsym{:}   \{  \nu  :   \TINT    \mid   \varphi  \}   \ottsym{]}   \vdash    \LET  x  =   \ALLOC  y   \ottsym{:}    \ottsym{(}   \tau^{-}  \TREF   \ottsym{)}  \TREF    \IN  e_{{\mathrm{0}}}   :  \tau   \produces   \Gamma' 
  }
\infrule[T-Deref]{
  \Gamma  \ottsym{,}   z \COL  \{  \nu  :   \TINT    \mid   \nu \,  =  \,  0   \}    \models  r \,  >  \,  \mathbf{0}  \andalso
  \Gamma  \ottsym{,}   z \COL  \{  \nu  :   \TINT    \mid   \nu \,  =  \,  0   \}    \vdash     \tau'  +  \tau  _{ x }    \approx    \tau _{ y }   \\
  \Gamma  \ottsym{,}   x \COL  \tau _{ x }    \ottsym{,}   z \COL  \{  \nu  :   \TINT    \mid   \nu \,  =  \,  0   \}    \vdash    \tau' _{ y }    \approx    \ottsym{(}  \tau'  \ottsym{)}  ^ {= x }   \andalso
  \Gamma  \ottsym{,}   x \COL  \tau _{ x }    \ottsym{,}   z \COL  \{  \nu  :   \TINT    \mid   \nu \, \neq \,  0   \}    \vdash    \tau' _{ y }    \approx    \tau _{ y }   \\
   \Theta   \mid    \Gamma  \left[  y \hookleftarrow  \Pi z .(  \tau' _{ y }   \TREF^{\hspace{0.5pt} r })   \right]   \ottsym{,}   x \COL  \tau _{ x }     \vdash   e_{{\mathrm{0}}}  :  \tau   \produces   \ottsym{(}  \Gamma'  \ottsym{,}   x \COL \tau''   \ottsym{)}  %
  }{
     \Theta   \mid   \Gamma  \ottsym{[}  y  \ottsym{:}   \Pi z .(  \tau _{ y }   \TREF^{\hspace{0.5pt} r })   \ottsym{]}   \vdash    \LET  x  =   \ast  y   \IN  e_{{\mathrm{0}}}   :  \tau   \produces   \Gamma' 
  }
\infrule[T-Assign]{
  \Gamma  \ottsym{,}   z \COL  \{  \nu  :   \TINT    \mid   \nu \,  =  \,  0   \}    \models  r \,  =  \,  \mathbf{1}  \andalso
  \Gamma  \vdash     \tau'  +  \tau'  _{ y }    \approx    \tau _{ y }   \\
  \Gamma  \ottsym{,}   z \COL  \{  \nu  :   \TINT    \mid   \nu \,  =  \,  0   \}    \vdash    \tau' _{\ast  x }    \approx    \ottsym{(}  \tau'  \ottsym{)}  ^ {= y }   \andalso
  \Gamma  \ottsym{,}   z \COL  \{  \nu  :   \TINT    \mid   \nu \, \neq \,  0   \}    \vdash    \tau' _{\ast  x }    \approx    \tau _{\ast  x }   \\
   \Theta   \mid     \Gamma  \left[  x \hookleftarrow  \Pi z .(  \tau' _{\ast  x }   \TREF^{\hspace{0.5pt} r })   \right]   \left[  y \hookleftarrow  \tau' _{ y }   \right]    \vdash   e_{{\mathrm{0}}}  :  \tau   \produces   \Gamma' 
  }{
     \Theta   \mid   \Gamma  \ottsym{[}  x  \ottsym{:}   \Pi z .(  \tau _{\ast  x }   \TREF^{\hspace{0.5pt} r })   \ottsym{]}  \ottsym{[}  y  \ottsym{:}   \tau _{ y }   \ottsym{]}   \vdash     x  \WRITE  y   \SEQ  e_{{\mathrm{0}}}   :  \tau   \produces   \Gamma' 
  }
\infrule[T-AddPtr]{
    \Gamma  \vdash     \Pi w .( \tau_{{\mathrm{1}}}  \TREF^{\hspace{0.5pt}  { r }_{ y_{{\mathrm{1}}} }  })   +   \Pi w .   [ (  w  -  z  ) /  w  ]   ( \tau_{{\mathrm{2}}}  \TREF^{\hspace{0.5pt}  { r }_{ x }  })     \approx    \Pi w .( \tau_{{\mathrm{3}}}  \TREF^{\hspace{0.5pt}  { r }_{ y }  })   \\
     \Theta   \mid    \Gamma  \ottsym{[}  z  \ottsym{:}   \{  \nu  :   \TINT    \mid   \varphi  \}   \ottsym{]}  \left[  y \hookleftarrow  \Pi w .( \tau_{{\mathrm{1}}}  \TREF^{\hspace{0.5pt}  { r }_{ y_{{\mathrm{1}}} }  })   \right]   \ottsym{,}   x \COL  \Pi w .( \tau_{{\mathrm{2}}}  \TREF^{\hspace{0.5pt}  { r }_{ x }  })     \vdash   e_{{\mathrm{0}}}  :  \tau   \produces   \ottsym{(}  \Gamma'  \ottsym{,}   x \COL \tau'   \ottsym{)}  %
  }{
     \Theta   \mid   \Gamma  \ottsym{[}  y  \ottsym{:}   \Pi w .( \tau_{{\mathrm{3}}}  \TREF^{\hspace{0.5pt}  { r }_{ y }  })   \ottsym{]}  \ottsym{[}  z  \ottsym{:}   \{  \nu  :   \TINT    \mid   \varphi  \}   \ottsym{]}   \vdash    \LET  x  =   y   \boxplus   z   \IN  e_{{\mathrm{0}}}   :  \tau   \produces   \Gamma' 
  }
\infrule[T-AliasAddPtr]{
    x \, \neq \, y\\
  \begin{array}{r}
    \Gamma  \vdash  \ottsym{(}    \Pi w' .   [ (  w'  -  z  ) /  w'  ]   (  \tau _{\ast  x }   \TREF^{\hspace{0.5pt}  { r }_{ x }  })   +   \Pi w .(  \tau _{\ast  y }   \TREF^{\hspace{0.5pt}  { r }_{ y }  })    \ottsym{)}  \approx  \qquad \qquad \\   \ottsym{(}   \Pi w' .   [ (  w'  -  z  ) /  w'  ]   (  \tau' _{\ast  x }   \TREF^{\hspace{0.5pt}  { r' }_{ x }  })   \ottsym{)}  +   \Pi w .(  \tau' _{\ast  y }   \TREF^{\hspace{0.5pt}  { r' }_{ y }  })  
    \end{array}\\
     \Theta   \mid     \Gamma  \ottsym{[}  z  \ottsym{:}   \{  \nu  :   \TINT    \mid   \varphi  \}   \ottsym{]}  \left[  x \hookleftarrow  \Pi w' .(  \tau' _{\ast  x }   \TREF^{\hspace{0.5pt}  { r' }_{ x }  })   \right]   \left[  y \hookleftarrow  \Pi w .(  \tau' _{\ast  y }   \TREF^{\hspace{0.5pt}  { r' }_{ y }  })   \right]    \vdash   e_{{\mathrm{0}}}  :  \tau   \produces   \Gamma' \\
  }{
     \Theta   \mid   \Gamma  \ottsym{[}  x  \ottsym{:}   \Pi w' .(  \tau _{\ast  x }   \TREF^{\hspace{0.5pt}  { r }_{ x }  })   \ottsym{]}  \ottsym{[}  y  \ottsym{:}   \Pi w .(  \tau _{\ast  y }   \TREF^{\hspace{0.5pt}  { r }_{ y }  })   \ottsym{]}  \ottsym{[}  z  \ottsym{:}   \{  \nu  :   \TINT    \mid   \varphi  \}   \ottsym{]}   \vdash     \ALIAS(  x  =  y   \boxplus   z  )   \SEQ  e_{{\mathrm{0}}}   :  \tau   \produces   \Gamma' 
  }
\infrule[T-AliasDeref]{
  \Gamma  \ottsym{,}   w \COL  \{  \nu  :   \TINT    \mid   \nu \,  =  \,  0   \}    \models  r \,  >  \,  \mathbf{0}  \\
\begin{array}{l}
  \Gamma  \ottsym{,}   w \COL  \{  \nu  :   \TINT    \mid   \nu \,  =  \,  0   \}    \vdash  \\ \qquad \ottsym{(}    \Pi z' .(  \tau _{\ast  x }   \TREF^{\hspace{0.5pt}  { r }_{ x }  })   +   \Pi z .(  \tau _{\ast \ast  y }   \TREF^{\hspace{0.5pt}  { r }_{ \ast  y }  })    \ottsym{)}  \approx  \ottsym{(}    \Pi z' .(  \tau' _{\ast  x }   \TREF^{\hspace{0.5pt}  { r' }_{ x }  })   +   \Pi z .(  \tau' _{\ast \ast  y }   \TREF^{\hspace{0.5pt}  { r' }_{ \ast  y }  })    \ottsym{)}
  \end{array}\\
  \Gamma  \ottsym{,}   w \COL  \{  \nu  :   \TINT    \mid   \nu \, \neq \,  0   \}    \vdash    \Pi z .(  \tau _{\ast \ast  y }   \TREF^{\hspace{0.5pt}  { r }_{ \ast  y }  })    \approx    \Pi z .(  \tau' _{\ast \ast  y }   \TREF^{\hspace{0.5pt}  { r' }_{ \ast  y }  })  \\
   \Theta   \mid     \Gamma  \left[  x \hookleftarrow  \Pi z' .(  \tau' _{\ast  x }   \TREF^{\hspace{0.5pt}  { r' }_{ x }  })   \right]   \left[  y \hookleftarrow  \Pi w .(  \Pi z .(  \tau' _{\ast \ast  y }   \TREF^{\hspace{0.5pt}  { r' }_{ \ast  y }  })   \TREF^{\hspace{0.5pt} r })   \right]    \vdash   e_{{\mathrm{0}}}  :  \tau   \produces   \Gamma' \\
}{
   \Theta   \mid   \Gamma  \ottsym{[}  x  \ottsym{:}   \Pi z' .(  \tau _{\ast  x }   \TREF^{\hspace{0.5pt}  { r }_{ x }  })   \ottsym{]}  \ottsym{[}  y  \ottsym{:}   \Pi w .(  \Pi z .(  \tau _{\ast \ast  y }   \TREF^{\hspace{0.5pt}  { r }_{ \ast  y }  })   \TREF^{\hspace{0.5pt} r })   \ottsym{]}   \vdash     \ALIAS(  x  = \ast  y  )   \SEQ  e_{{\mathrm{0}}}   :  \tau   \produces   \Gamma' 
}
\infrule[T-Assert]{
  \Gamma  \models  \varphi \andalso
   \Theta   \mid   \Gamma   \vdash   e_{{\mathrm{0}}}  :  \tau   \produces   \Gamma' 
}{
   \Theta   \mid   \Gamma   \vdash     \ASSERT( \varphi )   \SEQ  e_{{\mathrm{0}}}   :  \tau   \produces   \Gamma' 
}
\infrule[T-Sub]{
  \Gamma  \leq  \Gamma'
  \andalso  \Theta   \mid   \Gamma'   \vdash   e  :  \tau   \produces   \Gamma'' 
  \andalso \Gamma''  \ottsym{,}  \tau  \leq  \Gamma'''  \ottsym{,}  \tau'
}{
   \Theta   \mid   \Gamma   \vdash   e  :  \tau'   \produces   \Gamma''' 
}

The rules \rn{T-Int}, \rn{T-Minus}, and \rn{T-Let} type a binding
by extending the environment for its body.
In \rn{T-Int}, the body $e$ of $ \LET  x  =  n  \IN  e $ is typed
under $\Gamma  \ottsym{,}   x \COL  \{  \nu  :   \TINT    \mid   \nu \,  =  \,  n   \}  $, recording the value of $x$.
Similarly, \rn{T-Minus} gives $x$ the type
$ \{  \nu  :   \TINT    \mid   \nu \,  =  \, y  \ottsym{-}  z  \} $.
In \rn{T-Let}, the post-environment $\Gamma'$ of $e_{{\mathrm{0}}}$ is
extended with $ x \COL \tau' $, where $\tau'$ is the type of $e_{{\mathrm{0}}}$,
to type the body $e_{{\mathrm{1}}}$.
Since the scope of $x$ is the body, its binding is dropped from
the body's post-environment when typing the whole expression.
Well-formedness requires that neither the remaining post-environment
nor the expression's result type mention $x$; if necessary,
\rn{T-Sub} is used to weaken them to eliminate this dependency.

\subsection{Typing Rules for Functions and Programs}
\label{sec:functionTyping}

\leavevmode
\infrule[T-FunDef]{
  \Theta  \ottsym{(}  f  \ottsym{)} =  \tuple{ x_{{\mathrm{1}}}  \ottsym{:}  \tau_{{\mathrm{1}}}  \ottsym{,} \, ... \, \ottsym{,}  x_{\ottmv{n}}  \ottsym{:}  \tau_{\ottmv{n}} }\ra\tuple{ x_{{\mathrm{1}}}  \ottsym{:}  \tau'_{{\mathrm{1}}}  \ottsym{,} \, ... \, \ottsym{,}  x_{\ottmv{n}}  \ottsym{:}  \tau'_{\ottmv{n}} \mid \tau } 
  \\  \Theta   \mid    x_{{\mathrm{1}}} \COL \tau_{{\mathrm{1}}} ,\ldots, x_{\ottmv{n}} \COL \tau_{\ottmv{n}}    \vdash   e  :  \tau   \produces    x_{{\mathrm{1}}} \COL \tau'_{{\mathrm{1}}} ,\ldots, x_{\ottmv{n}} \COL \tau'_{\ottmv{n}}  
}{
   \Theta   \vdash   f  \mapsto   (  x_{{\mathrm{1}}}  , \ldots ,  x_{\ottmv{n}}  )  \, e 
}
\infrule[T-Funs]{
  dom(D) = dom(\Theta)
  \andalso \forall f  \mapsto   (  x_{{\mathrm{1}}}  , \ldots ,  x_{\ottmv{n}}  )  \, e  \in 
  D .\  \Theta   \vdash   f  \mapsto   (  x_{{\mathrm{1}}}  , \ldots ,  x_{\ottmv{n}}  )  \, e 
}{
   \Theta   \vdash   D 
}
\infrule[T-Prog]{
   \Theta   \vdash   D 
  \andalso  \vdash   \Theta  \mbox{ ok} 
  \andalso  \Theta   \mid    \bullet    \vdash   e  :  \tau   \produces   \Gamma 
}{
   \vdash    \tuple{ D ,  e }  
}

\section{Proofs}
\label{sec:proofs}

\subsection{Auxiliary Definitions and Rocq Representation}

The runtime invariants and configuration typing are defined in
\cref{sec:soundness}.  Here we introduce two auxiliary relations used in
their preservation proofs and explain their Rocq representations.

The Rocq annotations accompanying definitions and lemmas identify the
corresponding declarations; file paths are relative to \texttt{rocq/theories/}.
An annotation marked ``Supporting Rocq result'' identifies a related checked
result rather than a standalone formalization of the displayed statement.
Such a result may have additional hypotheses.  We explicitly note where
there is no standalone Rocq counterpart.

\begin{definition}[Ownership comparison and heap equivalence]
The domain of an ownership map denotes its support.  We write
\[
\begin{array}{l}
  \OWN(H,R,v,\tau)\le\OWN(H',R',v',\tau') \iff \\
  \qquad \DOM(H)\subseteq\DOM(H')\ \land\ \DOM(R)\subseteq\DOM(R')\\
  \qquad \land\ \forall(a,i)\in\DOM(\OWN(H,R,v,\tau)).\ \OWN(H,R,v,\tau)(a,i)\le\OWN(H',R',v',\tau')(a,i).
\end{array}
\]
Heap equivalence except at one location is defined by
\[
\begin{array}{l}
  \ottnt{H_{{\mathrm{1}}}}   \approx _{(a,i)}  \ottnt{H_{{\mathrm{2}}}}  \iff\quad \\
 \qquad \DOM(H_1)=\DOM(H_2) \land\ \forall(a',i')\in\DOM(H_1)\setminus\{(a,i)\}.\ H_1(a',i')=H_2(a',i').
\end{array}
\]
\end{definition}

\paragraph{Rocq representation.}
The declarations corresponding to \cref{def:satisfaction-ownership} are:
\rocqnote{Rocq: \rocqdecl{Logic/Ownership.v}{ownership_denote};
  \rocqdecl{Runtime/Model.v}{value_ownership},
  \rocqdecl{Runtime/Model.v}{env_ownership},
  \rocqdecl{Runtime/Model.v}{value_satisfies}, and
  \rocqdecl{Runtime/Model.v}{env_satisfies}.}
The ownership bound in \cref{def:configuration-typing} is formalized by
\rocqdecl{Runtime/Model.v}{ownership_consistent}, and configuration typing by
\rocqdecl{Metatheory/Progress.v}{configuration_typed}.

The Rocq counterparts of the value-level definitions recurse structurally
on the type $\tau$.\footnote{The Rocq definitions are
\texttt{value\_satisfies} (lines 18--39) and \texttt{value\_ownership}
(lines 68--87) in \texttt{Runtime/Model.v}.}
At a reference type, recursive calls use the structurally smaller element
type, while the recursive ownership sum is implemented as a fold over the
finite heap domain.

The Rocq predicate \texttt{consistent} packages the runtime conditions of
configuration typing.  It combines $ \mathbf{SAT}( \ottnt{H} ,  \ottnt{R} ,  \Gamma ) $ and
$ \mathbf{ConOwn}( \ottnt{H} ,  \ottnt{R} ,  \Gamma ) $ with the requirement that the heap, register file,
and type environment have no duplicate keys.

The Rocq definition of configuration typing additionally includes
$ \mathbf{AliasFail} $, with any result type and output type environment, via the
constructor \texttt{CT\_AliasFail}.  The theorem
\texttt{configuration\_\allowbreak preservation} therefore also covers
transitions to alias failure.  The predicate \texttt{terminal} includes both a configuration
returning a variable and $ \mathbf{AliasFail} $, so
\texttt{typed\_\allowbreak configuration\_\allowbreak progress} allows the same terminal outcomes
as \cref{thm:soundness}.

\subsection{Proof of \cref{thm:soundness}}

We recall \cref{thm:soundness}.

\soundness*

We restate the three lemmas used to establish this theorem.

\initializationlemma*
\begin{proof}
From $\vdash \tuple{D,e}$, we have $ \Theta   \vdash   D $ and
$ \Theta   \mid    \bullet    \vdash   e  :  \tau   \produces   \Gamma' $ for
some $\Theta, \tau, \Gamma'$.  It is enough to show
$ \mathbf{ConOwn}(  \emptyset  ,   \emptyset  ,   \bullet  ) $ and
$ \mathbf{SAT}(  \emptyset  ,   \emptyset  ,   \bullet  ) $, which are obvious from the
definitions of $\CONOWN$ and $\SAT$.
\end{proof}

\preservationlemma*
\begin{proof}
  See \cref{sec:preservation}.
\end{proof}

\progresslemma*
\begin{proof}
  See \cref{sec:progress}.
\end{proof}

To avoid clutter, we tacitly assume that all function environments, type environments, types, and judgments appearing in statements and proofs are well formed, unless stated otherwise.  In particular, an environment obtained by extension or update is assumed to be well formed whenever it appears in a judgment.

\subsubsection{Structural properties of typing}
\label{sec:proofs-structural-typing}

These lemmas establish the structural properties needed to reuse typing
derivations during reduction. After basic facts about subtyping and type
equivalence, we show how to retain bindings that an expression does not
use
(\cref{lem:typing-expression-weakening}).
The substitution results in
\cref{lem:auxiliary-substitution,lem:index-substitution,lem:typing-expression-substitution}
justify replacing program variables and instantiating array indices.
\textsc{Call preparation} combines these results for function application,
while \cref{lem:inversion-expression-typing} recovers the premises needed
for each reduction case despite
possible uses of \rn{T-Sub}.

\begin{lemma}
\label{lem:subtyping-relation-preorder}
The relations $\Gamma  \vdash   \tau   \leq   \tau' $ and $\Gamma  \leq  \Gamma'$ are reflexive and transitive.
Namely,
\begin{enumerate}
\item $\Gamma  \vdash   \tau   \leq   \tau $,
\item if $\Gamma  \vdash   \tau_{{\mathrm{1}}}   \leq   \tau_{{\mathrm{2}}} $ and $\Gamma  \vdash   \tau_{{\mathrm{2}}}   \leq   \tau_{{\mathrm{3}}} $, then $\Gamma  \vdash   \tau_{{\mathrm{1}}}   \leq   \tau_{{\mathrm{3}}} $,
\item $\Gamma  \leq  \Gamma$, and
\item if $\Gamma_{{\mathrm{1}}}  \leq  \Gamma_{{\mathrm{2}}}$ and $\Gamma_{{\mathrm{2}}}  \leq  \Gamma_{{\mathrm{3}}}$, then $\Gamma_{{\mathrm{1}}}  \leq  \Gamma_{{\mathrm{3}}}$.
\end{enumerate}
\rocqnote{Rocq: \rocqdecl{Typing/Relations.v}{subtype_refl},
  \rocqdecl{Typing/Relations.v}{subtype_trans}, and
  \rocqdecl{Typing/Relations.v}{env_subtype_refl}.
  Environment transitivity is not separately stated in Rocq.}
\end{lemma}
\begin{proof}
We first show $\Gamma  \vdash   \tau   \leq   \tau $ for any $\tau$ by structural induction on $\tau$.
The base case is when $\tau =  \{  \nu  :   \TINT    \mid   \varphi  \} $. 
Since $\varphi  \implies  \varphi$ is tautology, $ \models  \LLBRAKET  \Gamma  \RRBRAKET  \produces   \ottsym{(}  \varphi  \implies  \varphi  \ottsym{)} $ always holds. 
Thus, we get $\Gamma  \vdash    \{  \nu  :   \TINT    \mid   \varphi  \}    \leq    \{  \nu  :   \TINT    \mid   \varphi  \}  $ by \rn{S-Int}.

The inductive case is the case where $\tau =  \Pi z .( \tau'  \TREF^{\hspace{0.5pt} r }) $.
By \rn{S-Ref}, our goal is to show that $\Gamma  \ottsym{,}   z \COL  \TINT    \models  r \, \ge \, r$ and $\Gamma  \ottsym{,}   z \COL  \TINT    \vdash   \tau'   \leq   \tau' $.
The former is trivial, and the latter follows from the I.H.

Then, we prove transitivity of subtyping.
We show that, $\forall$ $\Gamma$, $\tau_{{\mathrm{1}}}$, $\tau_{{\mathrm{2}}}$, $\tau_{{\mathrm{3}}}$, if $\Gamma  \vdash   \tau_{{\mathrm{1}}}   \leq   \tau_{{\mathrm{2}}} $ and $ \Gamma  \vdash   \tau_{{\mathrm{2}}}   \leq   \tau_{{\mathrm{3}}} $, then $\Gamma  \vdash   \tau_{{\mathrm{1}}}   \leq   \tau_{{\mathrm{3}}}  $ by induction on $\Gamma  \vdash   \tau_{{\mathrm{1}}}   \leq   \tau_{{\mathrm{2}}} $.
The base case is when $\tau_{{\mathrm{1}}} =  \{  \nu  :   \TINT    \mid   \varphi_{{\mathrm{1}}}  \} $, $\tau_{{\mathrm{2}}} =  \{  \nu  :   \TINT    \mid   \varphi_{{\mathrm{2}}}  \} $, in which case $\tau_{{\mathrm{3}}}$ must be of the form $ \{  \nu  :   \TINT    \mid   \varphi_{{\mathrm{3}}}  \} $. 
By \rn{S-Int}, we have $\Gamma  \models  \varphi_{{\mathrm{1}}}  \implies  \varphi_{{\mathrm{2}}}$ and $\Gamma  \models  \varphi_{{\mathrm{2}}}  \implies  \varphi_{{\mathrm{3}}}$.
By transitivity of implication, $\Gamma  \models  \varphi_{{\mathrm{1}}}  \implies  \varphi_{{\mathrm{3}}}$ holds.
Thus, we get $\Gamma  \vdash   \tau_{{\mathrm{1}}}   \leq   \tau_{{\mathrm{3}}} $ by \rn{S-Int}.

The inductive case is the case where
$\tau_{{\mathrm{1}}} =  \Pi z .( \tau'_{{\mathrm{1}}}  \TREF^{\hspace{0.5pt} r_{{\mathrm{1}}} }) $,
$\tau_{{\mathrm{2}}} =  \Pi z .( \tau'_{{\mathrm{2}}}  \TREF^{\hspace{0.5pt} r_{{\mathrm{2}}} }) $,
$\tau_{{\mathrm{3}}} =  \Pi z .( \tau'_{{\mathrm{3}}}  \TREF^{\hspace{0.5pt} r_{{\mathrm{3}}} }) $ 
with
$\Gamma  \ottsym{,}   z \COL  \TINT    \vdash   \tau'_{{\mathrm{1}}}   \leq   \tau'_{{\mathrm{2}}} $,
$\Gamma  \ottsym{,}   z \COL  \TINT    \models  r_{{\mathrm{1}}} \, \ge \, r_{{\mathrm{2}}}$,
$\Gamma  \ottsym{,}   z \COL  \TINT    \vdash   \tau'_{{\mathrm{2}}}   \leq   \tau'_{{\mathrm{3}}} $, and
$\Gamma  \ottsym{,}   z \COL  \TINT    \models  r_{{\mathrm{2}}} \, \ge \, r_{{\mathrm{3}}}$.
By I.H., $\Gamma  \ottsym{,}   z \COL  \TINT    \vdash   \tau'_{{\mathrm{1}}}   \leq   \tau'_{{\mathrm{3}}} $. 
Also, $\Gamma  \ottsym{,}   z \COL  \TINT    \models  r_{{\mathrm{1}}} \, \ge \, r_{{\mathrm{3}}}$ holds.
Finally, \rn{S-Ref} shows $\Gamma  \vdash   \tau_{{\mathrm{1}}}   \leq   \tau_{{\mathrm{3}}} $.

Reflexivity of $\Gamma  \leq  \Gamma$ easily follows from the reflexivity of $\Gamma  \vdash   \tau_{{\mathrm{1}}}   \leq   \tau_{{\mathrm{2}}} $.

Finally, we prove transitivity of $\Gamma  \leq  \Gamma'$.  First observe that if $\Gamma_{{\mathrm{1}}}  \leq  \Gamma_{{\mathrm{2}}}$, then the assumptions represented by $\Gamma_{{\mathrm{1}}}$ imply those represented by $\Gamma_{{\mathrm{2}}}$.
Consequently, if $\Gamma_{{\mathrm{2}}}  \vdash   \tau   \leq   \tau' $, then $\Gamma_{{\mathrm{1}}}  \vdash   \tau   \leq   \tau' $.
This can be shown from the following two:
\begin{itemize}
\item If $\Gamma_{{\mathrm{1}}}  \leq  \Gamma_{{\mathrm{2}}}$, then $ \fml{ \Gamma_{{\mathrm{1}}} }  \implies  \fml{ \Gamma_{{\mathrm{2}}} } $.
\item If $ \fml{ \Gamma_{{\mathrm{1}}} }  \implies  \fml{ \Gamma_{{\mathrm{2}}} } $ and $\Gamma_{{\mathrm{2}}}  \vdash   \tau   \leq   \tau' $, then $\Gamma_{{\mathrm{1}}}  \vdash   \tau   \leq   \tau' $.
\end{itemize}
The former follows from \rn{S-TyEnv}, \rn{S-Int}, and the definition of $ \fml{ \Gamma } $.
The latter follows by induction on the derivation of $\Gamma_{{\mathrm{2}}}  \vdash   \tau   \leq   \tau' $.  

Then, given $\Gamma_{{\mathrm{1}}}  \leq  \Gamma_{{\mathrm{2}}}$ and $\Gamma_{{\mathrm{2}}}  \leq  \Gamma_{{\mathrm{3}}}$, consider any $ y \COL \tau_{{\mathrm{3}}} $ in $\Gamma_{{\mathrm{3}}}$.
We have $ y \COL \tau_{{\mathrm{2}}} $ in $\Gamma_{{\mathrm{2}}}$ with $\Gamma_{{\mathrm{2}}}  \vdash   \tau_{{\mathrm{2}}}   \leq   \tau_{{\mathrm{3}}} $.
Similarly, we have $ y \COL \tau_{{\mathrm{1}}} $ in $\Gamma_{{\mathrm{1}}}$ with $\Gamma_{{\mathrm{1}}}  \vdash   \tau_{{\mathrm{1}}}   \leq   \tau_{{\mathrm{2}}} $.
By the observation above, $\Gamma_{{\mathrm{1}}}  \vdash   \tau_{{\mathrm{2}}}   \leq   \tau_{{\mathrm{3}}} $.  By transitivity,
$\Gamma_{{\mathrm{1}}}  \vdash   \tau_{{\mathrm{1}}}   \leq   \tau_{{\mathrm{3}}} $.  Since $ y \COL \tau_{{\mathrm{3}}} $ was arbitrary, \rn{S-TyEnv} gives $\Gamma_{{\mathrm{1}}}  \leq  \Gamma_{{\mathrm{3}}}$.
\end{proof}

\begin{lemma}
\label{lem:own-term-equality}
If $\Gamma  \vdash    \Pi z .( \tau_{{\mathrm{1}}}  \TREF^{\hspace{0.5pt} r_{{\mathrm{1}}} })    \approx    \Pi z .( \tau_{{\mathrm{2}}}  \TREF^{\hspace{0.5pt} r_{{\mathrm{2}}} })  $, then $\Gamma  \ottsym{,}   z \COL  \TINT    \vdash   \tau_{{\mathrm{1}}}   \approx   \tau_{{\mathrm{2}}} $ and $\Gamma  \ottsym{,}   z \COL  \TINT    \models  r_{{\mathrm{1}}} \,  =  \, r_{{\mathrm{2}}}$.
\rocqref[By definition in Rocq]{Typing/Relations.v}{type_equiv}
\end{lemma}
\begin{proof}
  By inversion of \rn{TEq-Sub} and \rn{S-Ref}, we have
  $\Gamma  \ottsym{,}   z \COL  \TINT    \vdash   \tau_{{\mathrm{1}}}   \leq   \tau_{{\mathrm{2}}} $ and $\Gamma  \ottsym{,}   z \COL  \TINT    \vdash   \tau_{{\mathrm{2}}}   \leq   \tau_{{\mathrm{1}}} $ and $\Gamma  \ottsym{,}   z \COL  \TINT    \models  r_{{\mathrm{1}}} \, \ge \, r_{{\mathrm{2}}}$ and $\Gamma  \ottsym{,}   z \COL  \TINT    \models  r_{{\mathrm{2}}} \, \ge \, r_{{\mathrm{1}}}$.
  We have $\Gamma  \ottsym{,}   z \COL  \TINT    \vdash   \tau_{{\mathrm{1}}}   \approx   \tau_{{\mathrm{2}}} $ by \rn{TEq-Sub} and $\Gamma  \ottsym{,}   z \COL  \TINT    \models  r_{{\mathrm{1}}} \,  =  \, r_{{\mathrm{2}}}$ by pointwise antisymmetry
  of the ordering on rational numbers.
\end{proof}

\begin{lemma}[Preservation of type well-formedness]
\label{lem:type-well-formedness-integer-scope}
If $\Gamma  \vdash   \tau  \mbox{ ok} $ and every variable with an integer type in $\Gamma$ also
has an integer type in $\Gamma'$, then $\Gamma'  \vdash   \tau  \mbox{ ok} $.
Neither environment is required to be well formed, and the integer
refinements need not agree.
\rocqref{Metatheory/Scoping.v}{ty_wf_at_weaken_names}
\end{lemma}
\begin{proof}
Proceed by induction on type well-formedness.  In the \rn{WF-Int} case,
every free variable of $\varphi$ other than $ \nu $ has an integer type
in $\Gamma$.  By assumption, it also has an integer type in $\Gamma'$.
Hence $ \Gamma'   \vdash   \varphi  \mbox{ ok} $ by \rn{WF-Phi}, and
$\Gamma'  \vdash    \{  \nu  :   \TINT    \mid   \varphi  \}   \mbox{ ok} $ by \rn{WF-Int}.

In \rn{WF-Ref}, choose the bound index fresh for both environments and
apply I.H. to the element type.  Well-formedness of the ownership term is
preserved by induction on its structure, using the same condition on
integer-typed variables and the unchanged coefficient bounds.  The unconditional
zero-ownership/emptiness premise is unchanged.
\end{proof}

\begin{lemma}[Weakening]
\label{lem:typing-expression-weakening}
If $ \Theta   \mid   \Gamma   \vdash   e  :  \tau   \produces   \Gamma' $ and $\Gamma_{{\mathrm{3}}}$ is a type environment such that
\(
  dom(\Gamma) \cap dom(\Gamma_{{\mathrm{3}}}) = \emptyset,
\)
\(
  dom(\Gamma') \cap dom(\Gamma_{{\mathrm{3}}}) = \emptyset,
\)
\(
   \vdash   \Gamma  \ottsym{,}  \Gamma_{{\mathrm{3}}}  \mbox{ ok} ,
\)
and
\(
   \vdash   \Gamma'  \ottsym{,}  \Gamma_{{\mathrm{3}}}  \mbox{ ok} ,
\)
then
\(
  \Theta   \mid   \Gamma  \ottsym{,}  \Gamma_{{\mathrm{3}}}   \vdash   e  :  \tau   \produces   \Gamma'  \ottsym{,}  \Gamma_{{\mathrm{3}}} .
\)
\rocqref{Metatheory/Preservation.v}{has_type_frame_append}
\end{lemma}
\begin{proof}
By induction on the derivation of $ \Theta   \mid   \Gamma   \vdash   e  :  \tau   \produces   \Gamma' $ with case analysis
on the last typing rule used.

First, well-formed environment extension preserves contextual validity,
subtyping, type equivalence, and type addition: the added bindings only
strengthen the logical assumptions.  Moreover,
if $\Gamma  \leq  \Gamma'$, then $\Gamma  \ottsym{,}  \Gamma_{{\mathrm{3}}}  \leq  \Gamma'  \ottsym{,}  \Gamma_{{\mathrm{3}}}$, by weakening of subtyping and
reflexivity for the added types; the same holds for result subtyping.

\begin{rncase}{T-Sub}
The premises give
\[
 \Gamma  \leq  \Gamma_{{\mathrm{1}}},\qquad
  \Theta   \mid   \Gamma_{{\mathrm{1}}}   \vdash   e  :  \tau'   \produces   \Gamma_{{\mathrm{2}}} ,\qquad
 \Gamma_{{\mathrm{2}}}  \ottsym{,}  \tau'  \leq  \Gamma'  \ottsym{,}  \tau.
\]
Every integer binding of $\Gamma'$ is present with integer type in both
$\Gamma_{{\mathrm{2}}}$ and $\Gamma_{{\mathrm{1}}}$, since subtyping may forget bindings but preserves
their simple types, and typing preserves the integer types of retained
bindings.  Apply
\cref{lem:type-well-formedness-integer-scope} to each binding of $\Gamma_{{\mathrm{3}}}$,
starting from $ \vdash   \Gamma'  \ottsym{,}  \Gamma_{{\mathrm{3}}}  \mbox{ ok} $.  Together with the well-formedness of
$\Gamma_{{\mathrm{1}}}$ and $\Gamma_{{\mathrm{2}}}$ and disjointness of the frame, this establishes
$ \vdash   \Gamma_{{\mathrm{1}}}  \ottsym{,}  \Gamma_{{\mathrm{3}}}  \mbox{ ok} $ and $ \vdash   \Gamma_{{\mathrm{2}}}  \ottsym{,}  \Gamma_{{\mathrm{3}}}  \mbox{ ok} $, allowing I.H. to give
\[
  \Theta   \mid   \Gamma_{{\mathrm{1}}}  \ottsym{,}  \Gamma_{{\mathrm{3}}}   \vdash   e  :  \tau'   \produces   \Gamma_{{\mathrm{2}}}  \ottsym{,}  \Gamma_{{\mathrm{3}}} .
\]
Extend the two subtyping premises with $\Gamma_{{\mathrm{3}}}$ and reapply \rn{T-Sub}.
\end{rncase}

The remaining cases follow by I.H., using
\cref{lem:type-well-formedness-integer-scope} for the unchanged bindings
and the auxiliary-judgment properties above.

\end{proof}

We use without further mention the standard compatibility of
capture-avoiding substitution with type-environment extension, lookup,
update, and substitution composition.  Bound variables are alpha-renamed
when necessary.

\begin{lemma}[Substitution for auxiliary judgments]
\label{lem:auxiliary-substitution}
Let $\theta$ be a simultaneous capture-avoiding substitution of
program-variable names by program-variable names.  Suppose that the
substituted type environments appearing below are well formed.  Then:
\begin{enumerate}
\item If $\Gamma  \models  \varphi$, then $\theta \, \Gamma  \models  \theta \, \varphi$.  Here $\varphi$ may
  also be a formula encoding an ownership comparison, an ownership
  equality, or a type-emptiness condition.
\rocqref[Supporting Rocq result]{Metatheory/Substitution.v}{env_formula_instantiate_reindex}

\item If $\Gamma  \vdash   \tau  \mbox{ ok} $, then $\theta \, \Gamma  \vdash   \theta \, \tau  \mbox{ ok} $.  Similarly, if
  $ \Gamma   \vdash   r  \mbox{ ok} $, then $ \theta \, \Gamma   \vdash    \theta   r   \mbox{ ok} $.
\rocqnote{Rocq: \rocqdecl{Metatheory/Substitution.v}{ty_wf_at_instantiate},
  \rocqdecl{Metatheory/Substitution.v}{ownership_wf_substitute}, and
  \rocqdecl{Metatheory/Substitution.v}{ownership_scoped_substitute}.}

\item If $\Gamma  \vdash   \tau_{{\mathrm{1}}}   \leq   \tau_{{\mathrm{2}}} $, then
  \(
    \theta \, \Gamma  \vdash   \theta \, \tau_{{\mathrm{1}}}   \leq   \theta \, \tau_{{\mathrm{2}}} .
  \)
\rocqref{Metatheory/Substitution.v}{subtype_substitute}

\item If $\Gamma  \vdash   \tau_{{\mathrm{1}}}   \approx   \tau_{{\mathrm{2}}} $, then
  \(
    \theta \, \Gamma  \vdash   \theta \, \tau_{{\mathrm{1}}}   \approx   \theta \, \tau_{{\mathrm{2}}} .
  \)
\rocqref{Metatheory/Substitution.v}{type_equiv_substitute}

\item If $\Gamma  \vdash    \tau_{{\mathrm{1}}}  +  \tau_{{\mathrm{2}}}    \approx   \tau_{{\mathrm{3}}} $, then
  \(
    \theta \, \Gamma  \vdash    \theta \, \tau_{{\mathrm{1}}}  +  \theta \, \tau_{{\mathrm{2}}}    \approx   \theta \, \tau_{{\mathrm{3}}} .
  \)
\rocqref{Metatheory/Substitution.v}{type_add_substitute}

\item If $\Gamma_{{\mathrm{1}}}  \leq  \Gamma_{{\mathrm{2}}}$, then
  \(
    \theta \, \Gamma_{{\mathrm{1}}}  \leq  \theta \, \Gamma_{{\mathrm{2}}}.
  \)
\rocqref{Metatheory/Substitution.v}{env_subtype_substitute}

\item If $\Gamma_{{\mathrm{1}}}  \ottsym{,}  \tau_{{\mathrm{1}}}  \leq  \Gamma_{{\mathrm{2}}}  \ottsym{,}  \tau_{{\mathrm{2}}}$, then
  \(
    \theta \, \Gamma_{{\mathrm{1}}}  \ottsym{,}  \theta \, \tau_{{\mathrm{1}}}  \leq  \theta \, \Gamma_{{\mathrm{2}}}  \ottsym{,}  \theta \, \tau_{{\mathrm{2}}}.
  \)
\rocqref{Metatheory/Substitution.v}{env_result_subtype_substitute}
\end{enumerate}
In particular,
\(
  \models   \Empty{ \tau } 
  \quad\Longrightarrow\quad
  \models   \Empty{ \theta \, \tau } .
\)
\rocqref{Metatheory/Substitution.v}{type_empty_substitute}
\end{lemma}
\begin{proof}
The validity case is the standard substitution property of the underlying
first-order logic.  The well-formedness, subtyping, type-equivalence, and
type-addition cases follow by induction on the structure of the types
involved, choosing bound array indices fresh for $\theta$.  The environment
and environment-result cases follow pointwise from the type-subtyping case.
Finally, $\theta \,  \Empty{ \tau } = \Empty{ \theta \, \tau } $ follows by induction on
$\tau$.
\end{proof}

\begin{lemma}[Substitution of an array index]
\label{lem:index-substitution}
Let $j\in \mathbb{Z} $, and let $z$ be fresh for $\Gamma$.  Then:
\begin{enumerate}
\item If $\Gamma  \ottsym{,}   z \COL  \TINT    \models  \varphi$, then $\Gamma  \models   [  j  /  z  ]  \, \varphi$.  Here $\varphi$
  may also encode an ownership comparison, an ownership equality, or a
  type-emptiness condition.

\item If $\Gamma  \ottsym{,}   z \COL  \TINT    \vdash   \tau  \mbox{ ok} $, then $\Gamma  \vdash    [  j  /  z  ]  \, \tau  \mbox{ ok} $.  Similarly, if
  $ \Gamma  \ottsym{,}   z \COL  \TINT     \vdash   r  \mbox{ ok} $, then $ \Gamma   \vdash     [  j  /  z  ]    r   \mbox{ ok} $.

\item If $\Gamma  \ottsym{,}   z \COL  \TINT    \vdash   \tau_{{\mathrm{1}}}   \leq   \tau_{{\mathrm{2}}} $, then
  \(
    \Gamma  \vdash    [  j  /  z  ]  \, \tau_{{\mathrm{1}}}   \leq    [  j  /  z  ]  \, \tau_{{\mathrm{2}}} .
  \)

\item If $\Gamma  \ottsym{,}   z \COL  \TINT    \vdash   \tau_{{\mathrm{1}}}   \approx   \tau_{{\mathrm{2}}} $, then
  \(
    \Gamma  \vdash    [  j  /  z  ]  \, \tau_{{\mathrm{1}}}   \approx    [  j  /  z  ]  \, \tau_{{\mathrm{2}}} .
  \)

\item If $\Gamma  \ottsym{,}   z \COL  \TINT    \vdash    \tau_{{\mathrm{1}}}  +  \tau_{{\mathrm{2}}}    \approx   \tau_{{\mathrm{3}}} $, then
  \(
    \Gamma  \vdash     [  j  /  z  ]  \, \tau_{{\mathrm{1}}}  +   [  j  /  z  ]  \, \tau_{{\mathrm{2}}}    \approx    [  j  /  z  ]  \, \tau_{{\mathrm{3}}} .
  \)
\end{enumerate}
In particular, if $\Gamma  \ottsym{,}   z \COL  \TINT    \models  \varphi$ and $\models  \ottsym{[}  \ottnt{R}  \ottsym{]} \,  \fml{ \Gamma } $, then Item~1
implies
\(
  \models  \ottsym{[}  \ottnt{R}  \ottsym{]} \,  [  j  /  z  ]  \, \varphi.
\)
Thus an ownership comparison or equality valid under $\Gamma  \ottsym{,}   z \COL  \TINT  $ may be
evaluated pointwise at any integer $j$.  Also,
\[
  \models   \Empty{ \tau } 
  \quad\Longrightarrow\quad
  \models   \Empty{  [  j  /  z  ]  \, \tau } .
\]
\rocqnote{Rocq: no standalone counterpart.  Bound array indices are
  interpreted through \rocqdecl{Logic/Refinement.v}{index_stack}
  rather than substituted as program-variable names.}
\end{lemma}
\begin{proof}
Item~1 is the standard semantic substitution property of the underlying
first-order logic, instantiating $z$ with the integer $j$.  Item~2
follows by induction on the well-formedness derivation, and Items~3--5 by
induction on the corresponding derivations.  In each reference case, choose
the bound array index fresh before applying the induction hypothesis.  The
last consequence also uses
$ [  j  /  z  ]  \,  \Empty{ \tau } = \Empty{  [  j  /  z  ]  \, \tau } $, proved by induction on $\tau$.
\end{proof}

\begin{lemma}[Simultaneous substitution]
\label{lem:typing-expression-substitution}
If
\(
   \Theta   \mid   \Gamma_{{\mathrm{1}}}   \vdash   e  :  \tau   \produces   \Gamma_{{\mathrm{2}}} ,
\)
and $\theta$ is a simultaneous (capture-avoiding) substitution such that
$\theta \, \Gamma_{{\mathrm{1}}}$ and $\theta \, \Gamma_{{\mathrm{2}}}$ are well-formed type environments, then
\(
   \Theta   \mid   \theta \, \Gamma_{{\mathrm{1}}}   \vdash    \theta   e   :  \theta \, \tau   \produces   \theta \, \Gamma_{{\mathrm{2}}} .
\)
\rocqref{Metatheory/Substitution.v}{has_type_substitute}
\end{lemma}
\begin{proof}
By induction on the derivation of $ \Theta   \mid   \Gamma_{{\mathrm{1}}}   \vdash   e  :  \tau   \produces   \Gamma_{{\mathrm{2}}} $, using
\cref{lem:auxiliary-substitution} for the auxiliary premises of each typing
rule.  Bound variables are chosen fresh for $\theta$.  In the
\rn{T-Call} case, simultaneous substitutions are composed; in the
\rn{T-Sub} case, we use Items 6 and 7 of
\cref{lem:auxiliary-substitution}.
\end{proof}

\begin{lemma}[Call preparation]
\label{lem:call-preparation}
Suppose that
\[
\begin{array}{l}
  \Theta  \ottsym{(}  f  \ottsym{)} =
   \tuple{  x_{{\mathrm{1}}} \COL \tau_{{\mathrm{1}}} ,\ldots, x_{\ottmv{n}} \COL \tau_{\ottmv{n}}  }\ra\tuple{  x_{{\mathrm{1}}} \COL \tau'_{{\mathrm{1}}} ,\ldots, x_{\ottmv{n}} \COL \tau'_{\ottmv{n}}   \mid  \tau } ,\\
   \Theta   \vdash   f  \mapsto   (  x_{{\mathrm{1}}}  , \ldots ,  x_{\ottmv{n}}  )  \, e ,\\
  \theta  =   [  y_{{\mathrm{1}}}  /  x_{{\mathrm{1}}}  , \ldots,  y_{\ottmv{n}}  /  x_{\ottmv{n}}  ] .
\end{array}
\]
Suppose also that $y_{{\mathrm{1}}},\ldots,y_{\ottmv{n}}$ are pairwise distinct and that
\[
   \Gamma_{{\mathrm{1}}}  (  y_{\ottmv{i}}  )  = \theta \, \tau_{\ottmv{i}}
  \qquad\text{for every } 1 \leq i \leq n.
\]
Let $\Gamma_{{\mathrm{3}}}$ consist of the bindings of $\Gamma_{{\mathrm{1}}}$ whose names are not among
$y_{{\mathrm{1}}},\ldots,y_{\ottmv{n}}$, and define
\[
\begin{aligned}
 \Gamma_{{\mathrm{2}}} &=  y_{{\mathrm{1}}} \COL \theta \, \tau_{{\mathrm{1}}} ,\ldots, y_{\ottmv{n}} \COL \theta \, \tau_{\ottmv{n}}   \ottsym{,}  \Gamma_{{\mathrm{3}}},\\
 \Gamma_{{\mathrm{4}}} &=  y_{{\mathrm{1}}} \COL \theta \, \tau'_{{\mathrm{1}}} ,\ldots, y_{\ottmv{n}} \COL \theta \, \tau'_{\ottmv{n}}   \ottsym{,}  \Gamma_{{\mathrm{3}}}.
\end{aligned}
\]
Then $\Gamma_{{\mathrm{1}}}  \leq  \Gamma_{{\mathrm{2}}}$ and
\[
   \Theta   \mid   \Gamma_{{\mathrm{2}}}   \vdash    \theta   e   :  \theta \, \tau   \produces   \Gamma_{{\mathrm{4}}} .
\]
In particular, $\Gamma_{{\mathrm{2}}}$ and $\Gamma_{{\mathrm{4}}}$ are well formed.
\rocqnote{Rocq: \rocqdecl{Metatheory/Preservation.v}{call_frame_endpoints}
  and \rocqdecl{Metatheory/Preservation.v}{prepare_call_body}.}
\end{lemma}
\begin{proof}
Since $\Gamma_{{\mathrm{2}}}$ is a reordering of $\Gamma_{{\mathrm{1}}}$, it is well formed and
$\Gamma_{{\mathrm{1}}}  \leq  \Gamma_{{\mathrm{2}}}$ follows by \rn{S-TyEnv} and reflexivity.
The instantiated signature is well formed and has the same integer-typed
names at both endpoints.  For each binding $ x \COL  \tau _{ x }  $ in $\Gamma_{{\mathrm{3}}}$,
well-formedness of $\Gamma_{{\mathrm{2}}}$ gives $\Gamma_{{\mathrm{2}}}  \vdash    \tau _{ x }   \mbox{ ok} $.
Since every integer-typed variable of $\Gamma_{{\mathrm{2}}}$ remains integer typed in
$\Gamma_{{\mathrm{4}}}$, \cref{lem:type-well-formedness-integer-scope} gives
$\Gamma_{{\mathrm{4}}}  \vdash    \tau _{ x }   \mbox{ ok} $.  The instantiated output types
remain well formed under extension by $\Gamma_{{\mathrm{3}}}$, so \rn{WF-Env} gives
$ \vdash   \Gamma_{{\mathrm{4}}}  \mbox{ ok} $.

Invert the function-typing premise and apply
\cref{lem:typing-expression-substitution} with $\theta$ to obtain the
instantiated body typing.  Then apply
\cref{lem:typing-expression-weakening} with frame $\Gamma_{{\mathrm{3}}}$, obtaining the
required judgment.
\end{proof}

\begin{lemma}[Inversion for expression typing]
  \label{lem:inversion-expression-typing}
  Suppose $ \Theta   \mid   \Gamma   \vdash   e  :  \tau   \produces   \Gamma' $.
\begingroup
\renewcommand{\theenumi}{\arabic{enumi}}
\renewcommand{\labelenumi}{(\theenumi)}
\renewcommand{\theenumii}{\alph{enumii}}
\renewcommand{\labelenumii}{(\theenumii)}
\makeatletter
\renewcommand{\p@enumii}{}
\makeatother
\begin{enumerate}
\item\label{lem:inversion-expression-typing-var}
If $e = x$, then there exist
$\Gamma_{{\mathrm{1}}}$, $\tau_{{\mathrm{1}}}, \tau_{{\mathrm{2}}}$, and $\tau_{{\mathrm{3}}}$
such that:
\begin{enumerate}
  \item $\Gamma  \leq  \Gamma_{{\mathrm{1}}}$;\label{inversion-var-env-subtype}
  \item $ \Gamma_{{\mathrm{1}}}  (  x  )  = \tau_{{\mathrm{3}}}$;\label{inversion-var-x-type}
  \item $\Gamma_{{\mathrm{1}}}  \vdash    \tau_{{\mathrm{1}}}  +  \tau_{{\mathrm{2}}}    \approx   \tau_{{\mathrm{3}}} $; and\label{inversion-var-split}
  \item $ \Gamma_{{\mathrm{1}}}  \left[  x \hookleftarrow \tau_{{\mathrm{2}}}  \right]   \ottsym{,}   \ottsym{(}  \tau_{{\mathrm{1}}}  \ottsym{)}  ^ {= x }   \leq  \Gamma'  \ottsym{,}  \tau$.\label{inversion-var-result-subtype}
\end{enumerate}

\item\label{lem:inversion-expression-typing1}
If $e =  \LET  x  =  n  \IN  e_{{\mathrm{0}}} $, then
  there exist
  $\Gamma_{{\mathrm{1}}}$ and $\tau'$
such that:
\begin{enumerate}
  \item $\Gamma  \leq  \Gamma_{{\mathrm{1}}}$; and\label{inversion-int-env-subtype}
  \item $ \Theta   \mid   \Gamma_{{\mathrm{1}}}  \ottsym{,}   x \COL  \{  \nu  :   \TINT    \mid   \nu \,  =  \,  n   \}     \vdash   e_{{\mathrm{0}}}  :  \tau   \produces   \ottsym{(}  \Gamma'  \ottsym{,}   x \COL \tau'   \ottsym{)} $.\label{inversion-int-continuation}
\end{enumerate}

\item\label{lem:inversion-expression-typing-null}
If $e =  \LET  x  =  \ottkw{null}  \IN  e_{{\mathrm{0}}} $, then there exist
$\Gamma_{{\mathrm{1}}}$, $z$, $r$, $\tau_{{\mathrm{1}}}$, and $\tau'$
such that:
\begin{enumerate}
  \item $\Gamma  \leq  \Gamma_{{\mathrm{1}}}$;\label{inversion-null-env-subtype}
  \item $\models   \Empty{  \Pi z .( \tau_{{\mathrm{1}}}  \TREF^{\hspace{0.5pt} r })  } $; and\label{inversion-null-empty}
  \item $ \Theta   \mid   \Gamma_{{\mathrm{1}}}  \ottsym{,}   x \COL  \Pi z .( \tau_{{\mathrm{1}}}  \TREF^{\hspace{0.5pt} r })     \vdash   e_{{\mathrm{0}}}  :  \tau   \produces   \ottsym{(}  \Gamma'  \ottsym{,}   x \COL \tau'   \ottsym{)} $.\label{inversion-null-continuation}
\end{enumerate}

\item\label{lem:inversion-expression-typing-minus}
If $e=  \LET  x  =   y  \mathop{  -  }  z   \IN  e_{{\mathrm{0}}} $, then there exist
$\Gamma_{{\mathrm{1}}}$, $\varphi$, $\varphi'$, and $\tau'$
such that:
\begin{enumerate}
  \item $\Gamma  \leq  \Gamma_{{\mathrm{1}}}$;\label{inversion-minus-env-subtype}
  \item $ \Gamma_{{\mathrm{1}}}  (  y  )  =  \{  \nu  :   \TINT    \mid   \varphi  \} $;\label{inversion-minus-y-type}
  \item $ \Gamma_{{\mathrm{1}}}  (  z  )  =  \{  \nu  :   \TINT    \mid   \varphi'  \} $; and\label{inversion-minus-z-type}
  \item $ \Theta   \mid   \Gamma_{{\mathrm{1}}}  \ottsym{,}   x \COL  \{  \nu  :   \TINT    \mid   \nu \,  =  \, y  \ottsym{-}  z  \}     \vdash   e_{{\mathrm{0}}}  :  \tau   \produces   \ottsym{(}  \Gamma'  \ottsym{,}   x \COL \tau'   \ottsym{)} $.\label{inversion-minus-continuation}
\end{enumerate}

\item\label{lem:inversion-expression-typing3}
If $e =  \IFNP  x  \THEN  e_{{\mathrm{0}}}  \ELSE  e_{{\mathrm{1}}} $, then there exist $\Gamma_{{\mathrm{1}}}$,
$\varphi$ that satisfy:
\begin{enumerate}
  \item $\Gamma  \leq  \Gamma_{{\mathrm{1}}}$;\label{inversion-if-env-subtype}
  \item $ \Gamma_{{\mathrm{1}}}  (  x  )  =  \{  \nu  :   \TINT    \mid   \varphi  \} $;\label{inversion-if-x-type}
  \item $ \Theta   \mid    \Gamma_{{\mathrm{1}}}  \left[  x \hookleftarrow  \{  \nu  :   \TINT    \mid    \varphi  \wedge  \nu \, \le \,  0    \}   \right]    \vdash   e_{{\mathrm{0}}}  :  \tau   \produces   \Gamma' $; and\label{inversion-if-nonpos-branch}
  \item $ \Theta   \mid    \Gamma_{{\mathrm{1}}}  \left[  x \hookleftarrow  \{  \nu  :   \TINT    \mid    \varphi  \wedge  \nu \,  >  \,  0    \}   \right]    \vdash   e_{{\mathrm{1}}}  :  \tau   \produces   \Gamma' $.\label{inversion-if-positive-branch}
\end{enumerate}

\item\label{lem:inversion-expression-typing8}
If $e =  \LET  x  =   f (  y_{{\mathrm{1}}} ,\ldots, y_{\ottmv{n}}  )   \IN  e_{{\mathrm{0}}} $, then
there exist $\Gamma_{{\mathrm{1}}}$, $x_{{\mathrm{1}}}$, \ldots, $x_{\ottmv{n}}$, $\theta$, $\tau_{{\mathrm{1}}}$, \ldots $\tau_{\ottmv{n}}$,
$\tau'_{{\mathrm{1}}}$, \ldots, $\tau'_{\ottmv{n}}$, $\tau'$, and $\tau''$ such that:
\begin{enumerate}
  \item $\Gamma  \leq  \Gamma_{{\mathrm{1}}}$;\label{inversion-call-env-subtype}
  \item $ \Gamma_{{\mathrm{1}}}  (  y_{\ottmv{i}}  )  = \theta \, \tau_{\ottmv{i}}$ for $1 \leq i \leq n$;\label{inversion-call-argument-types}
  \item $\Theta  \ottsym{(}  f  \ottsym{)} =  \tuple{  x_{{\mathrm{1}}} \COL \tau_{{\mathrm{1}}} ,\ldots, x_{\ottmv{n}} \COL \tau_{\ottmv{n}}  }\ra\tuple{  x_{{\mathrm{1}}} \COL \tau'_{{\mathrm{1}}} ,\ldots, x_{\ottmv{n}} \COL \tau'_{\ottmv{n}}   \mid  \tau' } $;\label{inversion-call-signature}
  \item $\theta  =   [  y_{{\mathrm{1}}}  /  x_{{\mathrm{1}}}  , \ldots,  y_{\ottmv{n}}  /  x_{\ottmv{n}}  ] $; and\label{inversion-call-substitution}
  \item $ \Theta   \mid    \Gamma_{{\mathrm{1}}}  \left[  y_{\ottmv{i}} \hookleftarrow \theta \, \tau'_{\ottmv{i}}  \right]   \ottsym{,}   x \COL \theta \, \tau'    \vdash   e_{{\mathrm{0}}}  :  \tau   \produces   \ottsym{(}  \Gamma'  \ottsym{,}   x \COL \tau''   \ottsym{)} $.\label{inversion-call-continuation}
\end{enumerate}

\item\label{lem:inversion-expression-typing-let}
If $e =  \LET  x  =  e_{{\mathrm{1}}}  \IN  e_{{\mathrm{2}}} $, then there exist $\Gamma_{{\mathrm{1}}}$, $\Gamma_{{\mathrm{2}}}$, $\tau_{{\mathrm{1}}}$, and $\tau'$ that satisfy:
\begin{enumerate}
  \item $\Gamma  \leq  \Gamma_{{\mathrm{1}}}$;\label{inversion-let-env-subtype}
  \item $ \Theta   \mid   \Gamma_{{\mathrm{1}}}   \vdash   e_{{\mathrm{1}}}  :  \tau_{{\mathrm{1}}}   \produces   \Gamma_{{\mathrm{2}}} $;\label{inversion-let-bound-expression}
  \item $ \Theta   \mid   \Gamma_{{\mathrm{2}}}  \ottsym{,}   x \COL \tau_{{\mathrm{1}}}    \vdash   e_{{\mathrm{2}}}  :  \tau   \produces   \ottsym{(}  \Gamma'  \ottsym{,}   x \COL \tau'   \ottsym{)} $; and\label{inversion-let-continuation}
  \item $x \not \in (\Gamma', \tau)$.\label{inversion-let-fresh}
\end{enumerate}

\item\label{lem:inversion-expression-typing6}
If $e =  \LET  x  =   \ALLOC  y   \ottsym{:}     \TINT   \TREF    \IN  e_{{\mathrm{0}}} $, then
there exist $\Gamma_{{\mathrm{1}}}$, $\tau'$, $\varphi$, $r$, and $z$
such that:
\begin{enumerate}
  \item $\Gamma  \leq  \Gamma_{{\mathrm{1}}}$;\label{inversion-alloc-int-env-subtype}
  \item $ \Gamma_{{\mathrm{1}}}  (  y  )  =  \{  \nu  :   \TINT    \mid   \varphi  \} $;\label{inversion-alloc-int-length-type}
  \item $\Gamma_{{\mathrm{1}}}  \ottsym{,}   z \COL  \TINT    \models  r \,  =  \, \ottsym{(}    \ottsym{(}    0  \, \le \, z  \wedge  z \, \le \, y  \ottsym{-}   1    \ottsym{)}   \produces    1    ,   \mathbf{0}    \ottsym{)}$; and\label{inversion-alloc-int-ownership-shape}
  \item $ \Theta   \mid   \Gamma_{{\mathrm{1}}}  \ottsym{,}   x \COL  \Pi z .(  \{  \nu  :   \TINT    \mid     0  \, \le \, z  \wedge  z \, \le \, y  \ottsym{-}   1    \implies  \nu \,  =  \,  0   \}   \TREF^{\hspace{0.5pt} r })     \vdash   e_{{\mathrm{0}}}  :  \tau   \produces   \ottsym{(}  \Gamma'  \ottsym{,}   x \COL \tau'   \ottsym{)} $.\label{inversion-alloc-int-continuation}
\end{enumerate}

\item\label{lem:inversion-expression-typing7}
If $e =  \LET  x  =   \ALLOC  y   \ottsym{:}    \ottsym{(}   \tau^{-}  \TREF   \ottsym{)}  \TREF    \IN  e_{{\mathrm{0}}} $, then
there exist $\Gamma_{{\mathrm{1}}}$, $\varphi$, $\tau'$, $\tau_{{\mathrm{0}}}$, $z$, $r$ such that:
\begin{enumerate}
  \item $\Gamma  \leq  \Gamma_{{\mathrm{1}}}$;\label{inversion-alloc-nested-env-subtype}
  \item $ \Gamma_{{\mathrm{1}}}  (  y  )  =  \{  \nu  :   \TINT    \mid   \varphi  \} $;\label{inversion-alloc-nested-length-type}
  \item $\Gamma_{{\mathrm{1}}}  \ottsym{,}   z \COL  \TINT    \models  r \,  =  \, \ottsym{(}   \ottsym{(}    0  \, \le \, z  \wedge  z \, \le \, y  \ottsym{-}   1    \ottsym{)}   \produces    1    \ottsym{)}$;\label{inversion-alloc-nested-ownership-shape}
  \item $\models   \Empty{ \tau_{{\mathrm{0}}} } $;\label{inversion-alloc-nested-empty}
  \item $|\tau_{{\mathrm{0}}}| =  \tau^{-}  \TREF $; and\label{inversion-alloc-nested-element-shape}
  \item $ \Theta   \mid   \Gamma_{{\mathrm{1}}}  \ottsym{,}   x \COL  \Pi z .( \tau_{{\mathrm{0}}}  \TREF^{\hspace{0.5pt} r })     \vdash   e_{{\mathrm{0}}}  :  \tau   \produces   \ottsym{(}  \Gamma'  \ottsym{,}   x \COL \tau'   \ottsym{)} $.\label{inversion-alloc-nested-continuation}
\end{enumerate}

\item\label{lem:inversion-expression-typing4}
If $e =  \LET  x  =   \ast  y   \IN  e_{{\mathrm{0}}} $, then there exist $\Gamma_{{\mathrm{1}}}$,
$z$, $ \tau _{ y } $, $r$, $\tau'$,$\tau''$,  $ \tau _{ x } $, and $ \tau' _{ y } $ such that:
\begin{enumerate}
  \item $\Gamma  \leq  \Gamma_{{\mathrm{1}}}$;\label{inversion-deref-env-subtype}
  \item $ \Gamma_{{\mathrm{1}}}  (  y  )  =  \Pi z .(  \tau _{ y }   \TREF^{\hspace{0.5pt} r }) $;\label{inversion-deref-y-type}
  \item $\Gamma_{{\mathrm{1}}}  \ottsym{,}   z \COL  \{  \nu  :   \TINT    \mid   \nu \,  =  \,  0   \}    \vdash     \tau'  +  \tau  _{ x }    \approx    \tau _{ y }  $;\label{inversion-deref-split}
  \item $\Gamma_{{\mathrm{1}}}  \ottsym{,}   x \COL  \tau _{ x }    \ottsym{,}   z \COL  \{  \nu  :   \TINT    \mid   \nu \,  =  \,  0   \}    \vdash    \tau' _{ y }    \approx    \ottsym{(}  \tau'  \ottsym{)}  ^ {= x }  $;\label{inversion-deref-zero-equiv}
  \item $\Gamma_{{\mathrm{1}}}  \ottsym{,}   x \COL  \tau _{ x }    \ottsym{,}   z \COL  \{  \nu  :   \TINT    \mid   \nu \, \neq \,  0   \}    \vdash    \tau' _{ y }    \approx    \tau _{ y }  $;\label{inversion-deref-nonzero-equiv}
  \item $\Gamma_{{\mathrm{1}}}  \ottsym{,}   z \COL  \{  \nu  :   \TINT    \mid   \nu \,  =  \,  0   \}    \models  r \,  >  \,  \mathbf{0} $; and\label{inversion-deref-positive}
  \item $ \Theta   \mid    \Gamma_{{\mathrm{1}}}  \left[  y \hookleftarrow  \Pi z .(  \tau' _{ y }   \TREF^{\hspace{0.5pt} r })   \right]   \ottsym{,}   x \COL  \tau _{ x }     \vdash   e_{{\mathrm{0}}}  :  \tau   \produces   \ottsym{(}  \Gamma'  \ottsym{,}   x \COL \tau''   \ottsym{)} $.\label{inversion-deref-continuation}
\end{enumerate}

\item\label{lem:inversion-expression-typing-assign}
If $ e =   x  \WRITE  y   \SEQ  e_{{\mathrm{0}}} $, then
there exist
$\Gamma_{{\mathrm{1}}}$, $z$, $ \tau _{\ast  x } $, $r$, $ \tau' _{ y } $, $\tau'$, $ \tau _{ y } $, and $ \tau' _{\ast  x } $ such that:
\begin{enumerate}
  \item $\Gamma  \leq  \Gamma_{{\mathrm{1}}}$;\label{inversion-assign-env-subtype}
  \item $ \Gamma_{{\mathrm{1}}}  (  x  )  =  \Pi z .(  \tau _{\ast  x }   \TREF^{\hspace{0.5pt} r }) $;\label{inversion-assign-x-type}
  \item $ \Gamma_{{\mathrm{1}}}  (  y  )  =  \tau _{ y } $;\label{inversion-assign-y-type}
  \item $\Gamma_{{\mathrm{1}}}  \ottsym{,}   z \COL  \{  \nu  :   \TINT    \mid   \nu \,  =  \,  0   \}    \vdash    \tau' _{\ast  x }    \approx    \ottsym{(}  \tau'  \ottsym{)}  ^ {= y }  $;\label{inversion-assign-zero-equiv}
  \item $\Gamma_{{\mathrm{1}}}  \ottsym{,}   z \COL  \{  \nu  :   \TINT    \mid   \nu \, \neq \,  0   \}    \vdash    \tau' _{\ast  x }    \approx    \tau _{\ast  x }  $;\label{inversion-assign-nonzero-equiv}
  \item $ \Theta   \mid     \Gamma_{{\mathrm{1}}}  \left[  x \hookleftarrow  \Pi z .(  \tau' _{\ast  x }   \TREF^{\hspace{0.5pt} r })   \right]   \left[  y \hookleftarrow  \tau' _{ y }   \right]    \vdash   e_{{\mathrm{0}}}  :  \tau   \produces   \Gamma' $;\label{inversion-assign-continuation}
  \item $\Gamma_{{\mathrm{1}}}  \ottsym{,}   z \COL  \{  \nu  :   \TINT    \mid   \nu \,  =  \,  0   \}    \models  r \,  =  \,  \mathbf{1} $; and\label{inversion-assign-full-ownership}
  \item $\Gamma_{{\mathrm{1}}}  \vdash     \tau'  +  \tau'  _{ y }    \approx    \tau _{ y }  $.\label{inversion-assign-split}
\end{enumerate}

\item\label{lem:inversion-expression-typing5}
If $e =  \LET  x  =   y   \boxplus   z   \IN  e_{{\mathrm{0}}} $, then
there exist $\Gamma_{{\mathrm{1}}}$, $w$, $\tau_{{\mathrm{1}}}$, $\tau_{{\mathrm{2}}}$, $\tau_{{\mathrm{3}}}$, $\tau'$, $ { r }_{ y } $, $\varphi$, $ { r }_{ x } $, and $ { r }_{ y_{{\mathrm{1}}} } $
such that:
\begin{enumerate}
  \item $\Gamma  \leq  \Gamma_{{\mathrm{1}}}$;\label{inversion-addptr-env-subtype}
  \item $ \Gamma_{{\mathrm{1}}}  (  y  )  =  \Pi w .( \tau_{{\mathrm{3}}}  \TREF^{\hspace{0.5pt}  { r }_{ y }  }) $;\label{inversion-addptr-y-type}
  \item $ \Gamma_{{\mathrm{1}}}  (  z  )  =  \{  \nu  :   \TINT    \mid   \varphi  \} $;\label{inversion-addptr-offset-type}
  \item $\Gamma_{{\mathrm{1}}}  \vdash     \Pi w .( \tau_{{\mathrm{1}}}  \TREF^{\hspace{0.5pt}  { r }_{ y_{{\mathrm{1}}} }  })   +   \Pi w .   [ (  w  -  z  ) /  w  ]   ( \tau_{{\mathrm{2}}}  \TREF^{\hspace{0.5pt}  { r }_{ x }  })     \approx    \Pi w .( \tau_{{\mathrm{3}}}  \TREF^{\hspace{0.5pt}  { r }_{ y }  })  $; and\label{inversion-addptr-split}
  \item $ \Theta   \mid    \Gamma_{{\mathrm{1}}}  \left[  y \hookleftarrow  \Pi w .( \tau_{{\mathrm{1}}}  \TREF^{\hspace{0.5pt}  { r }_{ y_{{\mathrm{1}}} }  })   \right]   \ottsym{,}   x \COL  \Pi w .( \tau_{{\mathrm{2}}}  \TREF^{\hspace{0.5pt}  { r }_{ x }  })     \vdash   e_{{\mathrm{0}}}  :  \tau   \produces   \ottsym{(}  \Gamma'  \ottsym{,}   x \COL \tau'   \ottsym{)} $.\label{inversion-addptr-continuation}
\end{enumerate}

\item\label{lem:inversion-expression-typing10}
If $e =   \ALIAS(  x  =  y   \boxplus   z  )   \SEQ  e_{{\mathrm{0}}} $, then
there exist
$\Gamma_{{\mathrm{1}}}$, $w$, $ \tau _{\ast  x } $, $ { r }_{ x } $, $ \tau _{\ast  y } $, $ { r }_{ y } $,
 $ { r' }_{ x } $, $ { r' }_{ y } $, $w'$, $\varphi$, $ \tau' _{\ast  x } $, and $ \tau' _{\ast  y } $
such that:
\begin{enumerate}
  \item $\Gamma  \leq  \Gamma_{{\mathrm{1}}}$;\label{inversion-alias-addptr-env-subtype}
  \item $x \, \neq \, y$;\label{inversion-alias-addptr-distinct}
  \item $ \Gamma_{{\mathrm{1}}}  (  x  )  =  \Pi w' .(  \tau _{\ast  x }   \TREF^{\hspace{0.5pt}  { r }_{ x }  }) $;\label{inversion-alias-addptr-x-type}
  \item $ \Gamma_{{\mathrm{1}}}  (  y  )  =  \Pi w .(  \tau _{\ast  y }   \TREF^{\hspace{0.5pt}  { r }_{ y }  }) $;\label{inversion-alias-addptr-y-type}
  \item $ \Gamma_{{\mathrm{1}}}  (  z  )  =  \{  \nu  :   \TINT    \mid   \varphi  \} $;\label{inversion-alias-addptr-offset-type}
  \item\label{inversion-alias-addptr-combined-equiv}
    \(\begin{aligned}[t]
      &\Gamma_{{\mathrm{1}}}  \vdash  \ottsym{(}    \Pi w' .   [ (  w'  -  z  ) /  w'  ]   (  \tau _{\ast  x }   \TREF^{\hspace{0.5pt}  { r }_{ x }  })   +   \Pi w .(  \tau _{\ast  y }   \TREF^{\hspace{0.5pt}  { r }_{ y }  })    \ottsym{)}\\
      &\qquad  \approx   \ottsym{(}   \Pi w' .   [ (  w'  -  z  ) /  w'  ]   (  \tau' _{\ast  x }   \TREF^{\hspace{0.5pt}  { r' }_{ x }  })   \ottsym{)}  +   \Pi w .(  \tau' _{\ast  y }   \TREF^{\hspace{0.5pt}  { r' }_{ y }  })  
        \text{; and}
    \end{aligned}\)
  \item $  \Theta   \mid     \Gamma_{{\mathrm{1}}}  \left[  x \hookleftarrow  \Pi w' .(  \tau' _{\ast  x }   \TREF^{\hspace{0.5pt}  { r' }_{ x }  })   \right]   \left[  y \hookleftarrow  \Pi w .(  \tau' _{\ast  y }   \TREF^{\hspace{0.5pt}  { r' }_{ y }  })   \right]    \vdash   e_{{\mathrm{0}}}  :  \tau   \produces   \Gamma' $.\label{inversion-alias-addptr-continuation}
\end{enumerate}

\item\label{lem:inversion-expression-typing9}
If $ e =   \ALIAS(  x  = \ast  y  )   \SEQ  e_{{\mathrm{0}}} $, then
there exist $\Gamma_{{\mathrm{1}}}$, $z$, $ \tau _{\ast  x } $, $ { r }_{ x } $, $w$, $ \tau _{\ast \ast  y } $,
$ { r }_{ \ast  y } $, $r$, $ \tau' _{\ast  x } $, $ \tau' _{\ast \ast  y } $, $ { r' }_{ x } $, and $ { r' }_{ \ast  y } $
such that:
\begin{enumerate}
  \item $\Gamma  \leq  \Gamma_{{\mathrm{1}}}$;\label{inversion-alias-deref-env-subtype}
  \item $ \Gamma_{{\mathrm{1}}}  (  x  )  =  \Pi z .(  \tau _{\ast  x }   \TREF^{\hspace{0.5pt}  { r }_{ x }  }) $;\label{inversion-alias-deref-x-type}
  \item $ \Gamma_{{\mathrm{1}}}  (  y  )  =  \Pi w .(  \Pi z .(  \tau _{\ast \ast  y }   \TREF^{\hspace{0.5pt}  { r }_{ \ast  y }  })   \TREF^{\hspace{0.5pt} r }) $;\label{inversion-alias-deref-y-type}
  \item $\Gamma_{{\mathrm{1}}}  \ottsym{,}   w \COL  \{  \nu  :   \TINT    \mid   \nu \,  =  \,  0   \}    \vdash   \ottsym{(}    \Pi z .(  \tau _{\ast  x }   \TREF^{\hspace{0.5pt}  { r }_{ x }  })   +   \Pi z .(  \tau _{\ast \ast  y }   \TREF^{\hspace{0.5pt}  { r }_{ \ast  y }  })    \ottsym{)}   \approx   \ottsym{(}    \Pi z .(  \tau' _{\ast  x }   \TREF^{\hspace{0.5pt}  { r' }_{ x }  })   +   \Pi z .(  \tau' _{\ast \ast  y }   \TREF^{\hspace{0.5pt}  { r' }_{ \ast  y }  })    \ottsym{)} $;\label{inversion-alias-deref-zero-equiv}
  \item $\Gamma_{{\mathrm{1}}}  \ottsym{,}   w \COL  \{  \nu  :   \TINT    \mid   \nu \, \neq \,  0   \}    \vdash    \Pi z .(  \tau _{\ast \ast  y }   \TREF^{\hspace{0.5pt}  { r }_{ \ast  y }  })    \approx    \Pi z .(  \tau' _{\ast \ast  y }   \TREF^{\hspace{0.5pt}  { r' }_{ \ast  y }  })  $;\label{inversion-alias-deref-nonzero-equiv}
  \item $ \Theta   \mid     \Gamma_{{\mathrm{1}}}  \left[  x \hookleftarrow  \Pi z .(  \tau' _{\ast  x }   \TREF^{\hspace{0.5pt}  { r' }_{ x }  })   \right]   \left[  y \hookleftarrow  \Pi w .(  \Pi z .(  \tau' _{\ast \ast  y }   \TREF^{\hspace{0.5pt}  { r' }_{ \ast  y }  })   \TREF^{\hspace{0.5pt} r })   \right]    \vdash   e_{{\mathrm{0}}}  :  \tau   \produces   \Gamma' $; and\label{inversion-alias-deref-continuation}
  \item $\Gamma_{{\mathrm{1}}}  \ottsym{,}   w \COL  \{  \nu  :   \TINT    \mid   \nu \,  =  \,  0   \}    \models  r \,  >  \,  \mathbf{0} $.\label{inversion-alias-deref-positive}
\end{enumerate}

\item\label{lem:inversion-expression-typing11}
If $  e =   \ASSERT( \varphi )   \SEQ  e_{{\mathrm{0}}} $, then
there exists $\Gamma_{{\mathrm{1}}}$
such that:
\begin{enumerate}
  \item $\Gamma  \leq  \Gamma_{{\mathrm{1}}}$;\label{inversion-assert-env-subtype}
  \item $\Gamma_{{\mathrm{1}}}  \models  \varphi$; and\label{inversion-assert-validity}
  \item $ \Theta   \mid   \Gamma_{{\mathrm{1}}}   \vdash   e_{{\mathrm{0}}}  :  \tau   \produces   \Gamma' $.\label{inversion-assert-continuation}
\end{enumerate}
\end{enumerate}
\endgroup
\rocqnote{Rocq: no standalone inversion lemma.  The proofs invert
  \rocqdecl{Typing/Rules.v}{has_type}, notably in
  \rocqdecl{Metatheory/Progress.v}{progress} and
  \rocqdecl{Metatheory/Preservation.v}{source_preservation}.}
\end{lemma}
\begin{proof}
  By induction on the derivation of $ \Theta   \mid   \Gamma   \vdash   e  :  \tau   \produces   \Gamma' $ with case analysis on $e$.
We show only the proof for case $e =  \LET  x  =  n  \IN  e_{{\mathrm{0}}} $.
The other items are proved similarly.

The last typing rule used is either \rn{T-Int} or \rn{T-Sub}.
If it is \rn{T-Int}, choose $\Gamma_{{\mathrm{1}}} = \Gamma$; $\tau'$ is given by the premise of \rn{T-Int};
the reflexivity of type environment subtyping finishes the subcase.
Otherwise, if it is \rn{T-Sub}, it gives
\begin{gather*}
  \Gamma  \leq  \Gamma_{{\mathrm{0}}} \\
   \Theta   \mid   \Gamma_{{\mathrm{0}}}   \vdash    \LET  x  =  n  \IN  e_{{\mathrm{0}}}   :  \tau_{{\mathrm{0}}}   \produces   \Gamma'_{{\mathrm{0}}}  \\
  \Gamma'_{{\mathrm{0}}}  \ottsym{,}  \tau_{{\mathrm{0}}}  \leq  \Gamma'  \ottsym{,}  \tau
\end{gather*}
for some $\Gamma_{{\mathrm{0}}}$, $\Gamma'_{{\mathrm{0}}}$ and $\tau_{{\mathrm{0}}}$. By I.H.,
there exist $\Gamma_{{\mathrm{1}}}$ and $ \tau _{ x } $ such that
\begin{gather*}
  \Gamma_{{\mathrm{0}}}  \leq  \Gamma_{{\mathrm{1}}} \\
   \Theta   \mid   \Gamma_{{\mathrm{1}}}  \ottsym{,}   x \COL  \{  \nu  :   \TINT    \mid   \nu \,  =  \,  n   \}     \vdash   e_{{\mathrm{0}}}  :  \tau_{{\mathrm{0}}}   \produces   \Gamma'_{{\mathrm{0}}}  \ottsym{,}   x \COL  \tau _{ x }   
\end{gather*}
By transitivity of type environment subtyping, $\Gamma  \leq  \Gamma_{{\mathrm{1}}}$.
From $\Gamma'_{{\mathrm{0}}}  \ottsym{,}  \tau_{{\mathrm{0}}}  \leq  \Gamma'  \ottsym{,}  \tau$, it is easy to show, by induction on $ \tau _{ x } $,
there exists some $ \tau' _{ x } $ such that
$\ottsym{(}  \Gamma'_{{\mathrm{0}}}  \ottsym{,}   x \COL  \tau _{ x }    \ottsym{)}  \ottsym{,}  \tau_{{\mathrm{0}}}  \leq  \ottsym{(}  \Gamma'  \ottsym{,}   x \COL  \tau' _{ x }    \ottsym{)}  \ottsym{,}  \tau$ and $\Gamma'  \ottsym{,}   x \COL  \tau' _{ x }  $ is well formed.
By \rn{T-Sub},
\[
   \Theta   \mid   \Gamma_{{\mathrm{1}}}  \ottsym{,}   x \COL  \{  \nu  :   \TINT    \mid   \nu \,  =  \,  n   \}     \vdash   e_{{\mathrm{0}}}  :  \tau   \produces   \Gamma'  \ottsym{,}   x \COL  \tau' _{ x }   \ .
\]
Let $\tau'$ be $ \tau' _{ x } $, finishing the subcase.
\end{proof}

\subsubsection{Basic properties of satisfaction and ownership}
\label{sec:proofs-runtime-facts}

The lemmas in this section connects the assumptions recorded in types
to facts about the current heap and register file. Satisfaction
determines the forms of stored values, guarantees the presence of
positively owned cells, and justifies evaluating logical consequences
of the type environment. The accompanying ownership lemmas relate
symbolic addition to its interpretation and establish the bounds used
throughout the subsequent arguments.

\begin{lemma}[Runtime inversion from satisfaction]
\label{lem:sat-runtime-inversion}
If $ \mathbf{SAT}( \ottnt{H} ,  \ottnt{R} ,  \Gamma ) $ and $x \in dom(\Gamma)$, 
then $x \in dom(\ottnt{R})$ and
\(
   \mathbf{SATv} ( \ottnt{H} ,  \ottnt{R} ,  \ottnt{R}  \ottsym{(}  x  \ottsym{)} ,   \Gamma  (  x  )  ) .
\)
Moreover:
\begin{enumerate}
\item
  If
  \(
     \Gamma  (  x  )   =   \{  \nu  :   \TINT    \mid   \varphi  \} ,
  \)
  then there exists an integer $n$ such that
  \(
  \ottnt{R}  \ottsym{(}  x  \ottsym{)} = n
  \)
  and
  \(
    \models  \ottsym{[}  \ottnt{R}  \ottsym{]} \, \ottsym{[}   n   \ottsym{/}  \nu  \ottsym{]}  \varphi.
  \)

\item
  If
  \(
     \Gamma  (  x  )   =   \Pi z .( \tau  \TREF^{\hspace{0.5pt} r }) ,
  \)
  then either
  \(
    \ottnt{R}  \ottsym{(}  x  \ottsym{)} =  \NULL 
  \)
  or there exist $a$ and $i$ such that
  \(
    \ottnt{R}  \ottsym{(}  x  \ottsym{)} = \ottsym{(}  a  \ottsym{,}  i  \ottsym{)}.
  \)

\item
  If
  \(
   \Gamma  (  x  )   =   \Pi z .( \tau  \TREF^{\hspace{0.5pt} r }) 
  \)
  and
  \(
     \llbracket    [  j  /  z  ]    r   \rrbracket_{ \ottnt{R} }  > 0,
  \)
  then there exist $a$ and $i$ such that
  \(
  \ottnt{R}  \ottsym{(}  x  \ottsym{)} = \ottsym{(}  a  \ottsym{,}  i  \ottsym{)},
  \)
  \(    
    (a,i+j) \in dom(\ottnt{H}),
  \)
  and
  \(
     \mathbf{SATv} ( \ottnt{H} ,  \ottnt{R} ,  \ottnt{H}  \ottsym{(}  \ottsym{(}  a  \ottsym{,}   i   \ottsym{+}   j   \ottsym{)}  \ottsym{)} ,   [  j  /  z  ]  \, \tau ) .
  \)
\end{enumerate}
\rocqnote{Rocq: \rocqdecl{Runtime/Lemmas.v}{env_satisfies_lookup};
  \rocqdecl{Runtime/Model.v}{value_satisfies_int_inv},
  \rocqdecl{Runtime/Model.v}{value_satisfies_ref_shape}, and
  \rocqdecl{Runtime/Model.v}{satisfying_positive_pointer_is_allocated};
  \rocqdecl{Runtime/Model.v}{satisfying_null_has_zero_ownership}.}
\end{lemma}
\begin{proof}
The first conclusion follows by unfolding $ \mathbf{SAT}( \ottnt{H} ,  \ottnt{R} ,  \Gamma ) $ at the binding of
$x$.  If $ \Gamma  (  x  ) $ is an integer refinement type, unfolding
$ \mathbf{SATv} ( \ottnt{H} ,  \ottnt{R} ,  \ottnt{R}  \ottsym{(}  x  \ottsym{)} ,   \Gamma  (  x  )  ) $ shows that $\ottnt{R}  \ottsym{(}  x  \ottsym{)}$ is an integer satisfying its
refinement.  If it is a reference type, the runtime value is either
$ \NULL $ or an address.  In the null case, satisfaction requires the
ownership to be zero at every offset.  Thus positive ownership at offset
$j$ forces the address case, in which the definition of
$ \mathbf{SATv} ( \ottnt{H} ,  \ottnt{R} ,  \ottnt{R}  \ottsym{(}  x  \ottsym{)} ,   \Gamma  (  x  )  ) $ gives both the required heap cell and satisfaction
of its element type.
\end{proof}

\begin{lemma}
  \label{lem:sat-soundness}
  If $ \mathbf{SAT}( \ottnt{H} ,  \ottnt{R} ,  \Gamma ) $, $ \Gamma   \vdash   \varphi  \mbox{ ok} $,
  and $\Gamma  \models  \varphi$, then $\models  \ottsym{[}  \ottnt{R}  \ottsym{]} \, \varphi$.
\rocqref[Supporting Rocq result]{Runtime/Lemmas.v}{env_satisfaction_implies_formula}
\end{lemma}

\begin{proof}
For every $x\in dom(\Gamma)$ such that $ \Gamma  (  x  )   =   \{  \nu  :   \TINT    \mid   \varphi'  \} $,
$ \mathbf{SAT}( \ottnt{H} ,  \ottnt{R} ,  \Gamma ) $ gives
\[
   \mathbf{SATv} ( \ottnt{H} ,  \ottnt{R} ,  \ottnt{R}  \ottsym{(}  x  \ottsym{)} ,   \{  \nu  :   \TINT    \mid   \varphi'  \}  ) .
\]
Thus there exists an integer $n$ such that $\ottnt{R}  \ottsym{(}  x  \ottsym{)}=n$ and
$\models  \ottsym{[}  \ottnt{R}  \ottsym{]} \, \ottsym{[}   n   \ottsym{/}  \nu  \ottsym{]}  \varphi'$.  Since $\ottnt{R}  \ottsym{(}  x  \ottsym{)}=n$, this also gives
$\models  \ottsym{[}  \ottnt{R}  \ottsym{]} \, \ottsym{[}  x  \ottsym{/}  \nu  \ottsym{]}  \varphi'$.  Reference bindings contribute only $ \top $ to
$ \fml{ \Gamma } $.  Taking the conjunction of these formulae therefore yields
\[
  \models  \ottsym{[}  \ottnt{R}  \ottsym{]} \,  \fml{ \Gamma } .
\]
By the definition of $\Gamma  \models  \varphi$, we have
$\models   \fml{ \Gamma }   \implies  \varphi$.  Instantiating this validity with $\ottnt{R}$ gives
\[
  \models  \ottsym{[}  \ottnt{R}  \ottsym{]} \,  \fml{ \Gamma }   \implies  \ottsym{[}  \ottnt{R}  \ottsym{]} \, \varphi.
\]
The substitutions are well-defined: well-formedness of $\Gamma$ and
$ \Gamma   \vdash   \varphi  \mbox{ ok} $ permit only integer-typed variables in these formulae, and
$ \mathbf{SAT}( \ottnt{H} ,  \ottnt{R} ,  \Gamma ) $ supplies their integer values.  Using
$\models  \ottsym{[}  \ottnt{R}  \ottsym{]} \,  \fml{ \Gamma } $, we conclude $\models  \ottsym{[}  \ottnt{R}  \ottsym{]} \, \varphi$.
\end{proof}

\begin{lemma}
  \label{lem:add_own_term}
  If $\Gamma  \models  r_{{\mathrm{1}}}  \ottsym{+}  r_{{\mathrm{2}}} \,  =  \, r_{{\mathrm{3}}}$ and
  $\models  \ottsym{[}  \ottnt{R}  \ottsym{]} \,  \fml{ \Gamma } $, then
$  \llbracket  r_{{\mathrm{1}}}  \rrbracket_{ \ottnt{R} }  +  \llbracket  r_{{\mathrm{2}}}  \rrbracket_{ \ottnt{R} }  =  \llbracket  r_{{\mathrm{3}}}  \rrbracket_{ \ottnt{R} } $.
\rocqref[By definition in Rocq]{Logic/Ownership.v}{ownership_add}
\end{lemma}
\begin{proof}
\begin{align*}
      &\Gamma  \models  r_{{\mathrm{1}}}  \ottsym{+}  r_{{\mathrm{2}}} \,  =  \, r_{{\mathrm{3}}}&\\
      & \iff  \models   \fml{ \Gamma }   \implies   \fml{ r_{{\mathrm{1}}}   \ottsym{+}   r_{{\mathrm{2}}}   =   r_{{\mathrm{3}}} }  &\tag{by definition}\\
      & \implies  \models  \ottsym{[}  \ottnt{R}  \ottsym{]} \,  \fml{ \Gamma }   \implies  \ottsym{[}  \ottnt{R}  \ottsym{]} \,  \fml{ r_{{\mathrm{1}}}   \ottsym{+}   r_{{\mathrm{2}}}   =   r_{{\mathrm{3}}} }  &\tag{by instantiation}\\
      & \implies   \llbracket  r_{{\mathrm{1}}}  \rrbracket_{ \ottnt{R} }   \ottsym{+}   \llbracket  r_{{\mathrm{2}}}  \rrbracket_{ \ottnt{R} }  \,  =  \,  \llbracket  r_{{\mathrm{3}}}  \rrbracket_{ \ottnt{R} } .
\end{align*}
\end{proof}

\begin{lemma}[Bounds on ownership] \label{lem:ownership-bounds}\ 
\begin{enumerate}
\item If $ \Gamma   \vdash   r  \mbox{ ok} $, then, for any $\ottnt{R}$,
$
  0 \le  \llbracket  r  \rrbracket_{ \ottnt{R} }  \le 1.
$
\item If $\Gamma  \vdash   \tau  \mbox{ ok} $, then, for every address $(a,i)$, $ 0 \le  \mathbf{own} ( \ottnt{H} ,  \ottnt{R} ,  v ,  \tau ) (a,i).$
\end{enumerate}
\rocqnote{Rocq: \rocqdecl{Logic/Ownership.v}{ownership_wf_bounds} and
  \rocqdecl{Metatheory/Subtyping.v}{value_ownership_nonnegative}.}
\end{lemma}
\begin{proof}
For (1), proceed by induction on $r$.  Evaluation either selects the
coefficient of the first satisfied guard or continues with the remaining
ownership term.  Every coefficient lies in $[0,1]$ by well-formedness.

For (2), proceed by induction on $\tau$.  Integer types and
non-address values have zero ownership.  For a reference type, the
direct contribution is nonnegative by (1).  Every recursive
contribution is nonnegative by the induction hypothesis, using Item~2 of
\cref{lem:index-substitution}.  Their sum
is therefore nonnegative.
\end{proof}

\subsubsection{Register extension and integer bindings}
\label{sec:proofs-register-extension}

\cref{lem:own-preservation2,lem:sat-preservation-extend-register} mean that fresh register bindings leave existing ownership and satisfaction facts
unchanged when their names do not occur in the types concerned.
\cref{lem:conown-irrevalent-to-integer,lem:sat-extension}
then account for adding an integer-valued register
or changing its refinement: integer bindings contribute no ownership, but
satisfaction requires the stored integer to meet its refinement. Together,
these results handle the fresh names and integer bindings introduced by
reduction.

\begin{lemma}
  \label{lem:own-preservation2}
  If $x \not \in  \DOM( \ottnt{R} )  \cup \fv(\tau)$, then
$ \mathbf{own} ( \ottnt{H} ,  \ottnt{R}  \ottsym{\{}  x  \mapsto  v'  \ottsym{\}} ,  v ,  \tau )  =  \mathbf{own} ( \ottnt{H} ,  \ottnt{R} ,  v ,  \tau ) $.
\rocqref{Metatheory/Scoping.v}{value_ownership_scoped_from_scope}
\end{lemma}
\begin{proof}
    By induction on the size of $\tau$ with case analysis on the form of $\tau$.

    The case  $\tau =  \{  \nu  :   \TINT    \mid   \varphi  \} $ is trivial because both
    $ \mathbf{own} ( \ottnt{H} ,  \ottnt{R}  \ottsym{\{}  x  \mapsto  v'  \ottsym{\}} ,  v ,  \tau ) $ and 
    $ \mathbf{own} ( \ottnt{H} ,  \ottnt{R} ,  v ,  \tau ) $ are the ``everywhere zero'' mapping $ \emptyset $.

    Suppose $\tau =  \Pi z .( \tau'  \TREF^{\hspace{0.5pt} r })  $ for some $\tau'$ and $r$.
    We show only the subcase where $v = (a, k)$, because otherwise it is trivial.
    Since $x \not \in \fv(\tau)$, for any integer $j$, the evaluation of $  [  j  /  z  ]    r $ is unchanged by
    extending $\ottnt{R}$ with $x$.
    Since $x \not \in \fv( [  j  /  z  ]  \, \tau')$, by I.H. (notice the sizes of $\tau'$ and $ [  j  /  z  ]  \, \tau'$ are equal),
    \[
         \mathbf{own} ( \ottnt{H} ,  \ottnt{R}  \ottsym{\{}  x  \mapsto  v'  \ottsym{\}} ,  \ottnt{H}  \ottsym{(}  \ottsym{(}  a  \ottsym{,}   k   \ottsym{+}   j   \ottsym{)}  \ottsym{)} ,   [  j  /  z  ]  \, \tau' )  =  \mathbf{own} ( \ottnt{H} ,  \ottnt{R} ,  \ottnt{H}  \ottsym{(}  \ottsym{(}  a  \ottsym{,}   k   \ottsym{+}   j   \ottsym{)}  \ottsym{)} ,   [  j  /  z  ]  \, \tau' ) 
    \]
    for any $(a, k+j) \in  \DOM( \ottnt{H} ) $.
    Since $\sem{  [  j  /  z  ]    r }_{\ottnt{R}  \ottsym{\{}  x  \mapsto  v'  \ottsym{\}}}>0$ if and only if $\sem{  [  j  /  z  ]    r }_{\ottnt{R}}>0$,
    it easily follows that $ \mathbf{own} ( \ottnt{H} ,  \ottnt{R}  \ottsym{\{}  x  \mapsto  v'  \ottsym{\}} ,  v ,  \tau )  =  \mathbf{own} ( \ottnt{H} ,  \ottnt{R} ,  v ,  \tau ) $.
\end{proof}

\begin{lemma}[Satisfaction weakening by register extension]
\label{lem:sat-preservation-extend-register}
If $ \mathbf{SATv} ( \ottnt{H} ,  \ottnt{R} ,   \ottnt{v}  ,  \tau ) $ and
$x \notin \ottkw{FV} \, \ottsym{(}  \tau  \ottsym{)} \cup dom(R)$, then
$ \mathbf{SATv} ( \ottnt{H} ,  \ottnt{R}  \ottsym{\{}  x  \mapsto  v'  \ottsym{\}} ,   \ottnt{v}  ,  \tau ) $.
\rocqref{Metatheory/Scoping.v}{value_satisfies_scoped_from_scope}
\end{lemma}
\begin{proof}
Put $\ottnt{R'} = \ottnt{R}  \ottsym{\{}  x  \mapsto  v'  \ottsym{\}}$.  We proceed by induction on $\tau$.

Suppose first that $\tau  =   \{  \nu  :   \TINT    \mid   \varphi  \} $.  By inversion of the premise,
$ \ottnt{v}  \,  =  \,  n $ for some integer $n$ and
\[
 \models  \ottsym{[}  \ottnt{R}  \ottsym{]} \, \ottsym{[}   n   \ottsym{/}  \nu  \ottsym{]}  \varphi.
\]
Since $x\notin\fv(\varphi)$, extending $\ottnt{R}$ at $x$ does not change the
interpretation of this formula.  Hence
\[
 \models  \ottsym{[}  \ottnt{R'}  \ottsym{]} \, \ottsym{[}   n   \ottsym{/}  \nu  \ottsym{]}  \varphi,
\]
which is equivalent to $ \mathbf{SATv} ( \ottnt{H} ,  \ottnt{R'} ,   n  ,   \{  \nu  :   \TINT    \mid   \varphi  \}  ) $.

Now suppose that $\tau  =   \Pi z .( \tau'  \TREF^{\hspace{0.5pt} r }) $, alpha-renaming $z$ if necessary
so that it is distinct from $x$.  Freshness gives
$x\notin\fv(r)$ and $x\notin\fv(\tau')$.  Consequently, for every
integer $j$,
\begin{equation}
  \llbracket    [  j  /  z  ]    r   \rrbracket_{ \ottnt{R'} }  =  \llbracket    [  j  /  z  ]    r   \rrbracket_{ \ottnt{R} } .
 \label{sat-preservation-extend-register-permission}
\end{equation}
If $ \ottnt{v}  \,  =  \,  \NULL $, the premise says that the right-hand side of
\cref{sat-preservation-extend-register-permission} is zero for every $j$.
The equality therefore proves $ \mathbf{SATv} ( \ottnt{H} ,  \ottnt{R'} ,  \ottkw{null} ,   \Pi z .( \tau'  \TREF^{\hspace{0.5pt} r })  ) $.

Otherwise, inversion of the premise gives $ \ottnt{v}  \,  =  \, \ottsym{(}  a  \ottsym{,}   k   \ottsym{)}$ for some $a$ and
$k$.  Consider an arbitrary integer $j$ such that
\[
  \llbracket    [  j  /  z  ]    r   \rrbracket_{ \ottnt{R'} }  > 0.
\]
By \cref{sat-preservation-extend-register-permission}, the corresponding
ownership under $\ottnt{R}$ is also positive.  The premise then gives
\[
  \ottsym{(}  a  \ottsym{,}  k  \ottsym{+}  j  \ottsym{)}  \in   \DOM( \ottnt{H} )  
 \qquad\text{and}\qquad
  \mathbf{SATv} ( \ottnt{H} ,  \ottnt{R} ,  \ottnt{H}  \ottsym{(}  \ottsym{(}  a  \ottsym{,}   k   \ottsym{+}   j   \ottsym{)}  \ottsym{)} ,   [  j  /  z  ]  \, \tau' ) .
\]
Moreover, $x\notin\fv( [  j  /  z  ]  \, \tau')$.  The induction hypothesis therefore
gives
\[
  \mathbf{SATv} ( \ottnt{H} ,  \ottnt{R'} ,  \ottnt{H}  \ottsym{(}  \ottsym{(}  a  \ottsym{,}   k   \ottsym{+}   j   \ottsym{)}  \ottsym{)} ,   [  j  /  z  ]  \, \tau' ) .
\]
Since $j$ was arbitrary, the definition of $\SATV$ yields
$ \mathbf{SATv} ( \ottnt{H} ,  \ottnt{R'} ,  \ottsym{(}  a  \ottsym{,}   k   \ottsym{)} ,   \Pi z .( \tau'  \TREF^{\hspace{0.5pt} r })  ) $, as required.
\end{proof}

\begin{lemma}\ 
\label{lem:conown-irrevalent-to-integer}
\begin{enumerate}
\item\label{item:conown-irrevalent-to-integer-extension}
  \label{conown-irrevalent-to-integer1}
  If $ \mathbf{ConOwn}( \ottnt{H} ,  \ottnt{R} ,  \Gamma ) $ and $x \notin dom(\Gamma) \cup dom(R)$, then
  \[
     \mathbf{ConOwn}( \ottnt{H} ,  \ottnt{R}  \ottsym{\{}  x  \mapsto   n   \ottsym{\}} ,  \ottsym{(}  \Gamma  \ottsym{,}   x \COL  \{  \nu  :   \TINT    \mid   \varphi  \}    \ottsym{)} ) .
  \]
\item If $ \mathbf{ConOwn}( \ottnt{H} ,  \ottnt{R} ,  \Gamma ) $ and $ \Gamma  (  x  )   =   \{  \nu  :   \TINT    \mid   \varphi_{{\mathrm{2}}}  \} $, then
  \label{item:conown-irrevalent-to-integer2}
  \[
     \mathbf{ConOwn}( \ottnt{H} ,  \ottnt{R} ,   \Gamma  \left[  x \hookleftarrow  \{  \nu  :   \TINT    \mid   \varphi_{{\mathrm{1}}}  \}   \right]  ) .
  \]
\end{enumerate}
\rocqnote{Supporting Rocq results:
  \rocqdecl{Metatheory/Conservation.v}{env_ownership_names_extend_type_fresh}
  and \rocqdecl{Runtime/Model.v}{integer_ownership_zero}.}
\end{lemma}
\begin{proof}
  \begin{enumerate}
  \item %
    By $x\notin dom(\Gamma)$ and the implicit assumption that $\Gamma$ is well-formed,
    we have $x \not \in \fv( \Gamma  (  y  ) )$ for any $y \in dom (\Gamma)$.
    Moreover, $y \ne x$, so $(R\set{x \mapsto n})(y) = R(y)$ and
    $  \Gamma  \left[  x \hookleftarrow  \{  \nu  :   \TINT    \mid   \varphi  \}   \right]   (  y  )  =  \Gamma  (  y  ) $.
    Together with $x \not \in dom(R)$, \cref{lem:own-preservation2} gives, for any $y \in dom (\Gamma)$,
\begin{equation}
   \mathbf{own} ( \ottnt{H} ,  \ottnt{R} ,  \ottnt{R}  \ottsym{(}  y  \ottsym{)} ,   \Gamma  (  y  )  )  =  \mathbf{own} ( \ottnt{H} ,  \ottnt{R}  \ottsym{\{}  x  \mapsto   n   \ottsym{\}} ,  \ottnt{R}  \ottsym{\{}  x  \mapsto   n   \ottsym{\}}  \ottsym{(}  y  \ottsym{)} ,   \ottsym{(}  \Gamma  \ottsym{,}   x \COL  \{  \nu  :   \TINT    \mid   \varphi  \}    \ottsym{)}  (  y  )  ) . \label{eq:conown-irrevalent1}
\end{equation}
By the definition of $\OWN$,
\begin{equation}
   \mathbf{own} ( \ottnt{H} ,  \ottnt{R}  \ottsym{\{}  x  \mapsto   n   \ottsym{\}} ,  \ottnt{R}  \ottsym{\{}  x  \mapsto   n   \ottsym{\}}  \ottsym{(}  x  \ottsym{)} ,   \ottsym{(}  \Gamma  \ottsym{,}   x \COL  \{  \nu  :   \TINT    \mid   \varphi  \}    \ottsym{)}  (  x  )  )  =  \emptyset .\label{eq:conown-irrevalent2}
\end{equation}
From (\ref{eq:conown-irrevalent1}) and (\ref{eq:conown-irrevalent2}), $ \mathbf{Own}( \ottnt{H} ,  \ottnt{R} ,  \Gamma )  =  \mathbf{Own}( \ottnt{H} ,  \ottnt{R}  \ottsym{\{}  x  \mapsto   n   \ottsym{\}} ,  \ottsym{(}  \Gamma  \ottsym{,}   x \COL  \{  \nu  :   \TINT    \mid   \varphi  \}    \ottsym{)} ) $, that is,
 $\CONOWN(H,R\set{x \mapsto n},(\Gamma, x \COL \set{\nu \COL \TINT \mid \varphi}))$ holds if $\CONOWN(H,R,\Gamma)$.

\item For any $y \in dom (\Gamma \setminus \{x\})$, 
\begin{equation}
   \mathbf{own} ( \ottnt{H} ,  \ottnt{R} ,  \ottnt{R}  \ottsym{(}  y  \ottsym{)} ,    \Gamma  \left[  x \hookleftarrow  \{  \nu  :   \TINT    \mid   \varphi_{{\mathrm{1}}}  \}   \right]   (  y  )  )  =  \mathbf{own} ( \ottnt{H} ,  \ottnt{R} ,  \ottnt{R}  \ottsym{(}  y  \ottsym{)} ,   \Gamma  (  y  )  ) .\label{eq:conown-irrevalent3}
\end{equation}
because $  \Gamma  \left[  x \hookleftarrow  \{  \nu  :   \TINT    \mid   \varphi_{{\mathrm{1}}}  \}   \right]   (  y  )  =  \Gamma  (  y  ) $.
Furthermore, 
\begin{equation}
   \mathbf{own} ( \ottnt{H} ,  \ottnt{R} ,  \ottnt{R}  \ottsym{(}  x  \ottsym{)} ,    \Gamma  \left[  x \hookleftarrow  \{  \nu  :   \TINT    \mid   \varphi_{{\mathrm{1}}}  \}   \right]   (  x  )  )  =  \mathbf{own} ( \ottnt{H} ,  \ottnt{R} ,  \ottnt{R}  \ottsym{(}  x  \ottsym{)} ,   \Gamma  (  x  )  )  =  \emptyset .\label{eq:conown-irrevalent4}
\end{equation}
by definition.
From (\ref{eq:conown-irrevalent3}) and (\ref{eq:conown-irrevalent4}), 
$ \mathbf{Own}( \ottnt{H} ,  \ottnt{R} ,   \Gamma  \left[  x \hookleftarrow  \{  \nu  :   \TINT    \mid   \varphi_{{\mathrm{1}}}  \}   \right]  )  =  \mathbf{Own}( \ottnt{H} ,  \ottnt{R} ,  \Gamma ) $, that is,\\
 $ \mathbf{ConOwn}( \ottnt{H} ,  \ottnt{R} ,   \Gamma  \left[  x \hookleftarrow  \{  \nu  :   \TINT    \mid   \varphi_{{\mathrm{1}}}  \}   \right]  ) $.
  \end{enumerate}
\end{proof}

\begin{lemma}
\label{lem:sat-extension}
If $ \mathbf{SAT}( \ottnt{H} ,  \ottnt{R} ,  \Gamma ) $, $\models  \ottsym{[}  \ottnt{R}  \ottsym{]} \, \ottsym{[}   n   \ottsym{/}  \nu  \ottsym{]}  \varphi$,
$ \Gamma   \vdash   \varphi  \mbox{ ok} $, and $x \notin dom(R)$, then
\[
   \mathbf{SAT}( \ottnt{H} ,  \ottnt{R}  \ottsym{\{}  x  \mapsto   n   \ottsym{\}} ,  \ottsym{(}  \Gamma  \ottsym{,}   x \COL  \{  \nu  :   \TINT    \mid   \varphi  \}    \ottsym{)} ) .
\]
\rocqnote{Supporting Rocq results:
  \rocqdecl{Metatheory/Scoping.v}{consistent_extend_irrelevant} and
  \rocqdecl{Metatheory/Conservation.v}{consistent_extend_zero_ownership}.}
\end{lemma}
\begin{proof}
Put $\ottnt{R'} = \ottnt{R}  \ottsym{\{}  x  \mapsto   n   \ottsym{\}}$.
By $ \mathbf{SAT}( \ottnt{H} ,  \ottnt{R} ,  \Gamma ) $, $dom(\Gamma)\subseteq dom(\ottnt{R})$.
Thus $x\notin dom(\Gamma)$.  By well-formedness of $\Gamma$,
$x$ does not occur freely in any type in $\Gamma$.

First, let $y\in dom(\Gamma)$.  The satisfaction premise gives
$ \mathbf{SATv} ( \ottnt{H} ,  \ottnt{R} ,  \ottnt{R}  \ottsym{(}  y  \ottsym{)} ,   \Gamma  (  y  )  ) $.
By \cref{lem:sat-preservation-extend-register},
\[
   \mathbf{SATv} ( \ottnt{H} ,  \ottnt{R'} ,  \ottnt{R}  \ottsym{(}  y  \ottsym{)} ,   \Gamma  (  y  )  ) .
\]
Since $\ottnt{R'}(y)=\ottnt{R}(y)$, this establishes
$ \mathbf{SAT}( \ottnt{H} ,  \ottnt{R'} ,  \Gamma ) $.

Next, $ \Gamma   \vdash   \varphi  \mbox{ ok} $ and $x\notin dom(\Gamma)$ give
$x\notin\fv(\varphi)$.  Therefore, extending $\ottnt{R}$ at $x$
does not change the interpretation of $\ottsym{[}   n   \ottsym{/}  \nu  \ottsym{]}  \varphi$.
The refinement premise consequently gives
\[
  \models  \ottsym{[}  \ottnt{R'}  \ottsym{]} \, \ottsym{[}   n   \ottsym{/}  \nu  \ottsym{]}  \varphi.
\]
Together with $\ottnt{R'}  \ottsym{(}  x  \ottsym{)}=n$, this establishes
$ \mathbf{SATv} ( \ottnt{H} ,  \ottnt{R'} ,  \ottnt{R'}  \ottsym{(}  x  \ottsym{)} ,   \{  \nu  :   \TINT    \mid   \varphi  \}  ) $.

Combining these facts yields $ \mathbf{SAT}( \ottnt{H} ,  \ottnt{R'} ,  \ottsym{(}  \Gamma  \ottsym{,}   x \COL  \{  \nu  :   \TINT    \mid   \varphi  \}    \ottsym{)} ) $.
\end{proof}

\subsubsection{Subtyping and runtime invariants}
\label{sec:proofs-subtyping-invariants}

Subtyping allows a typing derivation to forget information without
invalidating its runtime interpretation. At the level of individual types,
passing to a supertype cannot increase the ownership attributed to a value
and preserves satisfaction. The corresponding environment-level results
preserve ownership consistency and satisfaction for the whole
configuration. These lemmas justify uses of \rn{T-Sub} and the environment
subtyping premises exposed by \cref{lem:inversion-expression-typing}.

\begin{lemma}
\label{lem:own-preservation-weakening-ty}
If $\Gamma  \vdash   \tau_{{\mathrm{1}}}   \leq   \tau_{{\mathrm{2}}} $ and $\models  \ottsym{[}  \ottnt{R}  \ottsym{]} \,  \fml{ \Gamma } $, then $ \mathbf{own} ( \ottnt{H} ,  \ottnt{R} ,  v ,  \tau_{{\mathrm{2}}} )  \le  \mathbf{own} ( \ottnt{H} ,  \ottnt{R} ,  v ,  \tau_{{\mathrm{1}}} ) $.
\rocqref[Supporting Rocq result (with a satisfaction premise)]{Metatheory/Subtyping.v}{value_ownership_subtype}
\end{lemma}
\begin{proof}
    By (strong) induction on the sum of the sizes of $\tau_{{\mathrm{1}}}$ and $\tau_{{\mathrm{2}}}$ with case analysis on the last rule used to derive $\Gamma  \vdash   \tau_{{\mathrm{1}}}   \leq   \tau_{{\mathrm{2}}} $.
    The case \rn{S-Int} where $\tau_{{\mathrm{1}}} =  \{  \nu  :   \TINT    \mid   \varphi_{{\mathrm{1}}}  \} $ and 
    $\tau_{{\mathrm{2}}} =  \{  \nu  :   \TINT    \mid   \varphi_{{\mathrm{2}}}  \} $ is trivial because $ \mathbf{own} ( \ottnt{H} ,  \ottnt{R} ,  v ,  \tau_{{\mathrm{1}}} )  =  \mathbf{own} ( \ottnt{H} ,  \ottnt{R} ,  v ,  \tau_{{\mathrm{2}}} )  =  \emptyset $.

    The case \rn{S-Ref} where $\tau_{{\mathrm{1}}} =  \Pi z .( \tau'_{{\mathrm{1}}}  \TREF^{\hspace{0.5pt} r_{{\mathrm{1}}} })  $ and $\tau_{{\mathrm{2}}} =  \Pi z .( \tau'_{{\mathrm{2}}}  \TREF^{\hspace{0.5pt} r_{{\mathrm{2}}} })  $ 
    with $\Gamma  \ottsym{,}   z \COL  \TINT    \models  r_{{\mathrm{1}}} \, \ge \, r_{{\mathrm{2}}}$ and $\Gamma  \ottsym{,}   z \COL  \TINT    \vdash   \tau'_{{\mathrm{1}}}   \leq   \tau'_{{\mathrm{2}}} $:
    If $v$ is not an address, $\mathbf{own}$ returns the ``everywhere-zero'' mapping, hence the conclusion is trivial.
    Otherwise, let $v=\ottsym{(}  a  \ottsym{,}  k  \ottsym{)}$. For every integer $j$, Item~3 of
    \cref{lem:index-substitution} gives
    $\Gamma  \vdash    [  j  /  z  ]  \, \tau'_{{\mathrm{1}}}   \leq    [  j  /  z  ]  \, \tau'_{{\mathrm{2}}} $. Since substitution does not change type size, I.H. gives
  \[
   \mathbf{own} ( \ottnt{H} ,  \ottnt{R} ,  \ottnt{H}  \ottsym{(}  \ottsym{(}  a  \ottsym{,}   k   \ottsym{+}   j   \ottsym{)}  \ottsym{)} ,   [  j  /  z  ]  \, \tau'_{{\mathrm{2}}} ) 
  \le
   \mathbf{own} ( \ottnt{H} ,  \ottnt{R} ,  \ottnt{H}  \ottsym{(}  \ottsym{(}  a  \ottsym{,}   k   \ottsym{+}   j   \ottsym{)}  \ottsym{)} ,   [  j  /  z  ]  \, \tau'_{{\mathrm{1}}} ) 
  \]
  whenever $\ottsym{(}  a  \ottsym{,}  k  \ottsym{+}  j  \ottsym{)} \in dom(\ottnt{H})$.
  
  By Item~1 of \cref{lem:index-substitution} and
  $\models  \ottsym{[}  \ottnt{R}  \ottsym{]} \,  \fml{ \Gamma } $,
  \[
     \llbracket    [  j  /  z  ]    r_{{\mathrm{1}}}   \rrbracket_{ \ottnt{R} }  \ge  \llbracket    [  j  /  z  ]    r_{{\mathrm{2}}}   \rrbracket_{ \ottnt{R} } 
  \]
  for every $j$.  Hence the direct ownership contribution for $r_{{\mathrm{2}}}$
  is bounded by that for $r_{{\mathrm{1}}}$, and every index occurring in the
  recursive sum for $r_{{\mathrm{2}}}$ also occurs in the sum for $r_{{\mathrm{1}}}$.
  By part (2) of \cref{lem:ownership-bounds}, all these ownership maps
  are pointwise nonnegative, so the
  recursive sum for $\tau'_{{\mathrm{2}}}$ is bounded by that for $\tau'_{{\mathrm{1}}}$. Adding
  the two inequalities proves the result.
\end{proof}

\begin{lemma}
\label{lem:sat-preservation-weakening}
If $ \mathbf{SATv} ( \ottnt{H} ,  \ottnt{R} ,  v ,  \tau' ) $, $\Gamma  \vdash   \tau'   \leq   \tau $,
and $\models  \ottsym{[}  \ottnt{R}  \ottsym{]} \,  \fml{ \Gamma } $, then
$ \mathbf{SATv} ( \ottnt{H} ,  \ottnt{R} ,  v ,  \tau ) $.
\rocqref{Metatheory/Subtyping.v}{value_satisfies_subtype}
\end{lemma}
\begin{proof}
By structural induction on $\tau$, generalizing $\tau'$ and $v$.

Suppose $\tau  =   \{  \nu  :   \TINT    \mid   \varphi  \} $.  Inversion of the subtyping premise gives
$\tau'  =   \{  \nu  :   \TINT    \mid   \varphi'  \} $ and
$ \Gamma  ,  \nu  \colon \TINT   \models  \varphi'  \implies  \varphi$.  Inversion of the satisfaction premise gives
$v \,  =  \,  n $ for some integer $n$ and
$\models  \ottsym{[}  \ottnt{R}  \ottsym{]} \, \ottsym{[}   n   \ottsym{/}  \nu  \ottsym{]}  \varphi'$.  Instantiating the validity premise of \rn{S-Int}
with $\ottnt{R}$ and $\nu \,  =  \,  n $, using $\models  \ottsym{[}  \ottnt{R}  \ottsym{]} \,  \fml{ \Gamma } $, yields
$\models  \ottsym{[}  \ottnt{R}  \ottsym{]} \, \ottsym{[}   n   \ottsym{/}  \nu  \ottsym{]}  \varphi$.  Hence
$ \mathbf{SATv} ( \ottnt{H} ,  \ottnt{R} ,   n  ,   \{  \nu  :   \TINT    \mid   \varphi  \}  ) $ by the definition of $\SATV$.

Suppose $\tau  =   \Pi z .( \tau_{{\mathrm{1}}}  \TREF^{\hspace{0.5pt} r_{{\mathrm{1}}} }) $.  Inversion of the subtyping premise gives
$\tau'  =   \Pi z .( \tau_{{\mathrm{2}}}  \TREF^{\hspace{0.5pt} r_{{\mathrm{2}}} }) $,
$\Gamma  \ottsym{,}   z \COL  \TINT    \models  r_{{\mathrm{2}}} \, \ge \, r_{{\mathrm{1}}}$, and
$\Gamma  \ottsym{,}   z \COL  \TINT    \vdash   \tau_{{\mathrm{2}}}   \leq   \tau_{{\mathrm{1}}} $.  By Item~1 of
\cref{lem:index-substitution} and $\models  \ottsym{[}  \ottnt{R}  \ottsym{]} \,  \fml{ \Gamma } $, for every integer
$j$,
\begin{equation}
  \llbracket    [  j  /  z  ]    r_{{\mathrm{2}}}   \rrbracket_{ \ottnt{R} }  \ge  \llbracket    [  j  /  z  ]    r_{{\mathrm{1}}}   \rrbracket_{ \ottnt{R} } .
 \label{sat-preservation-weakening1}
\end{equation}
Inversion of the satisfaction premise leaves two cases for $v$.

If $v \,  =  \,  \NULL $, then $ \llbracket    [  j  /  z  ]    r_{{\mathrm{2}}}   \rrbracket_{ \ottnt{R} } =0$ for every integer $j$.
Together with \cref{sat-preservation-weakening1} and the nonnegativity of
$ \llbracket    [  j  /  z  ]    r_{{\mathrm{1}}}   \rrbracket_{ \ottnt{R} } $ from part (1) of \cref{lem:ownership-bounds}, this gives
$ \llbracket    [  j  /  z  ]    r_{{\mathrm{1}}}   \rrbracket_{ \ottnt{R} } =0$ for every $j$.  Thus
$ \mathbf{SATv} ( \ottnt{H} ,  \ottnt{R} ,  \ottkw{null} ,   \Pi z .( \tau_{{\mathrm{1}}}  \TREF^{\hspace{0.5pt} r_{{\mathrm{1}}} })  ) $.

Otherwise, $v \,  =  \, \ottsym{(}  a  \ottsym{,}   k   \ottsym{)}$ for some $a$ and $k$.  Consider an
arbitrary integer $j$ such that $ \llbracket    [  j  /  z  ]    r_{{\mathrm{1}}}   \rrbracket_{ \ottnt{R} } >0$.
By \cref{sat-preservation-weakening1},
$ \llbracket    [  j  /  z  ]    r_{{\mathrm{2}}}   \rrbracket_{ \ottnt{R} } >0$.  The satisfaction premise therefore gives
\[
 (a,k+j) \in dom(\ottnt{H})
 \qquad\text{and}\qquad
  \mathbf{SATv} ( \ottnt{H} ,  \ottnt{R} ,  \ottnt{H}  \ottsym{(}  \ottsym{(}  a  \ottsym{,}   k   \ottsym{+}   j   \ottsym{)}  \ottsym{)} ,   [  j  /  z  ]  \, \tau_{{\mathrm{2}}} ) .
\]
Moreover, Item~3 of \cref{lem:index-substitution} gives
$\Gamma  \vdash    [  j  /  z  ]  \, \tau_{{\mathrm{2}}}   \leq    [  j  /  z  ]  \, \tau_{{\mathrm{1}}} $.  The induction hypothesis then yields
\[
  \mathbf{SATv} ( \ottnt{H} ,  \ottnt{R} ,  \ottnt{H}  \ottsym{(}  \ottsym{(}  a  \ottsym{,}   k   \ottsym{+}   j   \ottsym{)}  \ottsym{)} ,   [  j  /  z  ]  \, \tau_{{\mathrm{1}}} ) .
\]
Since $j$ was arbitrary, the definition of $\SATV$ gives
$ \mathbf{SATv} ( \ottnt{H} ,  \ottnt{R} ,  \ottsym{(}  a  \ottsym{,}   k   \ottsym{)} ,   \Pi z .( \tau_{{\mathrm{1}}}  \TREF^{\hspace{0.5pt} r_{{\mathrm{1}}} })  ) $, as required.
\end{proof}

As an immediate consequence of \cref{lem:sat-preservation-weakening} and
\rn{TEq-Sub}, if $\Gamma  \vdash   \tau'   \approx   \tau $ and $\models  \ottsym{[}  \ottnt{R}  \ottsym{]} \,  \fml{ \Gamma } $, then
$ \mathbf{SATv} ( \ottnt{H} ,  \ottnt{R} ,  v ,  \tau )   \iff   \mathbf{SATv} ( \ottnt{H} ,  \ottnt{R} ,  v ,  \tau' ) $.

\begin{lemma}
\label{lem:conown-preservation-weakening-tyenv}
if $ \mathbf{ConOwn}( \ottnt{H} ,  \ottnt{R} ,  \Gamma ) $ and $\Gamma  \leq  \Gamma'$ and $\models  \ottsym{[}  \ottnt{R}  \ottsym{]} \,  \fml{ \Gamma } $, then $ \mathbf{ConOwn}( \ottnt{H} ,  \ottnt{R} ,  \Gamma' ) $.
\rocqref[Supporting Rocq result (for the combined consistency invariant)]{Metatheory/Subtyping.v}{env_subtype_preserves_consistent}
\end{lemma}
\begin{proof}
  Fix any $ \ottsym{(}  a  \ottsym{,}  k  \ottsym{)}  \in   \DOM( \ottnt{H} )  $.  By inversion of \rn{S-TyEnv},
  $ \DOM( \Gamma' )  \subseteq  \DOM( \Gamma ) $ and, for every
  $ x  \in   \DOM( \Gamma' )  $,
  $
    \Gamma  \vdash    \Gamma  (  x  )    \leq    \Gamma'  (  x  )  .
  $
  Hence, by \cref{lem:own-preservation-weakening-ty},
  \begin{equation}
     \mathbf{own} ( \ottnt{H} ,  \ottnt{R} ,  \ottnt{R}  \ottsym{(}  x  \ottsym{)} ,   \Gamma'  (  x  )  ) (a,k)
    \le
     \mathbf{own} ( \ottnt{H} ,  \ottnt{R} ,  \ottnt{R}  \ottsym{(}  x  \ottsym{)} ,   \Gamma  (  x  )  ) (a,k)
    \qquad (x\in  \DOM( \Gamma' ) ).
    \label{eq:conown-env-subtype-binding}
  \end{equation}
  It follows that
  \begin{align*}
     \mathbf{Own}( \ottnt{H} ,  \ottnt{R} ,  \Gamma' ) (a,k)
    &=  \sum_{ x  \in dom( \Gamma' ) }   \mathbf{own} ( \ottnt{H} ,  \ottnt{R} ,  \ottnt{R}  \ottsym{(}  x  \ottsym{)} ,   \Gamma'  (  x  )  )  (a,k)\\
    &\le  \sum_{ x  \in dom( \Gamma' ) }   \mathbf{own} ( \ottnt{H} ,  \ottnt{R} ,  \ottnt{R}  \ottsym{(}  x  \ottsym{)} ,   \Gamma  (  x  )  )  (a,k)
      &&\text{by \cref{eq:conown-env-subtype-binding}}\\
    &\le  \sum_{ x  \in dom( \Gamma ) }   \mathbf{own} ( \ottnt{H} ,  \ottnt{R} ,  \ottnt{R}  \ottsym{(}  x  \ottsym{)} ,   \Gamma  (  x  )  )  (a,k)\\
    &=  \mathbf{Own}( \ottnt{H} ,  \ottnt{R} ,  \Gamma ) (a,k) \le 1.
  \end{align*}
  The second inequality uses $ \DOM( \Gamma' ) \subseteq  \DOM( \Gamma ) $ and
  pointwise nonnegativity of ownership from
  \cref{lem:ownership-bounds}.  The last inequality follows from
  $ \mathbf{ConOwn}( \ottnt{H} ,  \ottnt{R} ,  \Gamma ) $.  Since $\ottsym{(}  a  \ottsym{,}  k  \ottsym{)}$ was an arbitrary location in
  $ \DOM( \ottnt{H} ) $, we obtain $ \mathbf{ConOwn}( \ottnt{H} ,  \ottnt{R} ,  \Gamma' ) $.
\end{proof}

\begin{lemma}
\label{lem:sat-preservation-greater-tyenv}
If $ \mathbf{SAT}( \ottnt{H} ,  \ottnt{R} ,  \Gamma ) $, $\Gamma  \leq  \Gamma'$,
and $\models  \ottsym{[}  \ottnt{R}  \ottsym{]} \,  \fml{ \Gamma } $, then
$ \mathbf{SAT}( \ottnt{H} ,  \ottnt{R} ,  \Gamma' ) $.
\rocqref{Metatheory/Subtyping.v}{env_satisfies_subtype}
\end{lemma}
\begin{proof}
Let $x$ be any variable in $dom(\Gamma')$.  By inversion of
$\Gamma  \leq  \Gamma'$, we have $x\in dom(\Gamma)$ and
$\Gamma  \vdash    \Gamma  (  x  )    \leq    \Gamma'  (  x  )  $.  Hence $ \mathbf{SAT}( \ottnt{H} ,  \ottnt{R} ,  \Gamma ) $ gives
$ \mathbf{SATv} ( \ottnt{H} ,  \ottnt{R} ,  \ottnt{R}  \ottsym{(}  x  \ottsym{)} ,   \Gamma  (  x  )  ) $.  Applying
\cref{lem:sat-preservation-weakening} with $\models  \ottsym{[}  \ottnt{R}  \ottsym{]} \,  \fml{ \Gamma } $ yields
$ \mathbf{SATv} ( \ottnt{H} ,  \ottnt{R} ,  \ottnt{R}  \ottsym{(}  x  \ottsym{)} ,   \Gamma'  (  x  )  ) $.  Since $x$ was arbitrary, the definition of
$\SAT$ gives $ \mathbf{SAT}( \ottnt{H} ,  \ottnt{R} ,  \Gamma' ) $.
\end{proof}

\subsubsection{Refinement strengthening, emptiness, and type addition}
\label{sec:proofs-type-operations}

These lemmas describe how the operations used to redistribute types affect
ownership and satisfaction. Adding an equality refinement does not change
ownership (\cref{lem:own-preservation-equivalent-ty-strength}).
Empty types contribute no ownership (\cref{lem:empty-own}), and empty
reference types impose no further requirements on pointer values
(\cref{lem:sat-preservation-empty}). Type addition accounts for splitting
and recombining ownership (\cref{lem:own-preservation}), with satisfaction
of the sum equivalent to satisfaction of both components
(\cref{lem:sat-preservation-add-ty}). These properties are used
when bindings share a value or an operation transfers ownership between
types.

\begin{lemma}
\label{lem:own-preservation-equivalent-ty-strength}
$ \mathbf{own} ( \ottnt{H} ,  \ottnt{R} ,  v ,  \tau )  =  \mathbf{own} ( \ottnt{H} ,  \ottnt{R} ,  v ,   \tau  ^ {= x }  )  $ holds.
\rocqref{Metatheory/Conservation.v}{value_ownership_equal_to_var}
\end{lemma}
\begin{proof}
  By case analysis on $\tau$ and the definitions of strengthening and $\mathbf{own}$.
\end{proof}

\begin{lemma}
\label{lem:empty-own}
If $\models  \ottsym{[}  \ottnt{R}  \ottsym{]} \,  \Empty{ \tau } $, then $ \mathbf{own} ( \ottnt{H} ,  \ottnt{R} ,  v ,  \tau ) = \emptyset $.
\rocqref{Metatheory/Conservation.v}{value_ownership_empty}
\end{lemma}
\begin{proof}
By case analysis on $\tau$. If $\tau$ is an integer type, its ownership map
is $ \emptyset $ by definition. Suppose
$\tau  =   \Pi z .( \tau'  \TREF^{\hspace{0.5pt} r }) $. If $v$ is not an address, the result follows
immediately. Otherwise, write $v \,  =  \, \ottsym{(}  a  \ottsym{,}   k   \ottsym{)}$. From $\models  \ottsym{[}  \ottnt{R}  \ottsym{]} \,  \Empty{ \tau } $ and the
definition of $ \Empty{ \tau } $, for every integer $j$,
\[
 \llbracket    [  j  /  z  ]    r   \rrbracket_{ \ottnt{R} } =0.
\]
Therefore every direct ownership contribution is zero, and the guard of
every recursive contribution is false. Hence
$ \mathbf{own} ( \ottnt{H} ,  \ottnt{R} ,  v ,  \tau ) = \emptyset $.
\end{proof}

\begin{lemma}
\label{lem:sat-preservation-empty}
If $\models   \Empty{  \Pi z .( \tau'  \TREF^{\hspace{0.5pt} r })  } $, then
$ \mathbf{SATv} ( \ottnt{H} ,  \ottnt{R} ,  pv ,   \Pi z .( \tau'  \TREF^{\hspace{0.5pt} r })  ) $.
\rocqref{Runtime/Model.v}{empty_value_satisfies}
\end{lemma}
\begin{proof}
Instantiating the emptiness premise with $\ottnt{R}$ and each index gives
\[
  \forall j.\ \sem{r}_{R\{z \mapsto j\}}=0.
\]
If $pv \,  =  \,  \NULL $, this is exactly the corresponding clause in the definition
of $\SATV$.  Otherwise, $pv \,  =  \, \ottsym{(}  a  \ottsym{,}   i   \ottsym{)}$ for some $a$ and $i$.  For
every $j$, the antecedent $\sem{r}_{R\{z \mapsto j\}}>0$ in the definition
of $\SATV$ is false.  Hence $ \mathbf{SATv} ( \ottnt{H} ,  \ottnt{R} ,  pv ,   \Pi z .( \tau'  \TREF^{\hspace{0.5pt} r })  ) $ holds.
\end{proof}

\begin{lemma}
\label{lem:own-preservation}
If $\Gamma  \vdash    \tau_{{\mathrm{1}}}  +  \tau_{{\mathrm{2}}}    \approx   \tau_{{\mathrm{3}}} $ and $\models  \ottsym{[}  \ottnt{R}  \ottsym{]} \,  \fml{ \Gamma } $, then
$ \mathbf{own} ( \ottnt{H} ,  \ottnt{R} ,  v ,  \tau_{{\mathrm{3}}} )  =  \mathbf{own} ( \ottnt{H} ,  \ottnt{R} ,  v ,  \tau_{{\mathrm{1}}} )  +  \mathbf{own} ( \ottnt{H} ,  \ottnt{R} ,  v ,  \tau_{{\mathrm{2}}} ) $.
\rocqref[Supporting Rocq result (with a satisfaction premise)]{Metatheory/Conservation.v}{value_ownership_add}
\end{lemma}
\begin{proof}

  By strong induction on the size of $\tau_{{\mathrm{3}}}$, with case analysis on the
  last rule used to derive $\Gamma  \vdash    \tau_{{\mathrm{1}}}  +  \tau_{{\mathrm{2}}}    \approx   \tau_{{\mathrm{3}}} $.

  \begin{rncase}{TA-Int}
  By inversion, $\tau_{{\mathrm{1}}}$, $\tau_{{\mathrm{2}}}$, and $\tau_{{\mathrm{3}}}$ are integer types.
  Their ownership maps are all $ \emptyset $, so the conclusion follows
  immediately.

  \end{rncase}

  \begin{rncase}{TA-Ref}
  By inversion, there exist $\tau'_{{\mathrm{1}}}$, $\tau'_{{\mathrm{2}}}$, $\tau'_{{\mathrm{3}}}$,
  $r_{{\mathrm{1}}}$, $r_{{\mathrm{2}}}$, and $r_{{\mathrm{3}}}$ such that
  \[
  \begin{array}{lll}
   \tau_{{\mathrm{1}}}= \Pi z .( \tau'_{{\mathrm{1}}}  \TREF^{\hspace{0.5pt} r_{{\mathrm{1}}} }) ,&
   \tau_{{\mathrm{2}}}= \Pi z .( \tau'_{{\mathrm{2}}}  \TREF^{\hspace{0.5pt} r_{{\mathrm{2}}} }) ,&
   \tau_{{\mathrm{3}}}= \Pi z .( \tau'_{{\mathrm{3}}}  \TREF^{\hspace{0.5pt} r_{{\mathrm{3}}} }) ,
  \end{array}
  \]
  and
  \[
   \Gamma  \ottsym{,}   z \COL  \TINT    \models  r_{{\mathrm{1}}}  \ottsym{+}  r_{{\mathrm{2}}} \,  =  \, r_{{\mathrm{3}}}
   \qquad\text{and}\qquad
   \Gamma  \ottsym{,}   z \COL  \TINT    \vdash    \tau'_{{\mathrm{1}}}  +  \tau'_{{\mathrm{2}}}    \approx   \tau'_{{\mathrm{3}}} .
  \]

  If $v$ is not an address, all three ownership maps are
  $ \emptyset $.  Otherwise, let $v=\ottsym{(}  a  \ottsym{,}  k  \ottsym{)}$.

  By semantic instantiation of the ownership-addition premise and
  $\models  \ottsym{[}  \ottnt{R}  \ottsym{]} \,  \fml{ \Gamma } $, for every integer $j$,
  \begin{equation}
    \llbracket    [  j  /  z  ]    r_{{\mathrm{1}}}   \rrbracket_{ \ottnt{R} }   \ottsym{+}   \llbracket    [  j  /  z  ]    r_{{\mathrm{2}}}   \rrbracket_{ \ottnt{R} }  \,  =  \,  \llbracket    [  j  /  z  ]    r_{{\mathrm{3}}}   \rrbracket_{ \ottnt{R} } .
   \label{eq:own-add-coefficients}
  \end{equation}
  Consequently, the sum of the direct ownership contributions for
  $r_{{\mathrm{1}}}$ and $r_{{\mathrm{2}}}$ is the direct ownership contribution for
  $r_{{\mathrm{3}}}$.

  For every integer $j$, Item~5 of
  \cref{lem:index-substitution} gives
  \[
   \Gamma  \vdash     [  j  /  z  ]  \, \tau'_{{\mathrm{1}}}  +   [  j  /  z  ]  \, \tau'_{{\mathrm{2}}}    \approx    [  j  /  z  ]  \, \tau'_{{\mathrm{3}}} .
  \]
  Since substitution does not change type size, the induction hypothesis
  gives, whenever $\ottsym{(}  a  \ottsym{,}  k  \ottsym{+}  j  \ottsym{)}\in dom(\ottnt{H})$,
  \begin{equation}
  \begin{split}
   & \mathbf{own} ( \ottnt{H} ,  \ottnt{R} ,  \ottnt{H}  \ottsym{(}  \ottsym{(}  a  \ottsym{,}   k   \ottsym{+}   j   \ottsym{)}  \ottsym{)} ,   [  j  /  z  ]  \, \tau'_{{\mathrm{3}}} ) \\
   &\qquad =
    \mathbf{own} ( \ottnt{H} ,  \ottnt{R} ,  \ottnt{H}  \ottsym{(}  \ottsym{(}  a  \ottsym{,}   k   \ottsym{+}   j   \ottsym{)}  \ottsym{)} ,   [  j  /  z  ]  \, \tau'_{{\mathrm{1}}} ) 
   +
    \mathbf{own} ( \ottnt{H} ,  \ottnt{R} ,  \ottnt{H}  \ottsym{(}  \ottsym{(}  a  \ottsym{,}   k   \ottsym{+}   j   \ottsym{)}  \ottsym{)} ,   [  j  /  z  ]  \, \tau'_{{\mathrm{2}}} ) .
  \end{split}
  \label{eq:own-add-recursive}
  \end{equation}

  It remains to compare the guards of the three recursive sums.
  By part (1) of \cref{lem:ownership-bounds}, the values of $r_{{\mathrm{1}}}$
  and $r_{{\mathrm{2}}}$ are nonnegative.
  For each $j$ such that $\ottsym{(}  a  \ottsym{,}  k  \ottsym{+}  j  \ottsym{)}\in dom(\ottnt{H})$, we
  distinguish four cases.

  \begin{itemize}
  \item If
    $ \llbracket    [  j  /  z  ]    r_{{\mathrm{1}}}   \rrbracket_{ \ottnt{R} } >0$ and
    $ \llbracket    [  j  /  z  ]    r_{{\mathrm{2}}}   \rrbracket_{ \ottnt{R} } >0$,
    all three recursive sums contain the corresponding terms, and their
    equality follows from \eqref{eq:own-add-recursive}.

  \item If
    $ \llbracket    [  j  /  z  ]    r_{{\mathrm{1}}}   \rrbracket_{ \ottnt{R} } >0$ and
    $ \llbracket    [  j  /  z  ]    r_{{\mathrm{2}}}   \rrbracket_{ \ottnt{R} } =0$,
    the zero-ownership/Empty condition in the well-formedness of
    $\tau_{{\mathrm{2}}}$ gives
    \[
      \models  \ottsym{[}  \ottnt{R}  \ottsym{]} \,  \Empty{  [  j  /  z  ]  \, \tau'_{{\mathrm{2}}} } .
    \]
    Hence, by \cref{lem:empty-own}, 
    \[
       \mathbf{own} ( \ottnt{H} ,  \ottnt{R} ,  \ottnt{H}  \ottsym{(}  \ottsym{(}  a  \ottsym{,}   k   \ottsym{+}   j   \ottsym{)}  \ottsym{)} ,   [  j  /  z  ]  \, \tau'_{{\mathrm{2}}} ) = \emptyset .
    \]
    Equation~\eqref{eq:own-add-recursive} therefore identifies the term in the recursive sum for
    $\tau_{{\mathrm{3}}}$ with the term in the sum for $\tau_{{\mathrm{1}}}$.

  \item The case
    $ \llbracket    [  j  /  z  ]    r_{{\mathrm{1}}}   \rrbracket_{ \ottnt{R} } =0$ and
    $ \llbracket    [  j  /  z  ]    r_{{\mathrm{2}}}   \rrbracket_{ \ottnt{R} } >0$
    is symmetric.

  \item If both values are zero, \eqref{eq:own-add-coefficients} shows that
    $ \llbracket    [  j  /  z  ]    r_{{\mathrm{3}}}   \rrbracket_{ \ottnt{R} } =0$, so none of the recursive sums contains a term
    at $j$.
  \end{itemize}

  Thus, the sum of the recursive ownership contributions for $\tau_{{\mathrm{1}}}$
  and $\tau_{{\mathrm{2}}}$ equals the recursive ownership contribution for $\tau_{{\mathrm{3}}}$.
  Combining this equality with the equality of the direct contributions
  proves
  \(
    \mathbf{own} ( \ottnt{H} ,  \ottnt{R} ,  v ,  \tau_{{\mathrm{3}}} ) 
   =
    \mathbf{own} ( \ottnt{H} ,  \ottnt{R} ,  v ,  \tau_{{\mathrm{1}}} ) 
   +
    \mathbf{own} ( \ottnt{H} ,  \ottnt{R} ,  v ,  \tau_{{\mathrm{2}}} ) .
  \)
  \end{rncase}
\end{proof}

\begin{lemma}
  \label{lem:sat-preservation-add-ty}
  If $\Gamma  \vdash    \tau_{{\mathrm{1}}}  +  \tau_{{\mathrm{2}}}    \approx   \tau_{{\mathrm{3}}} $ and $\models  \ottsym{[}  \ottnt{R}  \ottsym{]} \,  \fml{ \Gamma } $, then
 $ \mathbf{SATv} ( \ottnt{H} ,  \ottnt{R} ,  v ,  \tau_{{\mathrm{1}}} )   \wedge   \mathbf{SATv} ( \ottnt{H} ,  \ottnt{R} ,  v ,  \tau_{{\mathrm{2}}} )   \iff   \mathbf{SATv} ( \ottnt{H} ,  \ottnt{R} ,  v ,  \tau_{{\mathrm{3}}} ) $.
\rocqref{Metatheory/Conservation.v}{value_satisfies_add_iff}
\end{lemma}
\begin{proof}
We use the standing convention that the types occurring in the judgment are
well formed.  We also use two elementary consequences of the definitions.
First, type addition relates types of the same simple shape.  Second, if
$\models  \ottsym{[}  \ottnt{R}  \ottsym{]} \,  \Empty{ \tau } $ and a runtime value has simple type $|\tau|$, then it
satisfies $\tau$.  The latter fact follows by induction on $\tau$: the
integer case uses the strengthened definition of $ \Empty{ \tau } $, and at a
reference type all outer ownerships are zero.

Proceed by induction on the size of $\tau_{{\mathrm{3}}}$, with the induction hypothesis
available for all strictly smaller result types, and invert the type-addition
derivation.

\begin{rncase}{TA-Int}
There are $\varphi_{{\mathrm{1}}}$, $\varphi_{{\mathrm{2}}}$, and $\varphi_{{\mathrm{3}}}$ such that
\[
 \tau_{{\mathrm{1}}}= \{  \nu  :   \TINT    \mid   \varphi_{{\mathrm{1}}}  \} ,\qquad
 \tau_{{\mathrm{2}}}= \{  \nu  :   \TINT    \mid   \varphi_{{\mathrm{2}}}  \} ,\qquad
 \tau_{{\mathrm{3}}}= \{  \nu  :   \TINT    \mid   \varphi_{{\mathrm{3}}}  \} ,
\]
and $\Gamma  \models  \varphi_{{\mathrm{1}}}  \wedge  \varphi_{{\mathrm{2}}}  \iff  \varphi_{{\mathrm{3}}}$.  If $v$ is not an integer, all
three satisfaction judgments are false.  Otherwise, write $v \,  =  \,  n $.
Semantic instantiation of the premise by $\ottnt{R}$ and $\nu \,  =  \,  n $ gives
\[
 \models  \ottsym{[}  \ottnt{R}  \ottsym{]} \, \ottsym{[}   n   \ottsym{/}  \nu  \ottsym{]}  \varphi_{{\mathrm{3}}}
 \quad \iff \quad
 \models  \ottsym{[}  \ottnt{R}  \ottsym{]} \, \ottsym{[}   n   \ottsym{/}  \nu  \ottsym{]}  \varphi_{{\mathrm{1}}}  \wedge  \models  \ottsym{[}  \ottnt{R}  \ottsym{]} \, \ottsym{[}   n   \ottsym{/}  \nu  \ottsym{]}  \varphi_{{\mathrm{2}}},
\]
which is exactly the required equivalence.

\end{rncase}

\begin{rncase}{TA-Ref}
After alpha-renaming the bound index if necessary, inversion gives
$\tau'_{{\mathrm{1}}}$, $\tau'_{{\mathrm{2}}}$, $\tau'_{{\mathrm{3}}}$, $r_{{\mathrm{1}}}$,
$r_{{\mathrm{2}}}$, and $r_{{\mathrm{3}}}$ such that
\[
\begin{array}{lll}
 \tau_{{\mathrm{1}}}= \Pi z .( \tau'_{{\mathrm{1}}}  \TREF^{\hspace{0.5pt} r_{{\mathrm{1}}} }) ,&
 \tau_{{\mathrm{2}}}= \Pi z .( \tau'_{{\mathrm{2}}}  \TREF^{\hspace{0.5pt} r_{{\mathrm{2}}} }) ,&
 \tau_{{\mathrm{3}}}= \Pi z .( \tau'_{{\mathrm{3}}}  \TREF^{\hspace{0.5pt} r_{{\mathrm{3}}} }) ,
\end{array}
\]
and
\[
 \Gamma  \ottsym{,}   z \COL  \TINT    \models  r_{{\mathrm{1}}}  \ottsym{+}  r_{{\mathrm{2}}} \,  =  \, r_{{\mathrm{3}}}
 \qquad\text{and}\qquad
 \Gamma  \ottsym{,}   z \COL  \TINT    \vdash    \tau'_{{\mathrm{1}}}  +  \tau'_{{\mathrm{2}}}    \approx   \tau'_{{\mathrm{3}}} .
\]
For an integer $j$, put
\[
 q_1(j)= \llbracket    [  j  /  z  ]    r_{{\mathrm{1}}}   \rrbracket_{ \ottnt{R} } ,\qquad
 q_2(j)= \llbracket    [  j  /  z  ]    r_{{\mathrm{2}}}   \rrbracket_{ \ottnt{R} } ,\qquad
 q_3(j)= \llbracket    [  j  /  z  ]    r_{{\mathrm{3}}}   \rrbracket_{ \ottnt{R} } .
\]
Item~1 of \cref{lem:index-substitution} and $\models  \ottsym{[}  \ottnt{R}  \ottsym{]} \,  \fml{ \Gamma } $ give
\begin{equation}
 q_1(j)+q_2(j)=q_3(j)
 \qquad(j\in\mathbb Z).
 \label{eq:sat-add-ownership}
\end{equation}
Moreover, Item~5 gives
\begin{equation}
 \Gamma  \vdash     [  j  /  z  ]  \, \tau'_{{\mathrm{1}}}  +   [  j  /  z  ]  \, \tau'_{{\mathrm{2}}}    \approx    [  j  /  z  ]  \, \tau'_{{\mathrm{3}}} .
 \label{eq:sat-add-elements}
\end{equation}
The \rn{WF-Ref} premises for $\tau_{{\mathrm{1}}}$, $\tau_{{\mathrm{2}}}$, and $\tau_{{\mathrm{3}}}$, together
with Item~2, give the well-formedness of the three instantiated element types
and ownership terms.  Hence the values $q_1(j)$ and $q_2(j)$ are nonnegative
by \cref{lem:ownership-bounds}.

If $v$ is an integer, all three satisfaction judgments are false.  If
$v \,  =  \,  \NULL $, satisfaction by $\tau_{{\mathrm{1}}}$, $\tau_{{\mathrm{2}}}$, and $\tau_{{\mathrm{3}}}$ means,
respectively, that every $q_1(j)$, every $q_2(j)$, and every $q_3(j)$ is
zero.  Equation~\eqref{eq:sat-add-ownership} and nonnegativity show that the
third condition holds if and only if both of the first two hold.

It remains to consider $v \,  =  \, \ottsym{(}  a  \ottsym{,}   k   \ottsym{)}$.  First suppose
$ \mathbf{SATv} ( \ottnt{H} ,  \ottnt{R} ,  \ottsym{(}  a  \ottsym{,}   k   \ottsym{)} ,  \tau_{{\mathrm{3}}} ) $.  For any $j$ such that $q_1(j)>0$,
\eqref{eq:sat-add-ownership} and nonnegativity imply $q_3(j)>0$.
Satisfaction by $\tau_{{\mathrm{3}}}$ therefore gives
\[
 (a,k+j)\in dom(\ottnt{H})
 \quad\text{and}\quad
  \mathbf{SATv} ( \ottnt{H} ,  \ottnt{R} ,  \ottnt{H}  \ottsym{(}  \ottsym{(}  a  \ottsym{,}   k   \ottsym{+}   j   \ottsym{)}  \ottsym{)} ,   [  j  /  z  ]  \, \tau'_{{\mathrm{3}}} ) .
\]
The induction hypothesis applied to
\eqref{eq:sat-add-elements} gives satisfaction of the same cell by
$ [  j  /  z  ]  \, \tau'_{{\mathrm{1}}}$.  This proves $ \mathbf{SATv} ( \ottnt{H} ,  \ottnt{R} ,  \ottsym{(}  a  \ottsym{,}   k   \ottsym{)} ,  \tau_{{\mathrm{1}}} ) $; the argument for
$\tau_{{\mathrm{2}}}$ is symmetric.

Conversely, suppose both $ \mathbf{SATv} ( \ottnt{H} ,  \ottnt{R} ,  \ottsym{(}  a  \ottsym{,}   k   \ottsym{)} ,  \tau_{{\mathrm{1}}} ) $ and
$ \mathbf{SATv} ( \ottnt{H} ,  \ottnt{R} ,  \ottsym{(}  a  \ottsym{,}   k   \ottsym{)} ,  \tau_{{\mathrm{2}}} ) $.  Fix $j$ such that $q_3(j)>0$.
Equation~\eqref{eq:sat-add-ownership} and nonnegativity imply that at least
one of $q_1(j)$ and $q_2(j)$ is positive, so the corresponding satisfaction
premise supplies the cell $(a,k+j)$ and satisfaction of its contents by the
corresponding element type.

If both values are positive, the two premises give satisfaction of this same
cell by both $ [  j  /  z  ]  \, \tau'_{{\mathrm{1}}}$ and $ [  j  /  z  ]  \, \tau'_{{\mathrm{2}}}$.  Suppose instead, without
loss of generality, that $q_1(j)>0$ and $q_2(j)=0$.  The \rn{WF-Ref} premise
for $\tau_{{\mathrm{2}}}$ is
\[
 \Gamma  \ottsym{,}   z \COL  \TINT    \models   \fml{ r_{{\mathrm{2}}}    =     \mathbf{0}  }   \implies   \Empty{ \tau'_{{\mathrm{2}}} } .
\]
Instantiating it with Item~1 of \cref{lem:index-substitution}, and using
$q_2(j)=0$ and $\models  \ottsym{[}  \ottnt{R}  \ottsym{]} \,  \fml{ \Gamma } $, gives
\[
 \models  \ottsym{[}  \ottnt{R}  \ottsym{]} \,  \Empty{  [  j  /  z  ]  \, \tau'_{{\mathrm{2}}} } .
\]
The supplied heap value satisfies $ [  j  /  z  ]  \, \tau'_{{\mathrm{1}}}$ and hence has its simple
type.  The same-shape property of \eqref{eq:sat-add-elements} gives it the
simple type of $ [  j  /  z  ]  \, \tau'_{{\mathrm{2}}}$ as well.  By the empty-type observation at the
start of the proof, it therefore satisfies $ [  j  /  z  ]  \, \tau'_{{\mathrm{2}}}$.  The case
$q_1(j)=0$ and $q_2(j)>0$ is symmetric.

Thus the heap cell satisfies both summand element types in every case.  The
reverse direction of the induction hypothesis applied to
\eqref{eq:sat-add-elements} shows
\[
  \mathbf{SATv} ( \ottnt{H} ,  \ottnt{R} ,  \ottnt{H}  \ottsym{(}  \ottsym{(}  a  \ottsym{,}   k   \ottsym{+}   j   \ottsym{)}  \ottsym{)} ,   [  j  /  z  ]  \, \tau'_{{\mathrm{3}}} ) .
\]
Since $j$ was arbitrary, the definition of $\SATV$ gives
$ \mathbf{SATv} ( \ottnt{H} ,  \ottnt{R} ,  \ottsym{(}  a  \ottsym{,}   k   \ottsym{)} ,  \tau_{{\mathrm{3}}} ) $, completing the proof.
\end{rncase}
\end{proof}

\subsubsection{Pointer shifting and shifted type addition}
\label{sec:proofs-pointer-shifting}

Pointer arithmetic changes the origin from which an array index is
measured. The lemmas here relate shifting that origin in a reference type to
shifting the pointer value itself, so that ownership continues to describe
the same heap locations (\cref{lem:own_shift}). Shifted type addition then
combines types whose pointers have different offsets
(\cref{def:shifted-type-addition}). Its ownership and satisfaction properties
provide the common reasoning needed for pointer addition and aliasing by
pointer addition (\cref{lem:own-pointer-split,lem:satv-pointer-split}).

\begin{lemma}
\label{lem:own_shift}
If $R(w) \in  \mathbb{Z} $, then
\[
   \mathbf{own} ( \ottnt{H} ,  \ottnt{R} ,  pv ,   \Pi z .   [ (  z  -  w  ) /  z  ]   ( \tau  \TREF^{\hspace{0.5pt} r })  ) 
  =
   \mathbf{own} ( \ottnt{H} ,  \ottnt{R} ,  pv  \boxplus  \ottnt{R}  \ottsym{(}  w  \ottsym{)} ,   \Pi z .( \tau  \TREF^{\hspace{0.5pt} r })  ) .
\]
\rocqnote{Rocq: no standalone shift-equality lemma.  The corresponding
  index alignment is performed directly in
  \rocqdecl{Metatheory/Conservation.v}{value_ownership_pointer_split_under}.}
\end{lemma}
\begin{proof}
If $pv=\NULL$, both sides are the zero ownership map.  Otherwise, write
$pv=(a,k)$ and put $n=R(w)$.  This address need not itself belong to
$dom(H)$.

For a target address $(a',k')$, the direct contributions on both sides are
zero when $a\ne a'$.  When $a=a'$, the right-hand side evaluates $r$ at
$k'-k-n$, while the left-hand side evaluates the shifted ownership term at
$k'-k$.  These values are equal by capture-avoiding substitution and
$R(w)=n$.

It remains to compare the recursive contributions.  A summand on the
right-hand side is indexed by an integer $j$ satisfying
\[
 (a,k+n+j)\in dom(H)
 \qquad\text{and}\qquad
 \sem{  [  j  /  z  ]    r }_R>0.
\]
Put $i=n+j$.  Then $(a,k+n+j)=(a,k+i)$, and semantic substitution gives
\[
 \sem{  [  i  /  z  ]  \,  [ (  z  -  w  ) /  z  ]    r }_R=\sem{  [  j  /  z  ]    r }_R.
\]
The analogous semantic substitution property for types, proved by induction
on $T$, identifies the corresponding recursive ownership maps:
\[
 \OWN(H,R,H((a,k+i)), [  i  /  z  ]  \,  [ (  z  -  w  ) /  z  ]  \, \tau)
 =
 \OWN(H,R,H((a,k+n+j)), [  j  /  z  ]  \, \tau).
\]
Thus this is exactly the summand indexed by $i$ on the left-hand side.
The change of variables $i=n+j$ is a bijection on the integers, so the
recursive sums, and hence the ownership maps, are equal.
\end{proof}

\begin{definition}[Shifted type addition]
\label{def:shifted-type-addition}
For a reference type, define
\[
  \mathsf{shift}_z\bigl( \Pi w .( \tau  \TREF^{\hspace{0.5pt} r }) \bigr)
  \defeq
   \Pi w .   [ (  w  -  z  ) /  w  ]   ( \tau  \TREF^{\hspace{0.5pt} r }) .
\]
Thus the substitution acts on both the element type and the ownership term.
For reference types $\tau_{{\mathrm{1}}}$, $\tau_{{\mathrm{2}}}$, and $\tau_{{\mathrm{3}}}$, we write
\[
  \Gamma\vdash \tau_{{\mathrm{1}}} \mathbin{+_z}\tau_{{\mathrm{2}}} \approx \tau_{{\mathrm{3}}}
\]
if
\[
  \Gamma\vdash
  \tau_{{\mathrm{1}}}+\mathsf{shift}_z(\tau_{{\mathrm{2}}}) \approx \tau_{{\mathrm{3}}}.
\]
\rocqref{Typing/Rules.v}{pointer_split}
\end{definition}

\begin{lemma}[Ownership under shifted type addition]
\label{lem:own-pointer-split}
Suppose
$\Gamma\vdash \tau_{{\mathrm{1}}}\mathbin{+_z}\tau_{{\mathrm{2}}} \approx \tau_{{\mathrm{3}}}$,
$\models  \ottsym{[}  \ottnt{R}  \ottsym{]} \,  \fml{ \Gamma } $, and $R(z)\in\mathbb{Z}$.
Then
\[
  \OWN(H,R,pv,\tau_{{\mathrm{1}}})
  +\OWN(H,R,pv  \boxplus  \ottnt{R}  \ottsym{(}  z  \ottsym{)},\tau_{{\mathrm{2}}})
  =\OWN(H,R,pv,\tau_{{\mathrm{3}}}).
\]
\rocqref[Supporting Rocq result (with a satisfaction premise)]{Metatheory/Conservation.v}{value_ownership_pointer_split}
\end{lemma}
\begin{proof}
By the definition of shifted type addition and
\cref{lem:own-preservation},
\[
  \OWN(H,R,pv,\tau_{{\mathrm{3}}})
  =
  \OWN(H,R,pv,\tau_{{\mathrm{1}}})
  +\OWN(H,R,pv,\mathsf{shift}_z(\tau_{{\mathrm{2}}})).
\]
After alpha-renaming the bound array index and instantiating the offset
variable with $z$, \cref{lem:own_shift} gives
\[
  \OWN(H,R,pv,\mathsf{shift}_z(\tau_{{\mathrm{2}}}))
  =
  \OWN(H,R,pv  \boxplus  \ottnt{R}  \ottsym{(}  z  \ottsym{)},\tau_{{\mathrm{2}}}).
\]
Substitution of this equality proves the result.
\end{proof}

\begin{lemma}[Satisfaction under shifted type addition]
\label{lem:satv-pointer-split}
If
$\Gamma\vdash \tau_{{\mathrm{1}}}\mathbin{+_z}\tau_{{\mathrm{2}}} \approx \tau_{{\mathrm{3}}}$,
$\models  \ottsym{[}  \ottnt{R}  \ottsym{]} \,  \fml{ \Gamma } $, and $R(z)\in\mathbb{Z}$, then
\[
  \SATV(H,R,pv,\tau_{{\mathrm{3}}})
  \quad\Longleftrightarrow\quad
  \SATV(H,R,pv,\tau_{{\mathrm{1}}})
  \ \land\
  \SATV(H,R,pv  \boxplus  \ottnt{R}  \ottsym{(}  z  \ottsym{)},\tau_{{\mathrm{2}}}).
\]
\rocqnote{Rocq: \rocqdecl{Metatheory/Conservation.v}{value_satisfies_pointer_split}
  and \rocqdecl{Metatheory/Conservation.v}{value_satisfies_pointer_merge_under}.}
\end{lemma}
\begin{proof}
Put $n=R(z)$.  By the definition of shifted type addition and
\cref{lem:sat-preservation-add-ty},
\[
\begin{aligned}
  \SATV(H,R,pv,\tau_{{\mathrm{3}}})
  \quad\Longleftrightarrow\quad&
  \SATV(H,R,pv,\tau_{{\mathrm{1}}})\\
  &{}\land\SATV(H,R,pv,\mathsf{shift}_z(\tau_{{\mathrm{2}}})).
\end{aligned}
\]
It therefore suffices to show that satisfaction of
$\mathsf{shift}_z(\tau_{{\mathrm{2}}})$ at $pv$ is equivalent to satisfaction of
$\tau_{{\mathrm{2}}}$ at $pv  \boxplus   n $.

If $pv \,  =  \,  \NULL $, pointer addition leaves it unchanged, and the equivalence
follows by reindexing the condition that all outer ownerships are zero.
Otherwise, write $pv \,  =  \, \ottsym{(}  a  \ottsym{,}  b  \ottsym{)}$ and change indices by $i=n+j$.
The corresponding offsets address the same heap cell, since
$(a,b+i)=(a,b+n+j)$.  Capture-avoiding substitution gives the same ownership
values and the same interpretation of the instantiated element types; the
latter property follows by induction on the element type.  Thus the
conditions defining satisfaction coincide under this bijection, proving
the required equivalence.
\end{proof}

\subsubsection{Heap updates}
\label{sec:proofs-heap-updates}

A heap update must preserve the types of unrelated values.
\Cref{lem:distinguish-address-by-own} distinguishes a location with zero
ownership from cells with positive outer ownership.
\Cref{lem:own-preservation-equivalent-heap,lem:sat-preservation-about-heap}
establish that changing only a
zero-owned cell preserves both a value's ownership map and its
satisfaction. In the assignment case, these facts protect the bindings
that do not own the cell being written.

\begin{lemma}
\label{lem:distinguish-address-by-own}
If $ \mathbf{own} ( \ottnt{H} ,  \ottnt{R} ,  \ottsym{(}  a'  \ottsym{,}    i '    \ottsym{)} ,   \Pi z .( \tau  \TREF^{\hspace{0.5pt} r })  ) (a, i) = 0$ and $ \llbracket    [  j  /  z  ]    r   \rrbracket_{ \ottnt{R} }  > 0$, then $(a,i) \neq (a',i'+j)$.
\rocqnote{Rocq: no standalone counterpart.  The address inequality is
  established inside the proof of
  \rocqdecl{Metatheory/HeapUpdate.v}{value_satisfies_heap_update_frame}.}
\end{lemma}
 \begin{proof}
  Suppose, for contradiction, that $(a,i)=(a',i'+j)$.  By the definition
  of $\mathbf{own}$,
  \[
    \mathbf{own} ( \ottnt{H} ,  \ottnt{R} ,  \ottsym{(}  a'  \ottsym{,}    i '    \ottsym{)} ,   \Pi z .( \tau  \TREF^{\hspace{0.5pt} r })  ) (a,i) =  \llbracket    [  j  /  z  ]    r   \rrbracket_{ \ottnt{R} }  + f(a,i),
  \]
  where $f$ is the recursive ownership contribution.  By part (2) of
  \cref{lem:ownership-bounds}, $f$ is pointwise nonnegative.  Since
  $ \llbracket    [  j  /  z  ]    r   \rrbracket_{ \ottnt{R} } >0$, the right-hand side is strictly positive,
  contradicting
  \[
    \mathbf{own} ( \ottnt{H} ,  \ottnt{R} ,  \ottsym{(}  a'  \ottsym{,}    i '    \ottsym{)} ,   \Pi z .( \tau  \TREF^{\hspace{0.5pt} r })  ) (a,i)=0.
  \]
  Therefore $(a,i)\neq(a',i'+j)$.
\end{proof}

\begin{lemma}
\label{lem:own-preservation-equivalent-heap}
If $ \ottnt{H}   \approx _{(a,i)}  \ottnt{H'} $ and $ \mathbf{own} ( \ottnt{H} ,  \ottnt{R} ,  v ,  \tau ) (a, i) = 0$, then $ \mathbf{own} ( \ottnt{H} ,  \ottnt{R} ,  v ,  \tau )  =  \mathbf{own} ( \ottnt{H'} ,  \ottnt{R} ,  v ,  \tau ) $.
\rocqref[Supporting Rocq result (for a heap update, with satisfaction)]{Metatheory/HeapUpdate.v}{value_ownership_heap_update_frame}
\end{lemma}
\begin{proof}
  We proceed by strong induction on the size of
  $\tau$.

  If $(a, i) \notin dom(\ottnt{H})$, then $ \ottnt{H}   \approx _{(a,i)}  \ottnt{H'} $ and equality of
  the heap domains imply $\ottnt{H}  =  \ottnt{H'}$, so the conclusion is immediate.

  If $\tau$ is an integer type, both ownership maps are $ \emptyset $.
  Suppose $\tau  =   \Pi z .( \tau'  \TREF^{\hspace{0.5pt} r }) $.  If $v$ is not an address, the
  conclusion is again immediate.  Otherwise, write $v \,  =  \, \ottsym{(}  a'  \ottsym{,}    k '    \ottsym{)}$.

  Consider any integer $j$ such that

  \[
     \ottsym{(}  a'  \ottsym{,}   k '   \ottsym{+}  j  \ottsym{)}  \in   \DOM( \ottnt{H} )  
    \qquad\text{and}\qquad
     \llbracket    [  j  /  z  ]    r   \rrbracket_{ \ottnt{R} } >0.
  \]
  By \cref{lem:ownership-bounds}, the direct contribution and every
  recursive ownership contribution are pointwise nonnegative.  Since
  \[
     \mathbf{own} ( \ottnt{H} ,  \ottnt{R} ,  \ottsym{(}  a'  \ottsym{,}    k '    \ottsym{)} ,   \Pi z .( \tau'  \TREF^{\hspace{0.5pt} r })  ) (a,i)=0,
  \]
  every summand of the recursive ownership contribution is zero.  In
  particular,
  \begin{equation}
     \mathbf{own} ( \ottnt{H} ,  \ottnt{R} ,  \ottnt{H}  \ottsym{(}  \ottsym{(}  a'  \ottsym{,}    k '    \ottsym{+}   j   \ottsym{)}  \ottsym{)} ,   [  j  /  z  ]  \, \tau' ) (a,i)=0.
    \label{eq:own-heap-frame-recursive-zero}
  \end{equation}

  Moreover, \cref{lem:distinguish-address-by-own} gives
  $(a',k'+j)\neq(a,i)$.  Consequently, $ \ottnt{H}   \approx _{(a,i)}  \ottnt{H'} $ gives
  \begin{equation}
    H((a',k'+j))=H'((a',k'+j)).
    \label{eq:own-heap-frame-lookup}
  \end{equation}
  Since index substitution does not change type size, the induction
  hypothesis applied to
  \cref{eq:own-heap-frame-recursive-zero} gives
  \[
  \begin{split}
    & \mathbf{own} ( \ottnt{H} ,  \ottnt{R} ,  \ottnt{H}  \ottsym{(}  \ottsym{(}  a'  \ottsym{,}    k '    \ottsym{+}   j   \ottsym{)}  \ottsym{)} ,   [  j  /  z  ]  \, \tau' ) \\
    &\qquad =
     \mathbf{own} ( \ottnt{H'} ,  \ottnt{R} ,  \ottnt{H}  \ottsym{(}  \ottsym{(}  a'  \ottsym{,}    k '    \ottsym{+}   j   \ottsym{)}  \ottsym{)} ,   [  j  /  z  ]  \, \tau' ) .
  \end{split}
  \]
  By \cref{eq:own-heap-frame-lookup}, the right-hand side is the
  corresponding recursive summand in
  $ \mathbf{own} ( \ottnt{H'} ,  \ottnt{R} ,  v ,  \tau ) $.

  Finally, $\ottnt{H}$ and $\ottnt{H'}$ have the same domain, the guards depend
  only on $\ottnt{R}$ and $r$, and the direct ownership contribution
  does not depend on the heap.  Hence the direct contributions and all
  recursive contributions agree, proving
  \[
     \mathbf{own} ( \ottnt{H} ,  \ottnt{R} ,  v ,  \tau ) = \mathbf{own} ( \ottnt{H'} ,  \ottnt{R} ,  v ,  \tau ) .\qedhere
  \]
\end{proof}

\begin{lemma}
\label{lem:sat-preservation-about-heap}
If $ \mathbf{SATv} ( \ottnt{H} ,  \ottnt{R} ,  v ,  \tau ) $, $ \ottnt{H}   \approx _{(a,i)}  \ottnt{H'} $,
and $ \mathbf{own} ( \ottnt{H} ,  \ottnt{R} ,  v ,  \tau ) (a,i)=0$, then
$ \mathbf{SATv} ( \ottnt{H'} ,  \ottnt{R} ,  v ,  \tau ) $.
\rocqref[Rocq (for a heap update)]{Metatheory/HeapUpdate.v}{value_satisfies_heap_update_frame}
\end{lemma}
\begin{proof}
Proceed by strong induction on the size of $\tau$, generalizing the other
parameters.  If $\tau  =   \{  \nu  :   \TINT    \mid   \varphi  \} $, inversion of the satisfaction premise
gives $v \,  =  \,  n $ for some integer $n$.  Satisfaction is then the
heap-independent assertion $\models  \ottsym{[}  \ottnt{R}  \ottsym{]} \, \ottsym{[}   n   \ottsym{/}  \nu  \ottsym{]}  \varphi$, so the conclusion follows.

Suppose $\tau  =   \Pi z .( \tau'  \TREF^{\hspace{0.5pt} r }) $.  If $v \,  =  \,  \NULL $, satisfaction depends only
on $r$ and $\ottnt{R}$, and is therefore unchanged.  Otherwise, write
$v \,  =  \, \ottsym{(}  a'  \ottsym{,}    i '    \ottsym{)}$.  Fix an integer $j$ such that
$ \llbracket    [  j  /  z  ]    r   \rrbracket_{ \ottnt{R} } >0$.  The satisfaction premise gives
\begin{equation}
   \ottsym{(}  a'  \ottsym{,}   i '   \ottsym{+}  j  \ottsym{)}  \in   \DOM( \ottnt{H} )  
  \quad\text{and}\quad
   \mathbf{SATv} ( \ottnt{H} ,  \ottnt{R} ,  \ottnt{H}  \ottsym{(}  \ottsym{(}  a'  \ottsym{,}    i '    \ottsym{+}   j   \ottsym{)}  \ottsym{)} ,   [  j  /  z  ]  \, \tau' ) .
  \label{eq:sat-heap-frame-old-cell}
\end{equation}

The recursive sum in the definition of
$ \mathbf{own} ( \ottnt{H} ,  \ottnt{R} ,  \ottsym{(}  a'  \ottsym{,}    i '    \ottsym{)} ,   \Pi z .( \tau'  \TREF^{\hspace{0.5pt} r })  ) (a,i)$ contains the contribution
$ \mathbf{own} ( \ottnt{H} ,  \ottnt{R} ,  \ottnt{H}  \ottsym{(}  \ottsym{(}  a'  \ottsym{,}    i '    \ottsym{+}   j   \ottsym{)}  \ottsym{)} ,   [  j  /  z  ]  \, \tau' ) (a,i)$.  By
\cref{lem:ownership-bounds}, the direct contribution and every recursive
contribution are nonnegative.  Since their sum is zero, this particular
contribution is zero:
\begin{equation}
   \mathbf{own} ( \ottnt{H} ,  \ottnt{R} ,  \ottnt{H}  \ottsym{(}  \ottsym{(}  a'  \ottsym{,}    i '    \ottsym{+}   j   \ottsym{)}  \ottsym{)} ,   [  j  /  z  ]  \, \tau' ) (a,i)=0.
  \label{eq:sat-heap-frame-recursive-zero}
\end{equation}

Moreover, \cref{lem:distinguish-address-by-own} gives
$(a',i'+j)\neq(a,i)$.  Hence $ \ottnt{H}   \approx _{(a,i)}  \ottnt{H'} $ and
\cref{eq:sat-heap-frame-old-cell} give
\begin{equation}
   \ottsym{(}  a'  \ottsym{,}   i '   \ottsym{+}  j  \ottsym{)}  \in   \DOM( \ottnt{H'} )  
  \quad\text{and}\quad
  H((a',i'+j))=H'((a',i'+j)).
  \label{eq:sat-heap-frame-cell-agreement}
\end{equation}
Index substitution preserves type size, so $ [  j  /  z  ]  \, \tau'$ is strictly
smaller than $\tau$.  Applying the induction hypothesis to
$ \ottnt{H}   \approx _{(a,i)}  \ottnt{H'} $ and
\cref{eq:sat-heap-frame-old-cell,eq:sat-heap-frame-recursive-zero} gives
\[
   \mathbf{SATv} ( \ottnt{H'} ,  \ottnt{R} ,  \ottnt{H}  \ottsym{(}  \ottsym{(}  a'  \ottsym{,}    i '    \ottsym{+}   j   \ottsym{)}  \ottsym{)} ,   [  j  /  z  ]  \, \tau' ) .
\]
Using \cref{eq:sat-heap-frame-cell-agreement}, this is the satisfaction
condition for the cell stored by $\ottnt{H'}$.  Since $j$ was arbitrary, the
definition of $\SATV$ gives
$ \mathbf{SATv} ( \ottnt{H'} ,  \ottnt{R} ,  \ottsym{(}  a'  \ottsym{,}    i '    \ottsym{)} ,   \Pi z .( \tau'  \TREF^{\hspace{0.5pt} r })  ) $, as required.
\end{proof}

The heap-update lemmas
\cref{lem:sat-preservation-about-heap,lem:own-preservation-equivalent-heap}
supply the framing argument for assignment.  If one typed variable has
full ownership of an allocated cell, \cref{lem:ownership-bounds} and
ownership consistency force every other typed variable's contribution
there to be zero.  The heap-update lemmas then show that updating the cell
preserves satisfaction and ownership for those other values.

\subsubsection{Rule-specific preservation of runtime invariants}
\label{sec:proofs-rule-invariants}

The final two lemmas assemble the preceding facts for the concrete changes
made by individual reduction rules.
\Cref{lem:conown-preservation,lem:sat-preservation} cover the same nine
operations, first establishing ownership consistency and then
satisfaction for the resulting heap, register file, and type environment.
Their parallel structure makes the two runtime obligations easy to locate
in each case of \textsc{Preservation}; the structural typing lemmas supply the
corresponding continuation-typing arguments.

\begin{lemma}[Preservation of $\CONOWN$]
\label{lem:conown-preservation}
The following items collect the operation-specific facts showing that the
heap, register, and type-environment transformations performed by the
reduction rules preserve $\CONOWN$.  Each item is used in the corresponding
case of the proof of Preservation.  Throughout this lemma, we assume that
$\models  \ottsym{[}  \ottnt{R}  \ottsym{]} \,  \fml{ \Gamma } $ holds for the register $\ottnt{R}$ and the type environment
$\Gamma$.
\begingroup
\makeatletter
\renewcommand{\p@enumii}{}
\makeatother
\begin{enumerate}
  \item\label{conown-preservation1}
    \emph{Variable binding (\rn{Rs-LetVar}).}
    If the following conditions hold:
    \begin{enumerate}
      \item $\CONOWN(H,R,\Gamma)$;\label{conown-preservation1-conown}
      \item $\Gamma \subt \Gamma_1, y \COL \tau_y, \Gamma_2$;\label{conown-preservation1-env-subtype}
      \item $\Gamma_{{\mathrm{1}}}  \ottsym{,}   y \COL  \tau _{ y }    \ottsym{,}  \Gamma_{{\mathrm{2}}}  \vdash    \tau _{ y }    \leq   \tau $;\label{conown-preservation1-source-subtype}
      \item $\Gamma_{{\mathrm{1}}}  \ottsym{,}   y \COL  \tau _{ y }    \ottsym{,}  \Gamma_{{\mathrm{2}}}  \vdash      \tau' _{ y }   +  \tau'  _{ x }    \approx   \tau $; and\label{conown-preservation1-split}
      \item $x' \notin dom(\Gamma_1, y \COL \tau_y, \Gamma_2)\cup dom(R)$,\label{conown-preservation1-fresh}
    \end{enumerate}
    then $\CONOWN(H,R\set{x' \mapsto R(y)},(\Gamma_1, y \COL \tau_y',
    \Gamma_2, x' \COL \tau_x'))$.
    \rocqref[Supporting Rocq result]{Metatheory/Conservation.v}{consistent_after_variable_bind}
  \item\label{conown-preservation2}
    \emph{Dereference (\rn{Rs-Deref}).}
    If the following conditions hold:
    \begin{enumerate}
      \item $\CONOWN(H,R,\Gamma)$;\label{conown-preservation2-conown}
      \item $\Gamma  \leq  \Gamma_{{\mathrm{1}}}  \ottsym{,}   y \COL  \Pi z .(  \tau _{ y }   \TREF^{\hspace{0.5pt} r })    \ottsym{,}  \Gamma_{{\mathrm{2}}}$;\label{conown-preservation2-env-subtype}
      \item $\Gamma_{{\mathrm{1}}}  \ottsym{,}   y \COL  \Pi z .(  \tau _{ y }   \TREF^{\hspace{0.5pt} r })    \ottsym{,}  \Gamma_{{\mathrm{2}}}  \ottsym{,}   z \COL  \{  \nu  :   \TINT    \mid   \nu \,  =  \,  0   \}    \vdash     \tau'  +  \tau  _{ x }    \approx    \tau _{ y }  $;\label{conown-preservation2-split}
      \item $\Gamma_{{\mathrm{1}}}  \ottsym{,}   y \COL  \Pi z .(  \tau _{ y }   \TREF^{\hspace{0.5pt} r })    \ottsym{,}  \Gamma_{{\mathrm{2}}}  \ottsym{,}   x' \COL  \tau _{ x }    \ottsym{,}   z \COL  \{  \nu  :   \TINT    \mid   \nu \,  =  \,  0   \}    \vdash    \tau' _{ y }    \approx    \ottsym{(}  \tau'  \ottsym{)}  ^ {= x' }  $;\label{conown-preservation2-zero-equiv}
      \item $\Gamma_{{\mathrm{1}}}  \ottsym{,}   y \COL  \Pi z .(  \tau _{ y }   \TREF^{\hspace{0.5pt} r })    \ottsym{,}  \Gamma_{{\mathrm{2}}}  \ottsym{,}   x' \COL  \tau _{ x }    \ottsym{,}   z \COL  \{  \nu  :   \TINT    \mid   \nu \, \neq \,  0   \}    \vdash    \tau' _{ y }    \approx    \tau _{ y }  $;\label{conown-preservation2-nonzero-equiv}
      \item $\Gamma_{{\mathrm{1}}}  \ottsym{,}   y \COL  \Pi z .(  \tau _{ y }   \TREF^{\hspace{0.5pt} r })    \ottsym{,}  \Gamma_{{\mathrm{2}}}  \ottsym{,}   z \COL  \{  \nu  :   \TINT    \mid   \nu \,  =  \,  0   \}    \models  r \,  >  \,  \mathbf{0} $;\label{conown-preservation2-positive}
      \item $R(y) = (a,i) \in dom(H)$; and\label{conown-preservation2-location}
      \item $x' \notin dom(\Gamma_{{\mathrm{1}}}  \ottsym{,}   y \COL  \Pi z .(  \tau _{ y }   \TREF^{\hspace{0.5pt} r })    \ottsym{,}  \Gamma_{{\mathrm{2}}}) \cup dom(R)$,\label{conown-preservation2-fresh}
    \end{enumerate}
    then $\CONOWN(H,R\set{x' \mapsto H(R(y))},(\Gamma_1, y \COL  \Pi z .(  \tau' _{ y }   \TREF^{\hspace{0.5pt} r }) ,
    \Gamma_2, x' \COL \tau_x))$.
    \rocqref[Supporting Rocq result]{Metatheory/Conservation.v}{consistent_after_deref_fresh}
  \item\label{conown-preservation3}
    \emph{Assignment (\rn{Rs-Assign}).}
    If the following conditions hold:
    \begin{enumerate}
      \item $\CONOWN(H,R,\Gamma)$;\label{conown-preservation3-conown}
      \item $\Gamma \subt \Gamma_1$;\label{conown-preservation3-env-subtype}
      \item $\Gamma_1(x) =  \Pi z .(  \tau _{\ast  x }   \TREF^{\hspace{0.5pt} r }) $;\label{conown-preservation3-x-type}
      \item $ \Gamma_{{\mathrm{1}}}  (  y  )   =   \tau _{ y } $;\label{conown-preservation3-y-type}
      \item $\Gamma_{{\mathrm{1}}}  \ottsym{,}   z \COL  \{  \nu  :   \TINT    \mid   \nu \,  =  \,  0   \}    \vdash    \tau' _{\ast  x }    \approx    \ottsym{(}  \tau'  \ottsym{)}  ^ {= y }  $;\label{conown-preservation3-zero-equiv}
      \item $\Gamma_{{\mathrm{1}}}  \ottsym{,}   z \COL  \{  \nu  :   \TINT    \mid   \nu \, \neq \,  0   \}    \vdash    \tau' _{\ast  x }    \approx    \tau _{\ast  x }  $;\label{conown-preservation3-nonzero-equiv}
      \item $\Gamma_{{\mathrm{1}}}  \ottsym{,}   z \COL  \{  \nu  :   \TINT    \mid   \nu \,  =  \,  0   \}    \models  r \,  =  \,  \mathbf{1} $;\label{conown-preservation3-full-ownership}
      \item $R(x) = (a,i) \in dom(H)$; and\label{conown-preservation3-location}
      \item $\Gamma_{{\mathrm{1}}}  \vdash     \tau' _{ y }   +  \tau'    \approx    \tau _{ y }  $,\label{conown-preservation3-split}
    \end{enumerate}
    then $\CONOWN(H\set{(a,i) \hookleftarrow R(y)},R,\Gamma_1[x \hookleftarrow
     \Pi z .(  \tau' _{\ast  x }   \TREF^{\hspace{0.5pt} r })  ][y \hookleftarrow  \tau' _{ y }  ])$.
    \rocqref[Supporting Rocq result]{Metatheory/Conservation.v}{env_ownership_after_assign_le}
  \item\label{conown-preservation4}
    \emph{Pointer addition (\rn{Rs-AddPtr}).}
    If the following conditions hold:
    \begin{enumerate}
      \item $\CONOWN(H,R,\Gamma)$;\label{conown-preservation4-conown}
      \item $\Gamma \subt \Gamma_1$;\label{conown-preservation4-env-subtype}
      \item $ \Gamma_{{\mathrm{1}}}  (  y  )   =   \Pi w .( \tau_{{\mathrm{3}}}  \TREF^{\hspace{0.5pt}  { r }_{ y }  }) $;\label{conown-preservation4-y-type}
      \item $\Gamma_{{\mathrm{1}}}  \vdash     \Pi w .( \tau_{{\mathrm{1}}}  \TREF^{\hspace{0.5pt}  { r }_{ y_{{\mathrm{1}}} }  })   +   \Pi w .   [ (  w  -  z  ) /  w  ]   ( \tau_{{\mathrm{2}}}  \TREF^{\hspace{0.5pt}  { r }_{ x }  })     \approx    \Pi w .( \tau_{{\mathrm{3}}}  \TREF^{\hspace{0.5pt}  { r }_{ y }  })  $;\label{conown-preservation4-split}
      \item $R(z)\in \mathbb{Z} $;\label{conown-preservation4-offset-value}
      \item $x' \notin dom(\Gamma_{{\mathrm{1}}}) \cup dom(R)$; and\label{conown-preservation4-fresh}
      \item $R(y) = pv$,\label{conown-preservation4-y-value}
    \end{enumerate}
    then $\CONOWN(H,R\set{x' \mapsto pv  \boxplus  \ottnt{R}  \ottsym{(}  z  \ottsym{)}},\Gamma_1[y
    \hookleftarrow  \Pi w .( \tau_{{\mathrm{1}}}  \TREF^{\hspace{0.5pt}  { r }_{ y_{{\mathrm{1}}} }  })  ], x' \COL
     \Pi w .( \tau_{{\mathrm{2}}}  \TREF^{\hspace{0.5pt}  { r }_{ x }  })  )$.
    \rocqref[Supporting Rocq result]{Metatheory/Conservation.v}{consistent_after_let_add_ptr}
  \item\label{conown-preservation5}
    \emph{Integer-array allocation (\rn{Rs-MkArrayIntref}).}
    If the following conditions hold:
    \begin{enumerate}
      \item $\CONOWN(H,R,\Gamma)$;\label{conown-preservation5-conown}
      \item $\SAT(H,R,\Gamma)$;\label{conown-preservation5-sat}
      \item $\Gamma \subt \Gamma_1$;\label{conown-preservation5-env-subtype}
      \item $\Gamma_1(y) = \set{\nu \COL \TINT \mid \varphi}$;\label{conown-preservation5-length-type}
      \item $\Gamma_{{\mathrm{1}}}  \ottsym{,}   z \COL  \TINT    \models  r \,  =  \, \ottsym{(}    \ottsym{(}    0  \, \le \, z  \wedge  z \, \le \, y  \ottsym{-}   1    \ottsym{)}   \produces    1    ,   \mathbf{0}    \ottsym{)}$;\label{conown-preservation5-ownership-shape}
      \item $x' \notin dom(\Gamma_{{\mathrm{1}}})\cup dom(R)$; and\label{conown-preservation5-fresh}
      \item $(a,0)\notin dom(H)$,\label{conown-preservation5-fresh-base}
    \end{enumerate}
    then
    \[
    \begin{aligned}
      &\CONOWN\bigl(H\set{(a,0) \mapsto 0, \dots, (a, R(y)-1) \mapsto 0},\\
      &\qquad R\set{x' \mapsto (a, 0)},
        (\Gamma_{{\mathrm{1}}}  \ottsym{,}   x' \COL  \Pi z .(  \{  \nu  :   \TINT    \mid     0  \, \le \, z  \wedge  z \, \le \, y  \ottsym{-}   1    \implies  \nu \,  =  \,  0   \}   \TREF^{\hspace{0.5pt} r })  )\bigr).
    \end{aligned}
    \]
    \rocqref[Supporting Rocq result]{Metatheory/Conservation.v}{consistent_after_alloc}
  \item\label{conown-preservation6}
    \emph{Nested-array allocation (\rn{Rs-MkArrayNestedref}).}
    If the following conditions hold:
    \begin{enumerate}
      \item $\CONOWN(H,R,\Gamma)$;\label{conown-preservation6-conown}
      \item $\SAT(H,R,\Gamma)$;\label{conown-preservation6-sat}
      \item $\Gamma \subt \Gamma_1$;\label{conown-preservation6-env-subtype}
      \item $\Gamma_1(y) = \set{\nu \COL \TINT \mid \varphi}$;\label{conown-preservation6-length-type}
      \item $\models   \Empty{ \tau' } $;\label{conown-preservation6-empty}
      \item $\Gamma_{{\mathrm{1}}}  \ottsym{,}   z \COL  \TINT    \models  r \,  =  \, \ottsym{(}    \ottsym{(}    0  \, \le \, z  \wedge  z \, \le \, y  \ottsym{-}   1    \ottsym{)}   \produces    1    ,   \mathbf{0}    \ottsym{)}$;\label{conown-preservation6-ownership-shape}
      \item $x' \notin dom(\Gamma_{{\mathrm{1}}})\cup dom(R)$; and\label{conown-preservation6-fresh}
      \item $(a,0)\notin dom(H)$,\label{conown-preservation6-fresh-base}
    \end{enumerate}
    then
    \[
    \begin{aligned}
      &\CONOWN\bigl(H\set{(a,0) \mapsto \NULL, \dots, (a, R(y)-1) \mapsto \NULL},\\
      &\qquad R\set{x' \mapsto (a, 0)}, (\Gamma_{{\mathrm{1}}}  \ottsym{,}   x' \COL  \Pi z .( \tau'  \TREF^{\hspace{0.5pt} r })  )\bigr).
    \end{aligned}
    \]
    \rocqref[Supporting Rocq result]{Metatheory/Conservation.v}{consistent_after_alloc}
  \item\label{conown-preservation7}
    \emph{Aliasing by dereference (\rn{Rs-AliasDeref}).}
    If the following conditions hold:
    \begin{enumerate}
      \item $\CONOWN(H,R,\Gamma)$;\label{conown-preservation7-conown}
      \item $\Gamma \subt \Gamma_1$;\label{conown-preservation7-env-subtype}
      \item $ \Gamma_{{\mathrm{1}}}  (  x  )   =    \Pi z'_{{\mathrm{1}}} .(  \tau _{\ast  x }   \TREF^{\hspace{0.5pt} r })  _{ x } $;\label{conown-preservation7-x-type}
      \item $ \Gamma_{{\mathrm{1}}}  (  y  )   =   \Pi z_{{\mathrm{2}}} .(  \Pi z_{{\mathrm{1}}} .(  \tau _{\ast \ast  y }   \TREF^{\hspace{0.5pt}  { r }_{ \ast  y }  })   \TREF^{\hspace{0.5pt} r }) $;\label{conown-preservation7-y-type}
      \item $\Gamma_{{\mathrm{1}}}  \ottsym{,}   z_{{\mathrm{2}}} \COL  \{  \nu  :   \TINT    \mid   \nu \,  =  \,  0   \}    \vdash   \ottsym{(}    \Pi z'_{{\mathrm{1}}} .(  \tau _{\ast  x }   \TREF^{\hspace{0.5pt}  { r }_{ x }  })   +   \Pi z_{{\mathrm{1}}} .(  \tau _{\ast \ast  y }   \TREF^{\hspace{0.5pt}  { r }_{ \ast  y }  })    \ottsym{)}   \approx   \ottsym{(}    \Pi z'_{{\mathrm{1}}} .(  \tau' _{\ast  x }   \TREF^{\hspace{0.5pt}  { r' }_{ x }  })   +   \Pi z_{{\mathrm{1}}} .(  \tau' _{\ast \ast  y }   \TREF^{\hspace{0.5pt}  { r' }_{ \ast  y }  })    \ottsym{)} $;\label{conown-preservation7-zero-equiv}
      \item $\Gamma_{{\mathrm{1}}}  \ottsym{,}   z_{{\mathrm{2}}} \COL  \{  \nu  :   \TINT    \mid   \nu \, \neq \,  0   \}    \vdash     \Pi z_{{\mathrm{1}}} .(  \tau _{\ast \ast  y }   \TREF^{\hspace{0.5pt} r })  _{\ast  y }    \approx     \Pi z_{{\mathrm{1}}} .(  \tau' _{\ast \ast  y }   \TREF^{\hspace{0.5pt} r' })  _{\ast  y }  $;\label{conown-preservation7-nonzero-equiv}
      \item $\Gamma_{{\mathrm{1}}}  \ottsym{,}   z_{{\mathrm{2}}} \COL  \{  \nu  :   \TINT    \mid   \nu \,  =  \,  0   \}    \models  r \,  >  \,  \mathbf{0} $; and\label{conown-preservation7-positive}
      \item $H(R(y)) = R(x)$,\label{conown-preservation7-alias}
    \end{enumerate}
    then $\CONOWN(H,R,\Gamma_1 [x \hookleftarrow  \Pi z'_{{\mathrm{1}}} .(  \tau' _{\ast  x }   \TREF^{\hspace{0.5pt}  { r' }_{ x }  })  ]
    [ y \hookleftarrow  \Pi z_{{\mathrm{2}}} .(  \Pi z_{{\mathrm{1}}} .(  \tau' _{\ast \ast  y }   \TREF^{\hspace{0.5pt}  { r' }_{ \ast  y }  })   \TREF^{\hspace{0.5pt} r })  ])$.
    \rocqref[Supporting Rocq result]{Metatheory/Conservation.v}{consistent_after_alias_deref}
  \item\label{conown-preservation8}
    \emph{Aliasing by pointer addition (\rn{Rs-AliasAddPtr}).}
    If the following conditions hold:
    \begin{enumerate}
      \item $\CONOWN(H,R,\Gamma)$;\label{conown-preservation8-conown}
      \item $R(x) = pv  \boxplus  \ottnt{R}  \ottsym{(}  z  \ottsym{)}$;\label{conown-preservation8-x-value}
      \item $R(y) = pv$;\label{conown-preservation8-y-value}
      \item $R(z)\in \mathbb{Z} $;\label{conown-preservation8-offset-value}
      \item $x \, \neq \, y$;\label{conown-preservation8-distinct}
      \item $\Gamma \subt \Gamma_1$;\label{conown-preservation8-env-subtype}
      \item $\Gamma_1(x) =  \Pi w' .(  \tau _{\ast  x }   \TREF^{\hspace{0.5pt}  { r }_{ x }  }) $;\label{conown-preservation8-x-type}
      \item $\Gamma_1(y) =  \Pi w .(  \tau _{\ast  y }   \TREF^{\hspace{0.5pt}  { r }_{ y }  }) $; and\label{conown-preservation8-y-type}
      \item $\begin{aligned}[t]
        &\Gamma_{{\mathrm{1}}}  \vdash  \ottsym{(}    \Pi w' .   [ (  w'  -  z  ) /  w'  ]   (  \tau _{\ast  x }   \TREF^{\hspace{0.5pt}  { r }_{ x }  })   +   \Pi w .(  \tau _{\ast  y }   \TREF^{\hspace{0.5pt}  { r }_{ y }  })    \ottsym{)}\\
        &\qquad  \approx   \ottsym{(}   \Pi w' .   [ (  w'  -  z  ) /  w'  ]   (  \tau' _{\ast  x }   \TREF^{\hspace{0.5pt}  { r' }_{ x }  })   \ottsym{)}  +   \Pi w .(  \tau' _{\ast  y }   \TREF^{\hspace{0.5pt}  { r' }_{ y }  })  ,
      \end{aligned}$\label{conown-preservation8-combined-equiv}
    \end{enumerate}
    then $\CONOWN(H,R,\Gamma_1\set{x \hookleftarrow  \Pi w' .(  \tau' _{\ast  x }   \TREF^{\hspace{0.5pt}  { r' }_{ x }  }) ,
    y \hookleftarrow  \Pi w .(  \tau' _{\ast  y }   \TREF^{\hspace{0.5pt}  { r' }_{ y }  }) })$.
    \rocqref[Supporting Rocq result]{Metatheory/Conservation.v}{consistent_after_alias_add}
  \item\label{conown-preservation9}
    \emph{Null binding (\rn{Rs-LetNull}).}
    If the following conditions hold:
    \begin{enumerate}
      \item $\CONOWN(H,R,\Gamma)$;\label{conown-preservation9-conown}
      \item $\Gamma \subt \Gamma_1$; and\label{conown-preservation9-env-subtype}
      \item $x' \notin dom(\Gamma_{{\mathrm{1}}})\cup dom(R)$,\label{conown-preservation9-fresh}
    \end{enumerate}
    then $\CONOWN(H,R\set{x' \mapsto \NULL}, (\Gamma_{{\mathrm{1}}}  \ottsym{,}   x' \COL  \Pi z .( \tau  \TREF^{\hspace{0.5pt} r })  ))$.
    \rocqref[Supporting Rocq result]{Runtime/Model.v}{null_ownership_zero}
 \end{enumerate}
\endgroup
\end{lemma}
\begin{proof}
\leavevmode
\proofitem{Item~\ref{conown-preservation1}: Variable binding
(\rn{Rs-LetVar}).}

Let $\Gamma_{{\mathrm{0}}}=\Gamma_{{\mathrm{1}}}  \ottsym{,}   y \COL  \tau _{ y }    \ottsym{,}  \Gamma_{{\mathrm{2}}}$.  By Lemma
\ref{lem:conown-preservation-weakening-tyenv},
premises~(\ref{conown-preservation1-conown}) and
(\ref{conown-preservation1-env-subtype}), and the standing validity
assumption, $ \mathbf{ConOwn}( \ottnt{H} ,  \ottnt{R} ,  \Gamma_{{\mathrm{0}}} ) $ holds.  By \rn{S-TyEnv}, \rn{S-Int}, and the definition of
$ \fml{ \Gamma } $, $\Gamma  \leq  \Gamma_{{\mathrm{0}}}$ implies
$ \fml{ \Gamma } \implies \fml{ \Gamma_{{\mathrm{0}}} } $.  Instantiating this implication by $\ottnt{R}$
and using the standing assumption $\models  \ottsym{[}  \ottnt{R}  \ottsym{]} \,  \fml{ \Gamma } $ gives
$\models  \ottsym{[}  \ottnt{R}  \ottsym{]} \,  \fml{ \Gamma_{{\mathrm{0}}} } $.
The total ownership is:
\begin{align*}
    & \mathbf{Own}( \ottnt{H} ,  \ottnt{R}  \ottsym{\{}  x'  \mapsto  \ottnt{R}  \ottsym{(}  y  \ottsym{)}  \ottsym{\}} ,  \ottsym{(}  \Gamma_{{\mathrm{1}}}  \ottsym{,}   y \COL  \tau' _{ y }    \ottsym{,}  \Gamma_{{\mathrm{2}}}  \ottsym{,}   x' \COL  \tau' _{ x }    \ottsym{)} ) \\
    &= \begin{aligned}
        &\sum_{z \in y  \ottsym{:}   \tau' _{ y }   \ottsym{,}  x'  \ottsym{:}   \tau' _{ x } }  \mathbf{own} ( \ottnt{H} ,  \ottnt{R}  \ottsym{\{}  x'  \mapsto  \ottnt{R}  \ottsym{(}  y  \ottsym{)}  \ottsym{\}} ,  \ottnt{R}  \ottsym{\{}  x'  \mapsto  \ottnt{R}  \ottsym{(}  y  \ottsym{)}  \ottsym{\}}  \ottsym{(}  z  \ottsym{)} ,   \ottsym{(}   y \COL  \tau' _{ y }    \ottsym{,}   x' \COL  \tau' _{ x }    \ottsym{)}  (  z  )  )  \\
        &\qquad + \sum_{z \in \Gamma_{{\mathrm{1}}}  \ottsym{,}  \Gamma_{{\mathrm{2}}}}  \mathbf{own} ( \ottnt{H} ,  \ottnt{R}  \ottsym{\{}  x'  \mapsto  \ottnt{R}  \ottsym{(}  y  \ottsym{)}  \ottsym{\}} ,  \ottnt{R}  \ottsym{\{}  x'  \mapsto  \ottnt{R}  \ottsym{(}  y  \ottsym{)}  \ottsym{\}}  \ottsym{(}  z  \ottsym{)} ,   \ottsym{(}  \Gamma_{{\mathrm{1}}}  \ottsym{,}  \Gamma_{{\mathrm{2}}}  \ottsym{)}  (  z  )  )   
    \end{aligned} \\
    &= \begin{aligned}
        & \mathbf{own} ( \ottnt{H} ,  \ottnt{R} ,  \ottnt{R}  \ottsym{(}  y  \ottsym{)} ,  \tau )  & \\
        &\qquad + \sum_{z \in \Gamma_{{\mathrm{1}}}  \ottsym{,}  \Gamma_{{\mathrm{2}}}}  \mathbf{own} ( \ottnt{H} ,  \ottnt{R} ,  \ottnt{R}  \ottsym{(}  z  \ottsym{)} ,   \ottsym{(}  \Gamma_{{\mathrm{1}}}  \ottsym{,}  \Gamma_{{\mathrm{2}}}  \ottsym{)}  (  z  )  )  & 
    \end{aligned} \tag{ by \cref{lem:own-preservation2}, \cref{lem:own-preservation}, and premise~(\ref{conown-preservation1-split})}
\end{align*}
From \cref{lem:own-preservation-weakening-ty} and
premise~(\ref{conown-preservation1-source-subtype}),
\[
   \mathbf{own} ( \ottnt{H} ,  \ottnt{R} ,  \ottnt{R}  \ottsym{(}  y  \ottsym{)} ,  \tau )  \le  \mathbf{own} ( \ottnt{H} ,  \ottnt{R} ,  \ottnt{R}  \ottsym{(}  y  \ottsym{)} ,   \tau _{ y }  ) .
\]
Furthermore,
\[
\begin{aligned}
  & \mathbf{own} ( \ottnt{H} ,  \ottnt{R} ,  \ottnt{R}  \ottsym{(}  y  \ottsym{)} ,   \tau _{ y }  ) 
    + \sum_{z \in \Gamma_{{\mathrm{1}}}  \ottsym{,}  \Gamma_{{\mathrm{2}}}}  \mathbf{own} ( \ottnt{H} ,  \ottnt{R} ,  \ottnt{R}  \ottsym{(}  z  \ottsym{)} ,   \ottsym{(}  \Gamma_{{\mathrm{1}}}  \ottsym{,}  \Gamma_{{\mathrm{2}}}  \ottsym{)}  (  z  )  ) \\
  &\qquad =  \mathbf{Own}( \ottnt{H} ,  \ottnt{R} ,  \Gamma_{{\mathrm{0}}} ) .
\end{aligned}
\]
Given the derived fact $ \mathbf{ConOwn}( \ottnt{H} ,  \ottnt{R} ,  \Gamma_{{\mathrm{0}}} ) $,
it follows that $ \mathbf{Own}( \ottnt{H} ,  \ottnt{R} ,  \Gamma_{{\mathrm{0}}} )  \ottsym{(}  a  \ottsym{,}   k   \ottsym{)}  \le 1$ for all $\ottsym{(}  a  \ottsym{,}  k  \ottsym{)} \in dom(H)$,
which implies
\[
   \mathbf{Own}( \ottnt{H} ,  \ottnt{R}  \ottsym{\{}  x'  \mapsto  \ottnt{R}  \ottsym{(}  y  \ottsym{)}  \ottsym{\}} ,  \ottsym{(}  \Gamma_{{\mathrm{1}}}  \ottsym{,}   y \COL  \tau' _{ y }    \ottsym{,}  \Gamma_{{\mathrm{2}}}  \ottsym{,}   x' \COL  \tau' _{ x }    \ottsym{)} )  \ottsym{(}  a  \ottsym{,}   k   \ottsym{)}  \le 1.
\]
Thus, $ \mathbf{ConOwn}( \ottnt{H} ,  \ottnt{R}  \ottsym{\{}  x'  \mapsto  \ottnt{R}  \ottsym{(}  y  \ottsym{)}  \ottsym{\}} ,  \ottsym{(}  \Gamma_{{\mathrm{1}}}  \ottsym{,}   y \COL  \tau' _{ y }    \ottsym{,}  \Gamma_{{\mathrm{2}}}  \ottsym{,}   x' \COL  \tau' _{ x }    \ottsym{)} ) $ holds.
 
\proofitem{Item~\ref{conown-preservation2}: Dereference
(\rn{Rs-Deref}).}

Let $\Gamma_{{\mathrm{0}}}$ be $(\Gamma_{{\mathrm{1}}}  \ottsym{,}   y \COL  \Pi z .(  \tau _{ y }   \TREF^{\hspace{0.5pt} r })    \ottsym{,}  \Gamma_{{\mathrm{2}}})$, $\Gamma'$ be $(\Gamma_{{\mathrm{1}}}  \ottsym{,}   y \COL  \Pi z .(  \tau' _{ y }   \TREF^{\hspace{0.5pt} r })    \ottsym{,}  \Gamma_{{\mathrm{2}}}  \ottsym{,}   x' \COL  \tau _{ x }  )$ and
 $R'$ be $\ottnt{R}  \ottsym{\{}  x'  \mapsto  \ottnt{H}  \ottsym{(}  \ottnt{R}  \ottsym{(}  y  \ottsym{)}  \ottsym{)}  \ottsym{\}}$.
By \cref{lem:conown-preservation-weakening-tyenv},
premises~(\ref{conown-preservation2-conown}) and
(\ref{conown-preservation2-env-subtype}), and the standing validity
assumption, $ \mathbf{ConOwn}( \ottnt{H} ,  \ottnt{R} ,  \Gamma_{{\mathrm{0}}} ) $ holds.
By \rn{S-TyEnv}, \rn{S-Int}, and the definition of $ \fml{ \Gamma } $,
$\Gamma  \leq  \Gamma_{{\mathrm{0}}}$ implies $ \fml{ \Gamma } \implies \fml{ \Gamma_{{\mathrm{0}}} } $.  Hence the standing
assumption $\models  \ottsym{[}  \ottnt{R}  \ottsym{]} \,  \fml{ \Gamma } $ gives $\models  \ottsym{[}  \ottnt{R}  \ottsym{]} \,  \fml{ \Gamma_{{\mathrm{0}}} } $.  Since $x'$ is
fresh, $R'$ agrees with $R$ on $dom(\Gamma_{{\mathrm{0}}})$, and therefore
$\models  \ottsym{[}  \ottnt{R'}  \ottsym{]} \,  \fml{ \Gamma_{{\mathrm{0}}} } $.
Since $x'$ is fresh, the update to $R'$ does not affect any existing types in $ \Gamma_{{\mathrm{1}}}, \Gamma_{{\mathrm{2}}}$. 
Therefore, we need only consider $ \mathbf{own} ( \ottnt{H} ,  \ottnt{R'} ,  \ottnt{R'}  \ottsym{(}  y  \ottsym{)} ,   \Pi z .(  \tau' _{ y }   \TREF^{\hspace{0.5pt} r })  ) $ and $ \mathbf{own} ( \ottnt{H} ,  \ottnt{R'} ,  \ottnt{R'}  \ottsym{(}  x'  \ottsym{)} ,   \tau _{ x }  ) $.
Moreover, premises~(\ref{conown-preservation2-fresh}) and
(\ref{conown-preservation2-location}) give
$R'(y)=R(y)=(a,i)$ and $R'(x')=H((a,i))$.
Semantic instantiation of premise~(\ref{conown-preservation2-positive}) using
$\models  \ottsym{[}  \ottnt{R'}  \ottsym{]} \,  \fml{ \Gamma_{{\mathrm{0}}} } $ gives
\begin{equation}\label{own-pos}
   \llbracket    [  0  /  z  ]    r   \rrbracket_{ \ottnt{R'} }  \,  >  \,  0 
\end{equation}
holds.
The type-addition premise~(\ref{conown-preservation2-split}) and the
zero-offset type-equivalence
premise~(\ref{conown-preservation2-zero-equiv}) are instantiated at $z=0$,
while the nonzero-offset type-equivalence
premise~(\ref{conown-preservation2-nonzero-equiv}) is instantiated at $z=j$
for $j\ne0$.  For the equivalence premises, consider
the common shape of the types.  If it is the integer shape, all the relevant
ownership maps are $ \emptyset $.  Otherwise $ \tau _{ x } $ is a reference type, so
$ \fmlT{  \tau _{ x }  }{ x' } = \top $ and $\models  \ottsym{[}  \ottnt{R'}  \ottsym{]} \,  \fml{ \Gamma_{{\mathrm{0}}}  \ottsym{,}   x' \COL  \tau _{ x }   } $.  Thus Lemma
\ref{lem:own-preservation-weakening-ty} may be applied in both directions.
\begingroup
\allowdisplaybreaks[1]
\begin{align*}
    & \mathbf{own} ( \ottnt{H} ,  \ottnt{R'} ,  \ottsym{(}  a  \ottsym{,}   i   \ottsym{)} ,   \Pi z .(  \tau' _{ y }   \TREF^{\hspace{0.5pt} r })  )  +  \mathbf{own} ( \ottnt{H} ,  \ottnt{R'} ,  \ottnt{H}  \ottsym{(}  \ottsym{(}  a  \ottsym{,}   i   \ottsym{)}  \ottsym{)} ,   \tau _{ x }  )  &\\
    &=  \sum_ { j  \in \mathbb{Z} }\{  \ottsym{(}  a  \ottsym{,}   i   \ottsym{+}   j   \ottsym{)}   \mapsto    \llbracket    [  j  /  z  ]    r   \rrbracket_{ \ottnt{R'} }    \} 
       + \sum_{\substack{j\in\mathbb Z\\(a,i+j)\in\DOM(H)\\\sem{[j/z]r}_{R'}>0}}
          \mathbf{own} ( \ottnt{H} ,  \ottnt{R'} ,  \ottnt{H}  \ottsym{(}  \ottsym{(}  a  \ottsym{,}   i   \ottsym{+}   j   \ottsym{)}  \ottsym{)} ,    [  j  /  z  ]  \, \tau' _{ y }  ) \\
    &\qquad +  \mathbf{own} ( \ottnt{H} ,  \ottnt{R'} ,  \ottnt{H}  \ottsym{(}  \ottsym{(}  a  \ottsym{,}   i   \ottsym{)}  \ottsym{)} ,   \tau _{ x }  ) 
    & \tag{by definition of $\OWN$}\\
&= \begin{aligned}
        & \sum_ { j  \in \mathbb{Z} }\{  \ottsym{(}  a  \ottsym{,}   i   \ottsym{+}   j   \ottsym{)}   \mapsto    \llbracket    [  j  /  z  ]    r   \rrbracket_{ \ottnt{R'} }    \} 
         + \sum_{\substack{j\in\mathbb Z,\ j\ne0\\(a,i+j)\in\DOM(H)\\\sem{[j/z]r}_{R'}>0}}
            \mathbf{own} ( \ottnt{H} ,  \ottnt{R'} ,  \ottnt{H}  \ottsym{(}  \ottsym{(}  a  \ottsym{,}   i   \ottsym{+}   j   \ottsym{)}  \ottsym{)} ,    [  j  /  z  ]  \, \tau' _{ y }  ) \\
        &\qquad +  \mathbf{own} ( \ottnt{H} ,  \ottnt{R'} ,  \ottnt{H}  \ottsym{(}  \ottsym{(}  a  \ottsym{,}   i   \ottsym{)}  \ottsym{)} ,    [  0  /  z  ]  \, \tau' _{ y }  )  +  \mathbf{own} ( \ottnt{H} ,  \ottnt{R'} ,  \ottnt{H}  \ottsym{(}  \ottsym{(}  a  \ottsym{,}   i   \ottsym{)}  \ottsym{)} ,   \tau _{ x }  ) 
    \end{aligned}\tag{by (\ref{own-pos})} &\\
&= \begin{aligned}
        & \sum_ { j  \in \mathbb{Z} }\{  \ottsym{(}  a  \ottsym{,}   i   \ottsym{+}   j   \ottsym{)}   \mapsto    \llbracket    [  j  /  z  ]    r   \rrbracket_{ \ottnt{R'} }    \} 
         + \sum_{\substack{j\in\mathbb Z,\ j\ne0\\(a,i+j)\in\DOM(H)\\\sem{[j/z]r}_{R'}>0}}
            \mathbf{own} ( \ottnt{H} ,  \ottnt{R'} ,  \ottnt{H}  \ottsym{(}  \ottsym{(}  a  \ottsym{,}   i   \ottsym{+}   j   \ottsym{)}  \ottsym{)} ,    [  j  /  z  ]  \, \tau' _{ y }  ) \\
        &\qquad +  \mathbf{own} ( \ottnt{H} ,  \ottnt{R'} ,  \ottnt{H}  \ottsym{(}  \ottsym{(}  a  \ottsym{,}   i   \ottsym{)}  \ottsym{)} ,   [  0  /  z  ]  \, \tau' )  +  \mathbf{own} ( \ottnt{H} ,  \ottnt{R'} ,  \ottnt{H}  \ottsym{(}  \ottsym{(}  a  \ottsym{,}   i   \ottsym{)}  \ottsym{)} ,   \tau _{ x }  ) 
    \end{aligned}\tag{by premise~(\ref{conown-preservation2-zero-equiv}), \cref{lem:own-preservation-weakening-ty} in both directions, and \cref{lem:own-preservation-equivalent-ty-strength}} &\\
&= \begin{aligned}
        & \sum_ { j  \in \mathbb{Z} }\{  \ottsym{(}  a  \ottsym{,}   i   \ottsym{+}   j   \ottsym{)}   \mapsto    \llbracket    [  j  /  z  ]    r   \rrbracket_{ \ottnt{R'} }    \} 
         + \sum_{\substack{j\in\mathbb Z,\ j\ne0\\(a,i+j)\in\DOM(H)\\\sem{[j/z]r}_{R'}>0}}
            \mathbf{own} ( \ottnt{H} ,  \ottnt{R'} ,  \ottnt{H}  \ottsym{(}  \ottsym{(}  a  \ottsym{,}   i   \ottsym{+}   j   \ottsym{)}  \ottsym{)} ,    [  j  /  z  ]  \, \tau' _{ y }  ) \\
        &\qquad +  \mathbf{own} ( \ottnt{H} ,  \ottnt{R'} ,  \ottnt{H}  \ottsym{(}  \ottsym{(}  a  \ottsym{,}   i   \ottsym{)}  \ottsym{)} ,    [  0  /  z  ]  \, \tau _{ y }  ) \\
    \end{aligned} \tag{by premise~(\ref{conown-preservation2-split}) and \cref{lem:own-preservation}}&\\
&= \begin{aligned}
        & \sum_ { j  \in \mathbb{Z} }\{  \ottsym{(}  a  \ottsym{,}   i   \ottsym{+}   j   \ottsym{)}   \mapsto    \llbracket    [  j  /  z  ]    r   \rrbracket_{ \ottnt{R'} }    \} 
         + \sum_{\substack{j\in\mathbb Z,\ j\ne0\\(a,i+j)\in\DOM(H)\\\sem{[j/z]r}_{R'}>0}}
            \mathbf{own} ( \ottnt{H} ,  \ottnt{R'} ,  \ottnt{H}  \ottsym{(}  \ottsym{(}  a  \ottsym{,}   i   \ottsym{+}   j   \ottsym{)}  \ottsym{)} ,    [  j  /  z  ]  \, \tau _{ y }  ) \\
        &\qquad +  \mathbf{own} ( \ottnt{H} ,  \ottnt{R'} ,  \ottnt{H}  \ottsym{(}  \ottsym{(}  a  \ottsym{,}   i   \ottsym{)}  \ottsym{)} ,    [  0  /  z  ]  \, \tau _{ y }  ) \\
    \end{aligned} \tag{by premise~(\ref{conown-preservation2-nonzero-equiv}) and \cref{lem:own-preservation-weakening-ty} in both directions}&\\
&=  \sum_ { j  \in \mathbb{Z} }\{  \ottsym{(}  a  \ottsym{,}   i   \ottsym{+}   j   \ottsym{)}   \mapsto    \llbracket    [  j  /  z  ]    r   \rrbracket_{ \ottnt{R'} }    \} 
   + \sum_{\substack{j\in\mathbb Z\\(a,i+j)\in\DOM(H)\\\sem{[j/z]r}_{R'}>0}}
      \mathbf{own} ( \ottnt{H} ,  \ottnt{R'} ,  \ottnt{H}  \ottsym{(}  \ottsym{(}  a  \ottsym{,}   i   \ottsym{+}   j   \ottsym{)}  \ottsym{)} ,    [  j  /  z  ]  \, \tau _{ y }  )  &\\
&=  \mathbf{own} ( \ottnt{H} ,  \ottnt{R'} ,  \ottsym{(}  a  \ottsym{,}   i   \ottsym{)} ,   \Pi z .(  \tau _{ y }   \TREF^{\hspace{0.5pt} r })  )  &\\
&=  \mathbf{own} ( \ottnt{H} ,  \ottnt{R} ,  \ottsym{(}  a  \ottsym{,}   i   \ottsym{)} ,   \Pi z .(  \tau _{ y }   \TREF^{\hspace{0.5pt} r })  )  \tag{by \cref{lem:own-preservation2}}&
\end{align*}
\endgroup
Together with the unchanged contributions from $\Gamma_{{\mathrm{1}}}  \ottsym{,}  \Gamma_{{\mathrm{2}}}$, this proves
\[
   \mathbf{Own}( \ottnt{H} ,  \ottnt{R'} ,  \Gamma' ) = \mathbf{Own}( \ottnt{H} ,  \ottnt{R} ,  \Gamma_{{\mathrm{0}}} ) .
\]
The derived fact $ \mathbf{ConOwn}( \ottnt{H} ,  \ottnt{R} ,  \Gamma_{{\mathrm{0}}} ) $ therefore gives
$ \mathbf{ConOwn}( \ottnt{H} ,  \ottnt{R'} ,  \Gamma' ) $, as required.

\proofitem{Item~\ref{conown-preservation3}: Assignment
(\rn{Rs-Assign}).}

Let $\Gamma_{{\mathrm{0}}}$ be $\Gamma_{{\mathrm{1}}}$, $\Gamma'$ be $(  \Gamma_{{\mathrm{1}}}  \left[  x \hookleftarrow  \Pi z .(  \tau' _{\ast  x }   \TREF^{\hspace{0.5pt} r })   \right]   \left[  y \hookleftarrow  \tau' _{ y }   \right] )$, and $\ottnt{H'}$ be $(\ottnt{H}  \ottsym{\{}  \ottsym{(}  a  \ottsym{,}   i   \ottsym{)}  \hookleftarrow  \ottnt{R}  \ottsym{(}  y  \ottsym{)}  \ottsym{\}})$.
By \cref{lem:conown-preservation-weakening-tyenv},
$ \mathbf{ConOwn}( \ottnt{H} ,  \ottnt{R} ,  \Gamma_{{\mathrm{0}}} ) $ follows from
premises~(\ref{conown-preservation3-conown}) and
(\ref{conown-preservation3-env-subtype}), and the standing validity
assumption.

As in the preceding cases, \rn{S-TyEnv}, \rn{S-Int}, and the definition
of $ \fml{ \Gamma } $ give
\[
   \fml{ \Gamma } \implies \fml{ \Gamma_{{\mathrm{0}}} } .
\]
Consequently, $\models  \ottsym{[}  \ottnt{R}  \ottsym{]} \,  \fml{ \Gamma_{{\mathrm{0}}} } $.  Semantic instantiation of
premise~(\ref{conown-preservation3-full-ownership}) therefore gives
\begin{equation}
   \llbracket    [  0  /  z  ]    r   \rrbracket_{ \ottnt{R} }  \,  =  \,  1 .
  \label{eq:assign-full-ownership}
\end{equation}

We first observe that $x\ne y$.  Otherwise,
premises~(\ref{conown-preservation3-x-type}) and
(\ref{conown-preservation3-y-type}) would give
$ \tau _{ y }   =   \Pi z .(  \tau _{\ast  x }   \TREF^{\hspace{0.5pt} r }) $.
Premises~(\ref{conown-preservation3-split}),
(\ref{conown-preservation3-zero-equiv}), and
(\ref{conown-preservation3-nonzero-equiv}) imply that $ \tau _{ y } $ and
$ \tau _{\ast  x } $ have the same finite type shape,
which is impossible when $ \tau _{ y } $ is a reference to $ \tau _{\ast  x } $.

By premise~(\ref{conown-preservation3-location}), write
$R(x)=(a,i)\in dom(H)$.  Let $\ell_0=(a,i)$ and let
\[
 J=\{j\in\mathbb Z
       \mid (a,i+j)\in dom(H)
       \text{ and }\sem{  [  j  /  z  ]    r }_{\ottnt{R}}>0\}.
\]
By \cref{eq:assign-full-ownership}, the direct contribution of $x$ at
$\ell_0$ is $1$.  Hence, by the definition of $\OWN$,
\[
\begin{split}
& \mathbf{own} ( \ottnt{H} ,  \ottnt{R} ,  \ottnt{R}  \ottsym{(}  x  \ottsym{)} ,   \Pi z .(  \tau _{\ast  x }   \TREF^{\hspace{0.5pt} r })  ) (\ell_0)\\
&\qquad =
1+\sum_{j\in J}
  \mathbf{own} ( \ottnt{H} ,  \ottnt{R} ,  \ottnt{H}  \ottsym{(}  \ottsym{(}  a  \ottsym{,}   i   \ottsym{+}   j   \ottsym{)}  \ottsym{)} ,    [  j  /  z  ]  \, \tau _{\ast  x }  ) (\ell_0).
\end{split}
\]
All ownership contributions are nonnegative by
\cref{lem:ownership-bounds}.  Since $ \mathbf{ConOwn}( \ottnt{H} ,  \ottnt{R} ,  \Gamma_{{\mathrm{0}}} ) $, the total
ownership at $\ell_0$ is at most $1$.  It follows that
\begin{equation*}
  \mathbf{own} ( \ottnt{H} ,  \ottnt{R} ,  \ottnt{R}  \ottsym{(}  x  \ottsym{)} ,   \Pi z .(  \tau _{\ast  x }   \TREF^{\hspace{0.5pt} r })  ) (\ell_0)=1,
\end{equation*}
that
\begin{equation}
  \mathbf{own} ( \ottnt{H} ,  \ottnt{R} ,  \ottnt{R}  \ottsym{(}  b  \ottsym{)} ,   \Gamma_{{\mathrm{0}}}  (  b  )  ) (\ell_0)=0
 \qquad
 (b\in dom(\Gamma_{{\mathrm{0}}})\setminus\{x\}),
 \label{eq:assign-other-zero}
\end{equation}
and that
\begin{equation}
  \mathbf{own} ( \ottnt{H} ,  \ottnt{R} ,  \ottnt{H}  \ottsym{(}  \ottsym{(}  a  \ottsym{,}   i   \ottsym{+}   j   \ottsym{)}  \ottsym{)} ,    [  j  /  z  ]  \, \tau _{\ast  x }  ) (\ell_0)=0
 \qquad (j\in J).
 \label{eq:assign-element-zero}
\end{equation}

In particular, since $x\ne y$, \cref{eq:assign-other-zero} gives
\begin{equation}
  \mathbf{own} ( \ottnt{H} ,  \ottnt{R} ,  \ottnt{R}  \ottsym{(}  y  \ottsym{)} ,   \tau _{ y }  ) (\ell_0)=0.
 \label{eq:assign-source-zero}
\end{equation}
By premise~(\ref{conown-preservation3-split}) and Lemma
\ref{lem:own-preservation},
\[
\begin{split}
  \mathbf{own} ( \ottnt{H} ,  \ottnt{R} ,  \ottnt{R}  \ottsym{(}  y  \ottsym{)} ,   \tau _{ y }  ) 
 &=  \mathbf{own} ( \ottnt{H} ,  \ottnt{R} ,  \ottnt{R}  \ottsym{(}  y  \ottsym{)} ,   \tau' _{ y }  ) \\
 &\qquad+ \mathbf{own} ( \ottnt{H} ,  \ottnt{R} ,  \ottnt{R}  \ottsym{(}  y  \ottsym{)} ,  \tau' ) .
\end{split}
\]
Together with \cref{eq:assign-source-zero} and nonnegativity, this gives
\begin{equation}
  \mathbf{own} ( \ottnt{H} ,  \ottnt{R} ,  \ottnt{R}  \ottsym{(}  y  \ottsym{)} ,   \tau' _{ y }  ) (\ell_0)=0
 \quad\text{and}\quad
  \mathbf{own} ( \ottnt{H} ,  \ottnt{R} ,  \ottnt{R}  \ottsym{(}  y  \ottsym{)} ,  \tau' ) (\ell_0)=0.
 \label{eq:assign-split-zero}
\end{equation}

We have $ \ottnt{H}   \approx _{(a,i)}  \ottnt{H'} $.  Using
premises~(\ref{conown-preservation3-zero-equiv}) and
(\ref{conown-preservation3-nonzero-equiv}) as appropriate to transfer the
zero equalities to the corresponding updated types, Lemma
\ref{lem:own-preservation-equivalent-heap}, applied using
\cref{eq:assign-other-zero,eq:assign-element-zero,eq:assign-split-zero},
shows that the heap update does not change:
\begin{itemize}
\item the ownership contributed by any binding other than $x$ and $y$;
\item the ownership of $R(y)$ at types $\tau'$ and $ \tau' _{ y } $; or
\item the ownership of any nonzero element reached from $x$.
\end{itemize}

Now fix an arbitrary address $\ell$.  For every $j\in J\setminus\{0\}$,
premise~(\ref{conown-preservation3-nonzero-equiv}), instantiated at $z=j$,
and Lemma
\ref{lem:own-preservation-weakening-ty} in both directions give
\begin{equation}
\begin{split}
& \mathbf{own} ( \ottnt{H'} ,  \ottnt{R} ,  \ottnt{H'}  \ottsym{(}  \ottsym{(}  a  \ottsym{,}   i   \ottsym{+}   j   \ottsym{)}  \ottsym{)} ,    [  j  /  z  ]  \, \tau' _{\ast  x }  ) (\ell)\\
&\qquad =
  \mathbf{own} ( \ottnt{H} ,  \ottnt{R} ,  \ottnt{H}  \ottsym{(}  \ottsym{(}  a  \ottsym{,}   i   \ottsym{+}   j   \ottsym{)}  \ottsym{)} ,    [  j  /  z  ]  \, \tau _{\ast  x }  ) (\ell).
\end{split}
\label{eq:assign-nonzero-elements}
\end{equation}
Here we also use $H'((a,i+j))=H((a,i+j))$, since $j\ne0$, and the
heap-frame equality above.

At offset zero, $H'((a,i))=R(y)$.
Premise~(\ref{conown-preservation3-zero-equiv}), Lemma
\ref{lem:own-preservation-weakening-ty} in both directions, Lemma
\ref{lem:own-preservation-equivalent-ty-strength}, and
\cref{eq:assign-split-zero} give
\begin{equation}
\begin{split}
& \mathbf{own} ( \ottnt{H'} ,  \ottnt{R} ,  \ottnt{H'}  \ottsym{(}  \ottsym{(}  a  \ottsym{,}   i   \ottsym{)}  \ottsym{)} ,    [  0  /  z  ]  \, \tau' _{\ast  x }  ) \\
&\qquad =  \mathbf{own} ( \ottnt{H'} ,  \ottnt{R} ,  \ottnt{R}  \ottsym{(}  y  \ottsym{)} ,  \tau' ) 
 =  \mathbf{own} ( \ottnt{H} ,  \ottnt{R} ,  \ottnt{R}  \ottsym{(}  y  \ottsym{)} ,  \tau' ) .
\end{split}
\label{eq:assign-zero-element}
\end{equation}

Let
\[
 D= \sum_ { j  \in \mathbb{Z} }\{  \ottsym{(}  a  \ottsym{,}   i   \ottsym{+}   j   \ottsym{)}   \mapsto    \llbracket    [  j  /  z  ]    r   \rrbracket_{ \ottnt{R} }    \} 
\]
be the direct ownership contribution of $x$.  Using
\cref{eq:assign-nonzero-elements,eq:assign-zero-element}, the heap-frame
equality for $ \tau' _{ y } $, and
premise~(\ref{conown-preservation3-split}), we obtain
\begin{align*}
& \mathbf{own} ( \ottnt{H'} ,  \ottnt{R} ,  \ottnt{R}  \ottsym{(}  x  \ottsym{)} ,   \Pi z .(  \tau' _{\ast  x }   \TREF^{\hspace{0.5pt} r })  ) (\ell)
 + \mathbf{own} ( \ottnt{H'} ,  \ottnt{R} ,  \ottnt{R}  \ottsym{(}  y  \ottsym{)} ,   \tau' _{ y }  ) (\ell)\\
&\quad =D(\ell)
 +\sum_{j\in J\setminus\{0\}}
     \mathbf{own} ( \ottnt{H} ,  \ottnt{R} ,  \ottnt{H}  \ottsym{(}  \ottsym{(}  a  \ottsym{,}   i   \ottsym{+}   j   \ottsym{)}  \ottsym{)} ,    [  j  /  z  ]  \, \tau _{\ast  x }  ) (\ell)
 + \mathbf{own} ( \ottnt{H} ,  \ottnt{R} ,  \ottnt{R}  \ottsym{(}  y  \ottsym{)} ,   \tau _{ y }  ) (\ell)\\
&\quad \le D(\ell)
 +\sum_{j\in J}
     \mathbf{own} ( \ottnt{H} ,  \ottnt{R} ,  \ottnt{H}  \ottsym{(}  \ottsym{(}  a  \ottsym{,}   i   \ottsym{+}   j   \ottsym{)}  \ottsym{)} ,    [  j  /  z  ]  \, \tau _{\ast  x }  ) (\ell)
 + \mathbf{own} ( \ottnt{H} ,  \ottnt{R} ,  \ottnt{R}  \ottsym{(}  y  \ottsym{)} ,   \tau _{ y }  ) (\ell)\\
&\quad = \mathbf{own} ( \ottnt{H} ,  \ottnt{R} ,  \ottnt{R}  \ottsym{(}  x  \ottsym{)} ,   \Pi z .(  \tau _{\ast  x }   \TREF^{\hspace{0.5pt} r })  ) (\ell)
 + \mathbf{own} ( \ottnt{H} ,  \ottnt{R} ,  \ottnt{R}  \ottsym{(}  y  \ottsym{)} ,   \tau _{ y }  ) (\ell).
\end{align*}
The inequality follows from \cref{lem:ownership-bounds}: its right-hand
side additionally contains the nonnegative zero-offset recursive contribution
of the old value of $x$.

Every remaining binding has the same ownership map before and after the heap
update, by \cref{eq:assign-other-zero} and Lemma
\ref{lem:own-preservation-equivalent-heap}.  Therefore,
\[
   \mathbf{Own}( \ottnt{H'} ,  \ottnt{R} ,  \Gamma' ) \le  \mathbf{Own}( \ottnt{H} ,  \ottnt{R} ,  \Gamma_{{\mathrm{0}}} ) .
\]
Since the heap domain is unchanged, the desired $ \mathbf{ConOwn}( \ottnt{H'} ,  \ottnt{R} ,  \Gamma' ) $ follows.

\proofitem{Item~\ref{conown-preservation4}: Pointer addition
(\rn{Rs-AddPtr}).}
Let
\[
\begin{array}{lll}
  \tau_{{\mathrm{4}}}= \Pi w .( \tau_{{\mathrm{1}}}  \TREF^{\hspace{0.5pt}  { r }_{ y_{{\mathrm{1}}} }  }) ,&
  \tau_{{\mathrm{5}}}= \Pi w .( \tau_{{\mathrm{2}}}  \TREF^{\hspace{0.5pt}  { r }_{ x }  }) ,&
  \tau_{{\mathrm{6}}}= \Pi w .( \tau_{{\mathrm{3}}}  \TREF^{\hspace{0.5pt}  { r }_{ y }  }) .
\end{array}
\]
Let $\Gamma'$ be $ \Gamma_{{\mathrm{1}}}  \left[  y \hookleftarrow  \Pi w .( \tau_{{\mathrm{1}}}  \TREF^{\hspace{0.5pt}  { r }_{ y_{{\mathrm{1}}} }  })   \right]   \ottsym{,}   x' \COL  \Pi w .( \tau_{{\mathrm{2}}}  \TREF^{\hspace{0.5pt}  { r }_{ x }  })  $ and let
$R'$ be $\ottnt{R}  \ottsym{\{}  x'  \mapsto  pv  \boxplus  \ottnt{R}  \ottsym{(}  z  \ottsym{)}  \ottsym{\}}$.
By \cref{lem:conown-preservation-weakening-tyenv},
premises~(\ref{conown-preservation4-conown}) and
(\ref{conown-preservation4-env-subtype}), and the standing validity
assumption give $ \mathbf{ConOwn}( \ottnt{H} ,  \ottnt{R} ,  \Gamma_{{\mathrm{1}}} ) $.
As in the preceding cases, $\Gamma  \leq  \Gamma_{{\mathrm{1}}}$ gives
$ \fml{ \Gamma } \implies \fml{ \Gamma_{{\mathrm{1}}} } $, so the standing validity assumption also
gives $\models  \ottsym{[}  \ottnt{R}  \ottsym{]} \,  \fml{ \Gamma_{{\mathrm{1}}} } $.

By the definition of shifted type addition,
premise~(\ref{conown-preservation4-split}) is
\[
  \Gamma_1\vdash\tau_{{\mathrm{4}}}\mathbin{+_z}\tau_{{\mathrm{5}}} \approx \tau_{{\mathrm{6}}}.
\]
Premises~(\ref{conown-preservation4-fresh}) and
(\ref{conown-preservation4-y-value}) give $R'(y)=R(y)=pv$ and ensure that
the fresh binding does not change the interpretation of the three types.
Consequently,
\begin{align*}
&\OWN(H,R',R'(y),\tau_{{\mathrm{4}}})+\OWN(H,R',R'(x'),\tau_{{\mathrm{5}}})\\
&\quad=\OWN(H,R,pv,\tau_{{\mathrm{4}}})
  +\OWN(H,R,pv  \boxplus  \ottnt{R}  \ottsym{(}  z  \ottsym{)},\tau_{{\mathrm{5}}})
  \tag{by \cref{lem:own-preservation2}}\\
&\quad=\OWN(H,R,pv,\tau_{{\mathrm{6}}})
  \tag{by \cref{lem:own-pointer-split} and
    premise~(\ref{conown-preservation4-offset-value})}\\
&\quad=\OWN(H,R,R(y),\Gamma_1(y))
  \tag{by premises~(\ref{conown-preservation4-y-type}) and
    (\ref{conown-preservation4-y-value}).}
\end{align*}
For every other binding of $\Gamma_{{\mathrm{1}}}$, the type and runtime value are unchanged,
and \cref{lem:own-preservation2} shows that its ownership map is also
unchanged.  Since $x'$ is fresh, decomposing total ownership at the bindings
$y$ and $x'$ therefore gives
\[
   \mathbf{Own}( \ottnt{H} ,  \ottnt{R'} ,  \Gamma' ) = \mathbf{Own}( \ottnt{H} ,  \ottnt{R} ,  \Gamma_{{\mathrm{1}}} ) .
\]
The heap is unchanged, so the derived fact $ \mathbf{ConOwn}( \ottnt{H} ,  \ottnt{R} ,  \Gamma_{{\mathrm{1}}} ) $ now gives
$ \mathbf{ConOwn}( \ottnt{H} ,  \ottnt{R'} ,  \Gamma' ) $, as required.

\proofitem{Item~\ref{conown-preservation5}: Integer-array allocation
(\rn{Rs-MkArrayIntref}).}

Let $\Gamma'$ be $(\Gamma_{{\mathrm{1}}}  \ottsym{,}   x' \COL  \Pi z .(  \{  \nu  :   \TINT    \mid     0  \, \le \, z  \wedge  z \, \le \, y  \ottsym{-}   1    \implies  \nu \,  =  \,  0   \}   \TREF^{\hspace{0.5pt} r })  )$, 
$\ottnt{H'}$ be $ \ottnt{H}  \{ (  a  ,   0   ) \mapsto   0  , \ldots, (  a  ,  \ottnt{R}  \ottsym{(}  y  \ottsym{)} \,  -  \,  1   ) \mapsto   0   \} $
and $R'$ be $\ottnt{R}  \ottsym{\{}  x'  \mapsto  \ottsym{(}  a  \ottsym{,}   0   \ottsym{)}  \ottsym{\}}$.
By \cref{lem:conown-preservation-weakening-tyenv},
premises~(\ref{conown-preservation5-conown}) and
(\ref{conown-preservation5-env-subtype}), and the standing validity
assumption give $ \mathbf{ConOwn}( \ottnt{H} ,  \ottnt{R} ,  \Gamma_{{\mathrm{1}}} ) $.
As in the preceding cases, $\Gamma  \leq  \Gamma_{{\mathrm{1}}}$ implies
$ \fml{ \Gamma } \implies \fml{ \Gamma_{{\mathrm{1}}} } $, so the standing validity assumption also
gives $\models  \ottsym{[}  \ottnt{R}  \ottsym{]} \,  \fml{ \Gamma_{{\mathrm{1}}} } $.
By \cref{lem:sat-preservation-greater-tyenv},
premises~(\ref{conown-preservation5-sat}) and
(\ref{conown-preservation5-env-subtype}), and the standing validity
assumption also give $ \mathbf{SAT}( \ottnt{H} ,  \ottnt{R} ,  \Gamma_{{\mathrm{1}}} ) $.  In particular,
premise~(\ref{conown-preservation5-length-type}) gives $R(y)\in\mathbb Z$.

We first record the effect of extending the heap at the fresh base.  For every
$b\in dom(\Gamma_{{\mathrm{1}}})$, satisfaction gives
$ \mathbf{SATv} ( \ottnt{H} ,  \ottnt{R} ,  \ottnt{R}  \ottsym{(}  b  \ottsym{)} ,   \Gamma_{{\mathrm{1}}}  (  b  )  ) $.  An induction on $ \Gamma_{{\mathrm{1}}}  (  b  ) $, using
\cref{lem:ownership-bounds}, proves the following two facts:
\begin{align}
  \mathbf{own} ( \ottnt{H'} ,  \ottnt{R} ,  \ottnt{R}  \ottsym{(}  b  \ottsym{)} ,   \Gamma_{{\mathrm{1}}}  (  b  )  ) 
   &=  \mathbf{own} ( \ottnt{H} ,  \ottnt{R} ,  \ottnt{R}  \ottsym{(}  b  \ottsym{)} ,   \Gamma_{{\mathrm{1}}}  (  b  )  ) ,
   \label{eq:alloc-int-old-frame}\\
  \mathbf{own} ( \ottnt{H} ,  \ottnt{R} ,  \ottnt{R}  \ottsym{(}  b  \ottsym{)} ,   \Gamma_{{\mathrm{1}}}  (  b  )  ) (a,j)&=0
   \qquad(j\in\mathbb Z).
   \label{eq:alloc-int-old-zero}
\end{align}
For an integer type both facts are immediate.  For a reference type, a
positive direct ownership coefficient at $(a,j)$ would, by satisfaction, imply
$(a,j)\in dom(H)$, contradicting
premise~(\ref{conown-preservation5-fresh-base}) and the heap-domain invariant.
Thus the direct contribution at $(a,j)$ is zero; the induction hypothesis
handles the recursive contributions.  The same observation shows that none
of the cells newly added to $H'$ contributes a new recursive summand, while
the summands over the old heap cells are unchanged.  This proves
\cref{eq:alloc-int-old-frame,eq:alloc-int-old-zero}.

Since $x'$ is fresh, it does not occur in the types in $\Gamma_{{\mathrm{1}}}$, and $R'$ agrees
with $R$ on $dom(\Gamma_{{\mathrm{1}}})$.  Hence \cref{lem:own-preservation2} and summation
of \cref{eq:alloc-int-old-frame,eq:alloc-int-old-zero} over $dom(\Gamma_{{\mathrm{1}}})$
give
\begin{align}
  \mathbf{Own}( \ottnt{H'} ,  \ottnt{R'} ,  \Gamma_{{\mathrm{1}}} ) &= \mathbf{Own}( \ottnt{H} ,  \ottnt{R} ,  \Gamma_{{\mathrm{1}}} ) ,
   \label{eq:alloc-int-env-frame}\\
  \mathbf{Own}( \ottnt{H'} ,  \ottnt{R'} ,  \Gamma_{{\mathrm{1}}} ) (a,j)&=0
   \qquad(j\in\mathbb Z).
   \label{eq:alloc-int-env-zero}
\end{align}

It remains to calculate the contribution of the new binding.  Put
\[
 \tau_a= \Pi z .(  \{  \nu  :   \TINT    \mid     0  \, \le \, z  \wedge  z \, \le \, y  \ottsym{-}   1    \implies  \nu \,  =  \,  0   \}   \TREF^{\hspace{0.5pt} r }) .
\]
The recursive part of $\OWN(H',R',(a,0),\tau_a)$ is the zero ownership map,
because the element type of $\tau_a$ is an integer type.  Moreover,
premise~(\ref{conown-preservation5-ownership-shape}),
Item~1 of \cref{lem:index-substitution} applied in both directions,
pointwise
antisymmetry, and
$\models  \ottsym{[}  \ottnt{R}  \ottsym{]} \,  \fml{ \Gamma_{{\mathrm{1}}} } $ give, for every $j\in\mathbb Z$,
\[
 \sem{  [  j  /  z  ]    r }_{R'}=
 \begin{cases}
  1 & \text{if }0\leq j<R(y),\\
  0 & \text{otherwise}.
 \end{cases}
\]
Here evaluation under $R'$ agrees with evaluation under $R$, since $x'$ is
fresh.  Therefore, for every address $(b,j)$,
\begin{equation}
 \OWN(H',R',(a,0),\tau_a)(b,j)=
 \begin{cases}
  1 & \text{if }b=a\text{ and }0\leq j<R(y),\\
  0 & \text{otherwise}.
 \end{cases}
 \label{eq:alloc-int-new-own}
\end{equation}

Finally, let $(b,j)\in dom(H')$.  If $(b,j)\in dom(H)$, then $b\ne a$ by
premise~(\ref{conown-preservation5-fresh-base}) and the heap-domain
invariant.  Thus the new binding contributes zero by
\cref{eq:alloc-int-new-own}, while
\cref{eq:alloc-int-env-frame} and the derived fact
$ \mathbf{ConOwn}( \ottnt{H} ,  \ottnt{R} ,  \Gamma_{{\mathrm{1}}} ) $ bound the old contribution by $1$.  Otherwise,
$(b,j)$ is newly allocated, so $b=a$ and $0\leq j<R(y)$.  The new binding
contributes exactly $1$ by \cref{eq:alloc-int-new-own}, and the old
environment contributes zero by \cref{eq:alloc-int-env-zero}.  Hence
$ \mathbf{Own}( \ottnt{H'} ,  \ottnt{R'} ,  \Gamma' ) (b,j)\leq1$ in either case, proving
$ \mathbf{ConOwn}( \ottnt{H'} ,  \ottnt{R'} ,  \Gamma' ) $.

\proofitem{Item~\ref{conown-preservation6}: Nested-array allocation
(\rn{Rs-MkArrayNestedref}).}

Let $\Gamma'$ be $(\Gamma_{{\mathrm{1}}}  \ottsym{,}   x' \COL  \Pi z .( \tau'  \TREF^{\hspace{0.5pt} r })  )$, $\ottnt{H'}$ be
$ \ottnt{H}  \{ (  a  ,   0   ) \mapsto  \ottkw{null} , \ldots, (  a  ,  \ottnt{R}  \ottsym{(}  y  \ottsym{)} \,  -  \,  1   ) \mapsto  \ottkw{null}  \} $, and $R'$ be
$\ottnt{R}  \ottsym{\{}  x'  \mapsto  \ottsym{(}  a  \ottsym{,}   0   \ottsym{)}  \ottsym{\}}$.
By \cref{lem:conown-preservation-weakening-tyenv},
premises~(\ref{conown-preservation6-conown}) and
(\ref{conown-preservation6-env-subtype}), and the standing validity
assumption give $ \mathbf{ConOwn}( \ottnt{H} ,  \ottnt{R} ,  \Gamma_{{\mathrm{1}}} ) $.
As in the preceding item, $\Gamma  \leq  \Gamma_{{\mathrm{1}}}$ gives
$ \fml{ \Gamma } \implies \fml{ \Gamma_{{\mathrm{1}}} } $, so the standing validity assumption gives
$\models  \ottsym{[}  \ottnt{R}  \ottsym{]} \,  \fml{ \Gamma_{{\mathrm{1}}} } $.  By
\cref{lem:sat-preservation-greater-tyenv},
premises~(\ref{conown-preservation6-sat}) and
(\ref{conown-preservation6-env-subtype}), and the standing validity
assumption also give $ \mathbf{SAT}( \ottnt{H} ,  \ottnt{R} ,  \Gamma_{{\mathrm{1}}} ) $.  In particular,
premise~(\ref{conown-preservation6-length-type}) gives $R(y)\in\mathbb Z$.

The old-environment frame argument from Item~\ref{conown-preservation5}
does not depend on the values used to initialize the new cells.  More
explicitly, for every $b\in dom(\Gamma_{{\mathrm{1}}})$, an induction on $ \Gamma_{{\mathrm{1}}}  (  b  ) $, using
$ \mathbf{SATv} ( \ottnt{H} ,  \ottnt{R} ,  \ottnt{R}  \ottsym{(}  b  \ottsym{)} ,   \Gamma_{{\mathrm{1}}}  (  b  )  ) $, \cref{lem:ownership-bounds},
premise~(\ref{conown-preservation6-fresh-base}), and the heap-domain
invariant, gives
\begin{align*}
  \mathbf{own} ( \ottnt{H'} ,  \ottnt{R} ,  \ottnt{R}  \ottsym{(}  b  \ottsym{)} ,   \Gamma_{{\mathrm{1}}}  (  b  )  ) 
   &=  \mathbf{own} ( \ottnt{H} ,  \ottnt{R} ,  \ottnt{R}  \ottsym{(}  b  \ottsym{)} ,   \Gamma_{{\mathrm{1}}}  (  b  )  ) ,\\
  \mathbf{own} ( \ottnt{H} ,  \ottnt{R} ,  \ottnt{R}  \ottsym{(}  b  \ottsym{)} ,   \Gamma_{{\mathrm{1}}}  (  b  )  ) (a,j)&=0
   \qquad(j\in\mathbb Z).
\end{align*}
Since $x'$ is fresh, $R'$ agrees with $R$ on $dom(\Gamma_{{\mathrm{1}}})$ and does not
affect the types in $\Gamma_{{\mathrm{1}}}$.  Hence \cref{lem:own-preservation2} and
summation over $dom(\Gamma_{{\mathrm{1}}})$ give
\begin{align}
  \mathbf{Own}( \ottnt{H'} ,  \ottnt{R'} ,  \Gamma_{{\mathrm{1}}} ) &= \mathbf{Own}( \ottnt{H} ,  \ottnt{R} ,  \Gamma_{{\mathrm{1}}} ) ,
   \label{eq:alloc-nested-env-frame}\\
  \mathbf{Own}( \ottnt{H'} ,  \ottnt{R'} ,  \Gamma_{{\mathrm{1}}} ) (a,j)&=0
   \qquad(j\in\mathbb Z).
   \label{eq:alloc-nested-env-zero}
\end{align}

It remains to calculate the contribution of the new binding.  Put
\[
 \tau_a= \Pi z .( \tau'  \TREF^{\hspace{0.5pt} r }) .
\]
The freshness of $x'$ also gives $\models  \ottsym{[}  \ottnt{R'}  \ottsym{]} \,  \fml{ \Gamma_{{\mathrm{1}}} } $.  Independently,
for every integer $j$, instantiating the unconditional emptiness
premise~(\ref{conown-preservation6-empty}) gives
\[
 \models  \ottsym{[}  \ottnt{R'}  \ottsym{]} \,  \Empty{  [  j  /  z  ]  \, \tau' } .
\]
By \cref{lem:empty-own}, every recursive ownership contribution of an
element of the new array is $ \emptyset $.  Premise
(\ref{conown-preservation6-ownership-shape}), instantiated using
$\models  \ottsym{[}  \ottnt{R'}  \ottsym{]} \,  \fml{ \Gamma_{{\mathrm{1}}} } $, gives the direct contribution.  Consequently, for
every address $(b,j)$,
\begin{equation}
 \OWN(H',R',(a,0),\tau_a)(b,j)=
 \begin{cases}
  1 & \text{if }b=a\text{ and }0\leq j<R(y),\\
  0 & \text{otherwise}.
 \end{cases}
 \label{eq:alloc-nested-new-own}
\end{equation}

Finally, let $(b,j)\in dom(H')$.  If $(b,j)\in dom(H)$, then $b\ne a$ by
premise~(\ref{conown-preservation6-fresh-base}) and the heap-domain
invariant.  Thus the new binding contributes zero by
\cref{eq:alloc-nested-new-own}, while
\cref{eq:alloc-nested-env-frame} and the derived fact
$ \mathbf{ConOwn}( \ottnt{H} ,  \ottnt{R} ,  \Gamma_{{\mathrm{1}}} ) $ bound the old contribution by $1$.  Otherwise,
$(b,j)$ is newly allocated, so $b=a$ and $0\leq j<R(y)$.  The new binding
contributes exactly $1$ by \cref{eq:alloc-nested-new-own}, and the old
environment contributes zero by \cref{eq:alloc-nested-env-zero}.  Hence
$ \mathbf{Own}( \ottnt{H'} ,  \ottnt{R'} ,  \Gamma' ) (b,j)\leq1$ in either case, proving
$ \mathbf{ConOwn}( \ottnt{H'} ,  \ottnt{R'} ,  \Gamma' ) $.

\proofitem{Item~\ref{conown-preservation7}: Aliasing by dereference
(\rn{Rs-AliasDeref}).}

Let
\[
 \Gamma' =   \Gamma_{{\mathrm{1}}}  \left[  x \hookleftarrow  \Pi z'_{{\mathrm{1}}} .(  \tau' _{\ast  x }   \TREF^{\hspace{0.5pt}  { r' }_{ x }  })   \right]   \left[  y \hookleftarrow  \Pi z_{{\mathrm{2}}} .(  \Pi z_{{\mathrm{1}}} .(  \tau' _{\ast \ast  y }   \TREF^{\hspace{0.5pt}  { r' }_{ \ast  y }  })   \TREF^{\hspace{0.5pt} r })   \right] .
\]
By \cref{lem:conown-preservation-weakening-tyenv},
premises~(\ref{conown-preservation7-conown}) and
(\ref{conown-preservation7-env-subtype}), and the standing validity
assumption give $ \mathbf{ConOwn}( \ottnt{H} ,  \ottnt{R} ,  \Gamma_{{\mathrm{1}}} ) $.
As in the preceding cases, $\Gamma  \leq  \Gamma_{{\mathrm{1}}}$ also gives
$ \fml{ \Gamma } \implies \fml{ \Gamma_{{\mathrm{1}}} } $, and hence
\begin{equation}
 \models  \ottsym{[}  \ottnt{R}  \ottsym{]} \,  \fml{ \Gamma_{{\mathrm{1}}} } .
 \label{eq:alias-deref-env-valid}
\end{equation}

For readability, put
\[
\begin{array}{ll}
 \tau_x=  \Pi z'_{{\mathrm{1}}} .(  \tau _{\ast  x }   \TREF^{\hspace{0.5pt} r })  _{ x } ,&
 \tau_p=  \Pi z_{{\mathrm{1}}} .(  \tau _{\ast \ast  y }   \TREF^{\hspace{0.5pt} r })  _{\ast  y } ,\\
 \tau_x'=  \Pi z'_{{\mathrm{1}}} .(  \tau' _{\ast  x }   \TREF^{\hspace{0.5pt} r' })  _{ x } ,&
 \tau_p'=  \Pi z_{{\mathrm{1}}} .(  \tau' _{\ast \ast  y }   \TREF^{\hspace{0.5pt} r' })  _{\ast  y } ,\\
 \tau_y= \Pi z_{{\mathrm{2}}} .(   \Pi z_{{\mathrm{1}}} .(  \tau _{\ast \ast  y }   \TREF^{\hspace{0.5pt} r })  _{\ast  y }   \TREF^{\hspace{0.5pt} r }) ,&
 \tau_y'= \Pi z_{{\mathrm{2}}} .(   \Pi z_{{\mathrm{1}}} .(  \tau' _{\ast \ast  y }   \TREF^{\hspace{0.5pt} r' })  _{\ast  y }   \TREF^{\hspace{0.5pt} r }) .
\end{array}
\]
We first observe that $x\ne y$.  Otherwise,
premises~(\ref{conown-preservation7-x-type}) and
(\ref{conown-preservation7-y-type}) would identify $\tau_x$ with a reference
whose element type is $\tau_p$.  On the other hand, the type additions in
premise~(\ref{conown-preservation7-zero-equiv}) require $\tau_x$ and
$\tau_p$ to have the same simple shape.  This would make a finite simple type
equal to a reference to itself, which is impossible.

Premise~(\ref{conown-preservation7-alias}) implies that, for some $a$ and
$k$,
\begin{equation}
 R(y)=(a,k)\in dom(H)
 \qquad\text{and}\qquad
 H((a,k))=R(x).
 \label{eq:alias-deref-runtime}
\end{equation}
Semantic instantiation of
premise~(\ref{conown-preservation7-positive}) using
\cref{eq:alias-deref-env-valid} gives
\begin{equation}
 \sem{  [  0  /  z_{{\mathrm{2}}}  ]    r }_{R}>0.
 \label{eq:alias-deref-positive}
\end{equation}

We now compare the ownership contributed by $x$ and $y$.  At offset zero,
the four-type type-addition notation in
premise~(\ref{conown-preservation7-zero-equiv}), instantiated with $z_2=0$,
provides a common total for the old pair and the new pair.  Applying
\cref{lem:own-preservation} to the two additions and
\cref{lem:own-preservation-weakening-ty} in both directions to their
equivalent totals gives
\begin{equation}
\begin{split}
 &\OWN(H,R,R(x),\tau_x')
   +\OWN(H,R,R(x),[0/z_2]\tau_p')\\
 &\qquad=
   \OWN(H,R,R(x),\tau_x)
   +\OWN(H,R,R(x),[0/z_2]\tau_p).
\end{split}
\label{eq:alias-deref-zero-own}
\end{equation}
Here $[0/z_2]\tau_p$ and $[0/z_2]\tau_p'$ denote the corresponding
index substitutions in the displayed types.

For every $j\ne0$, premise
(\ref{conown-preservation7-nonzero-equiv}), instantiated with $z_2=j$, and
\cref{lem:own-preservation-weakening-ty} in both directions give
\begin{equation}
\begin{split}
 &\OWN(H,R,H((a,k+j)),[j/z_2]\tau_p')\\
 &\qquad=
   \OWN(H,R,H((a,k+j)),[j/z_2]\tau_p)
\end{split}
\label{eq:alias-deref-nonzero-own}
\end{equation}
whenever $(a,k+j)\in dom(H)$ and
$\sem{  [  j  /  z_{{\mathrm{2}}}  ]    r }_{R}>0$.

By \cref{eq:alias-deref-positive,eq:alias-deref-runtime}, the zero-offset
recursive term in the ownership of $y$ is present and its heap value is
$R(x)$.  Expanding the definition of $\OWN$ for $y$, separating this term,
and applying \cref{eq:alias-deref-zero-own} to it and the contribution of
$x$, while applying \cref{eq:alias-deref-nonzero-own} to every remaining
term, proves
\begin{equation}
\begin{split}
 &\OWN(H,R,R(x),\tau_x')
  +\OWN(H,R,R(y),\tau_y')\\
 &\qquad=
  \OWN(H,R,R(x),\tau_x)
  +\OWN(H,R,R(y),\tau_y).
\end{split}
\label{eq:alias-deref-pair-own}
\end{equation}
The direct ownership map of $y$ is identical on the two sides because its
outer ownership term $r$ is unchanged.

Since $x\ne y$, all other bindings have unchanged types and values.
Equation~\ref{eq:alias-deref-pair-own} therefore gives
$ \mathbf{Own}( \ottnt{H} ,  \ottnt{R} ,  \Gamma' ) = \mathbf{Own}( \ottnt{H} ,  \ottnt{R} ,  \Gamma_{{\mathrm{1}}} ) $.  The derived fact
$ \mathbf{ConOwn}( \ottnt{H} ,  \ottnt{R} ,  \Gamma_{{\mathrm{1}}} ) $ proves $ \mathbf{ConOwn}( \ottnt{H} ,  \ottnt{R} ,  \Gamma' ) $, as required.

\proofitem{Item~\ref{conown-preservation8}: Aliasing by pointer addition
(\rn{Rs-AliasAddPtr}).}

Let
\[
 \Gamma' =   \Gamma_{{\mathrm{1}}}  \left[  x \hookleftarrow  \Pi w' .(  \tau' _{\ast  x }   \TREF^{\hspace{0.5pt}  { r' }_{ x }  })   \right]   \left[  y \hookleftarrow  \Pi w .(  \tau' _{\ast  y }   \TREF^{\hspace{0.5pt}  { r' }_{ y }  })   \right] .
\]
By \cref{lem:conown-preservation-weakening-tyenv},
premises~(\ref{conown-preservation8-conown}) and
(\ref{conown-preservation8-env-subtype}), and the standing validity
assumption give $ \mathbf{ConOwn}( \ottnt{H} ,  \ottnt{R} ,  \Gamma_{{\mathrm{1}}} ) $.
As in the preceding cases, $\Gamma  \leq  \Gamma_{{\mathrm{1}}}$ gives
$ \fml{ \Gamma } \implies \fml{ \Gamma_{{\mathrm{1}}} } $, and hence
\begin{equation}
 \models  \ottsym{[}  \ottnt{R}  \ottsym{]} \,  \fml{ \Gamma_{{\mathrm{1}}} } .
 \label{eq:alias-add-env-valid}
\end{equation}

For readability, put
\[
\begin{array}{ll}
 \tau_x=  \Pi w' .(  \tau _{\ast  x }   \TREF^{\hspace{0.5pt} r })  _{ x } ,&
 \tau_y=  \Pi w .(  \tau _{\ast  y }   \TREF^{\hspace{0.5pt} r })  _{ y } ,\\
 \tau_x'=  \Pi w' .(  \tau' _{\ast  x }   \TREF^{\hspace{0.5pt} r' })  _{ x } ,&
 \tau_y'=  \Pi w .(  \tau' _{\ast  y }   \TREF^{\hspace{0.5pt} r' })  _{ y } .
\end{array}
\]
By inversion of \rn{TA-PlusPlus} in
premise~(\ref{conown-preservation8-combined-equiv}), there is a type
$\tau_{{\mathrm{3}}}$ such that
\begin{equation}
\begin{split}
 &\Gamma_1\vdash \mathsf{shift}_z(\tau_x)+\tau_y \approx \tau_{{\mathrm{3}}},\\
 &\Gamma_1\vdash \mathsf{shift}_z(\tau_x')+\tau_y' \approx \tau_{{\mathrm{3}}}.
\end{split}
\label{eq:alias-add-common-total}
\end{equation}
Applying \cref{lem:own-preservation} to both judgments in
\cref{eq:alias-add-common-total}, using
\cref{eq:alias-add-env-valid}, and then applying \cref{lem:own_shift} using
premise~(\ref{conown-preservation8-offset-value}), gives
\begin{equation}
\begin{split}
 &\OWN(H,R,pv  \boxplus  \ottnt{R}  \ottsym{(}  z  \ottsym{)},\tau_x')+\OWN(H,R,pv,\tau_y')\\
 &\qquad=\OWN(H,R,pv,\tau_{{\mathrm{3}}})\\
 &\qquad=\OWN(H,R,pv  \boxplus  \ottnt{R}  \ottsym{(}  z  \ottsym{)},\tau_x)+\OWN(H,R,pv,\tau_y).
\end{split}
\label{eq:alias-add-pair-own}
\end{equation}
Premises~(\ref{conown-preservation8-x-value}) and
(\ref{conown-preservation8-y-value}) identify these four terms with the
ownership contributions of $x$ and $y$ before and after the update.
Premise~(\ref{conown-preservation8-distinct}) ensures that the two updates
refer to distinct bindings; every other binding has the same type and value.
Consequently, \cref{eq:alias-add-pair-own} gives
$ \mathbf{Own}( \ottnt{H} ,  \ottnt{R} ,  \Gamma' ) = \mathbf{Own}( \ottnt{H} ,  \ottnt{R} ,  \Gamma_{{\mathrm{1}}} ) $.  The derived fact
$ \mathbf{ConOwn}( \ottnt{H} ,  \ottnt{R} ,  \Gamma_{{\mathrm{1}}} ) $ proves $ \mathbf{ConOwn}( \ottnt{H} ,  \ottnt{R} ,  \Gamma' ) $, as required.

\proofitem{Item~\ref{conown-preservation9}: Null binding
(\rn{Rs-LetNull}).}

By \cref{lem:conown-preservation-weakening-tyenv},
premises~(\ref{conown-preservation9-conown}) and
(\ref{conown-preservation9-env-subtype}), and the standing validity
assumption give $ \mathbf{ConOwn}( \ottnt{H} ,  \ottnt{R} ,  \Gamma_{{\mathrm{1}}} ) $.
Put $\ottnt{R'} = \ottnt{R}  \ottsym{\{}  x'  \mapsto  \ottkw{null}  \ottsym{\}}$ and
$\Gamma' = \Gamma_{{\mathrm{1}}}  \ottsym{,}   x' \COL  \Pi z .( \tau  \TREF^{\hspace{0.5pt} r })  $.  For every $ y  \in   \DOM( \Gamma_{{\mathrm{1}}} )  $,
premise~(\ref{conown-preservation9-fresh}) gives $y \, \neq \, x'$ and
$\ottnt{R'}  \ottsym{(}  y  \ottsym{)} \,  =  \, \ottnt{R}  \ottsym{(}  y  \ottsym{)}$.  Moreover, well-formedness of $\Gamma_{{\mathrm{1}}}$ and
$x'\notin dom(\Gamma_{{\mathrm{1}}})$ give $x'\notin\fv(\Gamma_1(y))$.  Hence
\cref{lem:own-preservation2} gives
\[
  \mathbf{own} ( \ottnt{H} ,  \ottnt{R'} ,  \ottnt{R'}  \ottsym{(}  y  \ottsym{)} ,   \Gamma_{{\mathrm{1}}}  (  y  )  ) = \mathbf{own} ( \ottnt{H} ,  \ottnt{R} ,  \ottnt{R}  \ottsym{(}  y  \ottsym{)} ,   \Gamma_{{\mathrm{1}}}  (  y  )  ) .
\]
For the new binding, $\ottnt{R'}  \ottsym{(}  x'  \ottsym{)} \,  =  \,  \NULL $, and therefore the definition of
$\mathbf{own}$ gives
\[
  \mathbf{own} ( \ottnt{H} ,  \ottnt{R'} ,  \ottnt{R'}  \ottsym{(}  x'  \ottsym{)} ,   \Gamma'  (  x'  )  ) 
 = \mathbf{own} ( \ottnt{H} ,  \ottnt{R'} ,  \ottkw{null} ,   \Pi z .( \tau  \TREF^{\hspace{0.5pt} r })  ) 
 = \emptyset .
\]
Consequently,
$ \mathbf{Own}( \ottnt{H} ,  \ottnt{R'} ,  \Gamma' ) = \mathbf{Own}( \ottnt{H} ,  \ottnt{R} ,  \Gamma_{{\mathrm{1}}} ) $, so the derived fact
$ \mathbf{ConOwn}( \ottnt{H} ,  \ottnt{R} ,  \Gamma_{{\mathrm{1}}} ) $ proves $ \mathbf{ConOwn}( \ottnt{H} ,  \ottnt{R'} ,  \Gamma' ) $, as required.
\end{proof}

\begin{lemma}[Preservation of $\SAT$]
\label{lem:sat-preservation}
The following items collect the operation-specific facts showing that the
heap, register, and type-environment transformations performed by the
reduction rules preserve $\SAT$.  Each item is used in the corresponding
case of the proof of Preservation.  Throughout this lemma, we assume that
$\models  \ottsym{[}  \ottnt{R}  \ottsym{]} \,  \fml{ \Gamma } $ holds for the register $\ottnt{R}$ and the type environment
$\Gamma$.
\begingroup
\makeatletter
\renewcommand{\p@enumii}{}
\makeatother
\begin{enumerate}
  \item\label{sat-preservation1}
    \emph{Variable binding (\rn{Rs-LetVar}).}
    If the following conditions hold:
    \begin{enumerate}
      \item $\SAT(H,R,\Gamma)$;\label{sat-preservation1-sat}
      \item $\Gamma \subt \Gamma_1, y \COL \tau_y, \Gamma_2$;\label{sat-preservation1-env-subtype}
      \item $\Gamma_{{\mathrm{1}}}  \ottsym{,}   y \COL  \tau _{ y }    \ottsym{,}  \Gamma_{{\mathrm{2}}}  \vdash    \tau _{ y }    \leq   \tau $;\label{sat-preservation1-source-subtype}
      \item $\Gamma_{{\mathrm{1}}}  \ottsym{,}   y \COL  \tau _{ y }    \ottsym{,}  \Gamma_{{\mathrm{2}}}  \vdash      \tau' _{ y }   +  \tau'  _{ x }    \approx   \tau $; and\label{sat-preservation1-split}
      \item $x' \notin dom(\Gamma_1, y \COL \tau_y, \Gamma_2)\cup dom(R)$,\label{sat-preservation1-fresh}
    \end{enumerate}
    then $ \mathbf{SAT}( \ottnt{H} ,  \ottnt{R}  \ottsym{\{}  x'  \mapsto  \ottnt{R}  \ottsym{(}  y  \ottsym{)}  \ottsym{\}} ,  \ottsym{(}  \Gamma_{{\mathrm{1}}}  \ottsym{,}   y \COL  \tau' _{ y }    \ottsym{,}  \Gamma_{{\mathrm{2}}}  \ottsym{,}   x' \COL  \tau' _{ x }    \ottsym{)} ) $.
    \rocqref[Supporting Rocq result]{Metatheory/Conservation.v}{consistent_after_variable_bind}
  \item\label{sat-preservation2}
    \emph{Dereference (\rn{Rs-Deref}).}
    If the following conditions hold:
    \begin{enumerate}
      \item $\SAT(H,R,\Gamma)$;\label{sat-preservation2-sat}
      \item $\Gamma  \leq  \Gamma_{{\mathrm{1}}}  \ottsym{,}   y \COL  \Pi z .(  \tau _{ y }   \TREF^{\hspace{0.5pt} r })    \ottsym{,}  \Gamma_{{\mathrm{2}}}$;\label{sat-preservation2-env-subtype}
      \item $\ottsym{(}  \Gamma_{{\mathrm{1}}}  \ottsym{,}   y \COL  \Pi z .(  \tau _{ y }   \TREF^{\hspace{0.5pt} r })    \ottsym{,}  \Gamma_{{\mathrm{2}}}  \ottsym{,}   z \COL  \{  \nu  :   \TINT    \mid   \nu \,  =  \,  0   \}    \ottsym{)}  \vdash     \tau'  +  \tau  _{ x }    \approx    \tau _{ y }  $;\label{sat-preservation2-split}
      \item $\ottsym{(}  \Gamma_{{\mathrm{1}}}  \ottsym{,}   y \COL  \Pi z .(  \tau _{ y }   \TREF^{\hspace{0.5pt} r })    \ottsym{,}  \Gamma_{{\mathrm{2}}}  \ottsym{,}   x' \COL  \tau _{ x }    \ottsym{,}   z \COL  \{  \nu  :   \TINT    \mid   \nu \,  =  \,  0   \}    \ottsym{)}  \vdash    \tau' _{ y }    \approx    \ottsym{(}  \tau'  \ottsym{)}  ^ {= x' }  $;\label{sat-preservation2-zero-equiv}
      \item $\ottsym{(}  \Gamma_{{\mathrm{1}}}  \ottsym{,}   y \COL  \Pi z .(  \tau _{ y }   \TREF^{\hspace{0.5pt} r })    \ottsym{,}  \Gamma_{{\mathrm{2}}}  \ottsym{,}   x' \COL  \tau _{ x }    \ottsym{,}   z \COL  \{  \nu  :   \TINT    \mid   \nu \, \neq \,  0   \}    \ottsym{)}  \vdash    \tau' _{ y }    \approx    \tau _{ y }  $;\label{sat-preservation2-nonzero-equiv}
      \item $\Gamma_{{\mathrm{1}}}  \ottsym{,}   y \COL  \Pi z .(  \tau _{ y }   \TREF^{\hspace{0.5pt} r })    \ottsym{,}  \Gamma_{{\mathrm{2}}}  \ottsym{,}   z \COL  \{  \nu  :   \TINT    \mid   \nu \,  =  \,  0   \}    \models  r \,  >  \,  \mathbf{0} $;\label{sat-preservation2-positive}
      \item $R(y) = (a,i) \in dom(H)$; and\label{sat-preservation2-location}
      \item $x' \notin dom(\Gamma_{{\mathrm{1}}}  \ottsym{,}   y \COL  \Pi z .(  \tau _{ y }   \TREF^{\hspace{0.5pt} r })    \ottsym{,}  \Gamma_{{\mathrm{2}}}) \cup dom(R)$,\label{sat-preservation2-fresh}
    \end{enumerate}
    then $ \mathbf{SAT}( \ottnt{H} ,  \ottnt{R}  \ottsym{\{}  x'  \mapsto  \ottnt{H}  \ottsym{(}  \ottnt{R}  \ottsym{(}  y  \ottsym{)}  \ottsym{)}  \ottsym{\}} ,  \Gamma_{{\mathrm{1}}}  \ottsym{,}   y \COL  \Pi z .(  \tau' _{ y }   \TREF^{\hspace{0.5pt} r })    \ottsym{,}  \Gamma_{{\mathrm{2}}}  \ottsym{,}   x' \COL  \tau _{ x }   ) $.
    \rocqref[Supporting Rocq result]{Metatheory/Conservation.v}{consistent_after_deref_fresh}
  \item\label{sat-preservation3}
    \emph{Assignment (\rn{Rs-Assign}).}
    If the following conditions hold:
    \begin{enumerate}
      \item $\SAT(H,R,\Gamma)$;\label{sat-preservation3-sat}
      \item $\CONOWN(H,R,\Gamma)$;\label{sat-preservation3-conown}
      \item $\Gamma \subt \Gamma_1$;\label{sat-preservation3-env-subtype}
      \item $ \Gamma_{{\mathrm{1}}}  (  x  )   =   \Pi z .(  \tau _{\ast  x }   \TREF^{\hspace{0.5pt} r }) $;\label{sat-preservation3-x-type}
      \item $ \Gamma_{{\mathrm{1}}}  (  y  )   =   \tau _{ y } $;\label{sat-preservation3-y-type}
      \item $\Gamma_{{\mathrm{1}}}  \ottsym{,}   z \COL  \{  \nu  :   \TINT    \mid   \nu \,  =  \,  0   \}    \vdash    \tau' _{\ast  x }    \approx    \ottsym{(}  \tau'  \ottsym{)}  ^ {= y }  $;\label{sat-preservation3-zero-equiv}
      \item $\Gamma_{{\mathrm{1}}}  \ottsym{,}   z \COL  \{  \nu  :   \TINT    \mid   \nu \, \neq \,  0   \}    \vdash    \tau' _{\ast  x }    \approx    \tau _{\ast  x }  $;\label{sat-preservation3-nonzero-equiv}
      \item $\Gamma_{{\mathrm{1}}}  \ottsym{,}   z \COL  \{  \nu  :   \TINT    \mid   \nu \,  =  \,  0   \}    \models  r \,  =  \,  \mathbf{1} $;\label{sat-preservation3-full-ownership}
      \item $\Gamma_{{\mathrm{1}}}  \vdash     \tau' _{ y }   +  \tau'    \approx    \tau _{ y }  $; and\label{sat-preservation3-split}
      \item $R(x) = (a, i) \in dom(H)$,\label{sat-preservation3-location}
    \end{enumerate}
    then $ \mathbf{SAT}( \ottnt{H}  \ottsym{\{}  \ottsym{(}  a  \ottsym{,}   i   \ottsym{)}  \hookleftarrow  \ottnt{R}  \ottsym{(}  y  \ottsym{)}  \ottsym{\}} ,  \ottnt{R} ,    \Gamma_{{\mathrm{1}}}  \left[  x \hookleftarrow  \Pi z .(  \tau' _{\ast  x }   \TREF^{\hspace{0.5pt} r })   \right]   \left[  y \hookleftarrow  \tau' _{ y }   \right]  ) $.
    \rocqref{Metatheory/Conservation.v}{env_satisfies_after_assign}
  \item\label{sat-preservation4}
    \emph{Pointer addition (\rn{Rs-AddPtr}).}
    If the following conditions hold:
    \begin{enumerate}
      \item $\SAT(H,R,\Gamma)$;\label{sat-preservation4-sat}
      \item $\Gamma \subt \Gamma_1$;\label{sat-preservation4-env-subtype}
      \item $ \Gamma_{{\mathrm{1}}}  (  y  )   =   \Pi w .( \tau_{{\mathrm{3}}}  \TREF^{\hspace{0.5pt}  { r }_{ y }  }) $;\label{sat-preservation4-y-type}
      \item $ \Gamma_{{\mathrm{1}}}  (  z  )   =   \{  \nu  :   \TINT    \mid   \varphi  \} $;\label{sat-preservation4-offset-type}
      \item $R(y) = pv$;\label{sat-preservation4-y-value}
      \item $\Gamma_{{\mathrm{1}}}  \vdash     \Pi w .( \tau_{{\mathrm{1}}}  \TREF^{\hspace{0.5pt}  { r }_{ y_{{\mathrm{1}}} }  })   +   \Pi w .   [ (  w  -  z  ) /  w  ]   ( \tau_{{\mathrm{2}}}  \TREF^{\hspace{0.5pt}  { r }_{ x }  })     \approx    \Pi w .( \tau_{{\mathrm{3}}}  \TREF^{\hspace{0.5pt}  { r }_{ y }  })  $; and\label{sat-preservation4-split}
      \item $x' \notin dom ( \Gamma_{{\mathrm{1}}} ) \cup dom(R)$,\label{sat-preservation4-fresh}
    \end{enumerate}
    then $ \mathbf{SAT}( \ottnt{H} ,  \ottnt{R}  \ottsym{\{}  x'  \mapsto  pv  \boxplus  \ottnt{R}  \ottsym{(}  z  \ottsym{)}  \ottsym{\}} ,   \Gamma_{{\mathrm{1}}}  \left[  y \hookleftarrow  \Pi w .( \tau_{{\mathrm{1}}}  \TREF^{\hspace{0.5pt}  { r }_{ y_{{\mathrm{1}}} }  })   \right]   \ottsym{,}   x' \COL  \Pi w .( \tau_{{\mathrm{2}}}  \TREF^{\hspace{0.5pt}  { r }_{ x }  })   ) $.
    \rocqref[Supporting Rocq result]{Metatheory/Conservation.v}{consistent_after_let_add_ptr}
  \item\label{sat-preservation5}
    \emph{Integer-array allocation (\rn{Rs-MkArrayIntref}).}
    If the following conditions hold:
    \begin{enumerate}
      \item $\SAT(H,R,\Gamma)$;\label{sat-preservation5-sat}
      \item $\Gamma \subt \Gamma_1$;\label{sat-preservation5-env-subtype}
      \item $ \Gamma_{{\mathrm{1}}}  (  y  )   =   \{  \nu  :   \TINT    \mid   \varphi  \} $;\label{sat-preservation5-length-type}
      \item $\Gamma_{{\mathrm{1}}}  \ottsym{,}   z \COL  \TINT    \models  r \,  =  \, \ottsym{(}    \ottsym{(}    0  \, \le \, z  \wedge  z \, \le \, y  \ottsym{-}   1    \ottsym{)}   \produces    1    ,   \mathbf{0}    \ottsym{)}$;\label{sat-preservation5-ownership-shape}
      \item $x' \notin dom ( \Gamma_{{\mathrm{1}}} )\cup dom(R)$; and\label{sat-preservation5-fresh}
      \item $(a,0)\notin dom(H)$,\label{sat-preservation5-fresh-base}
    \end{enumerate}
    then $ \mathbf{SAT}(  \ottnt{H}  \{ (  a  ,   0   ) \mapsto   0  , \ldots, (  a  ,  \ottnt{R}  \ottsym{(}  y  \ottsym{)} \,  -  \,  1   ) \mapsto   0   \}  ,  \ottnt{R}  \ottsym{\{}  x'  \mapsto  \ottsym{(}  a  \ottsym{,}   0   \ottsym{)}  \ottsym{\}} ,  \ottsym{(}  \Gamma_{{\mathrm{1}}}  \ottsym{,}   x' \COL  \Pi z .(  \{  \nu  :   \TINT    \mid     0  \, \le \, z  \wedge  z \, \le \, y  \ottsym{-}   1    \implies  \nu \,  =  \,  0   \}   \TREF^{\hspace{0.5pt} r })    \ottsym{)} ) $.
    \rocqref[Supporting Rocq result]{Metatheory/Conservation.v}{consistent_after_alloc}
  \item\label{sat-preservation6}
    \emph{Nested-array allocation (\rn{Rs-MkArrayNestedref}).}
    If the following conditions hold:
    \begin{enumerate}
      \item $\SAT(H,R,\Gamma)$;\label{sat-preservation6-sat}
      \item $\Gamma \subt \Gamma_1$;\label{sat-preservation6-env-subtype}
      \item $\Gamma_1(y) = \set{\nu \COL \TINT \mid \varphi}$;\label{sat-preservation6-length-type}
      \item $\models   \Empty{ \tau' } $;\label{sat-preservation6-empty}
      \item $|\tau'|= \tau^{-}  \TREF $;\label{sat-preservation6-element-shape}
      \item $\Gamma_{{\mathrm{1}}}  \ottsym{,}   z \COL  \TINT    \models  r \,  =  \, \ottsym{(}    \ottsym{(}    0  \, \le \, z  \wedge  z \, \le \, y  \ottsym{-}   1    \ottsym{)}   \produces    1    ,   \mathbf{0}    \ottsym{)}$;\label{sat-preservation6-ownership-shape}
      \item $x' \notin dom ( \Gamma_{{\mathrm{1}}} )\cup dom(R)$; and\label{sat-preservation6-fresh}
      \item $(a,0)\notin dom(H)$,\label{sat-preservation6-fresh-base}
    \end{enumerate}
    then
    \[
    \begin{aligned}
      &\SAT\bigl( \ottnt{H}  \{ (  a  ,   0   ) \mapsto  \ottkw{null} , \ldots, (  a  ,  \ottnt{R}  \ottsym{(}  y  \ottsym{)} \,  -  \,  1   ) \mapsto  \ottkw{null}  \} ,\\
      &\qquad \ottnt{R}  \ottsym{\{}  x'  \mapsto  \ottsym{(}  a  \ottsym{,}   0   \ottsym{)}  \ottsym{\}}, (\Gamma_{{\mathrm{1}}}  \ottsym{,}   x' \COL  \Pi z .( \tau'  \TREF^{\hspace{0.5pt} r })  )\bigr).
    \end{aligned}
    \]
    \rocqref[Supporting Rocq result]{Metatheory/Conservation.v}{consistent_after_alloc}
  \item\label{sat-preservation7}
    \emph{Aliasing by dereference (\rn{Rs-AliasDeref}).}
    If the following conditions hold:
    \begin{enumerate}
      \item $\SAT(H,R,\Gamma)$;\label{sat-preservation7-sat}
      \item $\Gamma \subt \Gamma_1$;\label{sat-preservation7-env-subtype}
      \item $ \Gamma_{{\mathrm{1}}}  (  x  )   =   \Pi z' .(  \tau _{\ast  x }   \TREF^{\hspace{0.5pt}  { r }_{ x }  }) $;\label{sat-preservation7-x-type}
      \item $ \Gamma_{{\mathrm{1}}}  (  y  )   =   \Pi w .(  \Pi z .(  \tau _{\ast \ast  y }   \TREF^{\hspace{0.5pt}  { r }_{ \ast  y }  })   \TREF^{\hspace{0.5pt} r }) $;\label{sat-preservation7-y-type}
      \item $\Gamma_{{\mathrm{1}}}  \ottsym{,}   w \COL  \{  \nu  :   \TINT    \mid   \nu \,  =  \,  0   \}    \vdash   \ottsym{(}    \Pi z' .(  \tau _{\ast  x }   \TREF^{\hspace{0.5pt}  { r }_{ x }  })   +   \Pi z .(  \tau _{\ast \ast  y }   \TREF^{\hspace{0.5pt}  { r }_{ \ast  y }  })    \ottsym{)}   \approx   \ottsym{(}    \Pi z' .(  \tau' _{\ast  x }   \TREF^{\hspace{0.5pt}  { r' }_{ x }  })   +   \Pi z .(  \tau' _{\ast \ast  y }   \TREF^{\hspace{0.5pt}  { r' }_{ \ast  y }  })    \ottsym{)} $;\label{sat-preservation7-zero-equiv}
      \item $\Gamma_{{\mathrm{1}}}  \ottsym{,}   w \COL  \{  \nu  :   \TINT    \mid   \nu \, \neq \,  0   \}    \vdash    \Pi z .(  \tau _{\ast \ast  y }   \TREF^{\hspace{0.5pt}  { r }_{ \ast  y }  })    \approx    \Pi z .(  \tau' _{\ast \ast  y }   \TREF^{\hspace{0.5pt}  { r' }_{ \ast  y }  })  $;\label{sat-preservation7-nonzero-equiv}
      \item $\Gamma_{{\mathrm{1}}}  \ottsym{,}   w \COL  \{  \nu  :   \TINT    \mid   \nu \,  =  \,  0   \}    \models  r \,  >  \,  \mathbf{0} $; and\label{sat-preservation7-positive}
      \item $H(R(y)) = R(x)$,\label{sat-preservation7-alias}
    \end{enumerate}
    then $ \mathbf{SAT}( \ottnt{H} ,  \ottnt{R} ,    \Gamma_{{\mathrm{1}}}  \left[  x \hookleftarrow  \Pi z' .(  \tau' _{\ast  x }   \TREF^{\hspace{0.5pt}  { r' }_{ x }  })   \right]   \left[  y \hookleftarrow  \Pi w .(  \Pi z .(  \tau' _{\ast \ast  y }   \TREF^{\hspace{0.5pt}  { r' }_{ \ast  y }  })   \TREF^{\hspace{0.5pt} r })   \right]  ) $.
    \rocqref[Supporting Rocq result]{Metatheory/Conservation.v}{consistent_after_alias_deref}
  \item\label{sat-preservation8}
    \emph{Aliasing by pointer addition (\rn{Rs-AliasAddPtr}).}
    If the following conditions hold:
    \begin{enumerate}
      \item $\SAT(H,R,\Gamma)$;\label{sat-preservation8-sat}
      \item $R(x) = pv  \boxplus  \ottnt{R}  \ottsym{(}  z  \ottsym{)}$;\label{sat-preservation8-x-value}
      \item $R(y) = pv$;\label{sat-preservation8-y-value}
      \item $R(z)\in \mathbb{Z} $;\label{sat-preservation8-offset-value}
      \item $x \, \neq \, y$;\label{sat-preservation8-distinct}
      \item $\Gamma \subt \Gamma_1$;\label{sat-preservation8-env-subtype}
      \item $ \Gamma_{{\mathrm{1}}}  (  x  )   =   \Pi w' .(  \tau _{\ast  x }   \TREF^{\hspace{0.5pt}  { r }_{ x }  }) $;\label{sat-preservation8-x-type}
      \item $ \Gamma_{{\mathrm{1}}}  (  y  )   =   \Pi w .(  \tau _{\ast  y }   \TREF^{\hspace{0.5pt}  { r }_{ y }  }) $; and\label{sat-preservation8-y-type}
      \item\label{sat-preservation8-combined-equiv}
        \(\begin{aligned}[t]
          &\Gamma_{{\mathrm{1}}}  \vdash  \ottsym{(}    \Pi w' .   [ (  w'  -  z  ) /  w'  ]   (  \tau _{\ast  x }   \TREF^{\hspace{0.5pt}  { r }_{ x }  })   +   \Pi w .(  \tau _{\ast  y }   \TREF^{\hspace{0.5pt}  { r }_{ y }  })    \ottsym{)}\\
          &\qquad  \approx   \ottsym{(}   \Pi w' .   [ (  w'  -  z  ) /  w'  ]   (  \tau' _{\ast  x }   \TREF^{\hspace{0.5pt}  { r' }_{ x }  })   \ottsym{)}  +   \Pi w .(  \tau' _{\ast  y }   \TREF^{\hspace{0.5pt}  { r' }_{ y }  })  ,
        \end{aligned}\)
    \end{enumerate}
    then $ \mathbf{SAT}( \ottnt{H} ,  \ottnt{R} ,    \Gamma_{{\mathrm{1}}}  \left[  x \hookleftarrow  \Pi w' .(  \tau' _{\ast  x }   \TREF^{\hspace{0.5pt}  { r' }_{ x }  })   \right]   \left[  y \hookleftarrow  \Pi w .(  \tau' _{\ast  y }   \TREF^{\hspace{0.5pt}  { r' }_{ y }  })   \right]  ) $.
    \rocqref[Supporting Rocq result]{Metatheory/Conservation.v}{consistent_after_alias_add}
  \item\label{sat-preservation9}
    \emph{Null binding (\rn{Rs-LetNull}).}
    If the following conditions hold:
    \begin{enumerate}
      \item $\SAT(H,R,\Gamma)$;\label{sat-preservation9-sat}
      \item $\Gamma \subt \Gamma_1$;\label{sat-preservation9-env-subtype}
      \item $\models   \Empty{  \Pi z .( \tau'  \TREF^{\hspace{0.5pt} r })  } $; and\label{sat-preservation9-empty}
      \item $x' \notin dom ( \Gamma_{{\mathrm{1}}} )\cup dom(R)$,\label{sat-preservation9-fresh}
    \end{enumerate}
    then
    \[
       \mathbf{SAT}( \ottnt{H} ,  \ottnt{R}  \ottsym{\{}  x'  \mapsto  \ottkw{null}  \ottsym{\}} ,  \ottsym{(}  \Gamma_{{\mathrm{1}}}  \ottsym{,}   x' \COL  \Pi z .( \tau'  \TREF^{\hspace{0.5pt} r })    \ottsym{)} ) .
    \]
    \rocqref[Supporting Rocq result]{Metatheory/Conservation.v}{consistent_after_let_null}
\end{enumerate}
\endgroup
\end{lemma}
\begin{proof}
\leavevmode
\proofitem{Item~\ref{sat-preservation1}: Variable binding
(\rn{Rs-LetVar}).}

Put $\Gamma_{{\mathrm{0}}}=\Gamma_{{\mathrm{1}}}  \ottsym{,}   y \COL  \tau _{ y }    \ottsym{,}  \Gamma_{{\mathrm{2}}}$ and $\ottnt{R'} = \ottnt{R}  \ottsym{\{}  x'  \mapsto  \ottnt{R}  \ottsym{(}  y  \ottsym{)}  \ottsym{\}}$.
By \cref{lem:sat-preservation-greater-tyenv},
premises~(\ref{sat-preservation1-sat}) and
(\ref{sat-preservation1-env-subtype}), and the standing validity assumption,
we have
\begin{equation}
   \mathbf{SAT}( \ottnt{H} ,  \ottnt{R} ,  \Gamma_{{\mathrm{0}}} ) .
  \label{eq:sat-letvar-subtyped-env}
\end{equation}

Premise~(\ref{sat-preservation1-fresh}) and well-formedness of $\Gamma_{{\mathrm{0}}}$
ensure that $x'$ occurs free in none of its types.  Moreover, $R'$ agrees
with $R$ on $dom(\Gamma_{{\mathrm{0}}})$.  Applying
\cref{lem:sat-preservation-extend-register} to every binding in
\cref{eq:sat-letvar-subtyped-env} therefore gives
\begin{equation}
   \mathbf{SAT}( \ottnt{H} ,  \ottnt{R'} ,  \Gamma_{{\mathrm{0}}} ) .
  \label{eq:sat-letvar-extended-register}
\end{equation}

We also need the refinement assumptions of $\Gamma_{{\mathrm{0}}}$ under $R'$.  By
\rn{S-TyEnv}, \rn{S-Int}, and the definition of $ \fml{ \Gamma } $,
premise~(\ref{sat-preservation1-env-subtype}) implies
$ \fml{ \Gamma } \implies \fml{ \Gamma_{{\mathrm{0}}} } $.  The standing validity assumption thus
gives $\models  \ottsym{[}  \ottnt{R}  \ottsym{]} \,  \fml{ \Gamma_{{\mathrm{0}}} } $.  Since $R'$ and $R$ agree on $dom(\Gamma_{{\mathrm{0}}})$,
\begin{equation}
  \models  \ottsym{[}  \ottnt{R'}  \ottsym{]} \,  \fml{ \Gamma_{{\mathrm{0}}} } .
  \label{eq:sat-letvar-valid-env}
\end{equation}

From
\cref{eq:sat-letvar-extended-register} we obtain
$ \mathbf{SATv} ( \ottnt{H} ,  \ottnt{R'} ,  \ottnt{R'}  \ottsym{(}  y  \ottsym{)} ,   \tau _{ y }  ) $.  Premise
(\ref{sat-preservation1-source-subtype}),
\cref{eq:sat-letvar-valid-env}, and
\cref{lem:sat-preservation-weakening} give
\[
   \mathbf{SATv} ( \ottnt{H} ,  \ottnt{R'} ,  \ottnt{R'}  \ottsym{(}  y  \ottsym{)} ,  \tau ) .
\]
Premise~(\ref{sat-preservation1-split}),
\cref{eq:sat-letvar-valid-env}, and
\cref{lem:sat-preservation-add-ty} then give
\begin{equation}
   \mathbf{SATv} ( \ottnt{H} ,  \ottnt{R'} ,  \ottnt{R'}  \ottsym{(}  y  \ottsym{)} ,   \tau' _{ y }  ) 
  \quad\text{and}\quad
   \mathbf{SATv} ( \ottnt{H} ,  \ottnt{R'} ,  \ottnt{R'}  \ottsym{(}  y  \ottsym{)} ,   \tau' _{ x }  ) .
  \label{eq:sat-letvar-split-values}
\end{equation}

Finally, premise~(\ref{sat-preservation1-fresh}) gives $x'\ne y$ and hence
$R'(x')=R(y)=R'(y)$.  Thus the two facts in
\cref{eq:sat-letvar-split-values} establish satisfaction of the updated
binding for $y$ and the new binding for $x'$, respectively.  Every binding
in $\Gamma_{{\mathrm{1}}}  \ottsym{,}  \Gamma_{{\mathrm{2}}}$ is satisfied by
\cref{eq:sat-letvar-extended-register}.  Therefore
\[
   \mathbf{SAT}( \ottnt{H} ,  \ottnt{R'} ,  \Gamma_{{\mathrm{1}}}  \ottsym{,}   y \COL  \tau' _{ y }    \ottsym{,}  \Gamma_{{\mathrm{2}}}  \ottsym{,}   x' \COL  \tau' _{ x }   ) ,
\]
as required.

\proofitem{Item~\ref{sat-preservation2}: Dereference
(\rn{Rs-Deref}).}

Put
\[
 \Gamma_{{\mathrm{0}}}=\Gamma_{{\mathrm{1}}}  \ottsym{,}   y \COL  \Pi z .(  \tau _{ y }   \TREF^{\hspace{0.5pt} r })    \ottsym{,}  \Gamma_{{\mathrm{2}}},\qquad
 \Gamma' = \Gamma_{{\mathrm{1}}}  \ottsym{,}   y \COL  \Pi z .(  \tau' _{ y }   \TREF^{\hspace{0.5pt} r })    \ottsym{,}  \Gamma_{{\mathrm{2}}}  \ottsym{,}   x' \COL  \tau _{ x }  ,\qquad
 \ottnt{R'} = \ottnt{R}  \ottsym{\{}  x'  \mapsto  \ottnt{H}  \ottsym{(}  \ottnt{R}  \ottsym{(}  y  \ottsym{)}  \ottsym{)}  \ottsym{\}}.
\]
By \cref{lem:sat-preservation-greater-tyenv},
premises~(\ref{sat-preservation2-sat}) and
(\ref{sat-preservation2-env-subtype}), and the standing validity assumption,
we obtain
\begin{equation}
   \mathbf{SAT}( \ottnt{H} ,  \ottnt{R} ,  \Gamma_{{\mathrm{0}}} ) .
  \label{eq:sat-deref-subtyped-env}
\end{equation}
As in Item~\ref{sat-preservation1}, environment subtyping also gives
$ \fml{ \Gamma } \implies \fml{ \Gamma_{{\mathrm{0}}} } $, so
$\models  \ottsym{[}  \ottnt{R}  \ottsym{]} \,  \fml{ \Gamma_{{\mathrm{0}}} } $.  Premise~(\ref{sat-preservation2-fresh}) and
well-formedness of $\Gamma_{{\mathrm{0}}}$ ensure that $x'$ occurs free in none of its
types.  Since $R'$ agrees with $R$ on $dom(\Gamma_{{\mathrm{0}}})$,
\cref{lem:sat-preservation-extend-register} applied to
\cref{eq:sat-deref-subtyped-env} gives the first of the following facts; the
agreement of the two registers gives the second:
\begin{equation}
   \mathbf{SAT}( \ottnt{H} ,  \ottnt{R'} ,  \Gamma_{{\mathrm{0}}} ) 
  \qquad\text{and}\qquad
  \models  \ottsym{[}  \ottnt{R'}  \ottsym{]} \,  \fml{ \Gamma_{{\mathrm{0}}} } .
  \label{eq:sat-deref-framed-env}
\end{equation}

Premise~(\ref{sat-preservation2-location}) and freshness give
$R'(y)=R(y)=(a,i)$ and $R'(x')=H((a,i))$.  Semantic instantiation of
premise~(\ref{sat-preservation2-positive}) at $z=0$, using
\cref{eq:sat-deref-framed-env}, yields
\begin{equation}
   \llbracket    [  0  /  z  ]    r   \rrbracket_{ \ottnt{R'} }  \,  >  \,  0 .
  \label{eq:sat-deref-positive-zero}
\end{equation}
The satisfaction of $y$ in \cref{eq:sat-deref-framed-env} and
\cref{eq:sat-deref-positive-zero} therefore give
\begin{equation}
   \mathbf{SATv} ( \ottnt{H} ,  \ottnt{R'} ,  \ottnt{H}  \ottsym{(}  \ottsym{(}  a  \ottsym{,}   i   \ottsym{)}  \ottsym{)} ,    [  0  /  z  ]  \, \tau _{ y }  ) .
  \label{eq:sat-deref-old-zero}
\end{equation}

Instantiate the type-addition premise
(\ref{sat-preservation2-split}) with the valuation $\ottnt{R'}  \ottsym{\{}  z  \mapsto   0   \ottsym{\}}$.  Its
environment formula holds by \cref{eq:sat-deref-framed-env} and $0=0$.
The semantic substitution property used in \cref{lem:index-substitution},
\cref{eq:sat-deref-framed-env,eq:sat-deref-old-zero}, and
\cref{lem:sat-preservation-add-ty} give
\begin{equation}
   \mathbf{SATv} ( \ottnt{H} ,  \ottnt{R'} ,  \ottnt{H}  \ottsym{(}  \ottsym{(}  a  \ottsym{,}   i   \ottsym{)}  \ottsym{)} ,   [  0  /  z  ]  \, \tau' ) 
  \quad\text{and}\quad
   \mathbf{SATv} ( \ottnt{H} ,  \ottnt{R'} ,  \ottnt{H}  \ottsym{(}  \ottsym{(}  a  \ottsym{,}   i   \ottsym{)}  \ottsym{)} ,   \tau _{ x }  ) .
  \label{eq:sat-deref-split-zero}
\end{equation}
Here $z$ is not free in $ \tau _{ x } $ by well-formedness, so
$  [  0  /  z  ]  \, \tau _{ x }   =   \tau _{ x } $.  The second judgment in
\cref{eq:sat-deref-split-zero}, together with $R'(x')=H((a,i))$, establishes
satisfaction of the new binding $ x' \COL  \tau _{ x }  $.

We will use the refinement assumptions of the extended environment.  By the
definition of $ \fml{ \Gamma } $, $\models  \ottsym{[}  \ottnt{R'}  \ottsym{]} \,  \fml{ \Gamma_{{\mathrm{0}}} } $ from
\cref{eq:sat-deref-framed-env} and the second judgment in
\cref{eq:sat-deref-split-zero} imply
\begin{equation}
  \models  \ottsym{[}  \ottnt{R'}  \ottsym{]} \,  \fml{ \Gamma_{{\mathrm{0}}}  \ottsym{,}   x' \COL  \tau _{ x }   } .
  \label{eq:sat-deref-valid-extended-env}
\end{equation}
Indeed, if $ \tau _{ x } $ is an integer type, its value-satisfaction judgment is
exactly its refinement instantiated with $R'(x')$; if it is a reference type,
$ \fmlT{  \tau _{ x }  }{ x' } = \top $.

It remains to establish satisfaction of the updated type of $y$.  Consider an
arbitrary $  j   \in   \mathbb{Z}  $ such that $ \llbracket    [  j  /  z  ]    r   \rrbracket_{ \ottnt{R'} } >0$.  The satisfaction of
$y$ in \cref{eq:sat-deref-framed-env} gives
\begin{equation}
   \ottsym{(}  a  \ottsym{,}  i  \ottsym{+}  j  \ottsym{)}  \in   \DOM( \ottnt{H} )  
  \quad\text{and}\quad
   \mathbf{SATv} ( \ottnt{H} ,  \ottnt{R'} ,  \ottnt{H}  \ottsym{(}  \ottsym{(}  a  \ottsym{,}   i   \ottsym{+}   j   \ottsym{)}  \ottsym{)} ,    [  j  /  z  ]  \, \tau _{ y }  ) .
  \label{eq:sat-deref-old-element}
\end{equation}

If $j=0$, the first judgment of \cref{eq:sat-deref-split-zero} implies
\begin{equation}
   \mathbf{SATv} ( \ottnt{H} ,  \ottnt{R'} ,  \ottnt{H}  \ottsym{(}  \ottsym{(}  a  \ottsym{,}   i   \ottsym{)}  \ottsym{)} ,   [  0  /  z  ]  \, \ottsym{(}   \ottsym{(}  \tau'  \ottsym{)}  ^ {= x' }   \ottsym{)} ) .
  \label{eq:sat-deref-strengthened-zero}
\end{equation}
For a reference type, strengthening does not change the type; for an integer
type, the added equality holds because $R'(x')=H((a,i))$.  Instantiate the
zero-offset equivalence premise~(\ref{sat-preservation2-zero-equiv}) with
$\ottnt{R'}  \ottsym{\{}  z  \mapsto   0   \ottsym{\}}$.  Its environment formula is valid by
\cref{eq:sat-deref-valid-extended-env} and $0=0$, and hence type-equivalence
satisfaction applied to \cref{eq:sat-deref-strengthened-zero} gives
\[
   \mathbf{SATv} ( \ottnt{H} ,  \ottnt{R'} ,  \ottnt{H}  \ottsym{(}  \ottsym{(}  a  \ottsym{,}   i   \ottsym{)}  \ottsym{)} ,    [  0  /  z  ]  \, \tau' _{ y }  ) .
\]

If $j\ne0$, instantiate the nonzero-offset equivalence
premise~(\ref{sat-preservation2-nonzero-equiv}) with
$\ottnt{R'}  \ottsym{\{}  z  \mapsto   j   \ottsym{\}}$.  Again its
environment formula is valid by
\cref{eq:sat-deref-valid-extended-env}, now together with $j\ne0$.
Type-equivalence satisfaction and \cref{eq:sat-deref-old-element} yield
\[
   \mathbf{SATv} ( \ottnt{H} ,  \ottnt{R'} ,  \ottnt{H}  \ottsym{(}  \ottsym{(}  a  \ottsym{,}   i   \ottsym{+}   j   \ottsym{)}  \ottsym{)} ,    [  j  /  z  ]  \, \tau' _{ y }  ) .
\]
Thus, in either case, the cell at $(a,i+j)$ exists and satisfies
$  [  j  /  z  ]  \, \tau' _{ y } $.  Since $j$ was arbitrary, the definition of $\SATV$
gives
\begin{equation}
   \mathbf{SATv} ( \ottnt{H} ,  \ottnt{R'} ,  \ottsym{(}  a  \ottsym{,}   i   \ottsym{)} ,   \Pi z .(  \tau' _{ y }   \TREF^{\hspace{0.5pt} r })  ) .
  \label{eq:sat-deref-updated-pointer}
\end{equation}
The bindings in $\Gamma_{{\mathrm{1}}}  \ottsym{,}  \Gamma_{{\mathrm{2}}}$ remain satisfied by
\cref{eq:sat-deref-framed-env}; the new binding $ x' \COL  \tau _{ x }  $ is satisfied by
\cref{eq:sat-deref-split-zero}; and the updated binding for $y$ is satisfied
by \cref{eq:sat-deref-updated-pointer}.  Therefore $ \mathbf{SAT}( \ottnt{H} ,  \ottnt{R'} ,  \Gamma' ) $, as
required.
\proofitem{Item~\ref{sat-preservation3}: Assignment
(\rn{Rs-Assign}).}

Put
\[
 \ottnt{H'} = \ottnt{H}  \ottsym{\{}  \ottsym{(}  a  \ottsym{,}   i   \ottsym{)}  \hookleftarrow  \ottnt{R}  \ottsym{(}  y  \ottsym{)}  \ottsym{\}}
 \qquad\text{and}\qquad
 \Gamma' =   \Gamma_{{\mathrm{1}}}  \left[  x \hookleftarrow  \Pi z .(  \tau' _{\ast  x }   \TREF^{\hspace{0.5pt} r })   \right]   \left[  y \hookleftarrow  \tau' _{ y }   \right] .
\]
By \cref{lem:sat-preservation-greater-tyenv},
premises~(\ref{sat-preservation3-sat}) and
(\ref{sat-preservation3-env-subtype}), and the standing validity assumption,
we have
\begin{equation}
  \mathbf{SAT}( \ottnt{H} ,  \ottnt{R} ,  \Gamma_{{\mathrm{1}}} ) .
 \label{eq:sat-assign-subtyped-env}
\end{equation}
Similarly, \cref{lem:conown-preservation-weakening-tyenv},
premises~(\ref{sat-preservation3-conown}) and
(\ref{sat-preservation3-env-subtype}), and the standing validity assumption
give
\begin{equation}
  \mathbf{ConOwn}( \ottnt{H} ,  \ottnt{R} ,  \Gamma_{{\mathrm{1}}} ) .
 \label{eq:sat-assign-conown}
\end{equation}
As in the preceding items, environment subtyping also gives
$ \fml{ \Gamma } \implies \fml{ \Gamma_{{\mathrm{1}}} } $, and hence
\begin{equation}
 \models  \ottsym{[}  \ottnt{R}  \ottsym{]} \,  \fml{ \Gamma_{{\mathrm{1}}} } .
 \label{eq:sat-assign-valid-env}
\end{equation}

We first note that $x\ne y$.  Otherwise,
premises~(\ref{sat-preservation3-x-type}) and
(\ref{sat-preservation3-y-type}) give
$ \tau _{ y }   =   \Pi z .(  \tau _{\ast  x }   \TREF^{\hspace{0.5pt} r }) $.  The type-addition and two type-equivalence
premises force $ \tau _{ y } $ and $ \tau _{\ast  x } $ to have the same finite type shape,
which is impossible when the former is a reference to the latter.

Let $\ell_0=(a,i)$.  Semantic instantiation of
premise~(\ref{sat-preservation3-full-ownership}) using
\cref{eq:sat-assign-valid-env} gives
\begin{equation*}
  \llbracket    [  0  /  z  ]    r   \rrbracket_{ \ottnt{R} }  \,  =  \,  1 .
\end{equation*}
By premises~(\ref{sat-preservation3-x-type}) and
(\ref{sat-preservation3-location}), the direct contribution of $x$ to
$ \mathbf{Own}( \ottnt{H} ,  \ottnt{R} ,  \Gamma_{{\mathrm{1}}} ) (\ell_0)$ is therefore exactly $1$.  All direct and recursive
ownership contributions are nonnegative by \cref{lem:ownership-bounds}, while
\cref{eq:sat-assign-conown} bounds their sum by $1$.  Consequently,
\begin{equation}
  \mathbf{own} ( \ottnt{H} ,  \ottnt{R} ,  \ottnt{R}  \ottsym{(}  b  \ottsym{)} ,   \Gamma_{{\mathrm{1}}}  (  b  )  ) (\ell_0)=0
 \quad
 \bigl(b\in dom(\Gamma_{{\mathrm{1}}})\setminus\{x\}\bigr),
 \label{eq:sat-assign-other-zero}
\end{equation}
and, for every $j$ such that $(a,i+j)\in dom(H)$ and
$ \llbracket    [  j  /  z  ]    r   \rrbracket_{ \ottnt{R} } >0$,
\begin{equation}
  \mathbf{own} ( \ottnt{H} ,  \ottnt{R} ,  \ottnt{H}  \ottsym{(}  \ottsym{(}  a  \ottsym{,}   i   \ottsym{+}   j   \ottsym{)}  \ottsym{)} ,    [  j  /  z  ]  \, \tau _{\ast  x }  ) (\ell_0)=0.
 \label{eq:sat-assign-element-zero}
\end{equation}
These are precisely the binding and recursive summands for which the full
direct contribution at $\ell_0$ leaves no ownership available.

Updating the existing location $\ell_0$ gives $ \ottnt{H}   \approx _{(a,i)}  \ottnt{H'} $.  Hence,
for every $b\in dom(\Gamma_{{\mathrm{1}}})\setminus\{x\}$,
\cref{eq:sat-assign-subtyped-env,eq:sat-assign-other-zero} and
\cref{lem:sat-preservation-about-heap} give
\begin{equation}
  \mathbf{SATv} ( \ottnt{H'} ,  \ottnt{R} ,  \ottnt{R}  \ottsym{(}  b  \ottsym{)} ,   \Gamma_{{\mathrm{1}}}  (  b  )  ) .
 \label{eq:sat-assign-framed-binding}
\end{equation}
In particular, $x\ne y$ and premise~(\ref{sat-preservation3-y-type}) give
$ \mathbf{SATv} ( \ottnt{H'} ,  \ottnt{R} ,  \ottnt{R}  \ottsym{(}  y  \ottsym{)} ,   \tau _{ y }  ) $.  Applying
\cref{lem:sat-preservation-add-ty} with
premise~(\ref{sat-preservation3-split}) and
\cref{eq:sat-assign-valid-env} yields
\begin{equation}
  \mathbf{SATv} ( \ottnt{H'} ,  \ottnt{R} ,  \ottnt{R}  \ottsym{(}  y  \ottsym{)} ,   \tau' _{ y }  ) 
 \quad\text{and}\quad
  \mathbf{SATv} ( \ottnt{H'} ,  \ottnt{R} ,  \ottnt{R}  \ottsym{(}  y  \ottsym{)} ,  \tau' ) .
 \label{eq:sat-assign-split-source}
\end{equation}
The first judgment establishes satisfaction of the updated binding for $y$.

It remains to establish satisfaction of the updated reference type for $x$.
Fix an arbitrary integer $j$ such that $ \llbracket    [  j  /  z  ]    r   \rrbracket_{ \ottnt{R} } >0$.  We distinguish
whether $j$ is zero.

If $j=0$, then $H'((a,i))=R(y)$.  The second judgment in
\cref{eq:sat-assign-split-source} implies
\[
  \mathbf{SATv} ( \ottnt{H'} ,  \ottnt{R} ,  \ottnt{R}  \ottsym{(}  y  \ottsym{)} ,   \ottsym{(}  \tau'  \ottsym{)}  ^ {= y }  ) ,
\]
because strengthening is the identity on a reference type, while for an
integer type its added equality holds by definition.  Instantiate the
zero-offset equivalence premise~(\ref{sat-preservation3-zero-equiv}) with the
valuation $\ottnt{R}  \ottsym{\{}  z  \mapsto   0   \ottsym{\}}$.  Its environment formula holds by
\cref{eq:sat-assign-valid-env} and $0=0$.  Type-equivalence satisfaction then
gives
\begin{equation}
  \mathbf{SATv} ( \ottnt{H'} ,  \ottnt{R} ,  \ottnt{H'}  \ottsym{(}  \ottsym{(}  a  \ottsym{,}   i   \ottsym{)}  \ottsym{)} ,    [  0  /  z  ]  \, \tau' _{\ast  x }  ) .
 \label{eq:sat-assign-updated-zero}
\end{equation}

Suppose instead that $j\ne0$.  Satisfaction of the old binding for $x$ in
\cref{eq:sat-assign-subtyped-env} gives
\begin{equation}
  \ottsym{(}  a  \ottsym{,}  i  \ottsym{+}  j  \ottsym{)}  \in   \DOM( \ottnt{H} )  
 \quad\text{and}\quad
  \mathbf{SATv} ( \ottnt{H} ,  \ottnt{R} ,  \ottnt{H}  \ottsym{(}  \ottsym{(}  a  \ottsym{,}   i   \ottsym{+}   j   \ottsym{)}  \ottsym{)} ,    [  j  /  z  ]  \, \tau _{\ast  x }  ) .
 \label{eq:sat-assign-old-nonzero}
\end{equation}
By \cref{eq:sat-assign-element-zero} and
\cref{lem:sat-preservation-about-heap}, the latter judgment remains true with
$H'$ in place of $H$.  Moreover, $j\ne0$ gives
$H'((a,i+j))=H((a,i+j))$.  Instantiate the nonzero-offset equivalence
premise~(\ref{sat-preservation3-nonzero-equiv}) with
$\ottnt{R}  \ottsym{\{}  z  \mapsto   j   \ottsym{\}}$.  Its environment formula holds by
\cref{eq:sat-assign-valid-env} and $j\ne0$, so type-equivalence satisfaction
gives
\begin{equation}
  \mathbf{SATv} ( \ottnt{H'} ,  \ottnt{R} ,  \ottnt{H'}  \ottsym{(}  \ottsym{(}  a  \ottsym{,}   i   \ottsym{+}   j   \ottsym{)}  \ottsym{)} ,    [  j  /  z  ]  \, \tau' _{\ast  x }  ) .
 \label{eq:sat-assign-updated-nonzero}
\end{equation}

The heap update preserves the domain, so
\cref{eq:sat-assign-updated-zero,eq:sat-assign-old-nonzero,eq:sat-assign-updated-nonzero}
establish the required heap-cell condition at every offset with positive
ownership.  The definition of $\SATV$ therefore gives
\begin{equation}
  \mathbf{SATv} ( \ottnt{H'} ,  \ottnt{R} ,  \ottnt{R}  \ottsym{(}  x  \ottsym{)} ,   \Pi z .(  \tau' _{\ast  x }   \TREF^{\hspace{0.5pt} r })  ) .
 \label{eq:sat-assign-updated-pointer}
\end{equation}
Finally, the bindings other than $x$ and $y$ are satisfied by
\cref{eq:sat-assign-framed-binding}; the updated bindings for $x$ and $y$ are
satisfied by
\cref{eq:sat-assign-updated-pointer,eq:sat-assign-split-source}, respectively.
Since $x\ne y$, these cases cover $dom(\Gamma')$.  Thus
$ \mathbf{SAT}( \ottnt{H'} ,  \ottnt{R} ,  \Gamma' ) $, as required.

\proofitem{Item~\ref{sat-preservation4}: Pointer addition
(\rn{Rs-AddPtr}).}

Put
\[
 \tau_{{\mathrm{4}}}=  \Pi w .( \tau_{{\mathrm{1}}}  \TREF^{\hspace{0.5pt} r })  _{ y_{{\mathrm{1}}} } ,\qquad
 \tau_{{\mathrm{5}}}=  \Pi w .( \tau_{{\mathrm{2}}}  \TREF^{\hspace{0.5pt} r })  _{ x } ,\qquad
 \tau_{{\mathrm{6}}}=  \Pi w .( \tau_{{\mathrm{3}}}  \TREF^{\hspace{0.5pt} r })  _{ y } ,
\]
and let
\[
 \ottnt{R'}=\ottnt{R}  \ottsym{\{}  x'  \mapsto  pv  \boxplus  \ottnt{R}  \ottsym{(}  z  \ottsym{)}  \ottsym{\}},\qquad
 \Gamma'= \Gamma_{{\mathrm{1}}}  \left[  y \hookleftarrow   \Pi w .( \tau_{{\mathrm{1}}}  \TREF^{\hspace{0.5pt} r })  _{ y_{{\mathrm{1}}} }   \right]   \ottsym{,}   x' \COL   \Pi w .( \tau_{{\mathrm{2}}}  \TREF^{\hspace{0.5pt} r })  _{ x }  .
\]
By \cref{lem:sat-preservation-greater-tyenv},
premises~(\ref{sat-preservation4-sat}) and
(\ref{sat-preservation4-env-subtype}), and the standing validity assumption,
we obtain
\begin{equation}
  \mathbf{SAT}( \ottnt{H} ,  \ottnt{R} ,  \Gamma_{{\mathrm{1}}} ) .
 \label{eq:sat-addptr-subtyped-env}
\end{equation}
As in the preceding items, environment subtyping also gives
$ \fml{ \Gamma } \implies \fml{ \Gamma_{{\mathrm{1}}} } $, and hence
\begin{equation}
 \models  \ottsym{[}  \ottnt{R}  \ottsym{]} \,  \fml{ \Gamma_{{\mathrm{1}}} } .
 \label{eq:sat-addptr-valid-env}
\end{equation}

Premises~(\ref{sat-preservation4-y-type}) and
(\ref{sat-preservation4-y-value}), together with
\cref{eq:sat-addptr-subtyped-env}, give
\begin{equation}
 \SATV(H,R,pv,\tau_{{\mathrm{6}}}).
 \label{eq:sat-addptr-source}
\end{equation}
Similarly, premise~(\ref{sat-preservation4-offset-type}) and
\cref{eq:sat-addptr-subtyped-env} give $R(z)\in\mathbb Z$.
By the definition of shifted type addition,
premise~(\ref{sat-preservation4-split}) is precisely
\[
 \Gamma_1\vdash\tau_{{\mathrm{4}}}\mathbin{+_z}\tau_{{\mathrm{5}}} \approx \tau_{{\mathrm{6}}}.
\]
Thus \cref{lem:satv-pointer-split},
\cref{eq:sat-addptr-valid-env,eq:sat-addptr-source}, and $R(z)\in\mathbb Z$
give
\begin{equation}
 \SATV(H,R,pv,\tau_{{\mathrm{4}}})
 \quad\text{and}\quad
 \SATV(H,R,pv  \boxplus  \ottnt{R}  \ottsym{(}  z  \ottsym{)},\tau_{{\mathrm{5}}}).
 \label{eq:sat-addptr-split-values}
\end{equation}

By premise~(\ref{sat-preservation4-fresh}), well-formedness of
$\Gamma_1$, and the implicit well-formedness premises for the type-addition
judgment, the fresh variable $x'$ occurs free in none of the types in
$\Gamma_1$, nor in $\tau_{{\mathrm{4}}}$ or $\tau_{{\mathrm{5}}}$.  Applying
\cref{lem:sat-preservation-extend-register} to every binding in
\cref{eq:sat-addptr-subtyped-env}, and to the two facts in
\cref{eq:sat-addptr-split-values}, therefore gives
\begin{equation}
 \SAT(H,R',\Gamma_1),\qquad
 \SATV(H,R',pv,\tau_{{\mathrm{4}}}),\qquad
 \SATV(H,R',pv  \boxplus  \ottnt{R}  \ottsym{(}  z  \ottsym{)},\tau_{{\mathrm{5}}}).
 \label{eq:sat-addptr-framed}
\end{equation}
Finally, freshness gives $R'(y)=R(y)=pv$ and
$R'(x')=pv  \boxplus  \ottnt{R}  \ottsym{(}  z  \ottsym{)}$.  The last two facts in
\cref{eq:sat-addptr-framed} establish satisfaction of the updated binding
for $y$ and the new binding for $x'$, while its first fact establishes
satisfaction of every unchanged binding.  Consequently
$ \mathbf{SAT}( \ottnt{H} ,  \ottnt{R'} ,  \Gamma' ) $, as required.

\proofitem{Item~\ref{sat-preservation5}: Integer-array allocation
(\rn{Rs-MkArrayIntref}).}

Put
\[
 \tau_a= \Pi z .(  \{  \nu  :   \TINT    \mid     0  \, \le \, z  \wedge  z \, \le \, y  \ottsym{-}   1    \implies  \nu \,  =  \,  0   \}   \TREF^{\hspace{0.5pt} r }) ,
\]
and let
\[
\begin{aligned}
 \ottnt{R'}&=\ottnt{R}  \ottsym{\{}  x'  \mapsto  \ottsym{(}  a  \ottsym{,}   0   \ottsym{)}  \ottsym{\}},\\
 \ottnt{H'}&= \ottnt{H}  \{ (  a  ,   0   ) \mapsto   0  , \ldots, (  a  ,  \ottnt{R}  \ottsym{(}  y  \ottsym{)} \,  -  \,  1   ) \mapsto   0   \} ,\\
 \Gamma'&=\Gamma_{{\mathrm{1}}}  \ottsym{,}   x' \COL  \Pi z .(  \{  \nu  :   \TINT    \mid     0  \, \le \, z  \wedge  z \, \le \, y  \ottsym{-}   1    \implies  \nu \,  =  \,  0   \}   \TREF^{\hspace{0.5pt} r })  .
\end{aligned}
\]
By \cref{lem:sat-preservation-greater-tyenv},
premises~(\ref{sat-preservation5-sat}) and
(\ref{sat-preservation5-env-subtype}), and the standing validity assumption,
we have
\begin{equation}
  \mathbf{SAT}( \ottnt{H} ,  \ottnt{R} ,  \Gamma_{{\mathrm{1}}} ) .
 \label{eq:sat-alloc-int-subtyped-env}
\end{equation}
As in the preceding items, environment subtyping also gives
$ \fml{ \Gamma } \implies \fml{ \Gamma_{{\mathrm{1}}} } $, and hence
\begin{equation*}
 \models  \ottsym{[}  \ottnt{R}  \ottsym{]} \,  \fml{ \Gamma_{{\mathrm{1}}} } .
\end{equation*}
In particular, premise~(\ref{sat-preservation5-length-type}) and
\cref{eq:sat-alloc-int-subtyped-env} give $R(y)\in\mathbb Z$.

Premise~(\ref{sat-preservation5-fresh}) and well-formedness of
$\Gamma_1$ ensure that $x'$ occurs free in none of its types.  Moreover,
$R'$ agrees with $R$ on $dom(\Gamma_{{\mathrm{1}}})$.  Applying
\cref{lem:sat-preservation-extend-register} to every binding in
\cref{eq:sat-alloc-int-subtyped-env} gives the first fact below; the same
agreement of the two registers gives the second:
\begin{equation}
  \mathbf{SAT}( \ottnt{H} ,  \ottnt{R'} ,  \Gamma_{{\mathrm{1}}} ) 
 \qquad\text{and}\qquad
 \models  \ottsym{[}  \ottnt{R'}  \ottsym{]} \,  \fml{ \Gamma_{{\mathrm{1}}} } .
 \label{eq:sat-alloc-int-register-frame}
\end{equation}

We next use the following elementary allocation-frame property:
\begin{equation}
  \mathbf{SATv} ( \ottnt{H} ,  \ottnt{R'} ,  v ,  \tau ) 
 \quad\Longrightarrow\quad
  \mathbf{SATv} ( \ottnt{H'} ,  \ottnt{R'} ,  v ,  \tau ) .
 \label{eq:sat-alloc-int-heap-frame}
\end{equation}
It follows by induction on $\tau$.  The integer case is immediate.  At a
reference type, satisfaction in $\ottnt{H}$ supplies every cell required by a
positive ownership coefficient.  Since $\ottnt{H'}$ extends $\ottnt{H}$, that cell
has the same value in the two heaps, and the induction hypothesis applies to
its element type.  The null case is immediate because it refers only to the
ownership term.  Here premise~(\ref{sat-preservation5-fresh-base}) and the
heap-domain invariant ensure that no address with base $a$ was already in
$dom(\ottnt{H})$, so the displayed construction of $\ottnt{H'}$ is indeed a heap
extension.  Applying \cref{eq:sat-alloc-int-heap-frame} bindingwise to
\cref{eq:sat-alloc-int-register-frame} gives
\begin{equation}
  \mathbf{SAT}( \ottnt{H'} ,  \ottnt{R'} ,  \Gamma_{{\mathrm{1}}} ) .
 \label{eq:sat-alloc-int-old-bindings}
\end{equation}

It remains to establish satisfaction of the new binding.  By
premise~(\ref{sat-preservation5-ownership-shape}), Item~1 of
\cref{lem:index-substitution}, and
\cref{eq:sat-alloc-int-register-frame}, for every $j\in\mathbb Z$,
\begin{equation}
 \sem{  [  j  /  z  ]    r }_{R'}=
 \begin{cases}
  1 & \text{if }0\leq j<R(y),\\
  0 & \text{otherwise}.
 \end{cases}
 \label{eq:sat-alloc-int-permission}
\end{equation}
Evaluation under $R'$ agrees with evaluation under $R$ on the variables of
the ownership term because $x'$ is fresh.

Now fix $j\in\mathbb Z$ such that
$\sem{  [  j  /  z  ]    r }_{R'}>0$.  By
\cref{eq:sat-alloc-int-permission}, $0\leq j<R(y)$.  The definition of
$\ottnt{H'}$ therefore gives
\[
  \ottsym{(}  a  \ottsym{,}  j  \ottsym{)}  \in   \DOM( \ottnt{H'} )  
 \qquad\text{and}\qquad
 H'((a,j))=0.
\]
Under the valuation $\ottnt{R'}  \ottsym{\{}  z  \mapsto   j   \ottsym{\}}$, the initialized value $0$ satisfies
the refinement
$ \{  \nu  :   \TINT    \mid     0  \, \le \, z  \wedge  z \, \le \, y  \ottsym{-}   1    \implies  \nu \,  =  \,  0   \} $ because
$R'(y)=R(y)$ and $0\leq j<R(y)$.  Since $j$ was arbitrary, the definition
of $\SATV$ gives
\begin{equation}
 \SATV(H',R',(a,0),\tau_a).
 \label{eq:sat-alloc-int-new-binding}
\end{equation}
Finally, $R'(x')=(a,0)$.  Combining
\cref{eq:sat-alloc-int-old-bindings,eq:sat-alloc-int-new-binding} yields
$ \mathbf{SAT}( \ottnt{H'} ,  \ottnt{R'} ,  \Gamma' ) $, as required.

\proofitem{Item~\ref{sat-preservation6}: Nested-array allocation
(\rn{Rs-MkArrayNestedref}).}

Put
\[
 \tau_a= \Pi z .( \tau'  \TREF^{\hspace{0.5pt} r }) ,
\]
and let
\[
\begin{aligned}
 \ottnt{R'}&=\ottnt{R}  \ottsym{\{}  x'  \mapsto  \ottsym{(}  a  \ottsym{,}   0   \ottsym{)}  \ottsym{\}},\\
 \ottnt{H'}&= \ottnt{H}  \{ (  a  ,   0   ) \mapsto  \ottkw{null} , \ldots, (  a  ,  \ottnt{R}  \ottsym{(}  y  \ottsym{)} \,  -  \,  1   ) \mapsto  \ottkw{null}  \} ,\\
 \Gamma'&=\Gamma_{{\mathrm{1}}}  \ottsym{,}   x' \COL  \Pi z .( \tau'  \TREF^{\hspace{0.5pt} r })  .
\end{aligned}
\]
By \cref{lem:sat-preservation-greater-tyenv},
premises~(\ref{sat-preservation6-sat}) and
(\ref{sat-preservation6-env-subtype}), and the standing validity assumption,
we have
\begin{equation}
  \mathbf{SAT}( \ottnt{H} ,  \ottnt{R} ,  \Gamma_{{\mathrm{1}}} ) .
 \label{eq:sat-alloc-nested-subtyped-env}
\end{equation}
As in the preceding item, environment subtyping also gives
$ \fml{ \Gamma } \implies \fml{ \Gamma_{{\mathrm{1}}} } $, and hence
\begin{equation*}
 \models  \ottsym{[}  \ottnt{R}  \ottsym{]} \,  \fml{ \Gamma_{{\mathrm{1}}} } .
\end{equation*}
In particular, premise~(\ref{sat-preservation6-length-type}) and
\cref{eq:sat-alloc-nested-subtyped-env} give $R(y)\in\mathbb Z$.

Premise~(\ref{sat-preservation6-fresh}) and well-formedness of
$\Gamma_1$ ensure that $x'$ occurs free in none of its types.  Moreover,
$R'$ agrees with $R$ on $dom(\Gamma_{{\mathrm{1}}})$.  Applying
\cref{lem:sat-preservation-extend-register} to every binding in
\cref{eq:sat-alloc-nested-subtyped-env} gives the first fact below; register
agreement gives the second:
\begin{equation}
  \mathbf{SAT}( \ottnt{H} ,  \ottnt{R'} ,  \Gamma_{{\mathrm{1}}} ) 
 \qquad\text{and}\qquad
 \models  \ottsym{[}  \ottnt{R'}  \ottsym{]} \,  \fml{ \Gamma_{{\mathrm{1}}} } .
 \label{eq:sat-alloc-nested-register-frame}
\end{equation}

We next frame these old bindings across the allocation.  For every runtime
value $v$ and type $\tau$,
\begin{equation}
  \mathbf{SATv} ( \ottnt{H} ,  \ottnt{R'} ,  v ,  \tau ) 
 \quad\Longrightarrow\quad
  \mathbf{SATv} ( \ottnt{H'} ,  \ottnt{R'} ,  v ,  \tau ) .
 \label{eq:sat-alloc-nested-heap-frame}
\end{equation}
This follows by induction on $\tau$, exactly as in the preceding item.  At a
reference type, satisfaction in $\ottnt{H}$ supplies every cell required by a
positive ownership coefficient; $\ottnt{H'}$ contains that cell with the same
value, and the induction hypothesis applies to its element type.  Here
premise~(\ref{sat-preservation6-fresh-base}) and the heap-domain invariant
ensure that no address with base $a$ was already in $dom(\ottnt{H})$, so $\ottnt{H'}$
is a heap extension.  Applying
\cref{eq:sat-alloc-nested-heap-frame} bindingwise to
\cref{eq:sat-alloc-nested-register-frame} gives
\begin{equation}
  \mathbf{SAT}( \ottnt{H'} ,  \ottnt{R'} ,  \Gamma_{{\mathrm{1}}} ) .
 \label{eq:sat-alloc-nested-old-bindings}
\end{equation}

It remains to establish satisfaction of the new binding.  By
premise~(\ref{sat-preservation6-ownership-shape}), Item~1 of
\cref{lem:index-substitution}, and
\cref{eq:sat-alloc-nested-register-frame}, for every $j\in\mathbb Z$,
\begin{equation}
 \sem{  [  j  /  z  ]    r }_{R'}=
 \begin{cases}
  1 & \text{if }0\leq j<R(y),\\
  0 & \text{otherwise}.
 \end{cases}
 \label{eq:sat-alloc-nested-permission}
\end{equation}
Evaluation under $R'$ agrees with evaluation under $R$ on the variables of
the ownership term because $x'$ is fresh.

Now fix $j\in\mathbb Z$ such that
$\sem{  [  j  /  z  ]    r }_{R'}>0$.  By
\cref{eq:sat-alloc-nested-permission}, $0\leq j<R(y)$, and hence
\[
  \ottsym{(}  a  \ottsym{,}  j  \ottsym{)}  \in   \DOM( \ottnt{H'} )  
 \qquad\text{and}\qquad
 H'((a,j))=\NULL.
\]
Premise~(\ref{sat-preservation6-element-shape}) shows that $\tau'$ is a
reference type.  The emptiness consequence of
\cref{lem:index-substitution}, applied to
premise~(\ref{sat-preservation6-empty}), gives
\[
 \models   \Empty{  [  j  /  z  ]  \, \tau' } .
\]
\Cref{lem:sat-preservation-empty} therefore yields
\[
  \mathbf{SATv} ( \ottnt{H'} ,  \ottnt{R'} ,  \ottkw{null} ,   [  j  /  z  ]  \, \tau' ) .
\]
Since $j$ was arbitrary, the definition of $\SATV$ gives
\begin{equation}
 \SATV(H',R',(a,0),\tau_a).
 \label{eq:sat-alloc-nested-new-binding}
\end{equation}
Finally, $R'(x')=(a,0)$.  Combining
\cref{eq:sat-alloc-nested-old-bindings,eq:sat-alloc-nested-new-binding}
yields $ \mathbf{SAT}( \ottnt{H'} ,  \ottnt{R'} ,  \Gamma' ) $, as required.

\proofitem{Item~\ref{sat-preservation7}: Aliasing by dereference
(\rn{Rs-AliasDeref}).}

For readability, put
\[
\begin{array}{ll}
 \tau_x= \Pi z' .(  \tau _{\ast  x }   \TREF^{\hspace{0.5pt}  { r }_{ x }  }) ,&
 \tau_p= \Pi z .(  \tau _{\ast \ast  y }   \TREF^{\hspace{0.5pt}  { r }_{ \ast  y }  }) ,\\
 \tau_x'= \Pi z' .(  \tau' _{\ast  x }   \TREF^{\hspace{0.5pt}  { r' }_{ x }  }) ,&
 \tau_p'= \Pi z .(  \tau' _{\ast \ast  y }   \TREF^{\hspace{0.5pt}  { r' }_{ \ast  y }  }) .
\end{array}
\]
Let $\Gamma'$ be $  \Gamma_{{\mathrm{1}}}  \left[  x \hookleftarrow  \tau' _{ x }   \right]   \left[  y \hookleftarrow  \Pi w .(  \tau' _{ p }   \TREF^{\hspace{0.5pt} r })   \right] $.
By \cref{lem:sat-preservation-greater-tyenv},
premises~(\ref{sat-preservation7-sat}) and
(\ref{sat-preservation7-env-subtype}), and the standing validity assumption,
we have
\begin{equation}
  \mathbf{SAT}( \ottnt{H} ,  \ottnt{R} ,  \Gamma_{{\mathrm{1}}} ) .
 \label{eq:sat-alias-deref-subtyped-env}
\end{equation}
As in the preceding items, environment subtyping gives
$ \fml{ \Gamma } \implies \fml{ \Gamma_{{\mathrm{1}}} } $, so
\begin{equation}
 \models  \ottsym{[}  \ottnt{R}  \ottsym{]} \,  \fml{ \Gamma_{{\mathrm{1}}} } .
 \label{eq:sat-alias-deref-valid-env}
\end{equation}
Also, $x\ne y$: otherwise, the lookup premises would give
$|\tau_x|=|\tau_p|\ \mathtt{ref}$, whereas the type additions in
premise~(\ref{sat-preservation7-zero-equiv}) require
$|\tau_x|=|\tau_p|$.  A finite simple type cannot equal a reference to itself.

Premise~(\ref{sat-preservation7-alias}) implies that, for some $a$ and $k$,
\begin{equation}
 R(y)=(a,k)\in dom(H)
 \qquad\text{and}\qquad
 H((a,k))=R(x).
 \label{eq:sat-alias-deref-runtime}
\end{equation}
Instantiate premise~(\ref{sat-preservation7-positive}) at $w=0$ using
\cref{eq:sat-alias-deref-valid-env}.  This gives
$\sem{  [  0  /  w  ]    r }_R>0$.  Hence satisfaction of the old $y$-binding in
\cref{eq:sat-alias-deref-subtyped-env}, together with
\cref{eq:sat-alias-deref-runtime}, gives satisfaction of its selected cell.
The old $x$-binding is satisfied by the same environment judgment, so
\begin{equation}
 \SATV(H,R,R(x),\tau_x)
 \qquad\text{and}\qquad
 \SATV(H,R,R(x),[0/w]\tau_p).
 \label{eq:sat-alias-deref-old-zero}
\end{equation}

By \rn{TA-PlusPlus}, premise~(\ref{sat-preservation7-zero-equiv})
provides a common total type for the old pair and the new pair.  Instantiate
its two type-addition judgments at $w=0$ with the valuation
$\ottnt{R}  \ottsym{\{}  w  \mapsto   0   \ottsym{\}}$; its environment formula holds by
\cref{eq:sat-alias-deref-valid-env} and $0=0$.
Apply \cref{lem:sat-preservation-add-ty} first to merge the two judgments
in \cref{eq:sat-alias-deref-old-zero} into satisfaction of that total,
and then in the reverse direction to split it into satisfaction of the new
pair.  The semantic substitution property used in
\cref{lem:index-substitution} therefore gives
\begin{equation}
 \SATV(H,R,R(x),\tau_x')
 \qquad\text{and}\qquad
 \SATV(H,R,R(x),[0/w]\tau_p').
 \label{eq:sat-alias-deref-new-zero}
\end{equation}
Here $w$ is not free in $\tau_x$ or $\tau_x'$ by well-formedness of
the old and new environments, so substituting $0$ for $w$ leaves these
two types unchanged.

It remains to check the updated $y$-binding.  Fix an integer $j$ such that
$\sem{  [  j  /  w  ]    r }_R>0$.  Its outer ownership term $r$ is unchanged, so
\cref{eq:sat-alias-deref-subtyped-env} gives
\[
  \ottsym{(}  a  \ottsym{,}  k  \ottsym{+}  j  \ottsym{)}  \in   \DOM( \ottnt{H} )  
 \qquad\text{and}\qquad
 \SATV(H,R,H((a,k+j)),[j/w]\tau_p).
\]
If $j=0$, the second judgment in
\cref{eq:sat-alias-deref-new-zero} and
$H((a,k))=R(x)$ give satisfaction of the cell by $[0/w]\tau_p'$.
If $j\ne0$, instantiate premise~(\ref{sat-preservation7-nonzero-equiv})
with the valuation $\ottnt{R}  \ottsym{\{}  w  \mapsto   j   \ottsym{\}}$.  Its environment formula holds by
\cref{eq:sat-alias-deref-valid-env} and $j\ne0$.
Applying \cref{lem:sat-preservation-weakening} to the subtyping direction
from the old element type to the new one, and using semantic substitution,
gives
\[
 \SATV(H,R,H((a,k+j)),[j/w]\tau_p').
\]
Thus every cell required by the updated outer reference exists and satisfies
its new element type, establishing satisfaction of the updated $y$-binding.
The first judgment in \cref{eq:sat-alias-deref-new-zero} establishes
satisfaction of the updated $x$-binding.  Since $x\ne y$, every other binding
has its old type and value and is satisfied by
\cref{eq:sat-alias-deref-subtyped-env}.  Therefore $ \mathbf{SAT}( \ottnt{H} ,  \ottnt{R} ,  \Gamma' ) $,
as required.

\proofitem{Item~\ref{sat-preservation8}: Aliasing by pointer addition
(\rn{Rs-AliasAddPtr}).}

By \cref{lem:sat-preservation-greater-tyenv},
premises~(\ref{sat-preservation8-sat}) and
(\ref{sat-preservation8-env-subtype}), and the standing validity assumption,
we have
\begin{equation}
  \mathbf{SAT}( \ottnt{H} ,  \ottnt{R} ,  \Gamma_{{\mathrm{1}}} ) .
 \label{eq:sat-alias-add-subtyped-env}
\end{equation}
As in the preceding items, environment subtyping gives
$ \fml{ \Gamma } \implies \fml{ \Gamma_{{\mathrm{1}}} } $, so the standing validity assumption also gives
\begin{equation}
 \models  \ottsym{[}  \ottnt{R}  \ottsym{]} \,  \fml{ \Gamma_{{\mathrm{1}}} } .
 \label{eq:sat-alias-add-valid-env}
\end{equation}

Let $\Gamma'$ be $  \Gamma_{{\mathrm{1}}}  \left[  x \hookleftarrow  \Pi w' .(  \tau' _{\ast  x }   \TREF^{\hspace{0.5pt}  { r' }_{ x }  })   \right]   \left[  y \hookleftarrow  \Pi w .(  \tau' _{\ast  y }   \TREF^{\hspace{0.5pt}  { r' }_{ y }  })   \right] $.
Put $\tau_x$, $\tau_y$, $\tau_x'$, and $\tau_y'$ as in the proof of
item~\ref{conown-preservation8}.  By inversion of \rn{TA-PlusPlus} in
premise~(\ref{sat-preservation8-combined-equiv}), there is a type $\tau_{{\mathrm{3}}}$ satisfying
the two judgments in \cref{eq:alias-add-common-total}.  After commuting the
two summands, these judgments are precisely
\[
 \Gamma_1\vdash\tau_y\mathbin{+_z}\tau_x \approx \tau_{{\mathrm{3}}}
 \qquad\text{and}\qquad
 \Gamma_1\vdash\tau_y'\mathbin{+_z}\tau_x' \approx \tau_{{\mathrm{3}}}.
\]
By \cref{lem:satv-pointer-split},
\cref{eq:sat-alias-add-valid-env}, and
premise~(\ref{sat-preservation8-offset-value}),
\begin{align*}
 \SATV(H,R,pv,\tau_{{\mathrm{3}}})
 &\Longleftrightarrow
 \SATV(H,R,pv,\tau_y)\land
 \SATV(H,R,pv  \boxplus  \ottnt{R}  \ottsym{(}  z  \ottsym{)},\tau_x),\\
 \SATV(H,R,pv,\tau_{{\mathrm{3}}})
 &\Longleftrightarrow
 \SATV(H,R,pv,\tau_y')\land
 \SATV(H,R,pv  \boxplus  \ottnt{R}  \ottsym{(}  z  \ottsym{)},\tau_x').
\end{align*}
The old pair on the first line is satisfied by
\cref{eq:sat-alias-add-subtyped-env}, using
premises~(\ref{sat-preservation8-x-value}),
(\ref{sat-preservation8-y-value}),
(\ref{sat-preservation8-x-type}), and
(\ref{sat-preservation8-y-type}).  Hence the new pair on the second line is
satisfied.  By premise~(\ref{sat-preservation8-distinct}), the bindings are
distinct, and all other bindings are unchanged.  Therefore
$ \mathbf{SAT}( \ottnt{H} ,  \ottnt{R} ,  \Gamma' ) $ holds.

\proofitem{Item~\ref{sat-preservation9}: Null binding
(\rn{Rs-LetNull}).}

By \cref{lem:sat-preservation-greater-tyenv},
premises~(\ref{sat-preservation9-sat}) and
(\ref{sat-preservation9-env-subtype}), and the standing validity assumption,
we have $ \mathbf{SAT}( \ottnt{H} ,  \ottnt{R} ,  \Gamma_{{\mathrm{1}}} ) $.
Independently, premise~(\ref{sat-preservation9-empty}) and
\cref{lem:sat-preservation-empty} give
\[
  \mathbf{SATv} ( \ottnt{H} ,  \ottnt{R} ,  \ottkw{null} ,   \Pi z .( \tau'  \TREF^{\hspace{0.5pt} r })  ) .
\]
Put $\ottnt{R'}=\ottnt{R}  \ottsym{\{}  x'  \mapsto  \ottkw{null}  \ottsym{\}}$.  By well-formedness and
premise~(\ref{sat-preservation9-fresh}), $x'$ does not occur freely in the
types in $\Gamma_{{\mathrm{1}}}$ or in $ \Pi z .( \tau'  \TREF^{\hspace{0.5pt} r }) $.  Applying
\cref{lem:sat-preservation-extend-register} preserves satisfaction of the
old values and of $ \NULL $ under $\ottnt{R'}$.  Since $\ottnt{R'}(y)=R(y)$ for
every $y\in dom(\Gamma_{{\mathrm{1}}})$ and $\ottnt{R'}(x')= \NULL $, the definition of $\SAT$
gives $ \mathbf{SAT}( \ottnt{H} ,  \ottnt{R'} ,  \ottsym{(}  \Gamma_{{\mathrm{1}}}  \ottsym{,}   x' \COL  \Pi z .( \tau'  \TREF^{\hspace{0.5pt} r })    \ottsym{)} ) $, as required.
\end{proof}

\subsubsection{Proof of \cref{lem:preservation}}
\label{sec:preservation}
\begin{proof}
We prove the stronger assertion that the successor can be typed using the
same function type environment $\Theta$ as the source.  We proceed by
induction on the derivation of
$  \tuple{ \ottnt{H} ,  \ottnt{R} ,  e }     \longrightarrow _{  D  }     \tuple{ \ottnt{H'} ,  \ottnt{R'} ,  e' }  $, with
$\Theta$, $\Gamma$, $\tau$, and $\Gamma'$ arbitrary.
By configuration typing, choose $\Theta$ and $\Gamma$ such that
\[
\begin{gathered}
  \Theta   \mid   \Gamma   \vdash   e  :  \tau   \produces   \Gamma' , \qquad  \Theta   \vdash   D ,\\
  \mathbf{ConOwn}( \ottnt{H} ,  \ottnt{R} ,  \Gamma ) , \qquad  \mathbf{SAT}( \ottnt{H} ,  \ottnt{R} ,  \Gamma ) .
\end{gathered}
\]
These are the \emph{source typing and invariants} used below.  The definition
of $\SAT$ also gives $\models  \ottsym{[}  \ottnt{R}  \ottsym{]} \,  \fml{ \Gamma } $, supplying the standing validity
assumption of \cref{lem:conown-preservation,lem:sat-preservation}.
Whenever inversion gives $\Gamma  \leq  \Gamma_{{\mathrm{1}}}$, we may transfer the source invariants
to $\Gamma_{{\mathrm{1}}}$ by
\cref{lem:conown-preservation-weakening-tyenv,lem:sat-preservation-greater-tyenv}.

In each case we retain $\Theta$ and construct $\Gamma_p$ such that
\[
\begin{gathered}
 \Theta\mid\Gamma_p\vdash e'\COL\tau\Rightarrow\Gamma',\\
 \CONOWN(H',R',\Gamma_p), \qquad \SAT(H',R',\Gamma_p).
\end{gathered}
\]
These judgments give the required successor configuration typing.
We first consider the direct rules in the order of
\cref{sec:opsemFull}, and treat the recursive \rn{Rs-Context} case last.
The environments and types introduced by inversion below are existential
witnesses, and each case's premises are labeled locally.

For binding reductions, we write $\tau_{{\mathrm{0}}}$ for the residual type of the bound
variable in the continuation's output environment.  Renaming the bound
variable $x$ to the fresh register name $x'$ is capture-avoiding and may
initially give output $\Gamma'  \ottsym{,}   x' \COL \tau_{{\mathrm{0}}} $.  By
\cref{lem:typing-expression-substitution}, this renaming applies to the
expression and all types in the bound variable's scope that mention $x$,
including the residual type and updated element types.  After renaming, we
reuse their names for the renamed types.  Since the target environment $\Gamma'$
and type $\tau$ are well formed without $x'$, reflexivity and
\rn{S-TyEnv} give $\Gamma'  \ottsym{,}   x' \COL \tau_{{\mathrm{0}}}   \leq  \Gamma'$.  By \rn{S-Res},
\[
 \ottsym{(}  \Gamma'  \ottsym{,}   x' \COL \tau_{{\mathrm{0}}}   \ottsym{)}  \ottsym{,}  \tau  \leq  \Gamma'  \ottsym{,}  \tau.
\]
An application of \rn{T-Sub} therefore removes the fresh binding from the
output environment, yielding the target typing judgment above.

We also use the following freshness observation without repeating it in each
binding case.  From $\SAT(H,R,\Gamma)$, $dom(\Gamma)\subseteq dom(R)$.
Thus $x'\notin dom(R)$ implies $x'\notin dom(\Gamma)$; if
$\Gamma  \leq  \Gamma_{{\mathrm{1}}}$, then $dom(\Gamma_{{\mathrm{1}}})\subseteq dom(\Gamma)$, so
$x'\notin dom(\Gamma_{{\mathrm{1}}})\cup dom(R)$ as well.

\renewcommand{\theenumi}{\alph{enumi}}
\renewcommand{\labelenumi}{(\theenumi)}

\begin{rncase}{Rs-LetInt}
Here $e= \LET  x  =  n  \IN  e_{{\mathrm{0}}} $, $e'=  [  x'  /  x  ]    e_{{\mathrm{0}}} $,
$H'=H$, and $R'=\ottnt{R}  \ottsym{\{}  x'  \mapsto   n   \ottsym{\}}$, where $x'\notin dom(R)$.

By \cref{lem:inversion-expression-typing}, we have
\begin{enumerate}
\item $\Gamma  \leq  \Gamma_{{\mathrm{1}}}$;\label{preservation-let-int-subtype}
\item $\Theta \mid \Gamma_1, x \COL \set{\nu \COL \TINT \mid \nu = n} \vdash e_0 \COL \tau \Rightarrow \ottsym{(}  \Gamma'  \ottsym{,}   x \COL \tau_{{\mathrm{0}}}   \ottsym{)}$.\label{preservation-let-int-continuation}
\end{enumerate}

Put $\Gamma_p=\Gamma_{{\mathrm{1}}}  \ottsym{,}   x' \COL  \{  \nu  :   \TINT    \mid   \nu \,  =  \,  n   \}  $.
Renaming $x$ to $x'$ in premise~(\ref{preservation-let-int-continuation})
gives the continuation typing with output $\Gamma'  \ottsym{,}   x' \COL \tau_{{\mathrm{0}}} $.
By premise~(\ref{preservation-let-int-subtype}), transfer the source invariants to
$\Gamma_{{\mathrm{1}}}$.  Integer-register extension then gives
$\CONOWN(H',R',\Gamma_p)$ by
\cref{lem:conown-irrevalent-to-integer}(\ref{item:conown-irrevalent-to-integer-extension}).
Moreover, $\models  \ottsym{[}  \ottnt{R}  \ottsym{]} \, \ottsym{[}   n   \ottsym{/}  \nu  \ottsym{]}  \ottsym{(}  \nu \,  =  \,  n   \ottsym{)}$, so
\cref{lem:sat-extension} gives $\SAT(H',R',\Gamma_p)$.

\end{rncase}

\begin{rncase}{Rs-LetNull}
Here $e= \LET  x  =  \ottkw{null}  \IN  e_{{\mathrm{0}}} $, $e'=  [  x'  /  x  ]    e_{{\mathrm{0}}} $,
$H'=H$, and $R'=\ottnt{R}  \ottsym{\{}  x'  \mapsto  \ottkw{null}  \ottsym{\}}$, where $x'\notin dom(R)$.

By \cref{lem:inversion-expression-typing}, we have
\begin{enumerate}
\item $\Gamma  \leq  \Gamma_{{\mathrm{1}}}$;\label{preservation-let-null-subtype}
\item $\Theta \mid \Gamma_1, x \COL  \Pi z .( \tau'  \TREF^{\hspace{0.5pt} r })  \vdash e_0 \COL \tau \Rightarrow (\Gamma'  \ottsym{,}   x \COL \tau_{{\mathrm{0}}} )$;\label{preservation-let-null-continuation}
\item $\models   \Empty{  \Pi z .( \tau'  \TREF^{\hspace{0.5pt} r })  } $.\label{preservation-let-null-empty}
\end{enumerate}

Put $\Gamma_p=\Gamma_{{\mathrm{1}}}  \ottsym{,}   x' \COL  \Pi z .( \tau'  \TREF^{\hspace{0.5pt} r })  $.
Renaming $x$ to $x'$ in premise~(\ref{preservation-let-null-continuation})
gives the continuation typing with output $\Gamma'  \ottsym{,}   x' \COL \tau_{{\mathrm{0}}} $.
Premises~(\ref{preservation-let-null-subtype}) and
(\ref{preservation-let-null-empty}), together with freshness and the source
invariants, supply the hypotheses of the null-binding items.
Applying \cref{lem:conown-preservation}(\ref{conown-preservation9})
and \cref{lem:sat-preservation}(\ref{sat-preservation9}),
respectively, gives
\[
 \CONOWN(H',R',\Gamma_p)
 \qquad\text{and}\qquad
 \SAT(H',R',\Gamma_p).
\]

\end{rncase}

\begin{rncase}{Rs-Minus}
Here $e= \LET  x  =   y  \mathop{  -  }  z   \IN  e_{{\mathrm{0}}} $, $e'=  [  x'  /  x  ]    e_{{\mathrm{0}}} $,
$H'=H$, and $R'=R\{x'\mapsto R(y)-R(z)\}$.
The side conditions give $R(y),R(z)\in \mathbb{Z} $ and $x'\notin dom(R)$.

By \cref{lem:inversion-expression-typing}, we have
\begin{enumerate}
\item $\Gamma  \leq  \Gamma_{{\mathrm{1}}}$;\label{preservation-minus-subtype}
\item $\Theta \mid \Gamma_1, x \COL \set{\nu \COL \TINT \mid \nu = y - z} \vdash e_0 \COL \tau \Rightarrow (\Gamma'  \ottsym{,}   x \COL \tau_{{\mathrm{0}}} )$.\label{preservation-minus-continuation}
\end{enumerate}

Put $\Gamma_p=\Gamma_{{\mathrm{1}}}  \ottsym{,}   x' \COL  \{  \nu  :   \TINT    \mid   \nu \,  =  \, y  \ottsym{-}  z  \}  $.
Renaming $x$ to $x'$ in premise~(\ref{preservation-minus-continuation})
gives the continuation typing with output $\Gamma'  \ottsym{,}   x' \COL \tau_{{\mathrm{0}}} $.
By premise~(\ref{preservation-minus-subtype}), transfer the source invariants to
$\Gamma_{{\mathrm{1}}}$.  Integer-register extension gives
$\CONOWN(H',R',\Gamma_p)$ by
\cref{lem:conown-irrevalent-to-integer}(\ref{item:conown-irrevalent-to-integer-extension}).
Since $\models[R][R(y)-R(z)/\nu](\nu=y-z)$,
\cref{lem:sat-extension} also gives $\SAT(H',R',\Gamma_p)$.

\end{rncase}

\begin{rncase}{Rs-IfTrue}
Here $e= \IFNP  x  \THEN  e_{{\mathrm{0}}}  \ELSE  e_{{\mathrm{1}}} $, $e'=e_0$,
$H'=H$, $R'=R$, and $R(x)\leq0$.

By \cref{lem:inversion-expression-typing}, we have
\begin{enumerate}
\item $\Gamma  \leq  \Gamma_{{\mathrm{1}}}  \ottsym{,}   x \COL  \{  \nu  :   \TINT    \mid   \varphi  \}    \ottsym{,}  \Gamma_{{\mathrm{2}}}$;\label{preservation-if-true-subtype}
\item $\Theta \mid \Gamma_1, x \COL \set{\nu \COL \TINT \mid \varphi \land \nu \le 0}, \Gamma_2 \vdash e_0 \COL \tau \Rightarrow \Gamma'$.\label{preservation-if-true-continuation}
\end{enumerate}

Put $\Gamma_p=\Gamma_{{\mathrm{1}}}  \ottsym{,}   x \COL  \{  \nu  :   \TINT    \mid    \varphi  \wedge  \nu \, \le \,  0    \}    \ottsym{,}  \Gamma_{{\mathrm{2}}}$.
Premise~(\ref{preservation-if-true-continuation}) gives successor expression typing.
By premise~(\ref{preservation-if-true-subtype}), transfer the source invariants to
$\Gamma_{{\mathrm{1}}}  \ottsym{,}   x \COL  \{  \nu  :   \TINT    \mid   \varphi  \}    \ottsym{,}  \Gamma_{{\mathrm{2}}}$.
Changing an integer refinement preserves $\CONOWN$ by
\cref{lem:conown-irrevalent-to-integer}(\ref{item:conown-irrevalent-to-integer2}).
Satisfaction follows from the old refinement and $R(x)\leq0$, which give
$\models  \ottsym{[}  \ottnt{R}  \ottsym{]} \, \ottsym{[}  \ottnt{R}  \ottsym{(}  x  \ottsym{)}  \ottsym{/}  \nu  \ottsym{]}  \ottsym{(}  \varphi  \wedge  \nu \, \le \,  0   \ottsym{)}$.
Thus $\CONOWN(H',R',\Gamma_p)$ and $\SAT(H',R',\Gamma_p)$ hold.
The case \rn{Rs-IfFalse} is similar, using the refinement
$ \varphi  \wedge  \nu \,  >  \,  0  $ and $R(x)>0$.

\end{rncase}

\begin{rncase}{Rs-LetVar}
Here $e= \LET  x  =  y  \IN  e_{{\mathrm{0}}} $, $e'=  [  x'  /  x  ]    e_{{\mathrm{0}}} $,
$H'=H$, and $R'=\ottnt{R}  \ottsym{\{}  x'  \mapsto  \ottnt{R}  \ottsym{(}  y  \ottsym{)}  \ottsym{\}}$, where $x'\notin dom(R)$.
By alpha-renaming the bound variable, we may assume that $x\ne y$.

By the general-let item of \cref{lem:inversion-expression-typing}, we have
\begin{enumerate}
\item $\Gamma  \leq  \Gamma_{{\mathrm{1}}}$;\label{preservation-let-var-environment}
\item $ \Theta   \mid   \Gamma_{{\mathrm{1}}}   \vdash   y  :  \tau_{{\mathrm{1}}}   \produces   \Gamma_{{\mathrm{2}}} $;\label{preservation-let-var-variable}
\item $ \Theta   \mid   \Gamma_{{\mathrm{2}}}  \ottsym{,}   x \COL \tau_{{\mathrm{1}}}    \vdash   e_{{\mathrm{0}}}  :  \tau   \produces   \Gamma'  \ottsym{,}   x \COL \tau_{{\mathrm{0}}}  $.\label{preservation-let-var-continuation}
\end{enumerate}

Applying the variable item of the same lemma to
premise~(\ref{preservation-let-var-variable}) gives
$\Gamma_{{\mathrm{3}}}$, $ \tau _{ y } $, $ \tau' _{ y } $, and $ \tau _{ x } $ such that
\[
\begin{gathered}
 \Gamma_{{\mathrm{1}}}  \leq  \Gamma_{{\mathrm{3}}}, \qquad  \Gamma_{{\mathrm{3}}}  (  y  )   =   \tau _{ y } ,\\
 \Gamma_{{\mathrm{3}}}  \vdash      \tau' _{ y }   +  \tau  _{ x }    \approx    \tau _{ y }  , \qquad
 \Gamma_{{\mathrm{4}}}  \ottsym{,}   \ottsym{(}   \tau _{ x }   \ottsym{)}  ^ {= y }   \leq  \Gamma_{{\mathrm{2}}}  \ottsym{,}  \tau_{{\mathrm{1}}},
\end{gathered}
\]
where $\Gamma_{{\mathrm{4}}}= \Gamma_{{\mathrm{3}}}  \left[  y \hookleftarrow  \tau' _{ y }   \right] $; here we have exchanged the summands
using commutativity of type addition.
By transitivity, premise~(\ref{preservation-let-var-environment}) gives
$\Gamma  \leq  \Gamma_{{\mathrm{3}}}$.
Apply \cref{lem:conown-preservation}(\ref{conown-preservation1}) and
\cref{lem:sat-preservation}(\ref{sat-preservation1}), using the displayed
split and reflexivity for the source-type subtyping premise.  Together with
freshness and the source invariants, these give
\[
  \mathbf{ConOwn}( \ottnt{H'} ,  \ottnt{R'} ,  \Gamma_{{\mathrm{4}}}  \ottsym{,}   x' \COL  \tau _{ x }   ) 
 \qquad\text{and}\qquad
  \mathbf{SAT}( \ottnt{H'} ,  \ottnt{R'} ,  \Gamma_{{\mathrm{4}}}  \ottsym{,}   x' \COL  \tau _{ x }   ) .
\]
Since $R'(x')=R'(y)$, the new integer binding also satisfies the equality
refinement in $ \ottsym{(}   \tau _{ x }   \ottsym{)}  ^ {= y } $; for a reference, strengthening is the identity.
By \cref{lem:own-preservation-equivalent-ty-strength}, ownership is unchanged.
The two invariants therefore also hold under $\Gamma_{{\mathrm{4}}}  \ottsym{,}   x' \COL  \ottsym{(}   \tau _{ x }   \ottsym{)}  ^ {= y }  $.
By \rn{S-Res}, $\Gamma_{{\mathrm{4}}}  \ottsym{,}   \ottsym{(}   \tau _{ x }   \ottsym{)}  ^ {= y }   \leq  \Gamma_{{\mathrm{2}}}  \ottsym{,}  \tau_{{\mathrm{1}}}$ gives $\Gamma_{{\mathrm{4}}}  \leq  \Gamma_{{\mathrm{2}}}$ and
$\Gamma_{{\mathrm{4}}}  \vdash    \ottsym{(}   \tau _{ x }   \ottsym{)}  ^ {= y }    \leq   \tau_{{\mathrm{1}}} $.  Weakening their type-subtyping premises to
$\Gamma_{{\mathrm{4}}}  \ottsym{,}   x' \COL  \ottsym{(}   \tau _{ x }   \ottsym{)}  ^ {= y }  $ and applying \rn{S-TyEnv} gives
\[
 \Gamma_{{\mathrm{4}}}  \ottsym{,}   x' \COL  \ottsym{(}   \tau _{ x }   \ottsym{)}  ^ {= y }    \leq  \Gamma_{{\mathrm{2}}}  \ottsym{,}   x' \COL \tau_{{\mathrm{1}}} .
\]
Put $\Gamma_p=\Gamma_{{\mathrm{2}}}  \ottsym{,}   x' \COL \tau_{{\mathrm{1}}} $.
Environment subtyping preservation therefore gives
$\CONOWN(H',R',\Gamma_p)$ and $\SAT(H',R',\Gamma_p)$.
Renaming $x$ to $x'$ in premise~(\ref{preservation-let-var-continuation})
gives the continuation typing with output $\Gamma'  \ottsym{,}   x' \COL \tau_{{\mathrm{0}}} $.

\end{rncase}

\begin{rncase}{Rs-Deref}
Here $e= \LET  x  =   \ast  y   \IN  e_{{\mathrm{0}}} $, $e'=  [  x'  /  x  ]    e_{{\mathrm{0}}} $,
$H'=H$, and $R'=\ottnt{R}  \ottsym{\{}  x'  \mapsto  \ottnt{H}  \ottsym{(}  \ottnt{R}  \ottsym{(}  y  \ottsym{)}  \ottsym{)}  \ottsym{\}}$.
The side conditions give $R(y)=(a,i)\in dom(H)$ and $x'\notin dom(R)$.

By \cref{lem:inversion-expression-typing}, we have
\begin{enumerate}
\item $\Gamma  \leq  \Gamma_{{\mathrm{1}}}$;\label{preservation-deref-subtype}
\item $ \Gamma_{{\mathrm{1}}}  (  y  )  =  \Pi z .(  \tau _{ y }   \TREF^{\hspace{0.5pt} r }) $;\label{preservation-deref-pointer-type}
\item $\Gamma_{{\mathrm{1}}}  \ottsym{,}   z \COL  \{  \nu  :   \TINT    \mid   \nu \,  =  \,  0   \}    \vdash     \tau'  +  \tau  _{ x }    \approx    \tau _{ y }  $;\label{preservation-deref-split}
\item $\Gamma_{{\mathrm{1}}}  \ottsym{,}   x \COL  \tau _{ x }    \ottsym{,}   z \COL  \{  \nu  :   \TINT    \mid   \nu \,  =  \,  0   \}    \vdash    \tau' _{ y }    \approx    \ottsym{(}  \tau'  \ottsym{)}  ^ {= x }  $;\label{preservation-deref-zero-equiv}
\item $\Gamma_{{\mathrm{1}}}  \ottsym{,}   x \COL  \tau _{ x }    \ottsym{,}   z \COL  \{  \nu  :   \TINT    \mid   \nu \, \neq \,  0   \}    \vdash    \tau' _{ y }    \approx    \tau _{ y }  $;\label{preservation-deref-nonzero-equiv}
\item $\Gamma_{{\mathrm{1}}}  \ottsym{,}   z \COL  \{  \nu  :   \TINT    \mid   \nu \,  =  \,  0   \}    \models  r \,  >  \,  \mathbf{0} $;\label{preservation-deref-positive}
\item $ \Theta   \mid    \Gamma_{{\mathrm{1}}}  \left[  y \hookleftarrow  \Pi z .(  \tau' _{ y }   \TREF^{\hspace{0.5pt} r })   \right]   \ottsym{,}   x \COL  \tau _{ x }     \vdash   e_{{\mathrm{0}}}  :  \tau   \produces   \ottsym{(}  \Gamma'  \ottsym{,}   x \COL \tau_{{\mathrm{0}}}   \ottsym{)} $.\label{preservation-deref-continuation}
\end{enumerate}

Rename $x$ to $x'$ in premises
(\ref{preservation-deref-zero-equiv}),
(\ref{preservation-deref-nonzero-equiv}), and
(\ref{preservation-deref-continuation}), reusing $ \tau' _{ y } $ and $\tau_{{\mathrm{0}}}$
for the renamed types.
Put $\Gamma_p= \Gamma_{{\mathrm{1}}}  \left[  y \hookleftarrow  \Pi z .(  \tau' _{ y }   \TREF^{\hspace{0.5pt} r })   \right]   \ottsym{,}   x' \COL  \tau _{ x }  $.
The renamed premise~(\ref{preservation-deref-continuation})
gives the continuation typing with output $\Gamma'  \ottsym{,}   x' \COL \tau_{{\mathrm{0}}} $.
Premises~(\ref{preservation-deref-subtype})--(\ref{preservation-deref-positive}),
with the bound variable renamed as above, together with the operational
side conditions and the source invariants, supply the hypotheses of the
dereference items.
Applying \cref{lem:conown-preservation}(\ref{conown-preservation2})
and \cref{lem:sat-preservation}(\ref{sat-preservation2}),
respectively, gives
\[
 \CONOWN(H',R',\Gamma_p)
 \qquad\text{and}\qquad
 \SAT(H',R',\Gamma_p).
\]

\end{rncase}

\begin{rncase}{Rs-Assign}
Here $e=  x  \WRITE  y   \SEQ  e_{{\mathrm{0}}} $, $e'=e_0$, $R'=R$, and
$H'=\ottnt{H}  \ottsym{\{}  \ottsym{(}  a  \ottsym{,}   i   \ottsym{)}  \hookleftarrow  \ottnt{R}  \ottsym{(}  y  \ottsym{)}  \ottsym{\}}$, where $R(x)=(a,i)\in dom(H)$.

By \cref{lem:inversion-expression-typing}, we have
\begin{enumerate}
\item $\Gamma  \leq  \Gamma_{{\mathrm{1}}}$;\label{preservation-assign-subtype}
\item $\Gamma_1(x) =  \Pi z .(  \tau _{\ast  x }   \TREF^{\hspace{0.5pt} r }) $ and $\Gamma_1(y) =  \tau _{ y } $;\label{preservation-assign-operand-types}
\item $\Gamma_{{\mathrm{1}}}  \ottsym{,}   z \COL  \{  \nu  :   \TINT    \mid   \nu \,  =  \,  0   \}    \vdash    \tau' _{\ast  x }    \approx    \ottsym{(}  \tau'  \ottsym{)}  ^ {= y }  $;\label{preservation-assign-zero-equiv}
\item $\Gamma_{{\mathrm{1}}}  \ottsym{,}   z \COL  \{  \nu  :   \TINT    \mid   \nu \, \neq \,  0   \}    \vdash    \tau' _{\ast  x }    \approx    \tau _{\ast  x }  $;\label{preservation-assign-nonzero-equiv}
\item $\Gamma_{{\mathrm{1}}}  \ottsym{,}   z \COL  \{  \nu  :   \TINT    \mid   \nu \,  =  \,  0   \}    \models  r \,  =  \,  \mathbf{1} $;\label{preservation-assign-full-ownership}
\item $ \Theta   \mid     \Gamma_{{\mathrm{1}}}  \left[  x \hookleftarrow  \Pi z .(  \tau' _{\ast  x }   \TREF^{\hspace{0.5pt} r })   \right]   \left[  y \hookleftarrow  \tau' _{ y }   \right]    \vdash   e_{{\mathrm{0}}}  :  \tau   \produces   \Gamma' $;\label{preservation-assign-continuation}
\item $\Gamma_{{\mathrm{1}}}  \vdash     \tau' _{ y }   +  \tau'    \approx    \tau _{ y }  $.\label{preservation-assign-split}
\end{enumerate}

Put
\[
 \Gamma_p=  \Gamma_{{\mathrm{1}}}  \left[  x \hookleftarrow  \Pi z .(  \tau' _{\ast  x }   \TREF^{\hspace{0.5pt} r })   \right]   \left[  y \hookleftarrow  \tau' _{ y }   \right] .
\]
Premise~(\ref{preservation-assign-continuation}) gives successor expression typing.
Premises~(\ref{preservation-assign-subtype})--(\ref{preservation-assign-full-ownership})
and (\ref{preservation-assign-split}), together with $R(x)=(a,i)\in dom(H)$
and the source invariants, supply the hypotheses of the assignment items.
Applying \cref{lem:conown-preservation}(\ref{conown-preservation3})
and \cref{lem:sat-preservation}(\ref{sat-preservation3}),
respectively, gives
\[
 \CONOWN(H',R',\Gamma_p)
 \qquad\text{and}\qquad
 \SAT(H',R',\Gamma_p).
\]

\end{rncase}

\begin{rncase}{Rs-AddPtr}
Here $e= \LET  x  =   y   \boxplus   z   \IN  e_{{\mathrm{0}}} $, $e'=  [  x'  /  x  ]    e_{{\mathrm{0}}} $,
$H'=H$, and $R'=\ottnt{R}  \ottsym{\{}  x'  \mapsto  pv  \boxplus  \ottnt{R}  \ottsym{(}  z  \ottsym{)}  \ottsym{\}}$.
The side conditions give $R(y)=pv$, $R(z)\in \mathbb{Z} $, and
$x'\notin dom(R)$.

By \cref{lem:inversion-expression-typing}, we have
\begin{enumerate}
\item $\Gamma  \leq  \Gamma_{{\mathrm{1}}}$;\label{preservation-add-ptr-subtype}
\item $ \Gamma_{{\mathrm{1}}}  (  y  )  =  \Pi w .( \tau_{{\mathrm{3}}}  \TREF^{\hspace{0.5pt}  { r }_{ y }  }) $;\label{preservation-add-ptr-pointer-type}
\item $ \Gamma_{{\mathrm{1}}}  (  z  )  =  \{  \nu  :   \TINT    \mid   \varphi  \} $;\label{preservation-add-ptr-offset-type}
\item $\Gamma_{{\mathrm{1}}}  \vdash     \Pi w .( \tau_{{\mathrm{1}}}  \TREF^{\hspace{0.5pt}  { r }_{ y_{{\mathrm{1}}} }  })   +   \Pi w .   [ (  w  -  z  ) /  w  ]   ( \tau_{{\mathrm{2}}}  \TREF^{\hspace{0.5pt}  { r }_{ x }  })     \approx    \Pi w .( \tau_{{\mathrm{3}}}  \TREF^{\hspace{0.5pt}  { r }_{ y }  })  $;\label{preservation-add-ptr-split}
\item $ \Theta   \mid    \Gamma_{{\mathrm{1}}}  \left[  y \hookleftarrow  \Pi w .( \tau_{{\mathrm{1}}}  \TREF^{\hspace{0.5pt}  { r }_{ y_{{\mathrm{1}}} }  })   \right]   \ottsym{,}   x \COL  \Pi w .( \tau_{{\mathrm{2}}}  \TREF^{\hspace{0.5pt}  { r }_{ x }  })     \vdash   e_{{\mathrm{0}}}  :  \tau   \produces   \ottsym{(}  \Gamma'  \ottsym{,}   x \COL \tau_{{\mathrm{0}}}   \ottsym{)} $.\label{preservation-add-ptr-continuation}
\end{enumerate}

Put $\Gamma_p= \Gamma_{{\mathrm{1}}}  \left[  y \hookleftarrow  \Pi w .( \tau_{{\mathrm{1}}}  \TREF^{\hspace{0.5pt}  { r }_{ y_{{\mathrm{1}}} }  })   \right]   \ottsym{,}   x' \COL  \Pi w .( \tau_{{\mathrm{2}}}  \TREF^{\hspace{0.5pt}  { r }_{ x }  })  $.
Renaming $x$ to $x'$ in premise~(\ref{preservation-add-ptr-continuation})
gives the continuation typing with output $\Gamma'  \ottsym{,}   x' \COL \tau_{{\mathrm{0}}} $.
Premises~(\ref{preservation-add-ptr-subtype})--(\ref{preservation-add-ptr-split}),
together with the operational side conditions, freshness, and the source
invariants, supply the hypotheses of the pointer-addition items.
Applying \cref{lem:conown-preservation}(\ref{conown-preservation4})
and \cref{lem:sat-preservation}(\ref{sat-preservation4}),
respectively, gives
\[
 \CONOWN(H',R',\Gamma_p)
 \qquad\text{and}\qquad
 \SAT(H',R',\Gamma_p).
\]

\end{rncase}

\begin{rncase}{Rs-Call}
Here $ f  \mapsto   (  x_{{\mathrm{1}}}  , \ldots ,  x_{\ottmv{n}}  )  \, e_{{\mathrm{0}}}  \in  D $,
$e= \LET  x  =   f (  y_{{\mathrm{1}}} ,\ldots, y_{\ottmv{n}}  )   \IN  e_{{\mathrm{1}}} $,
$e'= \LET  x  =    [  y_{{\mathrm{1}}}  /  x_{{\mathrm{1}}}  , \ldots,  y_{\ottmv{n}}  /  x_{\ottmv{n}}  ]    e_{{\mathrm{0}}}   \IN  e_{{\mathrm{1}}} $, $H'=H$, and $R'=R$.

By \cref{lem:inversion-expression-typing}, we have
\begin{enumerate}
\item $\Gamma  \leq  \Gamma_{{\mathrm{1}}}$;\label{preservation-call-subtype}
\item $ \Gamma_{{\mathrm{1}}}  (  y_{\ottmv{i}}  )  =  \theta \, \tau_{\ottmv{i}}$ for each $i \in \set{1,\dots,n}$;\label{preservation-call-argument-types}
\item $\Theta  \ottsym{(}  f  \ottsym{)} =  \tuple{  x_{{\mathrm{1}}} \COL \tau_{{\mathrm{1}}} ,\ldots, x_{\ottmv{n}} \COL \tau_{\ottmv{n}}  }\ra\tuple{  x_{{\mathrm{1}}} \COL \tau'_{{\mathrm{1}}} ,\ldots, x_{\ottmv{n}} \COL \tau'_{\ottmv{n}}   \mid  \tau' } $;\label{preservation-call-function-type}
\item $\theta  =   [  y_{{\mathrm{1}}}  /  x_{{\mathrm{1}}}  , \ldots,  y_{\ottmv{n}}  /  x_{\ottmv{n}}  ] $;\label{preservation-call-substitution}
\item $ \Theta   \mid    \Gamma_{{\mathrm{1}}}  \left[  y_{\ottmv{i}} \hookleftarrow \theta \, \tau'_{\ottmv{i}}  \right]   \ottsym{,}   x \COL \theta \, \tau'    \vdash   e_{{\mathrm{1}}}  :  \tau   \produces   \ottsym{(}  \Gamma'  \ottsym{,}   x \COL \tau_{{\mathrm{0}}}   \ottsym{)} $.\label{preservation-call-continuation}
\end{enumerate}

By the function-type premise~(\ref{preservation-call-function-type}), the definition
of $f$, and $ \Theta   \vdash   D $, we have
$ \Theta   \vdash   f  \mapsto   (  x_{{\mathrm{1}}}  , \ldots ,  x_{\ottmv{n}}  )  \, e_{{\mathrm{0}}} $.
The argument-type and substitution premises
(\ref{preservation-call-argument-types}) and (\ref{preservation-call-substitution})
allow \cref{lem:call-preparation} to give
\[
 \Gamma_{{\mathrm{1}}}  \leq  \Gamma_{{\mathrm{2}}}
 \qquad\text{and}\qquad
  \Theta   \mid   \Gamma_{{\mathrm{2}}}   \vdash    \theta   e_{{\mathrm{0}}}   :  \theta \, \tau'   \produces   \Gamma_{{\mathrm{4}}} ,
\]
where $\Gamma_{{\mathrm{3}}}$ is the untouched caller frame and
\[
\begin{aligned}
 \Gamma_{{\mathrm{2}}} &=  y_{{\mathrm{1}}} \COL \theta \, \tau_{{\mathrm{1}}} ,\ldots, y_{\ottmv{n}} \COL \theta \, \tau_{\ottmv{n}}   \ottsym{,}  \Gamma_{{\mathrm{3}}},\\
 \Gamma_{{\mathrm{4}}} &=  y_{{\mathrm{1}}} \COL \theta \, \tau'_{{\mathrm{1}}} ,\ldots, y_{\ottmv{n}} \COL \theta \, \tau'_{\ottmv{n}}   \ottsym{,}  \Gamma_{{\mathrm{3}}}.
\end{aligned}
\]
Both environments are well formed.  Up to the order of bindings,
$\Gamma_{{\mathrm{4}}}$ is exactly $ \Gamma_{{\mathrm{1}}}  \left[  y_{\ottmv{i}} \hookleftarrow \theta \, \tau'_{\ottmv{i}}  \right] $, so the continuation
premise~(\ref{preservation-call-continuation}) gives
\[
  \Theta   \mid   \Gamma_{{\mathrm{4}}}  \ottsym{,}   x \COL \theta \, \tau'    \vdash   e_{{\mathrm{1}}}  :  \tau   \produces   \ottsym{(}  \Gamma'  \ottsym{,}   x \COL \tau_{{\mathrm{0}}}   \ottsym{)} .
\]
Rule \rn{T-Let} now gives
\[
  \Theta   \mid   \Gamma_{{\mathrm{2}}}   \vdash    \LET  x  =   \theta   e_{{\mathrm{0}}}   \IN  e_{{\mathrm{1}}}   :  \tau   \produces   \Gamma' .
\]
Put $\Gamma_p=\Gamma_{{\mathrm{2}}}$.  First use
premise~(\ref{preservation-call-subtype}) to transfer the source invariants
to $\Gamma_{{\mathrm{1}}}$, then use $\Gamma_{{\mathrm{1}}}  \leq  \Gamma_{{\mathrm{2}}}$ and environment subtyping preservation
to obtain $\CONOWN(H',R',\Gamma_p)$ and
$\SAT(H',R',\Gamma_p)$.

\end{rncase}

\begin{rncase}{Rs-MkArrayIntref}
Here $e= \LET  x  =   \ALLOC  y   \ottsym{:}     \TINT   \TREF    \IN  e_{{\mathrm{0}}} $,
$e'=  [  x'  /  x  ]    e_{{\mathrm{0}}} $, and
\[
 H'=H\{(a,0)\mapsto0,\ldots,(a,R(y)-1)\mapsto0\},
 \qquad R'=\ottnt{R}  \ottsym{\{}  x'  \mapsto  \ottsym{(}  a  \ottsym{,}   0   \ottsym{)}  \ottsym{\}}.
\]
The side conditions give $(a,0)\notin dom(H)$ and $x'\notin dom(R)$.

By \cref{lem:inversion-expression-typing}, we have
\begin{enumerate}
\item $\Gamma  \leq  \Gamma_{{\mathrm{1}}}$;\label{preservation-alloc-int-subtype}
\item $ \Gamma_{{\mathrm{1}}}  (  y  )   =   \{  \nu  :   \TINT    \mid   \varphi  \} $;\label{preservation-alloc-int-length-type}
\item $\Gamma_{{\mathrm{1}}}  \ottsym{,}   z \COL  \TINT    \models  r \,  =  \, \ottsym{(}    \ottsym{(}    0  \, \le \, z  \wedge  z \, \le \, y  \ottsym{-}   1    \ottsym{)}   \produces    1    ,   \mathbf{0}    \ottsym{)}$;\label{preservation-alloc-int-ownership-shape}
\item $ \Theta   \mid   \Gamma_{{\mathrm{1}}}  \ottsym{,}   x \COL \tau_{{\mathrm{1}}}    \vdash   e_{{\mathrm{0}}}  :  \tau   \produces   \Gamma'  \ottsym{,}   x \COL \tau_{{\mathrm{0}}}  $.\label{preservation-alloc-int-continuation}
\end{enumerate}

Here
\[
 \tau_{{\mathrm{1}}}= \Pi z .(  \{  \nu  :   \TINT    \mid     0  \, \le \, z  \wedge  z \, \le \, y  \ottsym{-}   1    \implies  \nu \,  =  \,  0   \}   \TREF^{\hspace{0.5pt} r }) .
\]
Put $\Gamma_p=\Gamma_{{\mathrm{1}}}  \ottsym{,}   x' \COL \tau_{{\mathrm{1}}} $.
Renaming $x$ to $x'$ in premise~(\ref{preservation-alloc-int-continuation})
gives the continuation typing with output $\Gamma'  \ottsym{,}   x' \COL \tau_{{\mathrm{0}}} $.
Premises~(\ref{preservation-alloc-int-subtype})--(\ref{preservation-alloc-int-ownership-shape}),
together with freshness of $x'$ and $(a,0)$ and the source invariants,
supply the hypotheses of the integer-array allocation items.
Applying \cref{lem:conown-preservation}(\ref{conown-preservation5})
and \cref{lem:sat-preservation}(\ref{sat-preservation5}),
respectively, gives
\[
 \CONOWN(H',R',\Gamma_p)
 \qquad\text{and}\qquad
 \SAT(H',R',\Gamma_p).
\]

\end{rncase}

\begin{rncase}{Rs-MkArrayNestedref}
Here $e= \LET  x  =   \ALLOC  y   \ottsym{:}    \ottsym{(}   \tau^{-}  \TREF   \ottsym{)}  \TREF    \IN  e_{{\mathrm{0}}} $,
$e'=  [  x'  /  x  ]    e_{{\mathrm{0}}} $, and
\[
 H'=H\{(a,0)\mapsto\NULL,\ldots,(a,R(y)-1)\mapsto\NULL\},
 \qquad R'=\ottnt{R}  \ottsym{\{}  x'  \mapsto  \ottsym{(}  a  \ottsym{,}   0   \ottsym{)}  \ottsym{\}}.
\]
The side conditions give $(a,0)\notin dom(H)$ and $x'\notin dom(R)$.

By \cref{lem:inversion-expression-typing}, we have
\begin{enumerate}
\item $\Gamma  \leq  \Gamma_{{\mathrm{1}}}$;\label{preservation-alloc-nested-subtype}
\item $ \Gamma_{{\mathrm{1}}}  (  y  )   =   \{  \nu  :   \TINT    \mid   \varphi  \} $;\label{preservation-alloc-nested-length-type}
\item $\models   \Empty{ \tau' } $;\label{preservation-alloc-nested-empty}
\item $\Gamma_{{\mathrm{1}}}  \ottsym{,}   z \COL  \TINT    \models  r \,  =  \, \ottsym{(}    \ottsym{(}    0  \, \le \, z  \wedge  z \, \le \, y  \ottsym{-}   1    \ottsym{)}   \produces    1    ,   \mathbf{0}    \ottsym{)}$;\label{preservation-alloc-nested-ownership-shape}
\item $ \Theta   \mid   \Gamma_{{\mathrm{1}}}  \ottsym{,}   x \COL \tau_{{\mathrm{1}}}    \vdash   e_{{\mathrm{0}}}  :  \tau   \produces   \Gamma'  \ottsym{,}   x \COL \tau_{{\mathrm{0}}}  $;\label{preservation-alloc-nested-continuation}
\item $|\tau'| =  \tau^{-}  \TREF $.\label{preservation-alloc-nested-element-shape}
\end{enumerate}

Here $\tau_{{\mathrm{1}}}= \Pi z .( \tau'  \TREF^{\hspace{0.5pt} r }) $.
Put $\Gamma_p=\Gamma_{{\mathrm{1}}}  \ottsym{,}   x' \COL \tau_{{\mathrm{1}}} $.
Renaming $x$ to $x'$ in premise~(\ref{preservation-alloc-nested-continuation})
gives the continuation typing with output $\Gamma'  \ottsym{,}   x' \COL \tau_{{\mathrm{0}}} $.
Premises~(\ref{preservation-alloc-nested-subtype})--(\ref{preservation-alloc-nested-ownership-shape})
and (\ref{preservation-alloc-nested-element-shape}), together with freshness
of $x'$ and $(a,0)$ and the source invariants, supply the hypotheses of the
nested-array allocation items.
Applying \cref{lem:conown-preservation}(\ref{conown-preservation6})
and \cref{lem:sat-preservation}(\ref{sat-preservation6}),
respectively, gives
\[
 \CONOWN(H',R',\Gamma_p)
 \qquad\text{and}\qquad
 \SAT(H',R',\Gamma_p).
\]

\end{rncase}

\begin{rncase}{Rs-AliasAddPtr}
Here $e=  \ALIAS(  x  =  y   \boxplus   z  )   \SEQ  e_{{\mathrm{0}}} $, $e'=e_0$, $H'=H$, and $R'=R$.
The side conditions give $R(y)=pv$, $R(z)\in \mathbb{Z} $, and
$R(x)=pv  \boxplus  \ottnt{R}  \ottsym{(}  z  \ottsym{)}$.

By \cref{lem:inversion-expression-typing}, we have
\begin{enumerate}
\item $\Gamma  \leq  \Gamma_{{\mathrm{1}}}$;\label{preservation-alias-add-subtype}
\item $x \, \neq \, y$;\label{preservation-alias-add-distinct}
\item $\Gamma_1(x) =  \Pi w' .(  \tau _{\ast  x }   \TREF^{\hspace{0.5pt}  { r }_{ x }  }) $;\label{preservation-alias-add-x-type}
\item $\Gamma_1(y) =  \Pi w .(  \tau _{\ast  y }   \TREF^{\hspace{0.5pt}  { r }_{ y }  }) $;\label{preservation-alias-add-y-type}
\item $\Gamma_1(z) =  \{  \nu  :   \TINT    \mid   \varphi  \} $;\label{preservation-alias-add-offset-type}
\item $\Gamma_{{\mathrm{1}}}  \vdash   \ottsym{(}    \Pi w' .   [ (  w'  -  z  ) /  w'  ]   (  \tau _{\ast  x }   \TREF^{\hspace{0.5pt}  { r }_{ x }  })   +   \Pi w .(  \tau _{\ast  y }   \TREF^{\hspace{0.5pt}  { r }_{ y }  })    \ottsym{)}   \approx    \ottsym{(}   \Pi w' .   [ (  w'  -  z  ) /  w'  ]   (  \tau' _{\ast  x }   \TREF^{\hspace{0.5pt}  { r' }_{ x }  })   \ottsym{)}  +   \Pi w .(  \tau' _{\ast  y }   \TREF^{\hspace{0.5pt}  { r' }_{ y }  })   $;\label{preservation-alias-add-combined-equiv}
\item $ \Theta   \mid     \Gamma_{{\mathrm{1}}}  \left[  x \hookleftarrow  \Pi w' .(  \tau' _{\ast  x }   \TREF^{\hspace{0.5pt}  { r' }_{ x }  })   \right]   \left[  y \hookleftarrow  \Pi w .(  \tau' _{\ast  y }   \TREF^{\hspace{0.5pt}  { r' }_{ y }  })   \right]    \vdash   e_{{\mathrm{0}}}  :  \tau   \produces   \Gamma' $.\label{preservation-alias-add-continuation}
\end{enumerate}

Put
$\Gamma_p=  \Gamma_{{\mathrm{1}}}  \left[  x \hookleftarrow  \Pi w' .(  \tau' _{\ast  x }   \TREF^{\hspace{0.5pt}  { r' }_{ x }  })   \right]   \left[  y \hookleftarrow  \Pi w .(  \tau' _{\ast  y }   \TREF^{\hspace{0.5pt}  { r' }_{ y }  })   \right] $.
Premise~(\ref{preservation-alias-add-continuation}) gives successor expression typing.
Premises~(\ref{preservation-alias-add-subtype})--(\ref{preservation-alias-add-combined-equiv}),
together with the operational side conditions and the source invariants,
supply the hypotheses of the pointer-aliasing items.
Applying \cref{lem:conown-preservation}(\ref{conown-preservation8})
and \cref{lem:sat-preservation}(\ref{sat-preservation8}),
respectively, gives
\[
 \CONOWN(H',R',\Gamma_p)
 \qquad\text{and}\qquad
 \SAT(H',R',\Gamma_p).
\]

\end{rncase}

\begin{rncase}{Rs-AliasDeref}
Here $e=  \ALIAS(  x  = \ast  y  )   \SEQ  e_{{\mathrm{0}}} $, $e'=e_0$, $H'=H$, and $R'=R$.
The side condition gives $H(R(y))=R(x)$.

By \cref{lem:inversion-expression-typing}, we have
\begin{enumerate}
\item $\Gamma  \leq  \Gamma_{{\mathrm{1}}}$;\label{preservation-alias-deref-subtype}
\item $ \Gamma_{{\mathrm{1}}}  (  x  )   =   \Pi z' .(  \tau _{\ast  x }   \TREF^{\hspace{0.5pt}  { r }_{ x }  }) $;\label{preservation-alias-deref-x-type}
\item $ \Gamma_{{\mathrm{1}}}  (  y  )   =   \Pi w .(  \Pi z .(  \tau _{\ast \ast  y }   \TREF^{\hspace{0.5pt}  { r }_{ \ast  y }  })   \TREF^{\hspace{0.5pt} r }) $;\label{preservation-alias-deref-y-type}
\item $\Gamma_{{\mathrm{1}}}  \ottsym{,}   w \COL  \{  \nu  :   \TINT    \mid   \nu \,  =  \,  0   \}    \vdash   \ottsym{(}    \Pi z' .(  \tau _{\ast  x }   \TREF^{\hspace{0.5pt}  { r }_{ x }  })   +   \Pi z .(  \tau _{\ast \ast  y }   \TREF^{\hspace{0.5pt}  { r }_{ \ast  y }  })    \ottsym{)}   \approx   \ottsym{(}    \Pi z' .(  \tau' _{\ast  x }   \TREF^{\hspace{0.5pt}  { r' }_{ x }  })   +   \Pi z .(  \tau' _{\ast \ast  y }   \TREF^{\hspace{0.5pt}  { r' }_{ \ast  y }  })    \ottsym{)} $;\label{preservation-alias-deref-zero-equiv}
\item $\Gamma_{{\mathrm{1}}}  \ottsym{,}   w \COL  \{  \nu  :   \TINT    \mid   \nu \, \neq \,  0   \}    \vdash    \Pi z .(  \tau _{\ast \ast  y }   \TREF^{\hspace{0.5pt}  { r }_{ \ast  y }  })    \approx    \Pi z .(  \tau' _{\ast \ast  y }   \TREF^{\hspace{0.5pt}  { r' }_{ \ast  y }  })  $;\label{preservation-alias-deref-nonzero-equiv}
\item $\Gamma_{{\mathrm{1}}}  \ottsym{,}   w \COL  \{  \nu  :   \TINT    \mid   \nu \,  =  \,  0   \}    \models  r \,  >  \,  \mathbf{0} $;\label{preservation-alias-deref-positive}
\item $ \Theta   \mid     \Gamma_{{\mathrm{1}}}  \left[  x \hookleftarrow  \Pi z' .(  \tau' _{\ast  x }   \TREF^{\hspace{0.5pt}  { r' }_{ x }  })   \right]   \left[  y \hookleftarrow  \Pi w .(  \Pi z .(  \tau' _{\ast \ast  y }   \TREF^{\hspace{0.5pt}  { r' }_{ \ast  y }  })   \TREF^{\hspace{0.5pt} r })   \right]    \vdash   e_{{\mathrm{0}}}  :  \tau   \produces   \Gamma' $.\label{preservation-alias-deref-continuation}
\end{enumerate}

Put
$\Gamma_p=  \Gamma_{{\mathrm{1}}}  \left[  x \hookleftarrow  \Pi z' .(  \tau' _{\ast  x }   \TREF^{\hspace{0.5pt}  { r' }_{ x }  })   \right]   \left[  y \hookleftarrow  \Pi w .(  \Pi z .(  \tau' _{\ast \ast  y }   \TREF^{\hspace{0.5pt}  { r' }_{ \ast  y }  })   \TREF^{\hspace{0.5pt} r })   \right] $.
Premise~(\ref{preservation-alias-deref-continuation}) gives successor expression typing.
Premises~(\ref{preservation-alias-deref-subtype})--(\ref{preservation-alias-deref-positive}),
together with $H(R(y))=R(x)$ and the source invariants, supply the hypotheses
of the dereference-aliasing items.
Applying \cref{lem:conown-preservation}(\ref{conown-preservation7})
and \cref{lem:sat-preservation}(\ref{sat-preservation7}),
respectively, gives
\[
 \CONOWN(H',R',\Gamma_p)
 \qquad\text{and}\qquad
 \SAT(H',R',\Gamma_p).
\]

\end{rncase}

\begin{rncase}{Rs-Assert}
Here $e=  \ASSERT( \varphi )   \SEQ  e_{{\mathrm{0}}} $, $e'=e_0$, $H'=H$, $R'=R$, and
$\models  \ottsym{[}  \ottnt{R}  \ottsym{]} \, \varphi$.

By \cref{lem:inversion-expression-typing}, we have
\begin{enumerate}
\item $\Gamma  \leq  \Gamma_{{\mathrm{1}}}$;\label{preservation-assert-subtype}
\item $\Gamma_{{\mathrm{1}}}  \models  \varphi$;\label{preservation-assert-validity}
\item $ \Theta   \mid   \Gamma_{{\mathrm{1}}}   \vdash   e_{{\mathrm{0}}}  :  \tau   \produces   \Gamma' $.\label{preservation-assert-continuation}
\end{enumerate}

Put $\Gamma_p=\Gamma_{{\mathrm{1}}}$.
Premise~(\ref{preservation-assert-continuation}) gives successor expression typing.
Premise~(\ref{preservation-assert-subtype}) and environment subtyping preservation
give $\CONOWN(H',R',\Gamma_p)$ and $\SAT(H',R',\Gamma_p)$.

\end{rncase}

\begin{rncase}{Rs-Context}
Here $e= \LET  x  =  e_{{\mathrm{0}}}  \IN  e_{{\mathrm{1}}} $,
$e'= \LET  x  =  e'_{{\mathrm{0}}}  \IN  e_{{\mathrm{1}}} $, and the premise of the reduction is
$  \tuple{ \ottnt{H} ,  \ottnt{R} ,  e_{{\mathrm{0}}} }     \longrightarrow _{  D  }     \tuple{ \ottnt{H'} ,  \ottnt{R'} ,  e'_{{\mathrm{0}}} }  $.

By \cref{lem:inversion-expression-typing}, we have
\begin{enumerate}
\item $\Gamma  \leq  \Gamma_{{\mathrm{1}}}$;\label{preservation-context-subtype}
\item $ \Theta   \mid   \Gamma_{{\mathrm{1}}}   \vdash   e_{{\mathrm{0}}}  :  \tau_{{\mathrm{1}}}   \produces   \Gamma_{{\mathrm{2}}} $;\label{preservation-context-inner}
\item $ \Theta   \mid   \Gamma_{{\mathrm{2}}}  \ottsym{,}   x \COL \tau_{{\mathrm{1}}}    \vdash   e_{{\mathrm{1}}}  :  \tau   \produces   \Gamma'  \ottsym{,}   x \COL \tau_{{\mathrm{0}}}  $.\label{preservation-context-continuation}
\end{enumerate}

By premise~(\ref{preservation-context-subtype}), transfer the source
invariants to $\Gamma_{{\mathrm{1}}}$.  Apply the strengthened induction hypothesis to
premise~(\ref{preservation-context-inner}) and the inner reduction.  It gives
$\Gamma_{{\mathrm{3}}}$ such that
\[
\begin{gathered}
  \Theta   \mid   \Gamma_{{\mathrm{3}}}   \vdash   e'_{{\mathrm{0}}}  :  \tau_{{\mathrm{1}}}   \produces   \Gamma_{{\mathrm{2}}} ,\\
  \mathbf{ConOwn}( \ottnt{H'} ,  \ottnt{R'} ,  \Gamma_{{\mathrm{3}}} ) , \qquad  \mathbf{SAT}( \ottnt{H'} ,  \ottnt{R'} ,  \Gamma_{{\mathrm{3}}} ) .
\end{gathered}
\]
The function type environment $\Theta$, the inner result type $\tau_{{\mathrm{1}}}$,
and the inner output environment $\Gamma_{{\mathrm{2}}}$ are unchanged.  Combining this
typing judgment with premise~(\ref{preservation-context-continuation}) by
\rn{T-Let} gives
\[
  \Theta   \mid   \Gamma_{{\mathrm{3}}}   \vdash    \LET  x  =  e'_{{\mathrm{0}}}  \IN  e_{{\mathrm{1}}}   :  \tau   \produces   \Gamma' .
\]
The binder side condition $x\notin(\Gamma_{{\mathrm{2}}},\tau)$ holds by well-formedness
and inversion, and is unchanged by the inner reduction.
Put $\Gamma_p=\Gamma_{{\mathrm{3}}}$; the displayed judgments give all three required
successor judgments.
\end{rncase}
\end{proof}

\subsubsection{Proof of \cref{lem:progress}}
\label{sec:progress}
\begin{proof}
From $\vdash_D  \tuple{ \ottnt{H} ,  \ottnt{R} ,  e }  \COL \tau \Rightarrow \Gamma'$, we
have
\begin{itemize}
\item $ \Theta   \mid   \Gamma   \vdash   e  :  \tau   \produces   \Gamma' $,
\item $ \Theta   \vdash   D $,
\item $ \mathbf{ConOwn}( \ottnt{H} ,  \ottnt{R} ,  \Gamma ) $,
\item $ \mathbf{SAT}( \ottnt{H} ,  \ottnt{R} ,  \Gamma ) $,
\end{itemize}
for some $\Theta$ and $\Gamma$.  We prove the full disjunction---$e$ is a
variable or $ \tuple{ \ottnt{H} ,  \ottnt{R} ,  e } $ can reduce---by induction on the derivation of
$ \Theta   \mid   \Gamma   \vdash   e  :  \tau   \produces   \Gamma' $, with $\ottnt{H}$, $\ottnt{R}$, and the runtime assumptions
arbitrary.  The case \rn{T-Var} is immediate.  The definition of $\SAT$
gives $\models  \ottsym{[}  \ottnt{R}  \ottsym{]} \,  \fml{ \Gamma } $, which we use below to instantiate validity premises.
Since $\ottnt{R}$ is finite, a fresh register variable can be chosen whenever
required.  We discuss the recursive cases and the cases requiring
satisfaction or allocation freshness.  The remaining direct cases follow
from \cref{lem:sat-runtime-inversion} and the corresponding operational
rules; in \rn{T-Call}, $ \Theta   \vdash   D $ supplies a definition of the called
function with the required arity.

\begin{namedcases}{\rn{T-MkIntArray} and \rn{T-MkNestedArray}}
By inversion of the typing derivation, $ \Gamma  (  y  ) $ is an integer refinement
type.  By \cref{lem:sat-runtime-inversion}, $\ottnt{R}  \ottsym{(}  y  \ottsym{)}$ is therefore an
integer.  Since $\ottnt{H}$ is finite, it contains cells at only finitely many
base addresses.  We can therefore choose a base address $a$ such that
\[
  \forall i \in  \mathbb{Z} .\;  \ottsym{(}  a  \ottsym{,}  i  \ottsym{)}  \not\in   \DOM( \ottnt{H} )  .
\]
Choosing a fresh register variable, we can apply
\rn{Rs-MkArrayIntref} when the element type is $ \TINT $.
Otherwise, we apply \rn{Rs-MkArrayNestedref}.

\end{namedcases}

\begin{rncase}{T-Assert}
we have
\begin{itemize}
\item $e =   \ASSERT( \varphi )   \SEQ  e_{{\mathrm{0}}} $,
\item $\Gamma  \models  \varphi$,
\item $ \Theta   \mid   \Gamma   \vdash   e_{{\mathrm{0}}}  :  \tau   \produces   \Gamma' $,
\end{itemize}
for some $\varphi$ and $e_0$.
We are to prove $\models  \ottsym{[}  \ottnt{R}  \ottsym{]} \, \varphi$, which follows from \cref{lem:sat-soundness}.
Thus \rn{Rs-Assert} applies.

\end{rncase}

\begin{rncase}{T-Deref}
we have
\begin{itemize}
\item $e =  \LET  x  =   \ast  y   \IN  e_{{\mathrm{0}}} $,
\item $ \Gamma  (  y  )   =   \Pi z .(  \tau _{ y }   \TREF^{\hspace{0.5pt} r }) $,
\item $\Gamma  \ottsym{,}   z \COL  \{  \nu  :   \TINT    \mid   \nu \,  =  \,  0   \}    \vdash     \tau'  +  \tau  _{ x }    \approx    \tau _{ y }  $,
\item $\Gamma  \ottsym{,}   x \COL  \tau _{ x }    \ottsym{,}   z \COL  \{  \nu  :   \TINT    \mid   \nu \,  =  \,  0   \}    \vdash    \tau' _{ y }    \approx    \ottsym{(}  \tau'  \ottsym{)}  ^ {= x }  $,
\item $\Gamma  \ottsym{,}   x \COL  \tau _{ x }    \ottsym{,}   z \COL  \{  \nu  :   \TINT    \mid   \nu \, \neq \,  0   \}    \vdash    \tau' _{ y }    \approx    \tau _{ y }  $,
\item $\Gamma  \ottsym{,}   z \COL  \{  \nu  :   \TINT    \mid   \nu \,  =  \,  0   \}    \models  r \,  >  \,  \mathbf{0} $,
\item $ \Theta   \mid    \Gamma  \left[  y \hookleftarrow  \Pi z .(  \tau' _{ y }   \TREF^{\hspace{0.5pt} r })   \right]   \ottsym{,}   x \COL  \tau _{ x }     \vdash   e_{{\mathrm{0}}}  :  \tau   \produces   \ottsym{(}  \Gamma'  \ottsym{,}   x \COL \tau''   \ottsym{)} $
\end{itemize}
for some $x$, $y$, $z$, $\tau'$,$\tau''$, $ \tau _{ x } $, $ \tau _{ y } $, $ \tau' _{ y } $, $r$ and $e_0$.
We are to prove $\ottnt{R}  \ottsym{(}  y  \ottsym{)} \in dom(H)$.
Using $\models  \ottsym{[}  \ottnt{R}  \ottsym{]} \,  \fml{ \Gamma } $, obtained from $ \mathbf{SAT}( \ottnt{H} ,  \ottnt{R} ,  \Gamma ) $, instantiate
$\Gamma  \ottsym{,}   z \COL  \{  \nu  :   \TINT    \mid   \nu \,  =  \,  0   \}    \models  r \,  >  \,  \mathbf{0} $ with $\ottnt{R}$ and $z=0$ to get
\[
   \llbracket    [  0  /  z  ]    r   \rrbracket_{ \ottnt{R} }  > 0.
\]
By Item~3 of \cref{lem:sat-runtime-inversion}, it follows that
$R(y) = (a, k) \in dom(H)$ for some address $(a,k)$.
Choosing a fresh register variable, we can apply \rn{Rs-Deref}.

\end{rncase}

\begin{rncase}{T-Assign}
The typing rule gives $e=  x  \WRITE  y   \SEQ  e_{{\mathrm{0}}} $ and
\[
\begin{gathered}
   \Gamma  (  x  )   =   \Pi z .(  \tau _{\ast  x }   \TREF^{\hspace{0.5pt} r }) , \qquad  \Gamma  (  y  )   =   \tau _{ y } ,\\
  \Gamma  \ottsym{,}   z \COL  \{  \nu  :   \TINT    \mid   \nu \,  =  \,  0   \}    \models  r \,  =  \,  \mathbf{1} .
\end{gathered}
\]
Using $\models  \ottsym{[}  \ottnt{R}  \ottsym{]} \,  \fml{ \Gamma } $, instantiate the last premise with $\ottnt{R}$ and
$z=0$.  This gives
\[
   \llbracket    [  0  /  z  ]    r   \rrbracket_{ \ottnt{R} }  = 1 > 0.
\]
By Item~3 of \cref{lem:sat-runtime-inversion}, there exist $a$ and
$i$ such that $\ottnt{R}  \ottsym{(}  x  \ottsym{)}=\ottsym{(}  a  \ottsym{,}  i  \ottsym{)}\in dom(\ottnt{H})$.  Since
$y\in dom(\Gamma)$, the first conclusion of the same lemma ensures that
$\ottnt{R}  \ottsym{(}  y  \ottsym{)}$ is defined.  We can therefore apply \rn{Rs-Assign}:
\[
    \tuple{ \ottnt{H} ,  \ottnt{R} ,    x  \WRITE  y   \SEQ  e_{{\mathrm{0}}}  }     \longrightarrow _{  D  }     \tuple{ \ottnt{H}  \ottsym{\{}  \ottsym{(}  a  \ottsym{,}   i   \ottsym{)}  \hookleftarrow  \ottnt{R}  \ottsym{(}  y  \ottsym{)}  \ottsym{\}} ,  \ottnt{R} ,  e_{{\mathrm{0}}} }  .
\]

\end{rncase}

\begin{rncase}{T-Let}
Here $e= \LET  x  =  e_{{\mathrm{0}}}  \IN  e_{{\mathrm{1}}} $, and the first typing premise is
$ \Theta   \mid   \Gamma   \vdash   e_{{\mathrm{0}}}  :  \tau_{{\mathrm{1}}}   \produces   \Gamma_{{\mathrm{1}}} $.  Apply the induction hypothesis to this
premise.  If
$  \tuple{ \ottnt{H} ,  \ottnt{R} ,  e_{{\mathrm{0}}} }     \longrightarrow _{  D  }     \tuple{ \ottnt{H'} ,  \ottnt{R'} ,  e'_{{\mathrm{0}}} }  $, then \rn{Rs-Context} gives a step
of the whole expression.  If $  \tuple{ \ottnt{H} ,  \ottnt{R} ,  e_{{\mathrm{0}}} }     \longrightarrow _{  D  }     \mathbf{AliasFail}  $, then
\rn{Rs-ContextFail} gives a step to $ \mathbf{AliasFail} $.  Otherwise,
$e_{{\mathrm{0}}}=y$ for some variable $y$.  Inversion of variable typing
by \cref{lem:inversion-expression-typing}, together with \rn{S-TyEnv},
gives $y\in dom(\Gamma)$.  By \cref{lem:sat-runtime-inversion},
$\ottnt{R}  \ottsym{(}  y  \ottsym{)}$ is defined.  Choosing a fresh register variable, we apply
\rn{Rs-LetVar}.

\end{rncase}

\begin{rncase}{T-Sub}
The premises include $\Gamma  \leq  \Gamma_{{\mathrm{1}}}$ and
$ \Theta   \mid   \Gamma_{{\mathrm{1}}}   \vdash   e  :  \tau_{{\mathrm{1}}}   \produces   \Gamma_{{\mathrm{2}}} $.  By
\cref{lem:conown-preservation-weakening-tyenv,lem:sat-preservation-greater-tyenv}
and $\models  \ottsym{[}  \ottnt{R}  \ottsym{]} \,  \fml{ \Gamma } $, we have $ \mathbf{ConOwn}( \ottnt{H} ,  \ottnt{R} ,  \Gamma_{{\mathrm{1}}} ) $ and
$ \mathbf{SAT}( \ottnt{H} ,  \ottnt{R} ,  \Gamma_{{\mathrm{1}}} ) $.  The induction hypothesis applied to the typing
premise therefore gives the required disjunction.

\end{rncase}

\begin{rncase}{T-AliasAddPtr}
Here $e=  \ALIAS(  x  =  y   \boxplus   z  )   \SEQ  e_{{\mathrm{0}}} $.  The typing rule assigns reference
types to $x$ and $y$ and an integer refinement type to $z$.  By
\cref{lem:sat-runtime-inversion}, $\ottnt{R}  \ottsym{(}  x  \ottsym{)}$ is defined,
$\ottnt{R}  \ottsym{(}  y  \ottsym{)}=pv$ for a pointer value $pv$, and $\ottnt{R}  \ottsym{(}  z  \ottsym{)}$ is an
integer.  Thus $pv  \boxplus  \ottnt{R}  \ottsym{(}  z  \ottsym{)}$ is defined, including when
$pv \,  =  \,  \NULL $.  According as $\ottnt{R}  \ottsym{(}  x  \ottsym{)}=pv  \boxplus  \ottnt{R}  \ottsym{(}  z  \ottsym{)}$ or not, apply
\rn{Rs-AliasAddPtr} or \rn{Rs-AliasAddPtrFail}.

\end{rncase}

\begin{rncase}{T-AliasDeref}
Here $e=  \ALIAS(  x  = \ast  y  )   \SEQ  e_{{\mathrm{0}}} $.  The typing premises give a reference
binding for $y$ and positive ownership at offset zero.  As in the
\rn{T-Deref} case, instantiating the ownership premise using
$\models  \ottsym{[}  \ottnt{R}  \ottsym{]} \,  \fml{ \Gamma } $ gives positive ownership under $\ottnt{R}$ at zero.
Item~3 of \cref{lem:sat-runtime-inversion} therefore shows that
$\ottnt{R}  \ottsym{(}  y  \ottsym{)}$ is an allocated address, so $\ottnt{H}  \ottsym{(}  \ottnt{R}  \ottsym{(}  y  \ottsym{)}  \ottsym{)}$ is defined.
The first conclusion of the same lemma ensures that $\ottnt{R}  \ottsym{(}  x  \ottsym{)}$ is
defined.  According as $\ottnt{H}  \ottsym{(}  \ottnt{R}  \ottsym{(}  y  \ottsym{)}  \ottsym{)}=\ottnt{R}  \ottsym{(}  x  \ottsym{)}$ or not, apply
\rn{Rs-AliasDeref} or \rn{Rs-AliasDerefFail}.

\end{rncase}
\end{proof}

\else
\fi

\end{document}